\documentclass[amsmath,amssymb,reprint,superscriptaddress,floatfix, nofootinbib]{revtex4-2}
\usepackage[dvipsnames]{xcolor}
\usepackage{graphicx}
\usepackage{changepage} 
\usepackage{multirow} 
\usepackage{hyperref}
\usepackage{float}
\usepackage{soul} 
\usepackage[maxfloats=256]{morefloats}
\usepackage{ulem}
\usepackage{url}
\usepackage{subcaption}
\usepackage[percent]{overpic} 

\usepackage[capitalise,noabbrev,nameinlink]{cleveref}
\hypersetup{
citecolor=blue,
 colorlinks=true,
 linkcolor=blue,
 filecolor=magenta,
 urlcolor=blue,
}

\usepackage{orcidlink}
\graphicspath{{./images/}}
\Crefname{equation}{Eq.}{Eqs.}

\begin{document}

\title{Phase Stability in the 3-Dimensional Open-source Code for the Chiral Mean Field Model}

\author{Nikolas Cruz-Camacho\,\orcidlink{0009-0004-7870-0039}}
\affiliation{University of Illinois Urbana-Champaign, Urbana, IL 61801, USA}
\author{Rajesh Kumar\,\orcidlink{0000-0003-2746-3956}}
\affiliation{Department of Physics, Kent State University, Kent, OH 44243, USA}
\author{Mateus Reinke Pelicer\,\orcidlink{0000-0002-2189-706X}}
\affiliation{Department of Physics, Kent State University, Kent, OH 44243, USA}
\author{Jeff Peterson\,\orcidlink{0000-0002-6703-418X}}
\affiliation{Department of Physics, Kent State University, Kent, OH 44243, USA}
\author{T. Andrew Manning\,\orcidlink{0000-0003-2545-9195}}
\affiliation{University of Illinois Urbana-Champaign, Urbana, IL 61801, USA}
\author{Roland Haas\,\orcidlink{0000-0003-1424-6178}}
\affiliation{University of Illinois Urbana-Champaign, Urbana, IL 61801, USA}
\author{Veronica Dexheimer\,\orcidlink{0000-0001-5578-2626}}
\affiliation{Department of Physics, Kent State University, Kent, OH 44243, USA}
\author{Jacquelyn Noronha-Hostler\,\orcidlink{0000-0003-3229-4958}}
\affiliation{University of Illinois Urbana-Champaign, Urbana, IL 61801, USA}

\collaboration{MUSES Collaboration}
\noaffiliation
\date{\today}

\begin{abstract}
{In this paper we explore independently for the first time three chemical potentials (baryon $\mu_B$, charged $\mu_Q$, and strange $\mu_S$) in the Chiral Mean Field (CMF) model. 
We designed and implemented \texttt{CMF++}, a new version of the CMF model rewritten in \texttt{C++} that is optimized, modular, and well-documented. \texttt{CMF++} has been integrated into the MUSES Calculation Engine as a free and open-source software module. 
The runtime improved in more than 4 orders of magnitude across all 3 chemical potentials, when compared to the legacy code.
Here we focus on the zero temperature case and study stable, as well as metastable and unstable, vacuum, hadronic, and quark phases, showing how phase boundaries vary with the different chemical potentials. Due to the significant numerical improvements in \texttt{CMF++}, we can now for the first time sweep the entire $\mu_B$, $\mu_S$, $\mu_Q$ phase space, investigate metastable phases, and calculate high-order susceptibilities within the CMF framework. This allows us to find phases of matter that include a light hadronic phase, a strangeness-dominated hadronic phase, and a quark phase. 
The numerical improvements also allow us to identify the order of the transitions among these phases, finding a first-order chiral symmetry restoration phase transition among the hadronic phases (favored for negative $\mu_Q$ and/or $\mu_S$ for some coupling schemes), in addition to third-order phase transitions. In particular, we identify for the first time triple points in the CMF model, where both chiral symmetry restoration and deconfinement phase transitions meet in the chemical potential phase space. Such points could potentially be identified in low-energy heavy-ion collisions.}
\end{abstract}

\maketitle

\section{INTRODUCTION} 
\label{sec:intro}

In the past decades, the increase of colliding energy in particle accelerators and the unprecedented accuracy in astronomical observations allowed us to grasp a better understanding of the building blocks of matter, the quarks, and gluons. In a way, this allows us to glimpse at the matter that existed in the first $10^{-6}$ s after the Big Bang. 
In the laboratory, it was shown that at extremely high temperatures the quark-gluon plasma created presents very low viscosity, behaving like an ideal fluid \cite{Heinz:2013th}. 
On the other hand, neutron stars reach extremely large baryon densities, the value being model dependent but attaining more than 14 times nuclear saturation density, $n_{\rm sat}$ in extreme cases~\cite{Lattimer:2004sa}, 10 $n_{\rm sat}$ for the heaviest neutron stars. Around these densities, several microphysical models have predicted deconfined quark matter within the core of neutron stars (starting with Ivanenko et. al in the 60's~\cite{Ivanenko:1969bs}), while being consistent with astrophysical data, see e.g.~\cite{Annala:2019puf}.
Finally, mergers of neutron stars provide both hot and dense environments, where deconfined quarks may be observed not only electromagnetically, but also gravitationally \cite{Most:2018eaw,Bauswein:2018bma}.

\begin{figure}[t!]
\centering
  \includegraphics[width=0.5\textwidth]{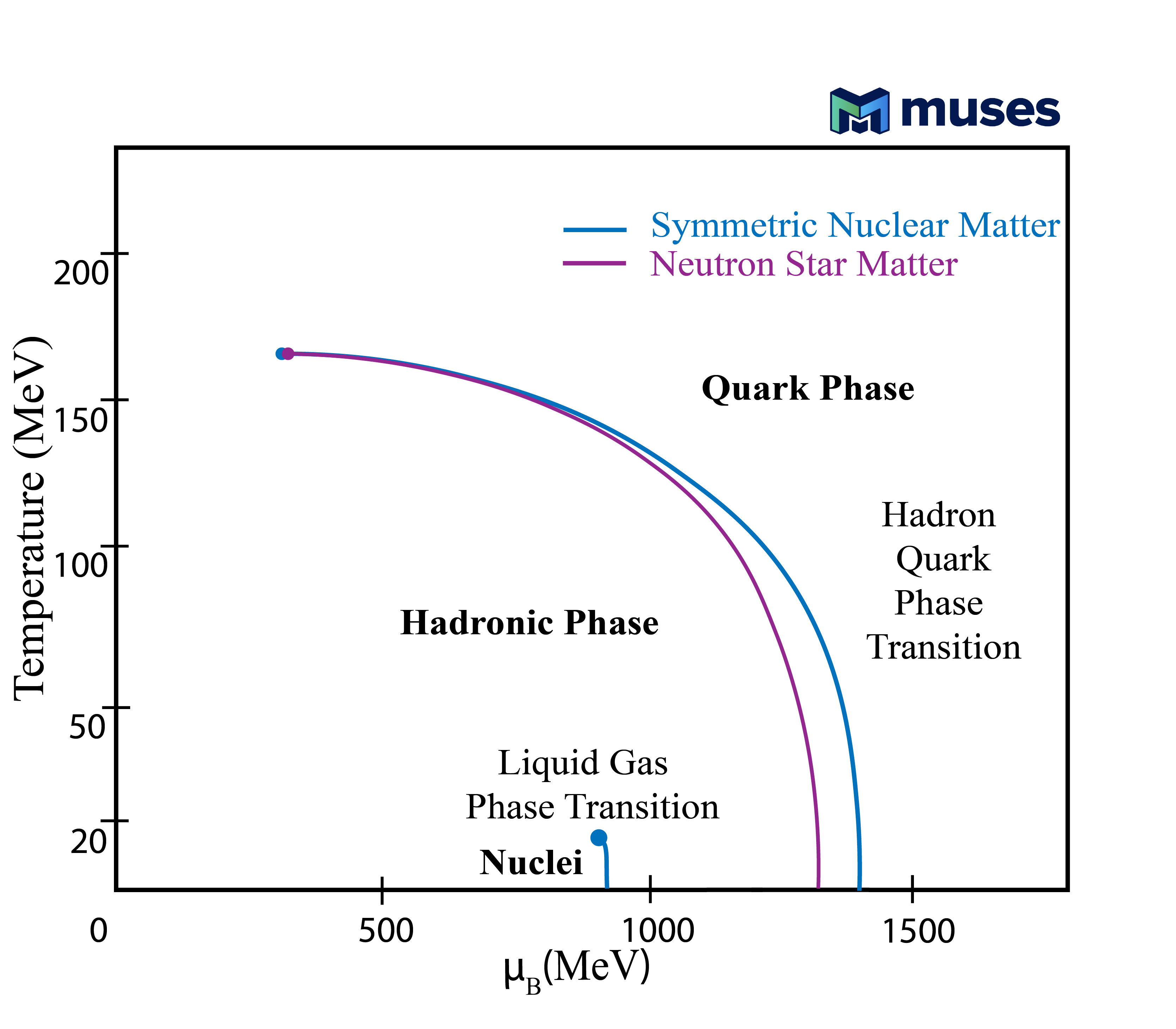}
  \caption{QCD Phase Diagram from the CMF model showing the nuclear liquid-gas and deconfinement phase transitions for symmetric matter (zero net strangeness and isospin) and neutron star matter (charge neutral in beta-equilibrium).}
  \label{fig:QCDPD}
\end{figure}

At lower energies, due to asymptotic freedom, quarks and gluons are confined within hadrons. At even lower energies baryons form atomic nuclei. These different ``phases'' of matter, which can be produced both in the laboratory and in the cosmos are usually depicted in a phase diagram, the Quantum chromodynamics (QCD) phase diagram, referring to the theory that describes quarks, gluons, and their interactions. 
The phase transition from nuclei to hadronic matter (composed of baryons with 3 quarks and mesons with one quark and one antiquark) is referred to as the nuclear Liquid-Gas phase transition, while the one from hadronic to deconfined quark matter is referred to as deconfinement. Both are expected to be first-order phase transitions in the low-temperature and high baryon density ($n_B$) regime and present a crossover region beyond a critical point~\cite{Stephanov:1998dy,Elliott:2013pna}. See \Cref{fig:QCDPD}, with different colors representing combinations of values for additional dimensions.

The thermodynamical description of equilibrated matter is done through the equation of state (EoS), usually given as the relation between pressure and energy density.
The dimensionality of the complete EoS depends on the characteristics of the system being described, such as temperature, number of conserved charges, and other effects (e.g. magnetic fields, spin, etc). 
In the case of QCD, the conserved charges typically considered are baryon number ($B$), electric charge ($Q$), and strangeness ($S$).
In principle, all quark flavors should be conserved on QCD time scales but not all quarks are produced in enough abundance to be considered in chemical equilibrium for the EoS (although some studies have considered charm \cite{Capellino:2022nvf}).
In this work, the dimensionality of our equation state is 3D $\vec{n}=\left\{n_B,n_Q,n_S\right\}$, where $n_x$ is the respective (number) density associated with the conserved charge $x=B,S,Q$. We plan to add finite temperature $T$ in future works. 

To describe fully evolved (beyond the protoneutron-star stage) cold neutron-star matter, one assumes charged neutrality and  
$\beta$-equilibrium with leptons. 
At $\beta$-equilibrium, the charge chemical potential is related to the electron and muon chemical potentials via $\mu_e=\mu_\mu=-\mu_Q$.
Electric charge neutrality is enforced i.e.  $n_Q=\sum_{i} Q_in_i=0$, where $i$ stands for all particles involved, $Q_i$ for electric charge, and $n_i$ for (number) density of each particle.
The time scales associated with neutron stars and their mergers allow for the creation of net strangeness through weak interactions, meaning that there is no strange chemical potential, $\mu_S=0$. Applying $\beta$-equilibrium and $\mu_S=0$, reduces the dimensionality of the system from 3 dimensions (at $T=0$) of $\mu_B,\mu_S,\mu_Q$ into 1 dimension, meaning that it only depends on baryon chemical potential, $\mu_B$.
Numerical relativity simulations of, e.g., neutron-star mergers typically require a 3D EoS, $T,n_B,Y_Q=\frac{n_Q}{n_B}$, although 1D EoS simulations of neutron star mergers are also performed \cite{Weih:2019xvw}. 
Additionally, out-of-equilibrium effects can be important (e.g. bulk viscosity) when there is a delay in the system to reach $\beta$-equilibrium, such that information about the EoS out of $\beta$ equilibrium is also required  \cite{Alford:2018lhf}.

On the other hand, the conserved charges in heavy-ion collisions are dictated by the choice of nuclei that are collided and the energy of the beam that collides. 
The colliding nuclei do not contain net strangeness, however, due to gluon splittings into quark anti-quark pairs, strangeness is produced over time in heavy-ion collision while preserving strangeness neutrality, i.e.,
$n_S=\sum_{i} S_i n_i=0$, where $S_i$ is the particle strangeness. This results in a non-zero $\mu_S$ when $\mu_B\neq 0$.
Heavy-ion collisions are close to the limit of isospin symmetric nuclear matter (SNM) where $Y_Q=Z/A=0.5$ (or $\mu_Q=0$), although data exist for nuclei with a range of $Y_Q=[0.38,0.5]$.
In the limit of symmetric nuclear matter, there is exactly the same number of protons and neutrons in the colliding nuclei.
Thus, one typically reduces the dimensionality of heavy-ion collisions from a 4D phase space into 2D ($T,\mu_B$).
Note that the temperature of heavy-ion collisions is always non-negligible, even at some of the lowest beam energies (estimates from HADES suggest an average temperature of $T\sim 70$ MeV \cite{HADES:2019auv}).
However, the $T=0$ EoS limit is still interesting to study as an input for theoretical models (see e.g. \cite{Sorensen:2020ygf,Oliinychenko:2022uvy} for heavy-ion transport simulations where the temperature is introduced through kinetic contributions and how they connect to neutron stars \cite{Yao:2023yda}). 

While the Lagrangian of QCD is well-known, solving QCD is far from easy.  The most common approach that has been extremely successful is lattice QCD, which represents space-time as a crystalline lattice with quarks at vertices connected by lines where the gluons travel \cite{Troyer:2004ge}. 
However, lattice QCD calculations can only be performed at vanishing densities due to the fermion sign problem~\cite{Muroya:2003qs,deForcrand:2010ys}.  
In order to circumvent the fermion sign problem, it is possible to perform calculations at either imaginary chemical potentials or perturb around vanishing chemical potentials in order to obtain susceptibilities, $\chi_{ijk}^{BSQ}$, of the EoS.  
Then these susceptibilities can be used to recreate the EoS in up to 4D ($T,\mu_B,\mu_S,\mu_Q$) through a Taylor series expanded in $\mu_x/T$, where $x=B,S,Q$ \cite{Monnai:2019hkn,Noronha-Hostler:2019ayj}. 
Given that lattice QCD results are only available at temperatures of $T\gtrsim 130$ MeV and the expansion is currently only valid up to $\mu_B/T\lesssim 3.5$ with the furthest reaching expansion scheme \cite{Borsanyi:2021sxv}, there is only a limited regime where lattice QCD results can be applied. For this reason, lattice QCD cannot be used to describe neutron stars, where $\mu_B>m_p\sim 938$ MeV at vanishing temperatures (in the MeV scale).

Outside the region covered by lattice QCD, perturbative QCD (pQCD) calculations are possible at extremely large $\mu_B$ and/or large $T$. They are performed using a perturbative expansion in the QCD coupling constant, which is small in this regime \cite{Andersen:2002jz, Fraga:2013qra,Kurkela:2016was}. However, near the deconfinement phase transition, the QCD coupling constant becomes large and the truncated sum from perturbation theory no longer approximates the infinite sum. On the other hand, chiral effective theory ($\chi$EFT) calculations are possible at nearly vanishing temperatures and baryon densities around nuclear saturation density \cite{Holt:2009ty,Tews:2012fj}. They include every allowed particle interaction and order them by the number of powers of mass and momentum \cite{Weinberg:1978kz}. However, even with the combination of lattice QCD, pQCD, and  $\chi$EFT the vast majority of the QCD phase diagram is not yet possible to map out from first principle calculations (see Fig.1 of \cite{MUSES:2023hyz}). 

As a result, one must turn to effective models,  utilizing phenomenological constraints to construct Lagrangians that contain the right degrees of freedom (quarks at high $T,\mu_B$, hadrons at intermediate $T,\mu_B$, and nuclei at very low $T,\mu_B$) to describe the entire space of 4D or 5D phase diagrams. These effective models should smoothly connect to first principle QCD calculations in their regime of validity and should also include known particles, their masses, and their known interactions.

In particular, the Chiral Mean Field (CMF) model is a very successful relativistic approach based on a non-linear realization of chiral symmetry \cite{Weinberg:1968de,Papazoglou:1997uw,Papazoglou:1998vr},
which allows for a very good agreement with experimental nuclear data. In addition, only the mean values of the mesons are used in the CMF model, as the mesonic field fluctuations are expected to be small at high densities. As an effective model, once it is calibrated to work on a certain regime of energies, it can produce reliable results for the matter EoS and particle compositions, which can then be used in e.g, hydrodynamical simulations of heavy-ion collisions \cite{Steinheimer:2009nn,Steinheimer:2007iy} or astrophysics \cite{Dexheimer:2008ax, Schurhoff:2010ph,Dexheimer:2015qha, Dexheimer:2018dhb,Roark:2018boj,Motornenko:2019arp}, including core-collapse supernova explosions \cite{Jakobus:2023fru}, stellar cooling \cite{Negreiros:2010hk,Dexheimer:2012eu}, and compact star mergers \cite{Most:2018eaw,Most:2019onn,Most:2022wgo}. See Ref.~\cite{Raduta:2021coc,Raduta:2022elz,Tsiopelas:2024ksy} for a recent reviews that compare the CMF with other multidimensional models available in the CompOSE repository \cite{CompOSECoreTeam:2022ddl,Dexheimer:2022qhn}.

The CMF model can be applied at $T=0$ as well as intermediate, and larger temperatures. It has also been extended to include the effects of strong magnetic fields \cite{Franzon:2015sya,Dexheimer:2021sxs,Marquez:2022fzh,Peterson:2021teb}.  It includes degrees of freedom (\textit{d.o.f}) expected to appear in different laboratory and astrophysical scenarios (leptons, baryons, mesons, and quarks) within a single framework. 
Both isospin asymmetry and strangeness (from hyperons and strange quarks) are included in the formalism, in order to describe the different environments. 
Inspired by unified approaches for the nuclear liquid-gas phase transition (between a phase with nuclei and a bulk hadronic one) \cite{Oertel:2016bki}, a unified approach for quark deconfinement was implemented in the CMF model~\cite{Dexheimer:2009hi}, as explained below.
All degrees of freedom are a priori included in the description, allowing deconfinement to appear as a first-order phase transition or crossover (in this case referring to higher than first-order phase transition), as expected at large temperatures \cite{Aoki:2006we}). 
The transition from baryons to quarks as the density and temperature increase is done utilizing an order parameter, a scalar field $\Phi$ named in analogy with the Polyakov loop \cite{Pisarski:2000eq,Fukushima:2003fw}. This order parameter is introduced in the effective mass of baryons and quarks. 
Within this approach, full QCD phase diagrams can be built, showing both the liquid gas and deconfinement phase transitions \cite{Dexheimer:2009hi,Steinheimer:2011ea,Hempel:2013tfa,Roark:2018uls,Aryal:2020ocm,Kumar:2024owe}.

The CMF model in its current form has been used for over 2 decades. 
However, the software developed for those calculations written in \texttt{Fortran 77}  was inefficient, not well-documented, proprietary, and had most variables hard-coded. 
Thus, the legacy CMF software placed many numerical limitations on the CMF model.  
For instance, while the theory allows for 4D or 5D calculations of the EoS, the legacy version of the CMF model was only calculated in maximum 3D \cite{Aryal:2020ocm} due to the very long run time.
In this paper, we report on a brand-new version of the CMF model in \texttt{C++} in collaboration with computer scientists through the MUSES collaboration (Modular Unified Solver of the Equation of State \cite{MUSES,ReinkePelicer:2025vuh}) that increases the efficiency of the code by several orders of magnitude. This allows for the first time a complete sweep of the independent chemical potentials. It also allows for more accurate solutions, such that high-order derivatives of the EoS are now possible for the first time, and captures not just the stable region of the phase diagram but also the metastable and unstable regions across first-order phase transitions.
In addition to the already known (in the CMF framework) deconfinement first-order phase transition, we now also find first order chiral symmetry phase transitions, in addition to second and third-order transitions within hadronic phases. We also encounter triple points across the $\mu_B$, $\mu_Q$, and $\mu_S$ phase space, where deconfinement and chiral symmetry restoration phase transitions meet.
For this work, we consider the vanishing temperature limit of the CMF model ($T=0$) and no magnetic fields ($B=0$).
However, future work is ongoing to extend the \texttt{C++} version of the CMF model both to finite $T$ and $B$.

The paper is organized as follows: in \Cref{sec:equations} the theoretical development of the CMF model is outlined.  First, the full chiral Lagrangian is built in \Cref{sec:chiralLagrangian}, then the mean-field Lagrangian is obtained in \Cref{sec:LCMF}, followed by the equations of motion in \Cref{sec:EOM}, the thermodynamical observables in \Cref{sec:thermo}, and the coupling constants used within CMF in \Cref{sec:coupling}. The numerical implementation of the CMF model in \texttt{C++} is outlined including a discussion of thermodynamical stability in \Cref{sec:CodeImplementation}, and the corresponding benchmark tests from this new code are presented.  Finally, the results from the upgraded version of CMF are shown in \Cref{sec:results} for different chemical potential combinations using different couplings. High-order derivatives of the EoS known as the susceptibilities are calculated and first-order phase transitions are explored, including the discussion of metastable and unstable regions. In \Cref{sec:outlook} concluding remarks and future plans are discussed. Appendices show particle properties, particle multiplets, Lagrangian calculations, equations for a free Fermi gas, instructions how to use \texttt{CMF++} software, discussions about thermodynamical stability, and additional results.

\section{CMF Formalism}\label{sec:equations}

This section outlines the equations used to calculate the thermodynamical properties of bulk hadronic and quark matter within the CMF model. For the first time, we show in detail in this paper the derivation of the entire mean-field Lagrangian, equations of motion, and the thermodynamic properties. Due to the large densities found in compact stars, the particles in their interior become relativistic, each with their momentum comparable to their mass. And so, a relativistic space-time metric must be adopted to describe them. 
The CMF model is relativistic, therefore, it respects causality, provided that the repulsive vector interactions are not too strong (see the footnote in \cite{Dexheimer:2020rlp}). 
Following the literature of relativistic models, we make use of natural units throughout the paper.

The CMF model is based on a non-linear realization of
the SU(3) sigma model \cite{Papazoglou:1998vr} in which hadrons interact by mesonic exchange:  $\sigma$,  $\zeta$,  $\delta$,  $\omega$,   $\phi$, and $\rho$. 
The scalar-isoscalar field $\sigma (u\bar{d})$ corresponds loosely to the light quark composed meson $f_0(500)$  and is the main driver for chiral symmetry restoration. 
A strange scalar-isoscalar field $\zeta$ ($s\bar{s}$) with hidden strangeness (which is assumed to couple with strange particles) is also introduced to provide necessary attraction and is associated with the $f_0(980)$ meson \cite{Schaffner:1993nn}.
The scalar-isovector field $\delta(\bar u u-\bar d d)$ couples differently to particles with different isospin, introducing a mass splitting between isospin multiples and making the EoS sensitive to isospin. It is associated with the $a_0(980)$ meson  \cite{Kubis:1997ew,Hofmann:2000vz}.
These fields mediate interactions among nucleons, hyperons, and quarks, contributing to attractive medium-range forces (scalar fields) and short-range repulsion (vector fields: vector-isoscalar $\omega$, strange vector-isoscalar $\phi$, and vector-isovector  $\rho$). 
The scalar dilaton field, $\chi$, representing the hypothetical glue ball field, is introduced to replicate QCD's trace anomaly \cite{Papazoglou:1998vr}.

While in reality the strong force is propagated by gluons, the CMF model approximates this interaction as an exchange of mesons. 
They are not the typical particle physics mesons, such as pions or kaons (bound states of a quark and an antiquark), instead they are virtual particles that serve as force carriers for the strong force, much like how the photon is the force carrier of the electromagnetic force. 
Unlike electromagnetism, the strong force changes sign (and therefore whether it is attractive or repulsive) based on the separation between particles. 
At low $T$, mesons do not contribute kinetically. 

The CMF model is built in a chirally invariant way, as the masses of the particles are built from interactions with the medium and, as a result, the masses decrease with the energy. Note that
the commonly used linear sigma model with linear realization approach in meson-baryon coupling leads to imbalanced hyperon potentials due to symmetric spin-0 and antisymmetric spin-1 meson interactions, and additional attraction from the $\zeta$ field without counterbalancing repulsion. This arises due to the symmetric coupling of spin-0 mesons (d-type) and the antisymmetric coupling of spin-1 mesons (f-type), combined with the additional attraction from the strange scalar field ($\zeta$) that lacks sufficient counterbalancing repulsion. Attempts to correct these imbalances by introducing explicit symmetry-breaking terms in the linear realization are problematic because they disrupt the partially conserved axial-vector current (PCAC) relations, which are fundamental to pseudoscalar mesons like pions and kaons~\cite{Papazoglou:1998vr}. In contrast, the non-linear realization of chiral symmetry treats pseudoscalar mesons as angular parameters of the chiral transformation, allowing explicit symmetry-breaking terms to be introduced without affecting PCAC relations. This approach also decouples the strange and non-strange condensates, resulting in a more balanced interaction and accurate values of hyperon potentials. Unlike the linear sigma model, where left- and right-handed chirality wave functions transform differently within the SU(3)$_{\rm L}\,\times$ SU(3)$_{\rm R}$ group, the non-linear realization applies a transformation that allows these wave functions to transform identically, ensuring consistency and resolving these issues (see Refs.\cite{Gerstein:1971fm, Coleman:1969sm,Callan:1969sn} for details).

The strength of the (confining) strong force changes with momentum transfer between particles, where the strong force becomes weaker with increased momentum transfer. 
This means that for high energies, temperature, or chemical potential, quarks become effectively deconfined \cite{Gross:1973id,Politzer:1973fx}. 
For this reason, quarks are also included in the CMF model (within the same description) but with different couplings~\cite{Dexheimer:2009hi}. 
The different phases, hadronic and quark, are characterized and distinguished from each other by the values of the condensates (such as $\sigma$) and the order parameter for deconfinement, $\Phi$. 

Although there are six known ``flavors" of quarks, effectively, only up, down, and strange quarks are present in the energy regime we are discussing in this work (given in \Cref{tab:particles} 
The gluons serve to carry both the attractive and repulsive attributes of the strong force, but in the CMF model, these attributes are split between scalar (spin-$0$) mesons mediating attractive interactions and vector (spin-$1$) mesons mediating repulsive interactions.
The baryons included in the CMF model are the baryonic octet (\Cref{tab:particles}) and the decuplet (\Cref{tab:particles}). 
An alternative version of the CMF model also exists that includes the chiral partners of the baryons, see Ref.~\cite{Steinheimer:2011ea,Motornenko:2019arp} but this approach is not included in \texttt{CMF++}.

\begin{figure}[htb]
    \centering
    \includegraphics[scale=0.65]{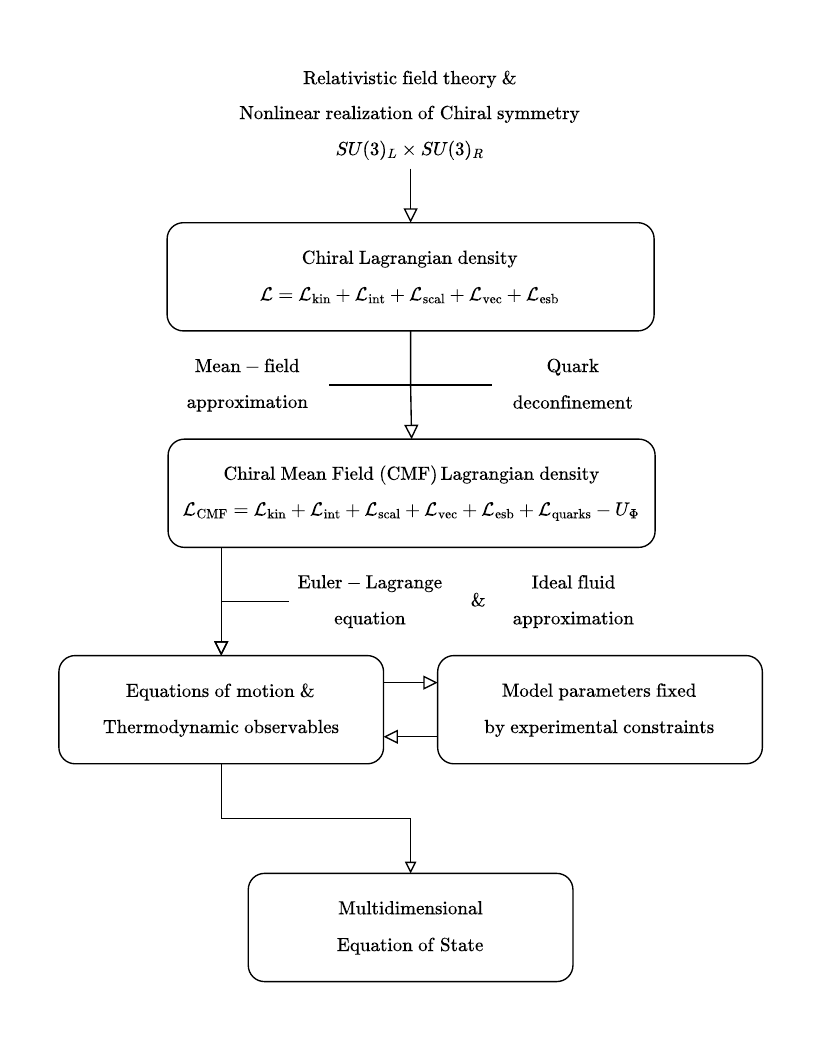}
    \caption{Flowchart depicting the steps involved in building the CMF model.}
    \label{fig:CMF_MFA}
\end{figure}

The construction of the CMF model is described in detail in the following subsections, however, the general procedure is shown in \Cref{fig:CMF_MFA}. One develops an effective theory by determining the relevant symmetry group and then constructing the appropriate Lagrangian that contains all the particles and their interactions.  
Once the Lagrangian is established, then the mean-field and no-sea approximations are made to simplify the Lagrangian to a form that can be solved straightforwardly. 
After the deconfinement mechanism is implemented, the model is named CMF. Next, the equations of motion are obtained from the Euler-Lagrange equation.  At this point, input from experimental and theoretical constraints for the model parameters are applied (e.g. mass and couplings of the baryons, etc.).  Then, for a given set of chemical potentials $\mu_B$, $\mu_S$, $\mu_Q$ (and $T$), the equations of motion for the mesons can be used to determine the particle population and to calculate the thermodynamic observables that allow one to obtain a multidimensional EoS. 

\subsection{Full chiral Lagrangian}\label{sec:chiralLagrangian}

For the non-linear realization of the sigma model, the full hadronic Lagrangian density reads \cite{Papazoglou:1998vr}
\begin{equation}
\mathcal{L}=\mathcal{L}_{\rm kin}+\mathcal{L}_{ \rm int}+\mathcal{L}_{ \rm scal}+\mathcal{L}_{ \rm  vec}+\mathcal{L}_{\rm  esb}\,,
\label{eq:L_before_MFA}
\end{equation}
where $\mathcal{L}_{\rm  kin}$ is the kinetic energy term, $\mathcal{L}_{\rm  int}$ is the baryon-meson interaction term, $\mathcal{L}_{ \rm scal}$ is the scalar meson self-interaction term, $\mathcal{L}_{\rm  vec}$ is the vector meson self-interaction term, and $\mathcal{L}_{\rm  esb}$ is the term for explicit breaking of chiral symmetry. We now cover each of these five terms in more detail. 

\subsubsection{\texorpdfstring{$\mathcal{L}_{\rm kin}$} - the kinetic-energy term}

The kinetic energy term expands as
\begin{align}
&\mathcal{L}_{\rm kin}=i\mathrm{Tr}(\bar{B}\gamma_\mu D^\mu B )+\frac{1}{2}\mathrm{Tr}(D_\mu X D^\mu X ) \nonumber \\
&+\mathrm{Tr}(u_\mu X u^\mu X + X u_\mu u^\mu X )+\frac{1}{2}\mathrm{Tr}(D_\mu Y D^\mu Y )\nonumber\\ &+\frac{1}{2}\mathrm{Tr}(D_\mu\chi D^\mu\chi)-\frac{1}{4}\mathrm{Tr}(V^{\mu\nu}V_{\mu\nu})-\frac{1}{4}\mathrm{Tr}(A^{\mu\nu}A_{\mu\nu})\,,
\label{eq:Lag_kinetic}
\end{align}
where $D_\mu$ is a covariant derivative defined by
\begin{equation}
D_\mu\diamond=\partial_\mu\diamond+i[\Gamma_\mu,\diamond]\,,
\end{equation}
with $\diamond$ being any of the following particle matrix: $B$ stands for the baryon octet matrix, $X$ for the scalar meson nonet and $Y$ for the pseudoscalar singlet. They are shown in the mean-field approximation in~\Cref{app:multiplets}. The $[\cdot,\diamond]$ represents the operator commutator and $\Gamma_\mu$ is a vector-type field that assures chiral invariance and is defined by
\begin{equation}
\Gamma_\mu=-\frac{i}{2}(u^\dagger\partial_\mu u+u\partial_\mu u^\dagger)\,.
\label{eq:Gamma_mu}
\end{equation}

The kinetic energy term of the pseudoscalar mesons is introduced (in analogy to \Cref{eq:Gamma_mu}) by defining the axial vector,
\begin{equation}
u_\mu=-\frac{i}{2}(u^\dagger\partial_\mu u-u\partial_\mu u^\dagger)\,,
\end{equation}
with
\begin{equation}
\begin{split}
u & =e^{\frac{i}{2\sigma_{0}}\pi^{a}\left(x\right)\lambda_{a}\gamma_{5}} =e^{\frac{i}{\sqrt{2}\sigma_{0}}P\gamma_{5}}\,,
\end{split}
\end{equation}
where $P=\frac{\pi^a\lambda_a}{\sqrt{2}}$ is the pseudoscalar octet matrix defined in \Cref{Pmatrix}. 
$\lambda_a$ are the Gell-mann matrices, $\gamma_5$ the fifth Dirac gamma matrix, which is the chirality operator, and $\pi^a$ are the components of the pseudoscalar meson octet.

The vector and axial-vector field tensors are $V^{\mu\nu}$=$\partial^\mu V^\nu-\partial^\nu V^\mu$ and $A^{\mu\nu}$=$\partial^\mu A^\nu-\partial^\nu A^\mu$, with the associated vector and axial field vectors $V_\mu$ and $A_\mu$. The vector meson nonet $V=V_\mu$ is shown in \Cref{app:multiplets} and $\chi$ represents the dilaton field, a.k.a. glueball field.

The first term in \Cref{eq:Lag_kinetic} is a Dirac term for the baryons, the second, fourth, and fifth terms are Klein-Gordon terms for their respective scalar, pseudo-scalar singlet, and dilaton fields, the sixth and seventh terms are Proca terms for the vector and axial-vector mesons, whereas the third term contains interaction between the scalar mesons and the pseudo-scalar meson nonet, including the pseudo-scalar kinetic term, $\mathrm{Tr}(u_\mu u^\mu)$. 

\subsubsection{\texorpdfstring{$\mathcal{L}_{\rm scal}$} - the scalar-meson self-interaction term}

The scalar meson self-interaction couplings are governed solely by SU(3)$_V$ symmetry, resulting in three lowest independent invariants,
\begin{equation}
\label{10}
I_0=\mathrm{det}( X )\,,~ I_1=\mathrm{Tr}( X )\,, ~\mathrm{and}~ I_2=\mathrm{Tr}( X ^2)\,.
\end{equation}
For integer $n>2$, $I_n=\mathrm{Tr}( X ^n)$ are invariant but not independent, as they can be written in terms of $I_0$, $I_1$, and $I_2$; for example, it can be shown using the matrix $X$ from \Cref{Pmatrix} that
\begin{equation}
\label{11}
I_4=\mathrm{Tr}( X ^4)\equiv I_1I_3+\frac{1}{2}\left(I_2-I_1^2\right)I_2+I_0I_1\,,
\end{equation}
where 
\begin{equation}
\label{12}
I_{3} = \operatorname{Tr} ( X ^3)=I_1 I_2+\frac{1}{2}\left(I_2-I_1^2\right) I_1+I_0\,.
\end{equation}

Using these invariants (excluding linear terms in the scalar mesonic fields that would generate a non-zero scalar density in vacuum), we define the scalar Lagrangian density up to order $4$ as 
\begin{equation}
\label{13s}
\mathcal{L}_0=-\frac{1}{2}k_0\chi^2I_2+k_1 I_2^2+k_2 I_4+2k_3\chi I_0+k_{3N}\chi I_3\,,
\end{equation}
where each term has been multiplied by an appropriate power of the dilaton field to allow the coupling constants $k$ to be dimensionless and thus make the Lagrangian scale invariant \cite{Schechter:1980ak,Callan:1969sn}. The parameter $k_{3N}$ is related to the nuclei-scalar meson interaction in the chiral model \cite{Papazoglou:1998vr}. It is not considered in the CMF, as it currently does not include nuclei as degrees of freedom, $k_{3N}=0$.

Whenever there are remaining dimensionful terms in the Lagrangian, the dilaton field $\chi^n$ must be multiplied with appropriate power to keep the coupling constants dimensionless and the Lagrangian scale invariant.
Additionally, to mimic the broken scale invariance property of QCD $\theta_\mu^\mu= \beta_{\rm QCD} G_{\mu \nu}G^{\mu\nu}/2g $,  a scale breaking term is added to the effective Lagrangian (with  $G^{\mu\nu}$ as a gluon field tensor)~\cite{Papazoglou:1997uw}
\begin{equation}
\mathcal{L}_{\rm scale\,break}= \frac{\chi^4}{4}\ln \bigg( \frac{\chi^4}{\chi_0^4}\bigg)   + \frac{\epsilon}{3}\chi^4\ln\bigg(\frac{I_0}{\mathrm{det}\langle X _0\rangle}\bigg) - k_4\chi^4\,.
\label{eq:L_broken_scale_invariance}
\end{equation}
In analogy to the scale-breaking term discussed in Ref.~\cite{Schechter:1980ak}, the first term is added to the effective Lagrangian at tree level, where $\chi$ represents a field associated with a spin $0^+$ glueball. This term disrupts scale invariance, resulting in the proportionality $\theta^\mu_\mu = \chi^4$, which follows from the definition of scale transformations \cite{Schechter:1971qa}. However, this form of the glueball potential is strictly applicable only to the effective low-energy theory of pure, quarkless QCD. To generalize the glueball potential for the case of massless quarks, a second term is introduced. Moreover, a third term is introduced to generate a phenomenological consistent finite vacuum expectation value~\cite{Detlef:Thesis}. The second and third terms extend the logarithmic term introduced in Ref.~\cite{Heide:1992tk} within the context of SU(3), ensuring that $\theta^\mu_\mu = \chi^4$ holds.
In \Cref{eq:L_broken_scale_invariance}, $\langle X _0\rangle$ is the vacuum expectation value of the scalar matrix, $\chi_0$ is the vacuum expectation value of the dilaton field, and we set $\epsilon=2/33$, which is related to the quark contribution to the QCD beta function~\cite{Schechter:1980ak}.
Adding these two pieces together gives us the full scalar mesonic self-interaction term

\begin{equation}
\begin{split}
\mathcal{L}_{\rm scal}=&-\frac{1}{2}k_0\chi^2I_2+k_1 I_2 ^2+k_2I_4+2k_3\chi I_0\\
&+k_{3N}\chi I_3 +\frac{\epsilon}{3}\chi^4\ln\bigg(\frac{I_0}{\mathrm{det}\langle X _0\rangle}\bigg) \\
&-k_4\chi^4+ \frac{\chi^4}{4}\ln \bigg( \frac{\chi^4}{\chi_0^4}\bigg)\,.
\end{split}
\end{equation}

\subsubsection{\texorpdfstring{$\mathcal{L}_{\rm vec}$} - the vector-meson interaction term}

The vector-meson interaction term is 

\begin{align}
\mathcal{L}_{\rm vec}=&\mathcal{L}_{\rm vec}^{m}+\mathcal{L}_{\rm vec}^{\rm SI}\,, \nonumber \\
&=\frac{1}{2}\frac{\chi^2}{\chi_0^2}m_{\rm V}^2\mathrm{Tr}({V}_\mu{V}^\mu)+\mathcal{L}_{\rm vec}^{\rm SI}\,,
\end{align}
where the first term is the mass term of each vector meson $\omega,\phi,\rho$. The second one presents different possibilities for the self-interaction terms of the vector mesons that are chiral invariant \cite{Dexheimer:2015qha}
\begin{align}
\text { C1: } \mathcal{L}_{\rm vec}^{\rm SI}=&2 g_{4} \mathrm{Tr}\left(V^4\right)\,, \nonumber \\
\text { C2: }\mathcal{L}_{\rm vec}^{\rm SI}=&g_{4}\Bigg[\frac{3}{2} \left[\mathrm{Tr}\left(V^2\right)\right]^2 -\mathrm{Tr}\left(V^4\right) \Bigg]\,, \nonumber \\
\text { C3: }\mathcal{L}_{\rm vec}^{\rm SI} =&g_{4} \left[\mathrm{Tr}\left(V^2\right)\right]^2\,, \nonumber \\
\text { C4: }\mathcal{L}_{\rm vec}^{\rm SI}
=&\frac{g_{4}}{4} [\mathrm{Tr}(V)]^4\,,
\end{align}
where C2 is a combination of other terms. Two more chiral invariant combinations can be used, but they were never studied in detail because they did not seem to produce physical results.

Note that the coupling scheme denoted as $\mathrm{C} 4$ for the self-interaction of vector mesons requires the introduction of a bare mass $m_0$ for the baryon octet,
\begin{equation}\label{eq:m0_lagrangian}
\begin{split}
    \mathcal{L}_{m_0} &=  - m_0 \mathrm{Tr} \left( \bar B B\right)\,,  \\
    &= -\sum_{i \, \in \, B}\bigg[\bar{\psi}_i m_0 \psi_i\bigg]\,,
\end{split}
\end{equation}
to properly fit the nuclear compressibility at saturation~\cite{Dexheimer:2008ax}.
We address parameter fitting in \Cref{sec:coupling}.

\subsubsection{\texorpdfstring{$\mathcal{L}_{\rm esb}$} - the explicit symmetry-breaking term}

As previously discussed in \Cref{sec:intro}, when chiral symmetry is spontaneously broken, Goldstone bosons emerge, which leads to large fluctuations that can lead to divergences or instabilities in the model. To remove the effects of these fluctuations, we add explicit symmetry-breaking terms to the Lagrangian density which also give rise to pseudoscalar-meson mass terms,
\begin{equation}\label{eq:LBS}
\mathcal{L}^u_{\rm esb}=-\frac{1}{2}\frac{\chi^2}{\chi_0^2}\mathrm{Tr}\left[A_p\left(u X u+u^\dagger X u^\dagger\right)\right]\,,
\end{equation}
where $A_p=\frac{1}{\sqrt{2}} \mathrm{diag}(m_\pi^2f_\pi, \, m_\pi^2f_\pi, \, 2m_K^2f_K-m_\pi^2f_\pi)$ is the matrix of explicit symmetry breaking parameters~\cite{Koch:1997ei}, with $f_\pi$ and $f_K$ being the decay constants of pions and kaons.
This term gives rise to a pion mass and leads to partially conserved axial current relations for $\pi$ and $K$ mesons. 
The choice of power for the dilaton field matches the dimension of the fields of the chiral condensates~\cite{Papazoglou:1997uw}.

In contrast to linear realization, a symmetry-breaking term can be explicitly introduced in the nonlinear realization to accurately reproduce the hyperon potentials without impacting the partially conserved axial current relations \cite{Papazoglou:1998vr}. We introduce the following term with a free parameter $m^{HO}_3$, which contributes to the bare mass of hyperons from the octet (with typical values given in \Cref{tab:vec_param})
\begin{equation}   
\mathcal{L}^{HO}_{\rm pot }=- m^{HO}_3 \operatorname{Tr}(\bar{B} B-\bar{B}[B, S]) \operatorname{Tr}\left(X-X_0\right)\,,
\label{eq:hyp_pot_Lag}
\end{equation}
where $S_b^a=-\frac{1}{3}\left[\sqrt{3}\left(\lambda_8\right)_b^a-\delta_b^a\right]$.
The net Lagrangian for the explicit symmetry-breaking contribution then reads
\begin{equation}
    \mathcal{L}_{\rm esb} = \mathcal{L}^u_{\rm esb}+\mathcal{L}^{HO}_{\rm pot}\,.
\end{equation}

\subsubsection{\texorpdfstring{$\mathcal{L}_{\rm int}$} - the baryon-meson interaction term}

This interaction term is similar for all mesons, with the only difference being the Lorentz space occupied by the mesons. Therefore we can write all the interactions with two compact terms, and any baryon ($B$) - meson ($M$) interaction expands as
\begin{align}
\mathcal{L}_{\rm int}=&-\sqrt{2}g_8^M \bigg(\alpha_M [{\bar{B}OB}M]_F +(1-\alpha_M )[{\bar{B}OB}M]_D \bigg) \nonumber \\
&-\frac{1}{\sqrt{3}}g_1^M \mathrm{Tr}({\bar{B}OB})\mathrm{Tr}(M )\,,
\label{eq:L_BM}
\end{align}
where $O$ depends on the specific interaction, with values listed in \Cref{tab:mesonint}, $g_1^M $ and $g_8^M $ are the coupling constants related to the singlet and octet (discussed in the following), and $\alpha_M $ controls the mixing between the $D$-type (symmetric) and $F$-type (anti-symmetric) terms, that read
\begin{equation}
[{\bar{B}OB}M]_D=\mathrm{Tr}({\bar{B}O}MB+{\bar{B}OB}M)-\frac{2}{3}\mathrm{Tr}({\bar{B}OB})\mathrm{Tr}(M )\,,
\label{symcouple}
\end{equation}
and
\begin{equation}
[{\bar{B}OB}M]_F=\mathrm{Tr}({\bar{B}O}MB-{\bar{B}OB}M)\,.
\label{antisymcouple}
\end{equation}
The last term in the $D$-type interaction term is added to cancel out the singlet contribution to the octet interaction when a nonet meson matrix is utilized.
\begin{table}[t!]
\centering
\def\arraystretch{1.8}
\begin{tabular}{ccc}
\hline
\hline
Interaction with baryons & $O$ & $M $\\
\hline
Scalar & $1$ & $ X $ \\
Pseudo-scalar & $\gamma_\mu\gamma_5$ & $u_\mu$ \\
Vector (vector interaction) & $\gamma_\mu$ & $V_\mu$ \\
Vector (tensor interaction) &$\sigma^{\mu\nu}$ & $V^{\mu\nu}$ \\
Axial-vector & $\gamma_\mu\gamma_5$ & $A_\mu$\\
\hline
\hline
\end{tabular}
\caption{Table with details for each baryon-meson interaction.}
\label{tab:mesonint}
\end{table}

\subsection{The mean-field Lagrangian}\label{sec:LCMF}

The full quantum operator fields in the Lagrangian (\Cref{eq:L_before_MFA}) lead to nonlinear quantum field equations with large couplings, making perturbative approaches infeasible and challenging to solve. Hence, reliable non-perturbative approximations are essential for solving these complex many-body interactions and achieving accurate comparisons between theory and experiment~\cite{Serot:1997xg}.
To describe dense matter, we apply the mean-field approximation, as first proposed in Ref.~\cite{Serot:1997xg}. Within the mean-field approximation, we assume homogeneous and isotropic infinite baryonic matter with defined parity (+) and charge (0). Thus, only mean-field mesons with positive parity (scalar mesons and time-like component of vector mesons) and zero third component of isospin (mesons along the diagonal of the matrices $X$ (see \Cref{Xmatrix}) and $V_\mu$ (see \Cref{Vmatrix}))  are non-vanishing. The mean-field mesons with negative parity (space-like component of vector mesons, time-like component of axial-vector mesons, and pseudoscalars) do not follow parity conservation, and there is no source term for them in mean-field infinite baryonic matter\footnote{The ground state expectation value of space-like components of axial-vector mesons, despite their positive parity is zero because of the homogeneous and isotropic medium assumption.}.

Furthermore, in this approximation, fluctuations around the constant ground state expectation values of the  scalar and vector field operators are neglected, for example,
\begin{align}
\sigma(x)&=\left\langle \sigma  \right\rangle + \delta \sigma \rightarrow \left\langle \sigma  \right\rangle \equiv \sigma\,, \nonumber \\
\omega^\mu(x)&=\left\langle \omega^\mu  \right\rangle \delta_{\mu 0} + \delta\omega^\mu  \rightarrow \left\langle \omega_0  \right\rangle \equiv \omega\,.
\end{align}
As a consequence, $\sigma$, $\delta$, $\zeta$, $\omega$, $\rho$, and $\phi$ are all reduced to time-and space-independent quantities. For simplicity, we omit the time index of vector mesons ($\omega$, $\rho$, and $\phi$) and also omit the third component of the isospin index of the isovector mesons ($\rho$ and $\delta$). 

We also work within the “no-sea” approximation, i.e. neglecting all contributions from antiparticles (section 2.1.2 of Ref.~\cite{Reinhard:1989zi}). A more elaborate evaluation of vacuum would include renormalization, with the vacuum contribution depending on the renormalization scale \cite{Skokov:2010sf,Chatterjee:2011jd}. This is will be the topic of a future extension of the finite-temperature CMF model.

We refer to the resulting Lagrangian, also including quark degrees of freedom \cite{Dexheimer:2009hi}, as the Chiral Mean Field (CMF) Lagrangian density
\begin{align}
\mathcal{L}_{\rm CMF}&=\mathcal{L}_{\rm kin}+\mathcal{L}_{\rm 
 int}+\mathcal{L}_{\rm  scal}+\mathcal{L}_{\rm  vec}+ \mathcal{L}_{m_0}+\mathcal{L}_{\rm esb} \nonumber \\ &+ \mathcal{L}_{\rm quarks} +  \mathcal{L}_{\Phi}\,.
 \label{eq:L_CMF}
\end{align}
In the regimes we examine, the $\chi$ field has a weak coupling to the baryons, resulting in little overall contribution to the baryon thermodynamic quantities regardless of the value of the $\chi$ field. 
Thus, for the remainder of this work, we set $\chi=\chi_0$ ($\chi$ remains ``frozen" at its vacuum value, $\chi_0$) and apply further simplifications.
For details regarding $\chi$, see Ref. \cite{Schechter:1980ak,Bonanno:2008tt,Sasaki:2011sd}.
As a result, no equation of motion for the $\chi$ field is shown. 

\subsubsection{\texorpdfstring{$\mathcal{L}_{\rm kin}$} - the kinetic-energy term}

In the mean-field approximation, $u=u^\dagger=1$ \footnote{ It may appear that, after the mean-field approximation, the Lagrangian or equations of motion for linear and non-linear realizations are similar. However, the non-linear realization allows explicit symmetry-breaking terms to be added without violating PCAC relations and decouples strange and non-strange condensates, enabling a more consistent treatment of hyperon potentials and meson interactions compared to the linear approach.}, thus the commutator $[\Gamma^\mu, \diamond ]\rightarrow0,$  meaning that the covariant derivative reduces to the partial derivative, $D^\mu\rightarrow\partial^\mu$. The mesons are taken as static, and thus no longer have kinetic terms ($\partial M = 0$, where $M$ is the matrix from \Cref{tab:mesonint}), such that all of \Cref{eq:Lag_kinetic} reduces to %
\begin{equation}
i\mathrm{Tr}(\bar{B}\gamma_{\mu}D^{\mu}B)=i\sum_{i\in B}\bigg(\bar{\psi}_i\gamma_\mu\partial^\mu\psi_i\bigg)\,.
\end{equation}
Quarks are discussed in the following.

\subsubsection{\texorpdfstring{$\mathcal{L}_{\rm scal}$} - the scalar-meson self-interaction term}

Applying the mean-field approximation to the scalar-meson self-interaction term and calculating the $I_n$ terms explicitly (see \Cref{app:CMFdetails} for a detailed calculation), we obtain
\begin{align}
\mathcal{L}_{\rm scal}=&-\frac{1}{2}k_0\chi_0^2(\sigma^2+\zeta^2+\delta^2)+k_1(\sigma^2+\zeta^2+\delta^2)^2\nonumber \\
&+k_2\left[\frac{\sigma^4+\delta^4}{2}+\zeta^4+3 \left(\sigma\delta \right)^2\right]+k_3\chi_0\left(\sigma^2-\delta^2\right)\zeta\nonumber\\ &+k_{3 N} \chi_0\left(\frac{\sigma^3}{\sqrt{2}}+\frac{3}{\sqrt{2}}\sigma\delta^2+\zeta^3\right)-k_4\chi_0^4\nonumber\\ &+\frac{\epsilon}{3}\chi_0^4\ln\left[\frac{\left(\sigma^2-\delta^2 \right)\zeta}{\sigma_0^2\zeta_0}\right]\,,
\end{align}
where $\sigma_0=-f_\pi$ and $\zeta_0=\dfrac{f_\pi}{\sqrt{2}} - \sqrt{2} f_K$ are the vacuum values of the $\sigma$ and $\zeta$ fields, respectively.

\subsubsection{\texorpdfstring{$\mathcal{L}_{\rm vec}$} - the vector-meson interaction term}
After applying the mean-field approximation, the total vector-meson Lagrangian can be expressed as a sum of the mass and  self-interaction terms
\begin{align}
\mathcal{L}_{\rm vec}=\frac{1}{2}\left(m_\omega^2\omega^2+m_\phi^2\phi^2+m_\rho^2\rho^2\right)+\mathcal{L}_{\rm vec}^{\rm SI}\,,
\end{align}
or explicitly
\begin{align}
&\mathcal{L}_{\rm vec}=\frac{1}{2}\left(m_\omega^2\omega^2+m_\phi^2\phi^2+m_\rho^2\rho^2\right)\nonumber \\
&+g_4\begin{cases}
\text { C1: } \left( \omega^{4}+6 \omega^2\rho^2+ \rho^{4} +2 \phi^{4}\right)\,,\\
\text { C2: } \left( \omega^4 + \rho^4+ \frac{ \phi^4}{2}+3 \rho^2 \phi^2 +3 \omega^2 \phi^2 \right)\,,\\
\text { C3: } \left( \omega^4+2 \omega^2 \rho^2+ \rho^4+2 \omega^2 \phi^2 + \phi^4 +2 \rho^2 \phi^2 \right)\,,\\
\text { C4: } \left(\omega^4+\frac{\phi^4}{4}+3\omega^2\phi^2+2\sqrt2\omega^3\phi+\sqrt2\omega\phi^3\right)\,. \\
\end{cases}
\label{eq:L_vec}
\end{align}

The coupling scheme $\mathrm{C} 2$ is a linear combination of $\mathrm{C} 1$ and $\mathrm{C} 3$ and which exhibits no $\omega \rho$ mixing. The coupling scheme denoted as $\mathrm{C} 4$ for the self-interaction of vector mesons is quite different from the other schemes, as it includes a term that exhibits a linear dependence on the strange vector meson $\phi$. Because of this linear dependence on $\phi$, the $\mathrm{C} 4$ scheme requires a different parametrization that includes a bare mass term, \Cref{eq:m0_lagrangian} to ensure that the compressibility of nucleons is in a better agreement with nuclear physics data \cite{Dexheimer:2007tn,Dexheimer:2008cv}.
See Ref.~\cite{Malik:2024qjw} for combinations of the couplings C1-C4 that allow one to separate each coupling term (in the non-strange case, $\phi=0$).

\subsubsection{\texorpdfstring{$\mathcal{L}_{\rm esb}$} - the explicit symmetry-breaking term}

In the mean-field approximation, the first explicit symmetry-breaking term \Cref{eq:LBS} (together with \Cref{Xmatrix}) simplifies to
\begin{equation}
\mathcal{L}^u_{\rm esb}=-\left[m_{\pi}^{2}f_{\pi}\sigma+\left(\sqrt{2}m_{K}^{2}f_{K}-\frac{1}{\sqrt{2}}m_{\pi}^{2}f_{\pi}\right)\zeta\right].
\label{eq:L_esb}
\end{equation}

From \Cref{eq:hyp_pot_Lag}, the expanded form of symmetry-breaking  Lagrangian related to the potential of the hyperons from the octet ($HO$) is given by 
\begin{equation}\label{eq:hyperon_potential_lagrangian}
\mathcal{L}^{HO}_{\rm pot}   = -\sum_{i \, \in \, H}\bigg[\bar{\psi}_i m^{HO}_3 \left( \sqrt{2} (\sigma-\sigma_0) +  (\zeta-\zeta_0) \right)  \psi_i\bigg]\,.
\end{equation}
It leads to an additional contribution to the coupling between the hyperons and the mesons $\sigma$ and $\zeta$ through the parameter $m^{HO}_3$, and to a constant (bare) mass term $\Delta m_i$. 

Note that $\Delta m_i$ also receives a contribution from the bare mass term (in the case of C4 vector-coupling),~\Cref{eq:m0_lagrangian} 
as follows
\begin{align}
    \Delta m_N &=m_0\,,\\
    \Delta m_{HO} &= m_0   - m^{HO}_3 \left( \sqrt{2} \sigma_0 + \zeta_0  \right)\,.
\end{align}

Similarly, a mass correction  due to an explicit breaking term with parameter $m^{HD}_3$ for the hyperons from the baryon decuplet ($HD$) is written as
\begin{align}
     \Delta m_\Delta &= 0\,,\\ 
     \Delta m_{\Sigma^*} &=  \Delta m_{\Xi^*}= -m^{HD}_{3}(\sqrt{2} \sigma_0+\zeta_0)\,,\\ 
     \Delta m_\Omega &= -\frac{3}{2}m^{HD}_{3}(\sqrt{2} \sigma_0+\zeta_0)\,,
\end{align}

\subsubsection{\texorpdfstring{$\mathcal{L}_{\rm int}$} - the baryon-meson interaction term}
\label{sec:the_baryon_meson_interaction_term}

Applying the mean-field approximation to the baryon-meson interaction term, we only get non-zero values for the cases where $M = X $ and $M =V$. This is due to the $A$-matrix for pseudovector mesons having vanishing expectation values and the pseudoscalar mesons only coupling to the baryons with a pseudovector coupling.  By doing the explicit calculation of the interaction term (see \Cref{app:interactions}), we can rewrite the interaction Lagrangian, \Cref{eq:L_BM}, as
\begin{equation}\label{hhh}
\begin{split}
    \mathcal{L}_{\rm int}=-\sum_{i \, \in \, {B}} \bar{\psi}_i&\big[\gamma_0 \big(g_{i\omega}\omega+g_{i\rho}\rho+g_{i\phi}\phi\big) \\
    &+ g_{i \sigma} \sigma + g_{i \zeta} \zeta + g_{i \delta} \delta \big]\psi_i\,,
    \end{split}
\end{equation}
where the couplings $g_{iM}$ are written in terms of $\alpha_M$ (from \Cref{eq:L_BM}), $g_8^M$, $g_1^M$, and $m^{HO}_3$, as shown in~\Cref{tab:scalar_couplings} for the scalar-mesons ($M=X$). From the curvature of the interaction potential (or minus the Lagrangian density), we can identify the effective mass terms for the baryons in terms of these couplings as
\begin{widetext}
\begin{flalign}
m_p^*=& \Delta m_N + \frac{1}{\sqrt3}g_1^X\left(\sqrt2\sigma+\zeta\right)-\frac{1}{3}g_8^ X \left(4\alpha_ X -1\right)\left(\sqrt2\zeta-\sigma\right)+g_8^ X \delta\,,\nonumber\\
m_n^*=&\Delta m_N +\frac{1}{\sqrt3}g_1^X\left(\sqrt2\sigma+\zeta\right)-\frac{1}{3}g_8^ X \left(4\alpha_ X -1\right)\left(\sqrt2\zeta-\sigma\right)-g_8^ X \delta\,,\nonumber\\
m_{\Lambda}^*=& \Delta m_\Lambda+ \left(m^{HO}_3 +\frac{1}{\sqrt3}g_1^X \right)\left(\sqrt2\sigma+\zeta\right)-\frac{2}{3}g_8^ X \left(\alpha_ X -1\right)(\sqrt2\zeta-\sigma)\,,\nonumber\\
m_{\Sigma^+}^*=&\Delta m_\Sigma +\left(m^{HO}_3 +\frac{1}{\sqrt3}g_1^X \right)\left(\sqrt2\sigma+\zeta\right)+\frac{2}{3}g_8^ X \left(\alpha_ X -1\right)\left(\sqrt2\zeta-\sigma\right)+2g_8^ X \alpha_ X \delta\,, \label{eq:effective_masses}\\
m_{\Sigma^0}^*=&\Delta m_\Sigma +\left(m^{HO}_3 +\frac{1}{\sqrt3}g_1^X \right)\left(\sqrt2\sigma+\zeta\right)+\frac{2}{3}g_8^ X \left(\alpha_ X -1\right)\left(\sqrt2\zeta-\sigma\right)\,,\nonumber\\
m_{\Sigma^-}^*=&\Delta m_\Sigma +\left(m^{HO}_3 +\frac{1}{\sqrt3}g_1^X \right)\left(\sqrt2\sigma+\zeta\right)+\frac{2}{3}g_8^ X \left(\alpha_ X -1\right)\left(\sqrt2\zeta-\sigma\right)-2g_8^ X \alpha_ X \delta\,,\nonumber\\
m_{\Xi^0}^*=&\Delta m_\Xi +\left(m^{HO}_3 +\frac{1}{\sqrt3}g_1^X \right)\left(\sqrt2\sigma+\zeta\right)+\frac{1}{3}g_8^ X (2\alpha_ X +1)\left(\sqrt2\zeta-\sigma\right)+g_8^ X \left(2\alpha_ X -1\right)\delta\,,\nonumber\\
m_{\Xi^-}^*=&\Delta m_\Xi +\left(m^{HO}_3 +\frac{1}{\sqrt3}g_1^X \right)\left(\sqrt2\sigma+\zeta\right)+\frac{1}{3}g_8^ X (2\alpha_ X +1)\left(\sqrt2\zeta-\sigma\right)-g_8^ X \left(2\alpha_ X -1\right)\delta\,,\nonumber
\end{flalign} 
\end{widetext}
which can be written compactly as
\begin{equation}\label{eq:eff_mass_nopol}   m_i^*=g_{i\sigma}\sigma+g_{i\zeta}\zeta+g_{i\delta}\delta+\Delta m_i\,.
\end{equation}
This derivation is shown in detail for the proton in \Cref{app:interactions}. It must be noted that an additional contribution to the effective mass must be accounted for when the deconfinement order parameter is introduced in the next section. If we disregard the $\delta$-meson contribution, the baryons masses of the nucleon doublet and hyperon triplets are degenerate. The inclusion of the isovector meson $\delta$ breaks this multiplet mass equality.

\begin{table*}[t!]
\centering
\def\arraystretch{1.8}
\begin{tabular}{cccc}
\hline
\hline
Particle & $g_\sigma$ & $g_\zeta$ & $g_\delta$ \\
\hline
$p$ & \multirow{2}{*}{$\sqrt{\frac{2}{3}} \, g^X_1 + \frac{1}{3}g^X_8 \, (4 \, \alpha_X - 1) $} & \multirow{2}{*}{$\sqrt{\frac{1}{3}} \, g^X_1 - \,\frac{\sqrt2 }{3}g^X_8 \, (4 \, \alpha_X - 1)  $} & $g^X_8$ \\
$n$ & & & $-g^X_8$ \\
\hline
$\Lambda$ & $\sqrt{\frac{2}{3}} \, g^X_1 + \frac{2}{3}g^X_8 (\alpha_X - 1)  + \sqrt{2} \, m^{HO}_3$ & $\sqrt{\frac{1}{3}} \, g^X_1 - \frac{2\sqrt2 }{3}g^X_8 (\, \alpha_X - 1) +m^{HO}_3 $ & 0 \\
\hline
$\Sigma^+$ & \multirow{3}{*}{$\sqrt{\frac{2}{3}} \, g^X_1 - \frac{2}{3} g^X_8 (\alpha_X - 1) +\sqrt{2} m^{HO}_3$} & \multirow{3}{*}{$\sqrt{\frac{1}{3}} \, g^X_1 +\frac{2\sqrt2}{3} g^X_8 (\alpha_X - 1)  + m^{HO}_3$} & $2 \, g^X_8 \, \alpha_X$ \\
$\Sigma^0$ & & & 0 \\
$\Sigma^-$ & & & $-2 \, g^X_8 \, \alpha_X$ \\
\hline
$\Xi^0$ & \multirow{2}{*}{$\sqrt{\frac{2}{3}} \, g^X_1 - \frac{1}{3}g^X_8 (2 \alpha_X + 1)  + \sqrt{2} \, m^{HO}_3 \, $} & \multirow{2}{*}{$\sqrt{\frac{1}{3}} \, g^X_1 +  \frac{\sqrt2}{3}g^X_8 (2 \alpha_X + 1)  + m^{HO}_3 \, $} & $g^X_8 \, (2 \, \alpha_X - 1)$ \\
$\Xi^-$ & & & $-g^X_8 \, (2 \, \alpha_X - 1)$ \\
\hline
$\Delta^{++}$ & \multirow{4}{*}{$g^X_D \, (3 - \alpha_{DX})$} & \multirow{4}{*}{$\sqrt{2} \,g^X_D \, \alpha_{DX}$} &
 \multirow{4}{*} 0 \\
$\Delta^{+}$ & & & \\
$\Delta^{0}$ & & & \\
$\Delta^{-}$ & & & \\
\cline{1-4}
$\Sigma^{*+}$ & \multirow{3}{*}{$2 \, g^X_D + \sqrt{2} \, m^{HD}_3$} & \multirow{3}{*}{$\sqrt{2} \, g^X_D + m^{HD}_3$} &  \multirow{3}{*}0 \\
$\Sigma^{*0}$ & & & \\
$\Sigma^{*-}$ & & & \\
\cline{1-4}
$\Xi^{*0}$ & \multirow{2}{*}{$g^X_D \, (1 + \alpha_{DX}) + \sqrt{2} \,  \, m^{HD}_3$} & \multirow{2}{*}{$\sqrt{2} \, g^X_D \, (2 - \alpha_{DX}) +  \, m^{HD}_3$} & \multirow{2}{*}0 \\
$\Xi^{*-}$ & & & \\
\cline{1-4}
$\Omega$ & $2 \, g^X_D \, \alpha_{DX} + \frac{3 \sqrt2}{2} m^{HD}_3$ & $\sqrt{2} \, g^X_D \, (3 - 2 \, \alpha_{DX}) + \frac{3}{2}m^{HD}_3 $ &0 \\
\hline
\hline
\end{tabular}
\caption{Table of scalar-meson coupling constants for the baryon octet and decuplet written in terms of the fundamental couplings $g_8^X$, $g_1^X$, and $\alpha_X$.}
\label{tab:scalar_couplings}
\end{table*}

For the baryon decuplet $D$, we follow \cite{Detlef:Thesis} and assume they are described by Dirac spinors such that, from the interactions between the baryon resonances and scalar mesons, we may extract the effective mass terms for the isospin degenerate baryon decuplet
\begin{widetext}
\begin{flalign}
m^*_{\Delta} & =\Delta m_\Delta + g_D^X\left[\left(3-\alpha_{D X}\right) \sigma+\alpha_{D X} \sqrt{2} \zeta\right]\,, \nonumber \\
m^*_{\Sigma^*} & = \Delta m_{\Sigma^*}+ m^{HD}_{3}(\sqrt{2} \sigma+\zeta)+g_D^X[2 \sigma+\sqrt{2} \zeta]\,,\nonumber \\
m^*_{\Xi^*} & =\Delta m_{\Xi^*}+ m^{HD}_{3}(\sqrt{2} \sigma+\zeta)+g_D^X\left[\left(1+\alpha_{D X}\right) \sigma+\left(2-\alpha_{D X}\right) \sqrt{2} \zeta\right]\,, \nonumber \\
m^*_{\Omega} & =\Delta m_{\Omega}+\frac{3}{2} m^{HD}_{3}(\sqrt{2} \sigma+\zeta)+g_D^X\left[2 \alpha_{D X} \sigma+\left(3-2\alpha_{D X}\right) \sqrt{2} \zeta\right]\,.
\label{eq:eff_mass_dec}
\end{flalign} 
\end{widetext}

\begin{table}[t!]
\centering
\def\arraystretch{1.8}
\begin{tabular}{cccc}
\hline
\hline
\hspace{1cm} & $\omega\ $ & $\phi\ $ & $\rho\ $\\
\hline
$n$ & $3$ & $0$ & $-1$\\
$p$ & $3$ & $0$ & $1$\\
$\Lambda$ & $2$ & $-\sqrt2$ & $0$\\
$\Sigma^+$ & $2$ & $-\sqrt2$ & $2$\\
$\Sigma^0$ & $2$ & $-\sqrt2$ & $0$\\
$\Sigma^-$ & $2$ & $-\sqrt2$ & $-2$\\
$\Xi^0$ & $1$ & $-2\sqrt2\ $ & $1$\\
$\Xi^-$ & $1$ & $-2\sqrt2\ $ & $-1$\\
$\Delta^{++}$ & $3$ & $0 $ & $3$\\
$\Delta^+$ & $3$ & $0 $ & $1$\\
$\Delta^0$ & $3$ & $0 $ & $-1$\\
$\Delta^-$ & $3$ & $0 $ & $-3$\\
$\Sigma^{*+}$ & $2$ & $-\sqrt2\ $ & $2$\\
$\Sigma^{*0}$ & $2$ & $-\sqrt2\ $ & $0$\\
$\Sigma^{*-}$ & $2$ & $-\sqrt2\ $ & $-2$\\
$\Xi^{*0}$ & $1$ & $-2\sqrt2\ $ & $1$\\
$\Xi^{*-}$ & $1$ & $-2\sqrt2\ $ & $-1$\\
$\Omega$ & $0$ & $-3\sqrt2\ $ & $0$\\
\hline
\hline
\end{tabular}
\caption{Table of SU(6) vector-meson coupling-constant coefficients $C_i$ with baryons (octet and decuplet), such that $g_{iV}=C_i\times g_8^V$.} 
\label{tab:vector_couplings}
\end{table}

Similarly to what has been done in \Cref{app:interactions} for the baryon-scalar meson coupling constants, we can calculate the baryon-vector meson coupling constants $g_{iV}$. 
Based on the vector dominance model (VDM) and the universality principle, it can be inferred that the $D$-type coupling is likely to be minimal~\cite{Sakurai:1969}. Therefore, in our analysis, we employ only $F$-type coupling by choosing $\alpha_V=1$ for all fits.
Additionally, we can decouple the nucleons from the strange vector meson $\phi$ by setting $g_1^V=\sqrt6g_8^V$ such that $g_{N\phi}=\sqrt{\frac{1}{3}} \, g^V_1 - \,\frac{\sqrt2 }{3}g^V_8 \, (4 \, \alpha_V - 1) \rightarrow 0$. Following a similar pattern, we assign $\alpha_{D V}$=0, resulting in the absence of coupling between the $\phi$ and the $\Delta$ baryons. 
The remaining couplings to the strange baryons are subsequently determined by symmetry relations (the quark model) \cite{Dover:1985ba} in terms of $g_8^V$ (the only free parameter for the baryon-vector mesonic coupling), such that the $\omega$ and $\phi$-meson couplings are given in \Cref{tab:vector_couplings}.
Note that the $\rho$-meson couplings follow the sign convention of the $\delta$-meson.
The scheme described is known as $F$-type or SU(6) as it includes SU(3) flavor symmetry and SU(2) spin symmetry~\cite{Weissenborn:2011ut,Lopes:2022vjx}. 
Nevertheless, in the CMF model we break this scheme and use, e.g. for C4 $g_{N\omega}/g_{N\rho}= 2.95$ (instead of $3$) to slightly modify $g_{N\rho}$ allowing a better fit of experimental data for the symmetry energy (as small differences matter).
Moreover, a parameter called $V_\Delta$ is introduced in the decuplet baryons' vector coupling ($g_{D V}=C_D \times g_8^V \times V_\Delta$) allowing a better fit of experimental data for the $\Delta$-nucleon potential. More general couplings will be explored in the future.

\subsubsection{\texorpdfstring{$\mathcal{L}_{\rm quarks}$}  - adding quarks to the model}
\label{here}

To reproduce quark deconfinement, we include up, down, and strange quarks in the model. 
We assume the same Lagrangian as the baryonic one, with kinetic, mass, and interaction terms given by
\begin{equation}
\begin{split}
    \mathcal{L}_{\rm quarks} &= \bar{\psi}_i\bigg[i\gamma_\mu\partial^\mu- \gamma_0 \big(g_{i\omega}\omega+g_{i\rho}\rho+g_{i\phi}\phi\big) \\
    & -\Delta m_i - g_{i \sigma} \sigma - g_{i \zeta} \zeta - g_{i \delta} \delta \bigg]\psi_i\,,
    \end{split}
\end{equation}
with $i=u, d, s$. We write the effective quark mass like the baryonic one,~\Cref{eq:eff_mass_nopol}, by defining
\begin{align}
    \Delta m_u =\Delta m_d=m_0^u\,,  \quad \quad \Delta m_s =&m_0^s\,,    
\end{align}
where $m_0^u=m_0^d = 5$ MeV for up and down quarks and $m_0^s=150$ MeV for the strange quark. CMF parameters associated with the quark sector have scalar couplings that are set to be roughly one-third of the nucleon scalar couplings, while the vector couplings are set to zero, in agreement with the findings of Ref.~\cite{Steinheimer:2014kka}. The coupling values are discussed in \Cref{sec:coupling}.

\subsubsection{\texorpdfstring{$\mathcal{L}_{\Phi}$} - the deconfinement Lagrangian}

The Lagrangian density of Polyakov-inspired field $\Phi$ can be written as

\begin{equation}\label{eq:Phi_lagrangian}
\begin{split}
    \mathcal{L}_{\Phi} &=  - g_{B\Phi}\Phi^2 \mathrm{Tr} \left( \bar B B\right) -g_{q\Phi}(1-\Phi) \mathrm{Tr} \left( \bar q q\right)-U_{\Phi}\,,
\end{split}
\end{equation}
where $g_{B\Phi}$ denotes the coupling of $\Phi$ field to the baryons whereas $g_{q\Phi}$  represents coupling of $\Phi$ field to the quarks. Moreover, $U_{\Phi}$ is implemented as a Polyakov-inspired potential term (referred to as the deconfinement potential)  to obtain a unified quark-hadron EoS~\cite{Dexheimer:2009hi}
\begin{align}
U_{\Phi}&=\left(a_0T^4+a_1\mu_B^4+a_2T^2\mu_B^2\right)\Phi^2\nonumber \\ &+a_3T_0^4\ln\left(1-6\Phi^2+8\Phi^3-3\Phi^4\right)\,,
\label{eq:Polyakov_Potential_temperature}
\end{align}
which at $T=0$ reduces to
\begin{align}
U_{\Phi}&=a_1\mu_B^4\Phi^2+a_3T_0^4\ln\left(1-6\Phi^2+8\Phi^3-3\Phi^4\right),
\label{eq:Polyakov_Potential}
\end{align}
where the $a$'s and $T_0$ are constants. 
Here we introduced a scalar field $\Phi\in[0,1]$, which serves as an order parameter for the quark-hadron phase transition. Note that this is not the Polyakov loop, which acts as an order parameter for confinement in pure gauge theories at nonzero temperatures only.
$U_{\Phi}$ was modified from its original form in the PNJL model \cite{Ratti:2005jh,Roessner:2006xn} to also contain baryon chemical potential dependent terms (of even order), to be used to study low-temperature and high-density environments, such as neutron stars. It has been shown that a $\mu_B^2$ term in $U_\Phi$ (instead of $\mu_B^4$) would significantly weaken the deconfinement phase transition at $T=0$ \cite{Dexheimer:2020rlp,Clevinger:2022xzl,Kumar:2023qcs}.
The form of $U_{\Phi}$ dictates the shape and location of the quark-hadron phase transition in the QCD phase diagram. 
If future information from the RHIC Beam Energy Scan and theoretical developments further constrain the QCD critical point, one could redefine $U_{\Phi}$ to reproduce these new constraints.

Explicitly, the first term in \Cref{eq:Phi_lagrangian} can we written as
\begin{equation}\label{eq:Phi_lagrangian_exp}
\begin{split}
    \mathcal{L}_{\Phi}^{(1)} &= -\sum_{i \, \in \, B}\bigg[\bar{\psi}_i  g_{B\Phi}\Phi^2  \psi_i \bigg] -\sum_{i \, \in \, q}\bigg[ \bar{\psi}_i  g_{q\Phi}(1-\Phi)   \psi_i\bigg]\,,
\end{split}
\end{equation}
which serves as an additional contribution to the effective masses of the baryons
\begin{equation}
m_i^*=g_{i\sigma}\sigma+g_{i\zeta}\zeta+g_{i\delta}\delta+\Delta m_{i}+g_{B\Phi}\Phi^2\,.
\label{eq:effective_mass_hadrons}
\end{equation}
and quarks
\begin{equation}
m_i^*=g_{i\sigma}\sigma+g_{i\zeta}\zeta+g_{i\delta}\delta + \Delta m_{i}+g_{q\Phi}\left(1-\Phi \right)\,.
\label{eq:effective_mass_quarks}
\end{equation}
Considering $g_{i\Phi}$ to be large, quark masses are large and baryon masses are small when $\Phi\sim0$ (and vice-versa when $\Phi\sim1.$)
The larger the mass of a particle, the more energy is required to create it. 
Therefore, when $\Phi$ is large it causes the baryon masses to be so large that it suppresses their influence and one is in a quark-dominated phase. 
On the other hand, when  $\Phi$ is small then the quark masses are large such that they are suppressed and one is in a hadron-dominated phase.
Putting this all together, $\Phi\sim0$ corresponds to having only hadrons and $\Phi\sim1$ corresponds to having only quarks, with intermediate values corresponding to having both hadrons and deconfined quarks (only reproduced at large temperatures).

\subsection{Equations of motion}\label{sec:EOM}

To derive the equations of motion for the six mean-field mesons and the deconfinement order parameter, $\Phi$, we apply the Euler-Lagrange equation to the CMF Lagrangian density
\begin{align}
\frac{\partial \mathcal{L_{\rm CMF}}}{\partial \varphi } - \partial_\mu\left(\frac{\partial \mathcal{L_{\rm CMF}}}{\partial (\partial_\mu \varphi)}\right) = 0\,,
\end{align}
with $\varphi =\sigma, \delta, \zeta, \omega, \phi, \rho$, and $\Phi$, resulting in
\\
\begin{widetext}
\begin{flalign}
&\sigma:~~\sum_i g_{i\sigma} n_{sc,i}=-k_0\chi_0^2\sigma+4k_1(\sigma^2+\zeta^2+\delta^2)\sigma+2k_2(\sigma^2+3\delta^2)\sigma+2k_3\chi_0\sigma\zeta+\frac{2\epsilon}{3}\chi_0^4\frac{\sigma}{\sigma^2-\delta^2}-m_\pi^2f_\pi\,,
\nonumber\\
&\zeta:~~\sum_ig_{i\zeta}n_{sc,i}=-k_0\chi_0^2\zeta+4k_1(\sigma^2+\zeta^2+\delta^2)\zeta+4k_2\zeta^3+k_3\chi_0(\sigma^2-\delta^2)+\frac{\epsilon}{3\zeta}\chi_0^4-\bigg(\sqrt2m_K^2f_K-\frac{1}{\sqrt2}m_\pi^2f_\pi\bigg)\,,
\nonumber\\
&\delta:~~\sum_ig_{i\delta}n_{sc,i}=-k_0\chi_0^2\delta+4k_1(\sigma^2+\zeta^2+\delta^2)\delta+2k_2(3\sigma^2+\delta^2)\delta-2k_3\chi_0\delta\zeta-\frac{2\epsilon}{3}\chi_0^4\frac{\delta}{\sigma^2-\delta^2}\,,
\nonumber\\
&\omega:~~\sum_i g_{i\omega}n_{i}=m_\omega^2\omega+2 g_4 \begin{cases}
\text { C1: }  2 \omega \left(\omega^2 + 3 \rho^2 \right)\,,\\
\text { C2: }  \omega \left(2 \omega^2 + 3 \phi^2 \right)\,,\\
\text { C3: }  2 \omega \left(\omega^2 + \rho^2 + \phi^2 \right)\,,\\
\text { C4: } \left(2\omega^3+3\phi^2\omega+3\sqrt2\phi\omega^2+\frac{\phi^3}{\sqrt{2}} \right)\,,\\
\end{cases}
\nonumber\\
&\phi:~~\sum_ig_{i\phi}n_{i}=m_\phi^2\phi+2 g_4\begin{cases}
\text { C1: }  4 \phi^3\,,\\
\text { C2: }  \phi \left( \phi^2 + 3 \left(\omega^2+\rho^2\right)\right)\,,\\
\text { C3: }  2 \phi \left( \omega^2 + \phi^2 + \rho^2 \right)\,,\\
\text { C4: }  \frac{\phi^3}{2} + 3 \omega^2 \phi + \sqrt{2} \omega^3 + \frac{3}{\sqrt{2}} \omega \phi^2\,,\\
\end{cases}
\nonumber\\
&\rho:~~\sum_ig_{i\rho}n_{i}=m_\rho^2\rho+2 g_4\begin{cases}
\text { C1: }  2 \rho ( 3\omega^2 + \rho^2)\,,\\
\text { C2: }  \rho ( 3 \phi^2 + 2\rho^2 )\,,\\
\text { C3: }  2 \rho ( \omega^2 + \phi^2 + \rho^2)\,,\\
\text { C4: }  0\,,
\end{cases}
\nonumber\\
&\Phi:~~\sum_{i\in q}
 g_{q\Phi}n_{sc,i} - 2 \Phi \sum_{i\in B} g_{B\Phi}n_{sc,i}=2 a_{1} \mu_{B}^{4} \Phi + a_3T_0^4\frac{12\Phi}{3\Phi^2-2\Phi-1}\,, 
\label{eq:Algebraic_System}
\end{flalign}
\end{widetext}
where $n_{sc,i}=\langle\bar{\psi}_i\psi_i\rangle$ is the scalar number density and $n_{i}=\langle{\psi}_i^\dagger\psi_i\rangle$  is the baryon (vector) number density. 
The index $i$ always indicates a summation of the baryon octet, decuplet, \emph{and} quark flavors.

Note that we do not derive equations of motion for fermions, as the expected value of their fields does not come from their equations of motion in our formalism, but instead directly from their effective chemical potentials and effective masses, which come from the chemical potentials looped over, $\mu_B, \mu_Q, \mu_S$. This is discussed in detail in the following.

\subsection{Thermodynamical observables}\label{sec:thermo}

The CMF Lagrangian density can be alternatively seen as consisting of a fermion part, a boson part, and a vector interaction term $\mathcal{L}\equiv\mathcal{L}_{\rm fermions}+\mathcal{L}_{\rm bosons} + \mathcal{L}_{\rm V,int}$.
For fermions, the Lagrangian density reads
\begin{align}
\mathcal{L}_{\rm fermions}=\sum_{i \, \in \, \rm fermions}\bigg[\bar{\psi}_i(i\gamma_\mu\partial^\mu-m_i^*)\psi_i\bigg]\,,
\end{align}
where the scalar-meson interactions are hiding within the effective mass $m_i^\ast$. This is the relativistic free Fermi gas Lagrangian but with effective masses $m_i^*$ (see \Cref{app:idealgas}).  The vector-meson interaction term is
\begin{equation}
 \mathcal{L}_{\rm V,int}=-\sum_{i \, \in \, {\rm fermions}} \bar{\psi}_i\big[\gamma_0 \big(g_{i\omega}\omega+g_{i\rho}\rho+g_{i\phi}\phi\big)\big]\psi_i\,,
 \end{equation}
and it leads to an effective chemical potential $\mu_i^*$ for the fermions, given by
\begin{equation}
\mu_i^*=\mu_i-g_{\omega i}\omega-g_{\rho i}\rho-g_{\phi i}\phi\,.
\label{eq:fermion_effective_chemical_potential}
\end{equation}
The individual particle chemical potentials $\mu_i$ are given by
\begin{align}
\mu_i=B_i\mu_B+S_i\mu_S+Q_i\mu_Q\,,
\label{eq:chemical_potential}
\end{align}
where $B_i$, the particle baryon number, is $1$ for baryons and $1/3$ for quarks. See \Cref{tab:particles} in \Cref{appendixA} for values for $B$, $Q$, and $S$ for the baryons and quarks described within CMF.

Within our formalism, bosons do not acquire effective masses.
There is also no contribution from the kinetic term, resulting in the bosonic Lagrangian as
\begin{align}
\mathcal{L}_{\rm bosons}&=\mathcal{L}_{\rm mesons}\,,
\end{align}
where $\mathcal{L}_{\rm mesons}=\mathcal{L}_{\rm scal}+\mathcal{L}_{\rm vec}+\mathcal{L}_{\rm esb} $. The total energy density, pressure, vector or baryon (number) density, and scalar density include the sum of contributions from individual fermions
\begin{equation}
\begin{split}
\varepsilon_B=&\sum_{i \, \in \, {\rm fermions}}\varepsilon_i\,,\\
P_B=&\sum_{i \, \in \, {\rm fermions}}P_i\,,\\
n_{B,{\rm no}\,\Phi}=&\sum_{i \, \in \, {\rm fermions}}B_i n_i\,,\\
n_{sc}=&\sum_{i \, \in \, {\rm fermions}}n_{sc,i}
\,,
\label{eq:Fermisums}
\end{split}
\end{equation}
with the individual particle contributions being calculated from the energy momentum-tensor as shown in~\Cref{app:idealgas}.
Here, $n_{B,{\rm no}\,\Phi}$ is the number density of the fermions without the scalar field $\Phi$  contribution
and, therefore, is different from the baryon density defined as $n_B=dP/d\mu_B$ (containing a $\Phi$ contribution) and discussed in the following. 
$B_i$ ensures that each quark counts as $1/3$ of a baryon. 

At vanishing temperature, the entropy density is identically zero in our framework.
Then the thermodynamic variables can be calculated directly using:
\begin{equation}
\begin{split}
\varepsilon_i=\frac{\gamma_i}{2\pi^2}\Bigg[ &\left(\frac{1}{8}m_i^{\ast 2}k_{F_i}+\frac{1}{4}k_{F_i}^3\right)\mu^*_i 
-\frac{1}{8}m_i^{\ast 4}\ln{\frac{k_{F_i}+\mu^*_i}{m_i}}\Bigg]\,,
\label{eps0}
\end{split}
\end{equation}
\begin{equation}
\begin{split}
P_i=\frac{1}{3}\frac{\gamma_i}{2\pi^2}\Bigg[&\left(\frac{1}{4}k_{F_i}^3-\frac{3}{8}m_i^{\ast 2}k_{F_i}\right)\mu^*_i +\frac{3}{8}m_i^{\ast 4}\ln{\frac{k_{F_i}+\mu^*_i}{m_i}}\Bigg]\,,
\end{split}
\end{equation}
\label{eq:press0}
\begin{equation}
n_i=\frac{\gamma_i}{6\pi^2}k_{F_i}^3\,,
\label{eq:vector_density}
\end{equation}
\begin{equation}
\begin{split}
n_{sc,i}=&\dfrac{\gamma_i \,{m_i^*}}{4\pi^{2}} \bigg[k_{F_i}\,\mu^*_i -{m_i^*}^{2}\ln\left(\dfrac{k_{F_i}+\mu^*_i}{{m_i^*}}\right)\bigg]\,,
\label{eq:scalar_density}
\end{split}
\end{equation}
where $\gamma_i$ is the total degeneracy (spin and color),  $k_{F_i}$ is the Fermi momentum of particle $i$, and at $T=0$ we can also write the effective chemical potential as the effective energy level
\begin{equation}
\mu^*_i=E^*_i=\sqrt{k_{F_i}^2+m_i^{\ast 2}}\,.
\label{pppp}
\end{equation}
The asterisks represent the influence of the strong interaction. For a given set of $\mu_B$, $\mu_Q$\, and $\mu_S$, once one determines the effective particle chemical potentials \Cref{eq:fermion_effective_chemical_potential} and masses \Cref{eq:effective_mass_hadrons,eq:effective_mass_quarks} (solving for the mean fields), at $T=0$ the Fermi momenta and thermodynamical properties easily follow.

The baryon-vector meson interactions modify the solution of the Dirac equation, \Cref{eq:psi_solution}, by modifying its energy in the plane wave exponential as $E_i \rightarrow E_i^\ast + g_{\omega i}\omega+g_{\rho i}\rho+g_{\phi i}\phi$. 
Then, the derivative of the Dirac spinor in \Cref{eq:energy_and_pressure}, applied to ~\Cref{eq:psi_solution} leads to a contribution to the energy density as
\begin{equation}
\varepsilon_{\rm int} = \sum_{i \in {\rm fermions}} \left( g_{i \omega} \omega +  g_{i \phi} \phi  + g_{i\rho} \rho  \right) n_{i}.
\end{equation}
We note here that the terms of the form $\bar{\psi}_ig_{i\omega}\gamma_0\omega\psi_i$ only contribute to the energy density due to $\gamma_0\omega$ being a simplification of $\gamma_\mu\omega^\mu$. The pressure, on the other hand, does not receive extra contributions from the mean-field mesons, since their spatial components are taken to be zero (see \Cref{eq:energy_and_pressure}). 

The $\Phi$ field has explicit temperature and chemical potential dependence (see \Cref{eq:Polyakov_Potential_temperature}). This means that to satisfy thermodynamic consistency, we must have $\varepsilon_\Phi=-P_\Phi+\mu_B n_\Phi$ (with $\Phi$ having no electric charge or strangeness), with $P_\Phi=-U_\Phi$ and $n_\Phi=\frac{\partial P_\Phi}{\partial \mu_B}.$ 
For the mesonic contribution to the thermodynamic quantities, we start from (\Cref{Tfull}) and acknowledge that $\partial_\mu M=0$ for all mesons and $\Phi$ due to the mean-field approximation.
This gives the energy density and pressure as
\begin{align}
\varepsilon_{\rm mesons}=&-\mathcal{L}_{\rm mesons}\,,\\
P_{\rm mesons}=&\mathcal{L}_{\rm mesons}\,.
\end{align}

Furthermore, in a vacuum, all thermodynamic quantities should be zero. 
This is not the case for the scalar meson contribution, which acts as self-energy. 
However, within the no-sea approximation, their vacuum values are constant, and we can always add a constant to the Lagrangian density. 
Accordingly, we make a final alteration to the CMF Lagrangian density by subtracting the constant vacuum state. We do not have the same issue with fermions in our approach because their thermodynamic variables are already zero in the vacuum. 
The net result is
\begin{align}
\mathcal{L}_{\rm mesons}\rightarrow\mathcal{L}_{\rm mesons}-\mathcal{L}_{\rm vacuum}\,,
\end{align}
where
\begin{equation}
\begin{split}
\mathcal{L}_{\rm vacuum}= & -\frac{1}{2}k_0\chi_0^2\left(\sigma_0^2+\zeta_0^2\right)+k_1\left(\sigma_0^2+\zeta_0^2\right)^2 \\
&+k_2\left(\frac{\sigma_0^4}{2}+\zeta_0^4\right) +k_3\chi_0\sigma_0^2\zeta_0-k_4\chi_0^4\,,
\end{split}
\end{equation}
is the aforementioned constant vacuum state value, achieved by taking $\sigma\rightarrow\sigma_0$ and $\zeta\rightarrow\zeta_0$ with all other meson fields vanishing, keeping in mind that we already have $\chi=\chi_0$ and $\delta_0=0$. 

Thus, we can write the expressions for the thermodynamic quantities (including mesonic  contributions)
\begin{align}
\varepsilon_{\rm bosons}=&-\left(\mathcal{L}_{\rm{mesons}}-\mathcal{L}_{\rm vacuum}\right)\,,\nonumber\\
P_{\rm bosons}=&{(\mathcal{L}_{\rm mesons}-\mathcal{L}_{\rm vacuum})}\,.\nonumber\\
\label{eq:boson_energy_pressure_density}
\end{align}
Finally, we get the total contribution to the thermodynamic quantities by adding the fermion, boson, and $\Phi$ contributions all together
\begin{align}
\varepsilon=&\varepsilon_B+\varepsilon_{\rm int}+\varepsilon_{\rm bosons}{-P_\Phi}+\frac{\partial P_\Phi}{\partial\mu_B}\mu_B\,,\nonumber\\
P=&P_B+P_{\rm bosons}+P_{ \Phi}\,,\nonumber\\
n_B=&n_{B,{\rm no}\,\Phi}+n_{\Phi}\,,
\label{eq:net_energy_pressure_density}
\end{align}
where the second term on the RHS of the above equation, $n_{\Phi}=\frac{\partial P_\Phi}{\partial\mu_B}=-4a_1\mu_B^3 \Phi^2$ is calculated analytically and would represent some sort of gluonic interaction contributing to the baryon density. 

Throughout this paper, we often refer to fractions instead of densities. They are defined as ratios of sums of quantum numbers (weighted by the number density of particles)
\begin{equation}
Y_Q = \frac{Q}{B}= \frac{\sum_i Q_i  n_{i}}{\sum_i B_i  n_{i}}= \frac{\sum_i Q_i  n_{i}}{n_{B,{\rm no}\,\Phi}}\,.
    \label{eq:Y_Q}
\end{equation} 
for the charge fraction and
\begin{equation}
Y_S = \frac{S}{B}= \frac{\sum_i S_i  n_{i}}{\sum_i B_i  n_{i}}=\frac{\sum_i S_i  n_{i}}{n_{B,{\rm no}\,\Phi}}\,,
    \label{eq:Y_S}
\end{equation} 
for the strangeness fraction. Since the field $\Phi$ possesses no quantum number, it does not contribute to these fractions.

\subsection{Coupling constants}\label{sec:coupling}

\begin{table}[t!]
\centering
\begin{tabular}{|c|c|c|}
\hline  Parameter  & Interaction & Constraint \\
\hline
$g_1^X, g_8^X, \alpha_X$  &  $\mathcal{L}_{B X}$ & $m_N$, $ m_{\Lambda}$, $m_{\Sigma}$  \\
\hline
$g_D^X,\alpha_{DX}$  &  
\footnote{\label{note1}For the decuplet the correct approach is to use the Rarita-Schwinger equation, which can be written as a Dirac equation with extra constraints~\cite{dePaoli:2012eq}. Here, we simply follow the results presented in Ref.~\cite{Detlef:Thesis}.} 
& $m_\Delta$, $ m_{\Sigma^*}$, $m_{\Omega}$  \\
\hline
$k_0$ & &$\left.\frac{\partial \mathcal{L}_{\rm scal}}{\partial \sigma}\right|_{\text {vac }}=0$ \\
$k_1$ &  & $m_\sigma$ \\
$k_2$ & & $\left.\frac{\partial \mathcal{L}_{\rm scal}}{\partial \zeta}\right|_{\text {vac }}=0$ \\
$k_3$  &  & $\eta, \eta^{\prime}$ splitting \\
$k_4$ &  $\mathcal{L}_{\rm scal}$  & $\left.\frac{\partial \mathcal{L}_{\rm scal}}{\partial \chi}\right|_{\text {vac }}=0$ \\
$\epsilon$  & & one-loop $\beta_{\mathrm{QCD}}$ function \\
$\chi_0$ & &$P\left(n_{\rm sat}\right)=0$\\
$\sigma_0$ & &$f_\pi$\\
$\zeta_0$ & &$f_\pi, f_K$\\
\hline 
$a_0$ & & $T^d_c$  \\
$a_1$ & & $n^d_{B,c}$\\
$a_2$ & &$T_c$, $\mu_{B,c}$ \\
$a_3$ &$U_{\Phi}$ & $\Phi\in{0,1}$  \\
$T^{\rm pure glue}_0$ & & $T^d_c$, $\Phi\in{0,1}$ \\
$T^{\rm crossover}_0$ & & $T^p_c$, $\Phi\in{0,1}$ \\
$g_{q\Phi}$ & & $T^p_c$ \\
\hline
$g_8^V$, $\alpha_V$, $g_4,$& $\mathcal{L}_{B V}$, $\mathcal{L}_{\rm vec }^{\rm SI}$& $g_{N \phi}=0,$ $g_1^V=\sqrt{6} g_8^V $, $ n_{\rm sat}$, $B^{\rm sat} / A,$  \\
 $m_0$ &  & $E^{\rm sat}_{\rm sym}$, $L^{\rm sat}$, $K$,  $f$-$d$ mixing (VDM) \\
\hline
$g_D^V, \alpha_{DV}$&  \footnotemark[1] &  $g_8^V, g_{\Delta\phi}=0$  \\
\hline
$m_V$ & $\mathcal{L}_{\rm vec }^{m}$  & $m_\omega$, $m_\rho$, $m_\phi$ \\
\hline
$m^{HO}_3$  &$\mathcal{L}^{HO}_{\rm pot}$ & $U_\Lambda$ \\
\hline
$V_\Delta$   &\footnotemark[1]& $U_\Delta$\\
\hline
$m^{HD}_3$  & \footnotemark[1]& $U_{\Sigma^*}, U_{\Xi^*}, U_\Omega$ \\
\hline 
\end{tabular}
\caption{Constraints imposed to fix model parameters and their corresponding terms in the Lagrangian or potential.}
\label{tab:param_constraints}
\end{table}

In \Cref{tab:param_constraints}, we list the free parameters of the CMF model and the corresponding constraints used to fix them. 
Note that the CMF couplings are constant, e.g., are not dependent on quantities like density. 
In the first set of rows, we present the scalar coupling constants concerning the interaction between scalar mesons and baryon octet (decuplet), which are determined based on the vacuum masses of baryons (See \Cref{tab:mass_comparison} in \Cref{appendixA}). 
For the octet, the following coupling relations are obtained through \Cref{eq:effective_masses} in the vacuum 
\begin{equation}
g_1^X = \frac{\sqrt{6}}{2} \frac{m_\Lambda + m_{\Sigma} - 2 m_{0}}{2 \sigma_0 + \sqrt{2} \zeta_0}\,,
\end{equation}
\begin{equation}
\alpha_X = \frac{-\frac{3}{2}m_{\Lambda}-\frac{1}{2}m_{\Sigma}+2m_{N}}{m_{\Sigma}-3m_{\Lambda}+2m_{N}}\,,
\label{eq:withoutgns}
\end{equation}
\begin{equation}
g_8^X = 3\frac{\frac{1}{2}m_{\Lambda}+\frac{1}{2}m_{\Sigma}-m_{N}}{(4\alpha_{X}-1)(\sqrt{2}\zeta_{0}-\sigma_{0})}\,,
\end{equation}
and for the decuplet, the coupling constants are obtained through \Cref{eq:eff_mass_dec} in the vacuum
\begin{equation}
\alpha_{DX} = \frac{-m_\Omega \sigma_0 + m_\Delta \sqrt{2} \zeta_0}{m_{\Sigma^*}(-\sigma_0 + \sqrt{2} \zeta_0)}\,,
\end{equation}
\begin{equation}
g_D^X = \frac{m_\Delta}{(3 - \alpha_{DX}) \sigma_0 + \alpha_{DX} \sqrt{2} \zeta_0}\,.
\end{equation}

Additionally, parameters like $k_0$, $k_2$, and $k_4$ governing scalar self-interactions are adjusted to the Lagrangian minima for $\sigma$, $\zeta$, and $\chi$ in the vacuum, while $k_1$ and $k_3$ are tuned to match the vacuum masses of $\sigma$ (which is uncertain) and the $\eta$, $\eta^\prime$ splitting, respectively.
The parameter $\epsilon$, linked to scalar scale breaking Lagrangian, is calibrated to the one-loop QCD beta function. 
Moreover, the vacuum value of $\chi_0$ is set to reproduce zero pressure at saturation. 
The vacuum value of the scalar field $\sigma_0$ is fitted to the decay constant of the $\pi$ meson, while $\zeta_0$ is fitted to the decay constants of the $\pi$ and $K$ mesons. See \Cref{tab:scalar_param} for a complete list.

The vector-baryon coupling constants have been fitted to reproduce nuclear saturation properties for isospin-symmetric matter and asymmetric matter, together with neutron-star observations. 
This includes $g_8^V$, $g_4$, and the bare mass of baryons $m_0$ fitting simultaneously saturation density $n_{\rm sat}=0.15$~fm$^{-3}$ and binding energy per nucleon $B^{\rm sat}/A=-16$~MeV (which results in compressibility of $K^{\rm sat}=300$~MeV) and the symmetry energy at saturation $E^{\rm sat}_{\rm sym}=30$~MeV (by using $g_{N\rho}\ne g_{N\omega}/3$) producing a slope $L^{\rm sat}=88$~MeV) separately for all for the vector couplings (C1-C4). 

There is also a requirement to reproduce $\sim2$ M$_\odot$ stars with radii consistent with observations. Reproducing these values requires a setting of vector coupling constants given in \Cref{tab:vec_param}. The remaining baryon-vector-meson coupling constants relate to the value of $g_8^V$ associated with $g_{N\omega}$.
Non-strange particles do not couple to $\phi$ and $\zeta$. 
Finally, parameter $m_V$ relates to experimental vector meson vacuum masses. 

We fit $m^{HO}_3$ to reproduce reasonable hyperon potentials (\mbox{$U_B=m^*_B-m_B+g_{B \omega}+g_{B \phi}+g_{B \rho}$}) \cite{Wang:2001jw} for symmetric matter at saturation, in particular $U_\Lambda\sim-28$~MeV (reproducing $U_\Sigma\sim5$~MeV and $U_\Xi\sim-18$~MeV). We fit $V_\Delta$ to reproduce a reasonable $\Delta$ baryon potential for symmetric matter at saturation, $U_\Delta\sim-76$~MeV (similar to the nucleon one $\sim70$ MeV). 
This procedure is done separately for each of the couplings C1-C4. 
We use a fixed value for $m^{HD}_3$, since there is little data available for the strange members of the baryon decuplet. Additionally, a full list of constants shared among coupling schemes is provided in \Cref{tab:scalar_param}, and a list of constants that are different in different coupling schemes is provided in \Cref{tab:vec_param}.

\begin{table}[t!]
\centering
\def\arraystretch{1.8}
\begin{tabular}{ccc}
\hline
\hline
$g_{N\sigma}=-9.83$ &  $g_{N\zeta}=1.22$ &
$g_{\Lambda\sigma}=-5.52$ \\
$g_{\Lambda\zeta}=-2.30$ &
$g_{\Sigma\sigma}=-4.01$ & 
$g_{\Sigma\zeta}=-4.44$\\
$g_{\Xi\sigma}=-1.67$ & 
$g_{\Xi\zeta}=-7.75$ &
$g_{\Delta\sigma}=-10.87$ \\
$g_{\Delta\zeta}=-2.03$ &
$g_{\Sigma^*\sigma}=-6.44$ &
 $g_{\Sigma^*\zeta}=-4.55$ \\
$g_{\Xi^*\sigma}=-3.78$ & 
$g_{\Xi^*\zeta}=-8.32$ &
$g_{\Omega\sigma}=-0.23$ \\ 
$g_{\Omega\zeta}=-11.47$ &
$g_{p\delta}=-2.34$&
$g_{n\delta}=2.34$\\
$g_{\Lambda\delta}=0$ & $g_{\Sigma^+\delta}=-6.95$ & 
$g_{\Sigma^0\delta}=0$ \\ 
$g_{\Sigma^-\delta}=6.95$ & 
$g_{\Xi^0\delta}=-4.61$ & 
$g_{\Xi^-\delta}=4.61$ \\
$g_{\Delta\delta}=0$ & 
$g_{\Sigma^*\delta}=0$ &
$g_{\Xi^*\delta}=0$ \\ $g_{\Omega\delta}=0$ & 
$\sigma_0=-93.3$~MeV&
$\mathbf{m_\pi}=139$~MeV\\
$\mathbf{f_\pi}=93.3$~MeV&
$\zeta_0=-106.56$~MeV&
$\mathbf{m_K}=498$~MeV\\
 $\mathbf{f_K}=122$~MeV&
$\mathbf{k_0}=2.37$ & $\mathbf{k_1}=1.4$ \\ $\mathbf{k_2}=-5.55$&
$\mathbf{k_3}=-2.65$ & $\mathbf{\chi_0}=401.93$~MeV \\
$\mathbf{ m^{HD}_{3}}=1.25$ &  $\delta_0 = 0$  &
$\mathbf{\epsilon}=0.060606$  \\
\hline
\hline
\end{tabular}
\caption{A table of constants shared among the couplings schemes used in the CMF model. Only some of these are independent. The variables in bold can be freely changed by the user of the \texttt{CMF++} code.}
\label{tab:scalar_param}
\end{table}

\begin{table}[t!]
\centering
\begin{tabular}{cccccccc}
\hline \hline
 Coupling&$ \mathbf{g_4}$& $ \mathbf{g_{N\omega}}$  &   $\mathbf{g_{p\rho}}$ &   $\mathbf{g_{n\rho}}$ &$ \mathbf{m_{0}}$ &$ \mathbf{m^{HO}_3}$& $\mathbf{V_\Delta}$\\ \hline
C1&58.40 & 13.66 & 11.06 & -11.06 & 0 & 1.24&1.07 \\ 
C2&58.40 & 13.66 & 3.51 & -3.51  & 0 & 1.24&1.07 \\ 
C3&58.40 & 13.66 & 3.82 & -3.82 & 0 & 1.24&1.07 \\ 
C4&38.90 & 11.90 & 4.03  & -4.03 & 150 & 0.86&1.2 \\
  \hline \hline
\end{tabular}
\caption{Table of constants that are different among the coupling schemes of CMF. These include the nucleon-vector coupling constants and the bare mass contributions for the different vector self-interaction terms~\cite{Dexheimer:2015qha}. The variables in bold can be freely changed by the user of the \texttt{CMF++} code.}
\label{tab:vec_param}
\end{table}

\begin{table}[t!]
\centering
\begin{tabular}{lcc}
\hline \hline
$\mathbf{g_{q \omega}}=0$ & $\mathbf{g_{q \phi}}=0$ & $\mathbf{g_{q \rho}}=0$ \\
$\mathbf{g_{u \sigma}}=-3.00$ & $\mathbf{g_{u \delta}}=0$ & $\mathbf{g_{u \zeta}}=0$ \\
$\mathbf{g_{d \sigma}}=-3.00$ & $\mathbf{g_{d \delta}}=0$ & $\mathbf{g_{d \zeta}}=0$ \\
$\mathbf{g_{s \sigma}}=0$ & $\mathbf{g_{s \delta}}=0$ & $\mathbf{g_{s \zeta}}=-3.00$ \\
$\mathbf{m^u_0}=5$ MeV &  $\mathbf{m^d_0}=5$ MeV&$\mathbf{m^s_0}=150$ MeV  \\
$\mathbf{a_1}=-1.443 \times 10^{-3}$ & $\mathbf{a_3}=-0.396$ &  $\mathbf{g_{B \Phi}}=1500~\mathrm{MeV}$  \\
$\mathbf{g_{q \Phi}}=500~\mathrm{MeV}$ &
$\mathbf{T_0^{\rm crossover}}$=200~MeV &
$\mathbf{T_0^{\rm pureglue}}=270$~MeV \\
\hline \hline
\end{tabular}
\caption{Parameters and coupling constants for the quark sector in the CMF model with the C4 coupling scheme~\cite{Dexheimer:2009hi}, where `$q$' stands for $u$, $d$, and $s$ quarks.  Variables in bold can be freely changed by the user of the \texttt{CMF++} code. }
\label{tab:quark_param}
\end{table}

Following this, we detail the parameters related to the deconfinement potential $U_\Phi$ (not including the decuplet) and quarks.
For the C4 coupling scheme, the quark and $\Phi$ coupling constants (listed in \Cref{tab:quark_param}) have been fitted to reproduce lattice results at zero and small chemical potential and known physics of the phase diagram. Lattice QCD predicts the first-order deconfinement phase transition (for pure glue Yang-Mills) observed at a temperature of $T^d_c=270$ MeV~\cite{Roessner:2006xn}. At $\mu_B=0$, we fit the parameter $a_0$ and $T^{\rm pureglue}_0$ together to $T^d_c$ as well as the pressure function $P(T)$ which mirrors patterns seen in previous works (\cite{Ratti:2005jh,Roessner:2006xn}) for pure glue Yang-Mills theories.
At vanishing chemical potential, when including fermions, the hadron to quark phase change is a crossover rather than a sharp transition, as well as chiral symmetry restoration. The mid value of the crossover band is known as the pseudo-critical temperature of chiral symmetry restoration, marked by a transition temperature  $T^p_c$. In the CMF model, this temperature is identified through the peak change in the condensate $\sigma$ and field $\Phi$. The parameters $T^{\rm crossover}_0$ and $g^q_\Phi$ (coupling between quarks and $\Phi$) are fitted together to reproduce the chiral crossover temperature $T^p_c$=171 MeV in agreement with results from 2001~\cite{Fodor:2001pe}. Note that, in the absence of our deconfinement mechanism, previous investigations within the CMF always predicted a crossover for the chiral symmetry restoration within the studied range of $\mu_B$ and $T$, while fixing $\mu_Q$ and $\mu_S$ \cite{Dexheimer:2008ax}. We come back to this discussion  in \Cref{sec:results}.
 
Furthermore, $a_1$ is fitted to the critical number density ($n^d_{B,c}=4$ $n_{\rm sat}$) at the onset of deconfinement transition at $T=0$ for neutron stars, and $a_2$ is constrained by the critical temperature ($T_c$=167 MeV) and critical baryon chemical potential ($\mu_{B,c}$=354 MeV) for isospin symmetric matter, aligned with findings from 2004~\cite{Fodor:2004nz}. Modifying $a_2$, one could easily reproduce a different critical point (without modifying $\mu_B=0$ or $T=0$ results).
Additionally, $a_3$ is tuned to maintain $\Phi$ value within 0 and 1. 
It is noteworthy that parameters from the Polyakov-inspired potential (\Cref{eq:Polyakov_Potential}) and quark couplings (\Cref{eq:effective_mass_quarks}) have been fitted solely for C4, determining the location of the deconfinement phase transition at specific $\mu_B$ and EoS behavior in the quark regime. Since the paper aims to compare \texttt{C++} and \texttt{Fortran} solutions while also analyzing stability, we employ the quark sector parameters of C4 (refer to \Cref{tab:quark_param}) for all other coupling schemes. Adjusting the $\Phi$ parameters to C1-C3 coupling schemes would lead to shifts in the location of the deconfinement phase transition as well as the behavior of  EoS post-deconfinement transition. See Ref.~\cite{Kumar:2024owe} for a recent work in which we broke the mass degeneracy of vector mesons in the CMF model using their field redefinition. This required us to fit the C1-C4 coupling schemes to the up-to-date constraints coming from lattice QCD, low-energy nuclear, and astrophysics.

\section{Code Implementation}\label{sec:CodeImplementation}

\subsection{Code Overview}

\begin{figure*}
    \centering
    \includegraphics[scale=0.55]{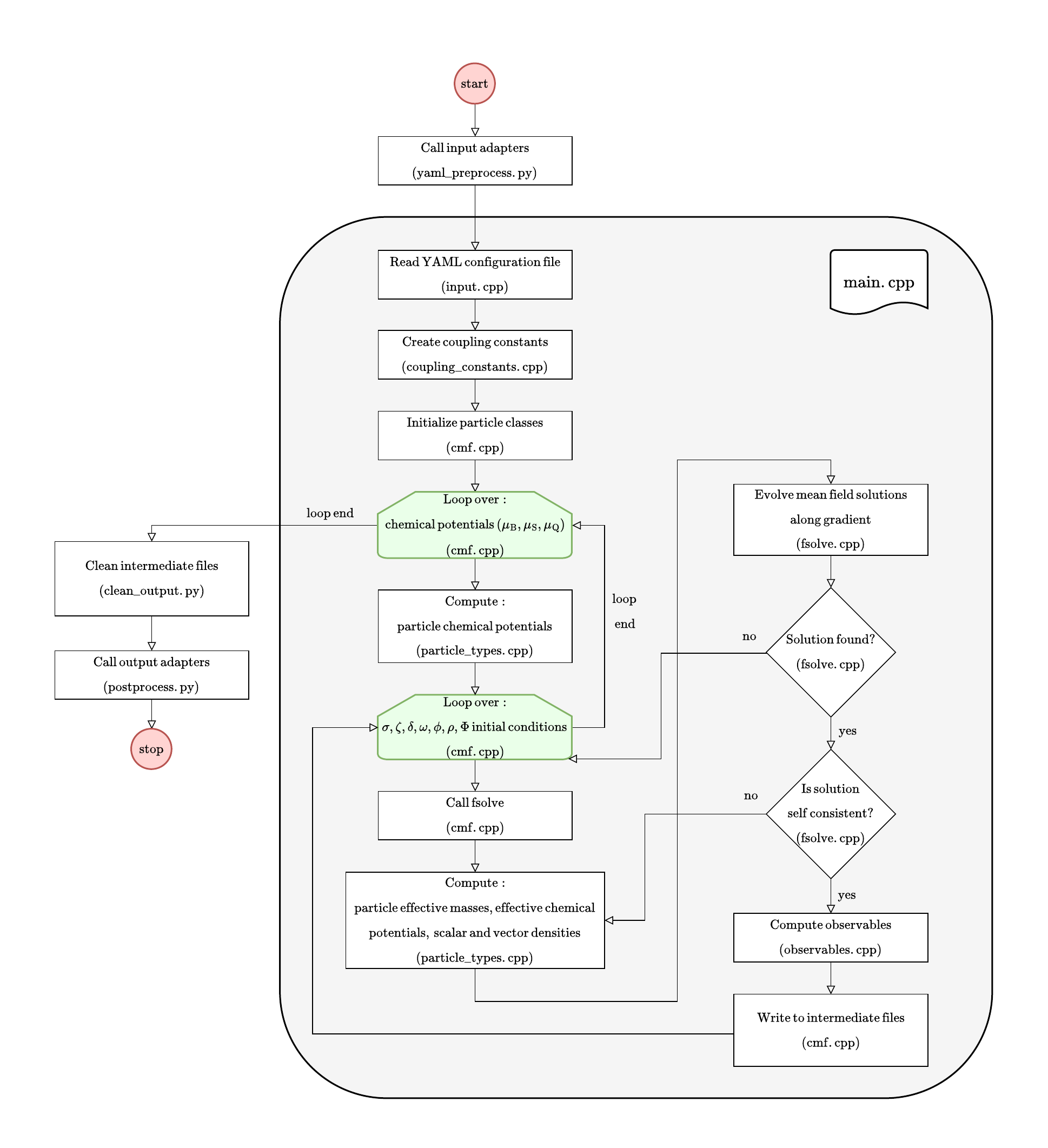}
    \caption{\texttt{CMF++} general algorithm flowchart detailing the procedures inside \texttt{Algorithm} (gray enclosed section).}
    \label{fig:General_algorithm}
\end{figure*}

\Cref{fig:General_algorithm} shows the \texttt{CMF++} code flowchart with the \texttt{Python} and \texttt{C++} layers, where the shaded gray region highlights the \texttt{C++} main driver routine. The code can be divided into three sections: input preprocessing, main algorithm, and output postprocessing. In the first section, \texttt{yaml\_preprocess.py} validates the YAML input configuration file required for the main execution. Details about the validation procedure are illustrated in \Cref{sec:Input_preprocessing}. In the second section, the main routine is responsible for finding solutions for \Cref{eq:Algebraic_System} and computing derived thermodynamic quantities (see \Cref{eq:net_energy_pressure_density}) for the valid solutions found. More details about this section are described in  \Cref{sec:C++Implementation}. The last section covers postprocessing and output. In the postprocessing section, the solutions found in the main algorithm are cleaned and classified as stable, metastable, or unstable. Additionally, the output is divided by the underlying degrees of freedom, i.e., quarks vs. baryons. The criteria and procedure for separating the solutions are detailed in  \Cref{sec:stability}. Finally, the output adapters are called in \texttt{postprocess.py}, which transforms the final output files into either CSV or HDF5 format via the MUSES Porter library for the consumption of other MUSES modules.
Details on how to run \texttt{CMF++} can be found in \Cref{sec:Run_Software}.

\subsection{Input preprocessing}
\label{sec:Input_preprocessing}

The only input required to execute the code is a YAML-formatted configuration file to ensure human and machine readability, which we named \texttt{config.yaml}. The YAML file contains all the computational options and the physical parameters required to run. The computational options detailed in \Cref{tab:input_parameters_computational} encompass the model hyperparameters like the name of the run, 
see \Cref{tab:input_parameters_physical}. The file structure is detailed in the OpenAPI specifications for the model version. 

The \texttt{config.yaml} file is processed by \texttt{yaml\_preprocess.py} which validates it via the \texttt{openapi-core} library and flattens it for the ingestion of the main algorithm. 

\begin{table*}[t!]
\centering
\begin{tabular}{cccp{8cm}}
\hline
\hline
\multirow{2}{*}{\textbf{Category}} & \multirow{2}{*}{\textbf{Variable}} & \multirow{2}{*}{\textbf{Value}} & \multirow{2}{*}{\textbf{Description}} \\
& & & \\
\hline
\multirow{5}{*}{computational\_parameters} & run\_name & default & name of the run \\
& solution\_resolution & 1.e-15 & resolution for mean-field solutions \\
& maximum\_for\_residues & 1.e-4 & threshold for solution residues \\
& production\_run & true & Is this a production run? \\
\hline
\multirow{15}{*}{options} & baryon\_mass\_coupling & 1 & baryon-meson coupling scheme \\
& use\_ideal\_gas & false & use ideal gas? \\
& use\_quarks & true & use quarks? \\
& use\_octet & true & use baryon octet? \\
& use\_decuplet & true & use baryon decuplet? \\
& use\_pure\_glue & false & use gluons only (no baryons nor quarks)? \\
& use\_hyperons & true & are hyperons included? \\
& use\_constant\_sigma\_mean\_field & false & fix sigma mean-field to chosen value \\
& use\_delta\_mean\_field & true & is delta mean-field included? \\
& use\_Phi\_order & true & use Polyakov-inspired potential? \\
& use\_constant\_Phi\_order & false & fix Phi field  value to chosen value \\
& vector\_potential & 4 & vector coupling scheme C1-C4 \\
& use\_default\_vector\_couplings & true & use default vector couplings? \\
\hline
\multirow{6}{*}{output\_files} & output\_Lepton & true & create output file for Lepton module \\
& output\_debug & false & create output file for debugging \\
& output\_flavor\_equilibration & true & create output file for Flavor equilibration module \\
& output\_format & CSV & create output files either in CSV or HDF5 format \\
& output\_particle\_properties & true & create output file for particle populations and properties \\
\hline
\multirow{11}{*}{chemical\_optical\_potentials} & muB\_begin & 900.0 & initial baryon chemical potential (MeV) \\
& muB\_end & 1800.0 & final baryon chemical potential (MeV) \\
& muB\_step & 1.0 & step for baryon chemical potential (MeV) \\
& muS\_begin & 0.0 & initial strange chemical potential (MeV) \\
& muS\_end & 1.0 & final strange chemical potential (MeV) \\
& muS\_step & 5.0 & step for strange chemical potential (MeV) \\
& muQ\_begin & 0.0 & initial charge chemical potential (MeV) \\
& muQ\_end & 1.0 & final charge chemical potential (MeV) \\
& muQ\_step & 5.0 & step for charge chemical potential (MeV) \\
\hline
\multirow{25}{*}{mean\_fields\_and\_Phi\_field} & sigma0\_begin & -100.0 & initial $\sigma$ mean-field (MeV) \\
& sigma0\_end & -10.0 & final $\sigma$ mean-field (MeV) \\
& sigma0\_step & 30.0 & step for $\sigma$ mean-field (MeV) \\
& zeta0\_begin & -110.0 & initial $\zeta$ mean-field (MeV) \\
& zeta0\_end & -40.0 & final $\zeta$ mean-field (MeV) \\
& zeta0\_step & 23.333 & step for $\zeta$ mean-field (MeV) \\
& delta0\_begin & 0.0 & initial $\delta$ mean-field (MeV) \\
& delta0\_end & 1.0 & final $\delta$ mean-field (MeV) \\
& delta0\_step & 10.0 & step for $\delta$ mean-field (MeV) \\
& omega0\_begin & 0.0 & initial $\omega$ mean-field (MeV) \\
& omega0\_end & 100.0 & final $\omega$ mean-field (MeV) \\
& omega0\_step & 33.333 & step for $\omega$ mean-field (MeV) \\
& phi0\_begin & -40.0 & initial $\phi$ mean-field (MeV) \\
& phi0\_end & 0.0 & final $\phi$ mean-field (MeV) \\
& phi0\_step & 13.333 & step for $\phi$ mean-field (MeV) \\
& rho0\_begin & 0.0 & initial $\rho$ mean-field (MeV) \\
& rho0\_end & 1.0 & final $\rho$ mean-field (MeV) \\
& rho0\_step & 10.0 & step for$\rho$ mean-field (MeV) \\
& Phi0\_begin & 0.0 & initial $\Phi$ mean-field (MeV) \\
& Phi0\_end & 0.9999 & final $\Phi$ mean-field (MeV) \\
& Phi0\_step & 0.333 & step for $\Phi$ mean-field (MeV) \\
\hline
\hline
\end{tabular}
\caption{\textit{config.yaml} Computational parameters and descriptions.}
\label{tab:input_parameters_computational}
\end{table*}

\squeezetable
\begin{table*}[t!]
\centering
\begin{tabular}{cccl}
\hline
\hline
\multirow{2}{*}{\textbf{Category}} & \multirow{2}{*}{\textbf{Variable} (Symbol)} & \multirow{2}{*}{\textbf{Value}} & \multirow{2}{*}{\textbf{Description}} \\
& & & \\
\hline
\multirow{6}{*}{physical\_parameters} & d\_betaQCD ($\epsilon$) & 0.0606060606 & fit parameter for beta QCD function\\
& f\_K ($f_K$) & 122.0 & $K$ decay constant (MeV) \\
& f\_pi ($f_\pi$) & 93.3000031 & $\pi$ decay constant (MeV) \\
& hbarc ($\hbar c$) & 197.3269804 & $\hbar c$ (MeV) \\
& chi\_field\_vacuum\_value ($\chi_0$) & 401.933763 & $\chi$ vacuum value (MeV) \\
\hline
\multirow{5}{*}{Phi\_order\_optical\_potential} & a\_1 ($a_1$) & -0.001443 & fit parameter for deconfinement phase transition  \\
& a\_3 ($a_3$) & -0.396 & fit parameter to keep $\Phi$ between 0 and 1 \\
& T0 (crossover) ($T^{\rm crossover}_0$) & 200 &fit parameter for pseudo critical transition temperature (MeV)\\
& T0 (pureglue) ($T^{\rm pureglue}_0$)  & 270 & fit parameter for deconfinement critical temperature (MeV) \\
\hline
\multirow{5}{*}{scalar\_mean\_field\_equation} & k\_0  ($k_0$) & 2.37321880 & fit parameter to minimize scalar Lagrangian with respect to $\sigma$\\
& k\_1 ($k_1$) & 1.39999998 & fit parameter for mass of $\sigma$ meson \\
& k\_2 ($k_2$) & -5.54911336 & fit parameter to minimize scalar Lagrangian with respect to $\zeta$\\
& k\_3 ($k_3$) & -2.65241888 & fit parameter to account $\eta-\eta^\prime$ splitting \\
\hline
\multirow{4}{*}{explicit\_symmetry\_breaking} & m\_3HO ($m^{HO}_3$) & 0.85914584 & fit parameter for potential of strange octet baryons \\
& m\_3HD  ($m^{HD}_3$) & 1.25 & fit parameter for  potential of strange decuplet baryons  \\
& V\_Delta ($V_\Delta$) & 1.2 & fit parameter for potential of decuplet $\Delta$ particles \\
\hline
\multirow{4}{*}{vector\_nucleon\_couplings} & gN\_omega ($g_{N\omega}$) & 11.90 & Nucleon coupling to $\omega$ field\\
& gN\_rho ($g_{N\rho}$) & 4.03 & Nucleon coupling to $\rho$ field\\
& g\_4 ($g_{4}$) & 38.90 & Self-coupling of the vector mesons \\
\hline
\multirow{4}{*}{mean\_field\_vacuum\_masses} & omega\_mean\_field\_vacuum\_mass ($m_\omega$) & 780.562988 & $\omega$ mean-field vacuum mass (MeV)\\
& phi\_mean\_field\_vacuum\_mass ($m_\phi$) & 1019. & $\phi$ mean-field vacuum mass (MeV)\\
& rho\_mean\_field\_vacuum\_mass ($m_\rho$) & 761.062988 & $\rho$ mean-field vacuum mass (MeV)\\
\hline
\multirow{4}{*}{quark\_bare\_masses} & up\_quark\_bare\_mass  ($m^u_0$) & 5.0 & up quark bare mass (MeV)\\
& down\_quark\_bare\_mass ($m^d_0$) & 5.0 & down quark bare mass (MeV)\\
& strange\_quark\_bare\_mass ($m^s_0$) & 150.0 & strange quark bare mass (MeV)\\
\hline
\multirow{12}{*}{vacuum\_masses} & Delta\_vacuum\_mass ($m_\Delta$) & 1232. & $\Delta$ vacuum mass (MeV)\\
& Lambda\_vacuum\_mass ($m_\Lambda$) & 1115. & $\Lambda$ vacuum mass (MeV)\\
& Sigma\_vacuum\_mass ($m_\Sigma$) & 1202. & $\Sigma$ vacuum mass (MeV)\\
& Sigma\_star\_vacuum\_mass ($m_\Sigma{^*}$) & 1385. & $\Sigma^*$ vacuum mass (MeV)\\
& Omega\_vacuum\_mass ($m_\Omega$) & 1691. & $\Omega$ vacuum mass (MeV)\\
& Kaon\_vacuum\_mass ($m_K$) & 498. & $K$ vacuum mass (MeV)\\
& Nucleon\_vacuum\_mass ($m_N$) & 937.242981 & Nucleon vacuum mass (MeV)\\
& Pion\_vacuum\_mass ($m_\pi$) & 139. & $\pi$ vacuum mass (MeV)\\
& mass0 ($m_0$) & 150. &  Bare vacuum mass (MeV) \\
\hline
\multirow{22}{*}{quark\_to\_fields\_couplings} & gu\_sigma ($g_{u \sigma}$) & -3.0 & up quark coupling for $\sigma$ mean-field \\
& gd\_sigma ($g_{d \sigma}$) & -3.0 & down quark coupling for $\sigma$ mean-field \\
& gs\_sigma ($g_{s \sigma}$) & 0 & strange quark coupling for $\sigma$ mean-field \\
& gu\_zeta ($g_{u \zeta}$) & 0 & up quark coupling for $\zeta$ mean-field \\
& gd\_zeta ($g_{d \zeta}$) & 0 & down quark coupling for $\zeta$ mean-field \\
& gs\_zeta ($g_{s \zeta}$) & -3.0 & strange quark coupling for $\zeta$ mean-field \\
& gu\_delta ($g_{u \delta}$) & 0.0 & up quark coupling for $\delta$ mean-field \\
& gd\_delta ($g_{d \delta}$) & 0.0 & down quark coupling for $\delta$ mean-field \\
& gs\_delta ($g_{s \delta}$) & 0.0 & strange quark coupling for $\delta$ mean-field \\
& gu\_omega ($g_{u \omega}$) & 0.0 & up quark coupling for $\omega$ mean-field \\
& gd\_omega ($g_{d \omega}$) & 0.0 & down quark coupling for $\omega$ mean-field \\
& gs\_omega ($g_{s \omega}$) & 0.0 & strange quark coupling for $\omega$ mean-field \\
& gu\_phi ($g_{u \phi}$) & 0.0 & up quark coupling for $\phi$ mean-field \\
& gd\_phi ($g_{d \phi}$) & 0.0 & down quark coupling for $\phi$ mean-field \\
& gs\_phi ($g_{s \phi}$) & 0.0 & strange quark coupling for $\phi$ mean-field \\
& gu\_rho ($g_{u \rho}$) & 0.0 & up quark coupling for $\rho$ mean-field \\
& gd\_rho ($g_{d \rho}$) & 0.0 & down quark coupling for $\rho$ mean-field \\
& gs\_rho ($g_{s \rho}$) & 0.0 & strange quark coupling for $\rho$ mean-field \\
& gq\_Phi ($g_{q \Phi}$) & 500.0 & quark coupling for $\Phi$ field (MeV)\\
\hline
baryon\_to\_Phi\_field\_coupling & gbar\_Phi ($g_{B \Phi}$) & 1500.0 & baryon coupling to $\Phi$ field (MeV)\\
\hline
\hline
\end{tabular}
\caption{Default \textit{config.yaml} physical parameters and descriptions related to the C4 coupling scheme.}
\label{tab:input_parameters_physical}
\end{table*}

\subsection{\texttt{Algorithm}}
\label{sec:C++Implementation}

In computational terms, the CMF model is a coupled system of nonlinear algebraic equations for the mean-field mesons $\sigma,\,\delta,\,\zeta,\,\omega,\,\rho,\,\phi$, and  $\Phi$ field (see \Cref{eq:Algebraic_System}), therefore, a root solver algorithm is required. 
In our implementation, we adopted the numerical root solver \texttt{fsolve} \cite{Burkardt:fsolve}, which is inspired by the \texttt{fsolve} function from MATLAB and is based on \texttt{MINPACK} \cite{More:1980, More:1984}. \texttt{MINPACK} is a \texttt{Fortran} library designed to solve systems of nonlinear equations by residual's least-squares minimization employing a pseudo-Gauss-Newton algorithm in conjunction with gradient descent. 

This validated \texttt{config.yaml} file is read by the \texttt{C++} layer via an input class and stored within a structure. The coupling constants for each particle respective to every mean-field (\Cref{tab:scalar_couplings,tab:vector_couplings}) are computed.
The different particle classes (quarks, baryons from octet, and/or decuplet) are initialized and filled with their quantum numbers read from the PDG table 2021+ \cite{SanMartin:2023zhv} and the couplings just computed.

The code loops over desired $\mu_B$, $\mu_S$, $\mu_Q$, so the chemical potential per particle is computed via \Cref{eq:chemical_potential}, then loops over every mean-field initial guesses ($\sigma$, $\zeta$, $\delta$, $\omega$, $\phi$, $\rho$, $\Phi$) follow. 
The \texttt{fsolve} routine is called where these initial values, in conjunction with the input parameters provided by the user, are used to compute the right hand side (RHS) of \Cref{eq:Algebraic_System}. To compute the left hand side (LHS) of \Cref{eq:Algebraic_System}, the scalar \Cref{eq:scalar_density} and vector \Cref{eq:vector_density} densities must be obtained for each particle involved, which implies the calculation of the effective chemical potential via \Cref{eq:fermion_effective_chemical_potential}, the effective masses (see \Cref{eq:effective_mass_hadrons} for hadrons  and see \Cref{eq:effective_mass_quarks} for quarks), and the Fermi momentum. 

The \texttt{fsolve} routine then computes the gradient for every field equation involved and updates the mean fields and $\Phi$ field to an improved guess that minimizes the difference in RHS and LHS of \Cref{eq:Algebraic_System}. 
If the new solution for the fields lies outside of the domains, then the code skips to the next initial conditions guess. If the solution found is not self-consistent (LHS not equal to RHS), recompute the effective masses, chemical potentials, and scalar and vector densities using the new guesses and evolve the field solutions along the gradient.

The previous procedure is performed until self-consistency is achieved, which means that LHS is equal to RHS within a certain threshold and that the solution has not been achieved before. 
Given that a valid solution has been found, the code now proceeds to compute a collection of thermodynamic observables like pressure, energy density, density (see \Cref{eq:net_energy_pressure_density}), and other relevant quantities like strangeness density, charge density, density without $\Phi$, and densities per particle sector (quarks, baryon octet, baryon decuplet). This data is written to an intermediate file, and the \texttt{C++} layer continues its execution into the next field's initial condition guess. 

Once all the $\mu_B, \mu_S, \mu_Q$ domains of interest have been exhausted, the main algorithm execution finishes. 

\subsection{Stability and phase transition criteria}
\label{sec:stability}

Let us begin the discussion by defining susceptibilities of the pressure:
\begin{equation}
    \chi_{ijk}^{{BSQ}}=\frac{\partial^{i+j+k}P}{(\partial\mu_B)^i(\partial\mu_S)^j(\partial\mu_Q)^k}\bigg|_{T}\,,
    \label{72}
\end{equation}
where whatever chemical potential is not being varied is kept constant as well. Due to the symmetries in QCD, the ordering of the derivatives does not matter, i.e. 
\begin{equation}\label{eqn:chi_symmetry}
    \chi_{ij}^{xy}=\chi_{ji}^{yx}\,,
\end{equation}
where $x$ and $y$ are any $B,S,Q$ combinations.
The first susceptibilities relate to the respective density of each conserved charge i.e.
\begin{equation}
\chi_1^B=n_B\,,\quad\chi_1^S=n_S\,,\quad \chi_1^Q=n_Q\,.
\end{equation}
Additionally, the second-order susceptibilities are then equivalent to 
\begin{align}
    \chi_2^B&=\frac{\partial n_B}{\partial \mu_B}\Big|_{T,\mu_S,\mu_Q}\label{75}\,,\\
    \chi_2^S&=\frac{\partial n_S}{\partial \mu_S}\Big|_{T,\mu_B,\mu_Q}\,,\\
    \chi_2^Q&=\frac{\partial n_Q}{\partial \mu_Q}\Big|_{T,\mu_B,\mu_Q}\,,
\end{align}
which have been shown to have interesting connections to the speed of sound in neutron stars \cite{McLerran:2018hbz}.
The susceptibilities are also important to provide connections to the search for the QCD critical point at finite $T$ and understanding the deconfinement phase transition \cite{Stephanov:2008qz,Stephanov:2011pb,Karsch:2012wm,Critelli:2017oub,Borsanyi:2018grb,Pradeep:2019ccv,Bazavov:2020bjn,Mroczek:2020rpm,Dore:2022qyz,Borsanyi:2023wno}.

A first-order phase transition is defined as a jump in $\chi_1^X$ at a specific $\mu_X$. 
Higher-order phase transitions appear as jumps in the higher-order susceptibilities. 
Thus, an $i^{th}$-order phase transition occurs at the point $\mu_X$ if $\chi_i^X(\mu_X)$ diverges.
When an $i^{th}$-order phase transition occurs then all higher order derivatives also diverge i.e. $\chi_l^X(\mu_X)$ diverges where $l>i$ at $\mu_X$.
However, we only determine the order of the phase transition by the first derivative where either a jump or divergence occurs.

In the grand canonical ensemble, in the infinite volume limit,  stability corresponds to minimizing the grand potential density or maximizing the pressure (see \Cref{Ensembles}). For this case, and assuming BSQ conserved charges, the 4-dimensional Hessian matrix is shown in \Cref{Ensembles5}. In the following, we show results only for $T=0$. In this case, the Hessian matrix is 3D:
\begin{equation}\label{eqn:M_matrix}
    M=\begin{bmatrix}
    \chi_2^B & \chi_{11}^{BS} & \chi_{11}^{BQ} \\
     \chi_{11}^{SB} & \chi_2^S & \chi_{11}^{SQ} \\
     \chi_{11}^{QB} & \chi_{11}^{QS}  & \chi_2^Q
    \end{bmatrix}\,,
\end{equation}
where the matrix is symmetric due to \Cref{eqn:chi_symmetry}.
Then, the determinant of each submatrix must be zero or positive. 
Thus, for the $1\times 1$ matrix 
\begin{equation}
    \det\left[M_{1\times1}\right]= \chi_2^B\geq 0\,,
\end{equation}
and for the $2\times 2$ matrix 
\begin{align}
    \det\left[M_{2\times2}\right]&= \chi_2^B\chi_2^S-\left(\chi_{11}^{BS}\right)^2\geq 0\,,\\
    \chi_2^B\chi_2^S&\geq \left(\chi_{11}^{BS}\right)^2\,.\label{eqn:22mat}
\end{align}
Using \Cref{eqn:22mat}, then it also implies that $\chi_2^S\geq0$ because $\chi_{11}^{BS}$ is real.
Finally, the $3\times 3$ matrix gives the condition that we show later on in \Cref{eqn:3by3mat}.

The matrix defined in \Cref{eqn:M_matrix} was somewhat arbitrarily built in that one could also have ordered it as SQB or QBS (or any other ordering). 
Thus, when considering all perturbations of the matrix we then arrive at the following independent conditions:
\begin{align}
    \chi_2^B\geq 0\,, \quad \chi_2^S&\geq 0\,, \quad \chi_2^Q\geq 0\,,\label{eqn:11mat_all}\\
\chi_2^B\chi_2^S&\geq \left(\chi_{11}^{BS}\right)^2\,, \label{eqn:chiBS_con} \\
\chi_2^S\chi_2^Q&\geq \left(\chi_{11}^{SQ}\right)^2\,, \label{eqn:chiSQ_con}\\
\chi_2^B\chi_2^Q&\geq \left(\chi_{11}^{BQ}\right)^2\,, \label{eqn:chiBQ_con}
\end{align}
\begin{align}
\chi_2^B\chi_2^S\chi_2^Q+2\left(\chi_{11}^{BS}\chi_{11}^{BQ}\chi_{11}^{SQ}\right)\geq \nonumber\\
\chi_2^B\left(\chi_{11}^{SQ}\right)^2+\chi_2^S\left(\chi_{11}^{BQ}\right)^2+\chi_2^Q\left(\chi_{11}^{BS}\right)^2\,.
\label{eqn:3by3mat}
\end{align}
The susceptibilities can be thought of as moments of the net-BSQ distributions (again recalling that the first moment implies the respective BSQ charge densities). 
Then, \Cref{eqn:11mat_all} implies that the variance of each net-BSQ distribution is positive and \Cref{eqn:chiBS_con,eqn:chiSQ_con,eqn:chiBQ_con} imply that the covariances must also be semi-negative definite. 
Finally, we note that the matrix in \Cref{eqn:M_matrix} becomes more complicated at finite temperatures. 
However, we leave finite $T$ studies to future work.

For multiple solutions of the EoS, if more than one solution obeys the stability criteria, then the one with the highest pressure (or lowest grand potential density) at fixed a $\vec{\mu}$ value is denoted as the stable EoS. 
The other EoS' that obey \Cref{eqn:11mat_all,eqn:chiBS_con,eqn:chiSQ_con,eqn:chiBQ_con} are called metastable~\footnote{If we have two mixtures, I and II, I is stable and II is metastable if $P^I(\mu_B) > P^{II}(\mu_B)$ and both satisfy \Cref{eqn:11mat_all,eqn:chiBS_con,eqn:chiSQ_con,eqn:chiBQ_con}.}. 
Additionally, if the pressure of an EoS is negative, but it obeys the stability criteria, it is also considered metastable. 
If there is a unique EoS with $P<0$ that obeys \Cref{eqn:11mat_all,eqn:chiBS_con,eqn:chiSQ_con,eqn:chiBQ_con}, then vacuum solutions are considered stable. 
We summarize our stability criteria in \Cref{tab:stability_criteria}. 

\begin{table*}[t!]
    \centering
    \begin{tabular}{c|c|c}
      stability label   &  thermodynamic criteria & multiple solution criteria for phase $i$\\
      \hline
      stable   & $P>0\  \land $ \Cref{eqn:11mat_all,eqn:chiBS_con,eqn:chiSQ_con,eqn:chiBQ_con,eqn:3by3mat} &  single solution $\lor\; P_i>P_{j\neq i}\forall j$\\
      metastable   & \Cref{eqn:11mat_all,eqn:chiBS_con,eqn:chiSQ_con,eqn:chiBQ_con,eqn:3by3mat} & $ \exists  P_{j\neq i} > P_i \lor P< 0$ \\
      unstable   & At least 1 fails: \Cref{eqn:11mat_all,eqn:chiBS_con,eqn:chiSQ_con,eqn:chiBQ_con,eqn:3by3mat} & \\
      stable vacuum   &  & $\exists_{=1} \land P_i<0$ \\
    \end{tabular}
    \caption{List of criteria for labeling the stability of a solution $i$ at a given point in the phase diagram.  We assume there is at least one solution $i$ that is a subset of $j$ i.e. $i\subset j$ (all possible solutions at that point in the phase diagram).}
    \label{tab:stability_criteria}
\end{table*}

The variables related to the stability of the system also dictate the occurrence of a phase transition. For example, a first-order phase transition, such as the quark deconfinement transition at low temperatures, occurs when 
\begin{equation}
    P^I = P^{II}, \qquad \vec{\mu}^I = \vec{\mu}^{II}.
\end{equation}
where the superindex $I$ indicates the hadronic phase and the superindex $II$ indicates the quark phase.
The presence of strangeness and charge chemical potentials offers different possibilities for the phase transition, such as making the phase transition at fixed charge fraction or strangeness fraction~\cite{Hempel:2013tfa, Aryal:2020ocm}. In this paper, we assume all charges are conserved during the phase transition (a non-congruent transition), where   
\begin{equation}
\mu_B^I = \mu_B^{II}, \qquad\mu_Q^I = \mu_Q^{II}, \qquad \mu_S^{I} = \mu_S^{II}.
\end{equation}

\begin{figure*}[h!]
    \centering
    \begin{subfigure}[b]{0.45\textwidth}
        \centering
\includegraphics[scale=0.1135]{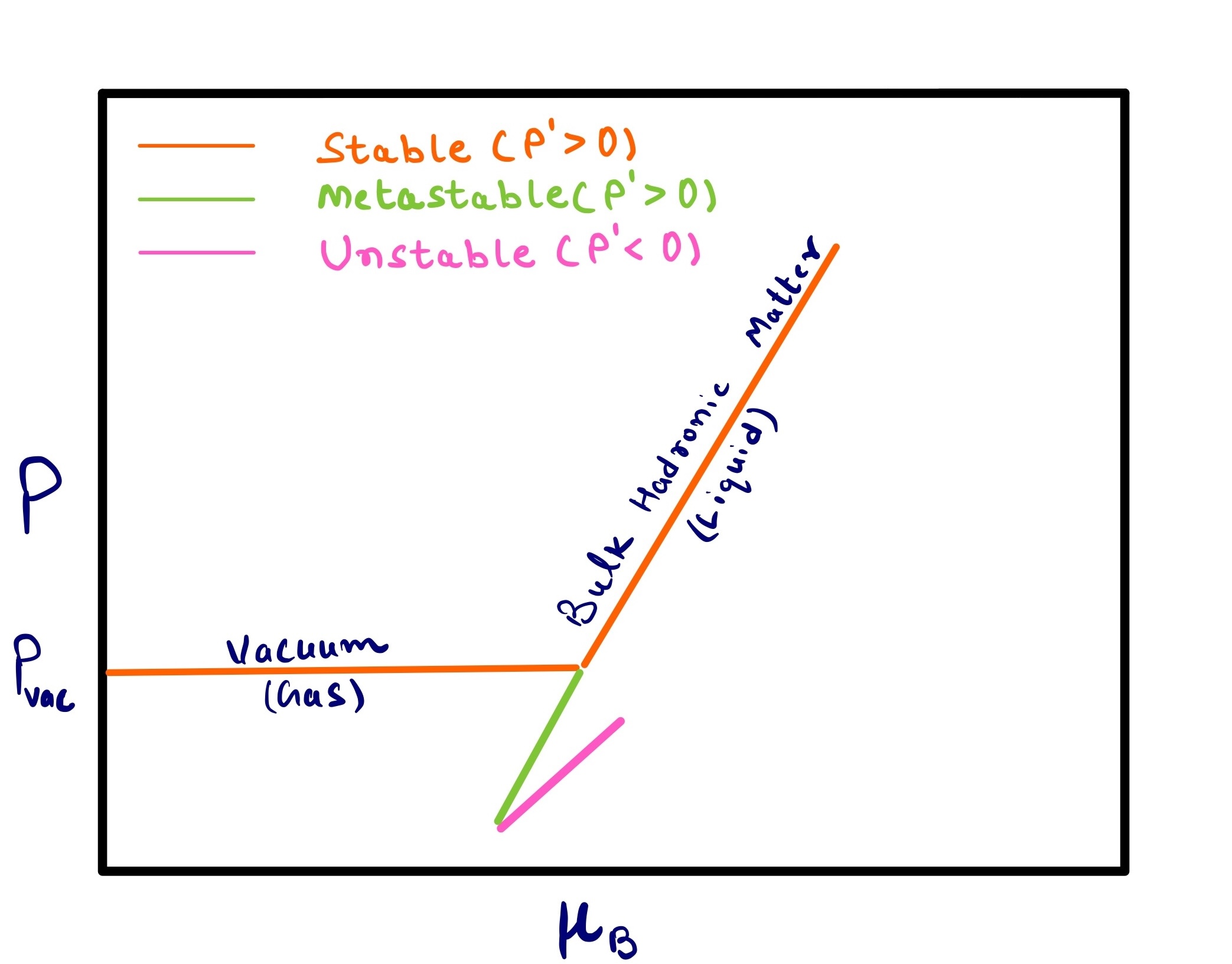}
        \label{fig:LGPT_mu}
    \end{subfigure}
    \hfill
    \begin{subfigure}[b]{0.45\textwidth}
        \centering
    \includegraphics[scale=0.1135]{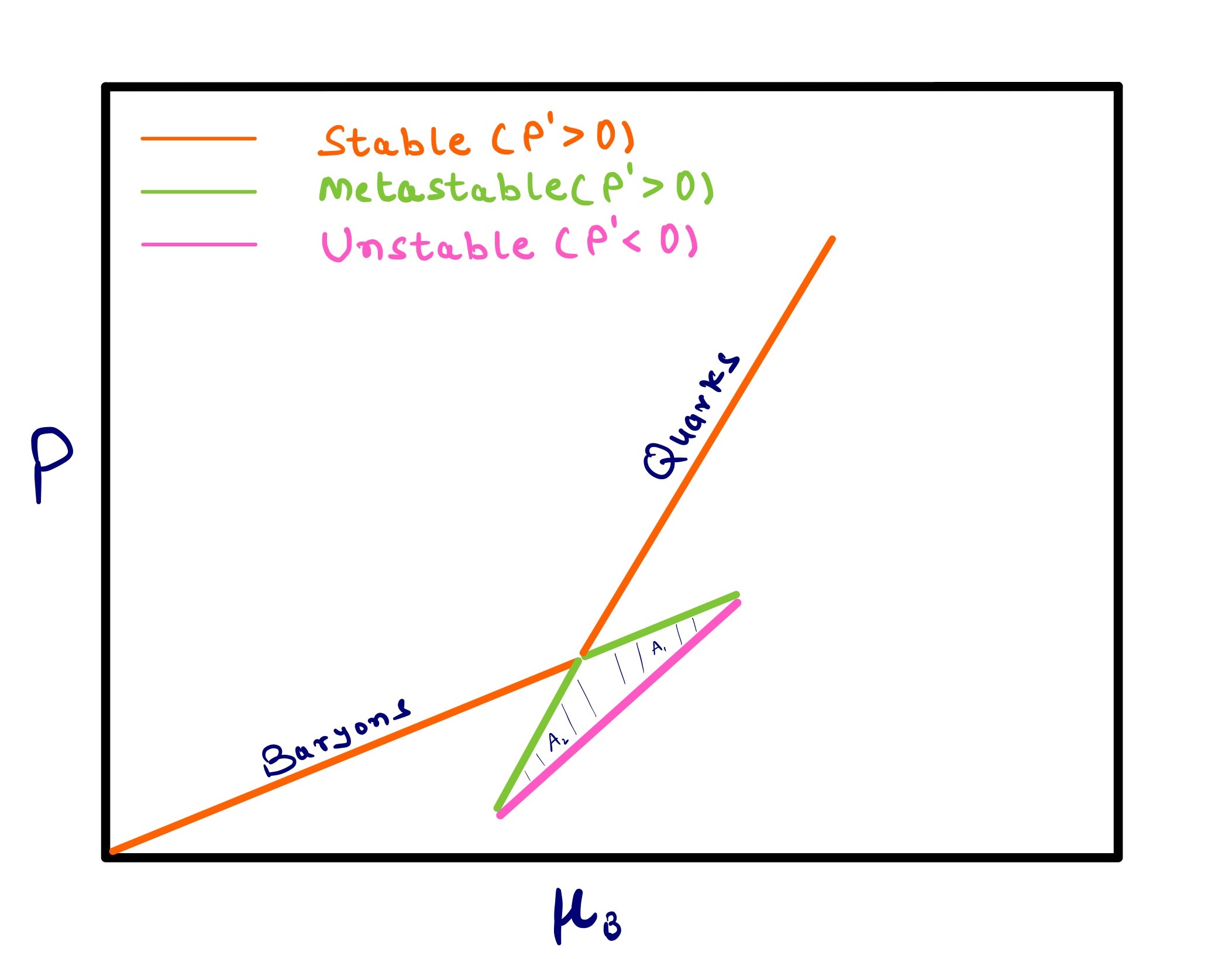}
        \label{fig:HQPT_mu}
    \end{subfigure}
     \begin{subfigure}[b]{0.45\textwidth}
        \centering
\includegraphics[scale=0.1135]{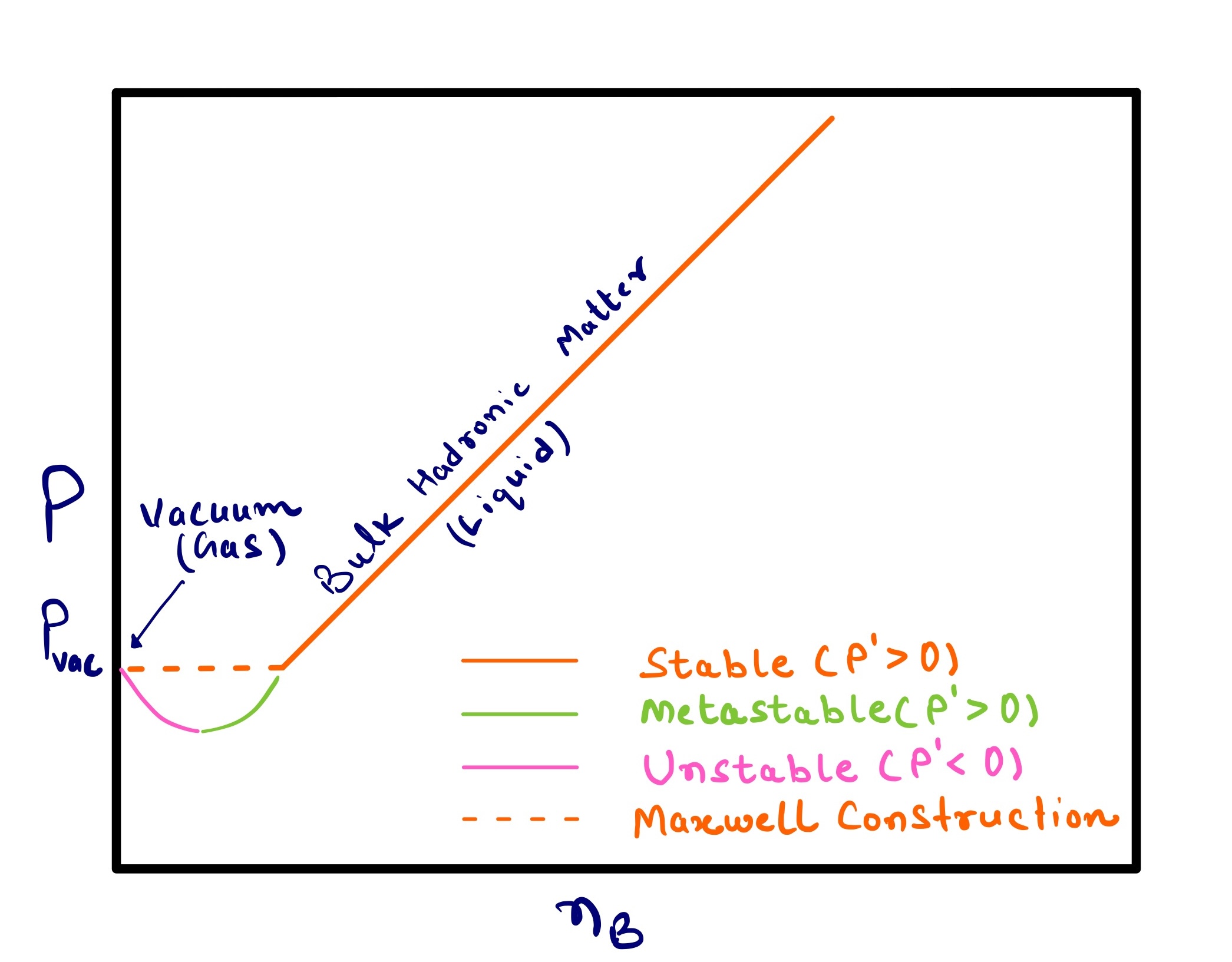}
        \label{fig:LGPT_nB}
    \end{subfigure}
    \hfill
    \begin{subfigure}[b]{0.45\textwidth}
        \centering
    \includegraphics[scale=0.1135]{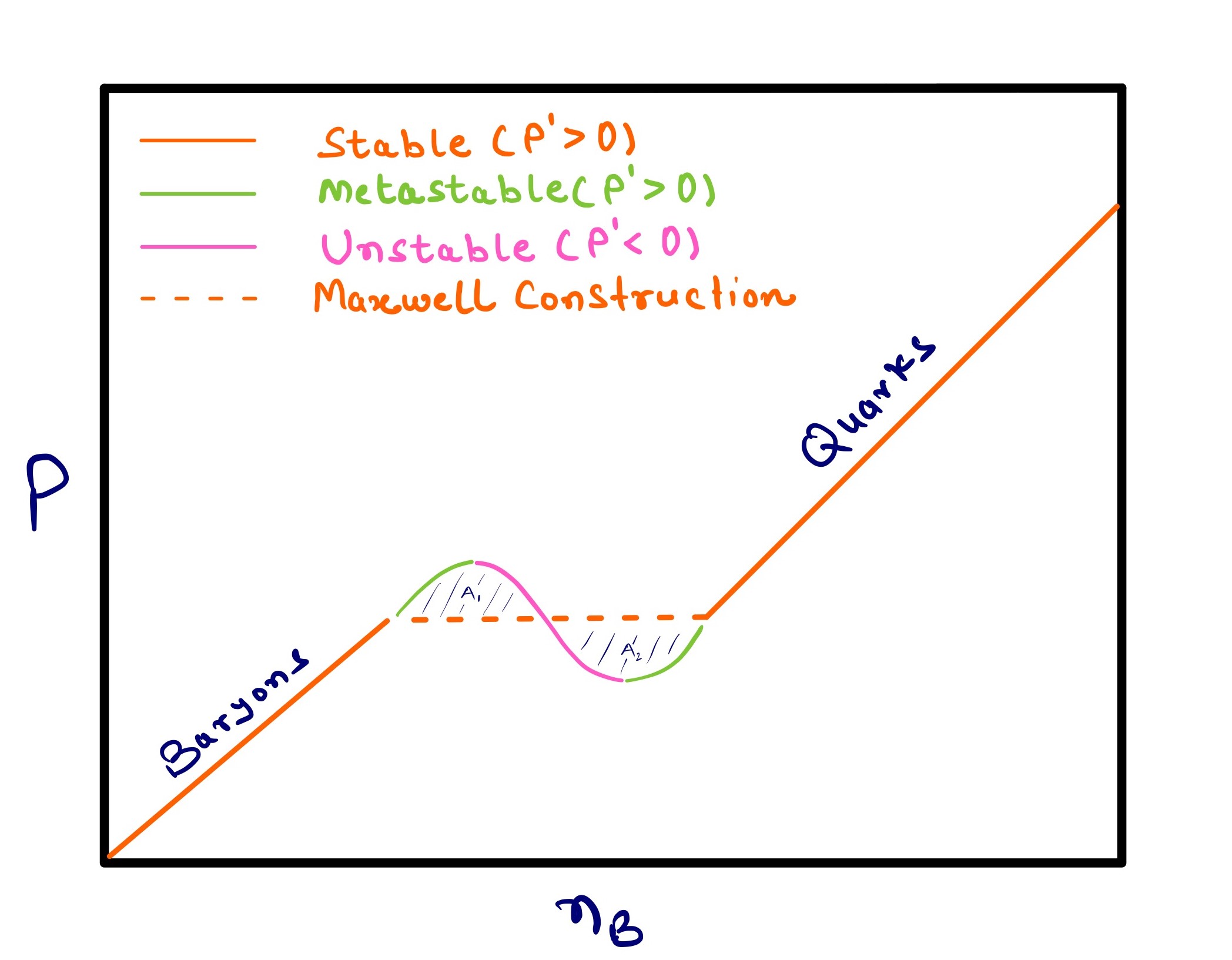}
        \label{fig:HQPT_nB}
    \end{subfigure}
    \caption{
    Stable, metastable, and unstable phases in pressure vs. baryon chemical potential (top panels) and pressure vs. baryon density (bottom panels) plane with $P^\prime={dP}/{dn_B}$.
   On the left, the liquid-gas phase transition is shown with a vacuum phase at low densities and a bulk hadronic phase at large densities.
   On the right, the deconfinement phase transition is shown with a bulk hadronic phase at low densities and a deconfined quark phase at high densities. A Maxwell construction is shown with examples of the equal area method.}
    \label{fig:stability}
\end{figure*}

In \Cref{fig:stability} different phases relevant, e.g., for the description of the core of neutron stars, are shown: vacuum, hadronic matter, and quark matter. 
The top panel shows a first-order phase transition in $P$ vs $\mu_x$ space and the bottom shows the first-order phase transition in $P$ vs $n_x$ space. 
Because $P$ vs $\mu_x$ must be continuous, we see a clear maximum solution at each point in $\mu_x$. 
In contrast, the stable solution for the first-order phase transition demonstrated here has a jump in $n_x$ such that across a range of $n_x$ we see only metastable and unstable solutions. 

As the chiral model in its current version does not include nuclei, we reproduce the liquid-gas phase transition as being from vacuum to bulk baryonic matter.
Depending on how they are connected and which ones are present, these phases can be unstable, metastable, or stable (see \Cref{tab:stability_criteria}). 
If a system is in equilibrium, then a Maxwell construction can be performed across the metastable/unstable regime, such that the EoS remains stable even across the phase transition. 
However, dynamical simulations often require metastable/unstable regimes in order to accurately describe the time spent in each phase of matter (see e.g. \cite{Sorensen:2020ygf,Oliinychenko:2022uvy}). 
Thus, in \texttt{CMF++} we build Maxwell constructions, but also preserve the metastable/unstable regimes.

Given an EoS with a metastable/unstable regime across a phase transition, one can obtain the Maxwell construction in one of two ways:
\begin{itemize}
    \item using the equal area method in $P(\vec{n})$ space, in which one finds the line (dashed line) such that the two areas in \Cref{fig:stability} d) are equal i.e. $A_1=A_2$. See \cite{Vovchenko:2015vxa} for examples and discussion in a van der Waals model for the liquid gas phase transition. This method is more typical for models within the canonical ensemble;
    \item choosing the maximum pressure (minimizing the grand potential density) at a specific point in $\vec{\mu}$ (one can also do the same at a specific point in $T$), which is demonstrated in \Cref{fig:stability} (a-b). 
    This method is more typical for models within the grand canonical ensemble.
\end{itemize}
In this work, we follow the procedure depicted in \Cref{tab:stability_criteria} and apply the second method to find the (most) stable phase such that at each point in our $\vec{\mu}$ phase diagram, we choose the $P_i(\vec{\mu})=\texttt{max}\left[P_j(\vec{\mu})\right]$ given multiple solutions $j$ where $i\subset j$.
The second method is significantly easier in \texttt{CMF++} because the metastable/unstable regime in \texttt{CMF++} can become significantly more complicated than the sinusoidal appearance shown in \Cref{fig:stability} d). 
Rather, depending on the degrees of freedom one may find more than 2 solutions or even solutions that cross each other. 
Thus, it is not always obvious what the definition of the areas is with so many solutions present, such that the equal area method would be impractical. To differentiate between unstable and metastable phases, we also follow the procedure depicted in \Cref{tab:stability_criteria}.

\subsection{Benchmark} 
\label{sec:Code_Benchmark}

\begin{figure}[H]
\centering
\includegraphics[width=.47\textwidth]{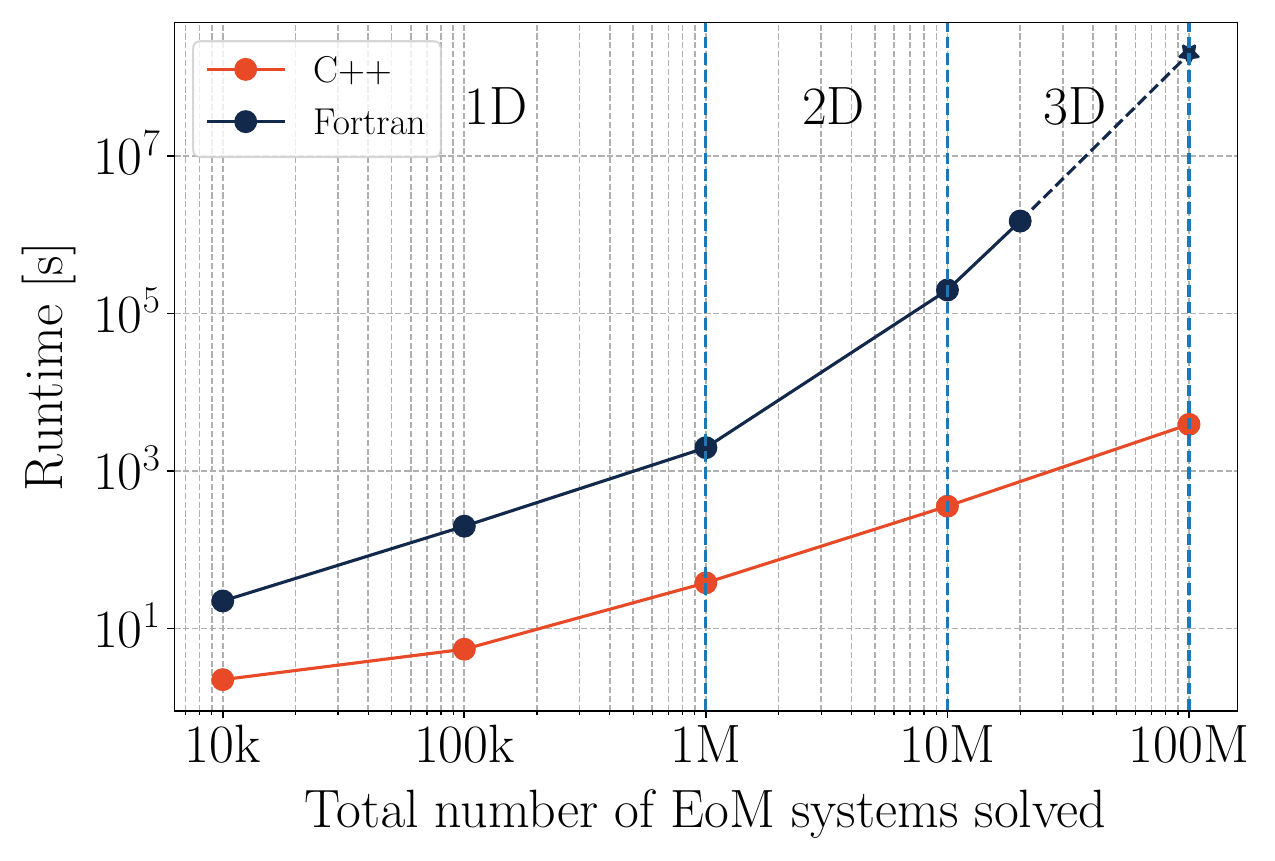}
\caption{Runtime comparison for \texttt{Fortran} vs \text{C++}.}
\label{fig:Runtime_test}
\end{figure} 

In \Cref{fig:Runtime_test} a benchmark of the time it takes to run \texttt{CMF++} compared to the legacy CMF in \texttt{Fortran} is shown. 
Along the x-axis, we demonstrate the typical total number of EoM systems solved (the set of equations in \Cref{eq:Algebraic_System}) for 1D, 2D, and 3D EoSs. For the 1D case, $\mu_S$ and $\mu_Q$ were kept at zero and $\mu_B$ was varied from 10 points (10k EoM systems) to 100 points (100k) and finally 1000 points (1M). For the 2D case, $\mu_S$ was kept at zero, 1000 points were used in $\mu_B$ and 10 points in $\mu_Q$. Finally, for the 3D case, 1000 points were used in $\mu_B$, 10 points in $\mu_Q$, and 10 points in $\mu_S$. Along the y-axis, we find the average runtime for 16 different particle configuration combinations (4 vector potentials, decuplet on/off, quarks on/off). The dashed line represents an extrapolation given Fortran's extreme runtime, where the star at the end is an educated guess based on the behavior at 20M. It is important to note that the time complexity for \texttt{CMF++} is $\mathcal{O}(n)$ whereas \texttt{Fortran} has an $\mathcal{O}(n \ln(n))$ one. We find that \texttt{CMF++} significantly improves the performance of the calculations of the EoS by at least an order of magnitude (for the simplest calculations) and up to 4 orders of magnitude for the 3-Dimensional case.

\section{Results}
\label{sec:results}

In this section, we present our numerical findings, exploring all four vector couplings across various combinations of degrees of freedom  (\textit{d.o.f.}) within the CMF model, using different combinations of independent chemical potentials.

Our general approach in the following subsections is to demonstrate our results for the mean fields, $\Phi$, and certain thermodynamic variables. 
Then, we show population plots for the individual species of hadrons and/or quarks. 
 Finally, the charge fractions and susceptibilities are shown. 
 Initially, we demonstrate that the new \texttt{CMF++} can both reproduce the legacy \texttt{Fortran} version of CMF and also obtain more precise results in 1D. 
 Later, new results across the 3D phase space of $\vec{\mu}$ are only shown for \texttt{CMF++} due to the extremely long run times that they would take in the legacy \texttt{Fortran} code.

\subsection{$\left\{\mu_S = \mu_Q = 0\right\}$}
\label{sec:muBN0_muQ0_muS0}

We begin by examining various sets of \textit{d.o.f}, considering the simplest case where $\mu_S=\mu_Q=0$.

\subsubsection{C3 and C4 with baryon octet + quarks}\label{sec:C3C4_octet_quarks}


We start by exploring the behavior of mean-field mesons, the deconfinement phase transition order parameter, $\Phi$, and thermodynamical properties, including the baryon octet plus quarks as \textit{d.o.f} under the influence of C3 and C4 vector couplings (see  \Cref{tab:scalar_param,tab:vec_param,tab:quark_param} for related parameters). 
Displaying all coupling schemes would involve an excessive amount of quantitative detail; therefore, we only present the results for C3 and C4 couplings, as C3 behaves similarly to C2 and C1 (see \Cref{sec:C1_C4_baryon_octet_quarks} for details on the other couplings).

\begin{figure*}
\centering
\includegraphics[width=0.8\textwidth]{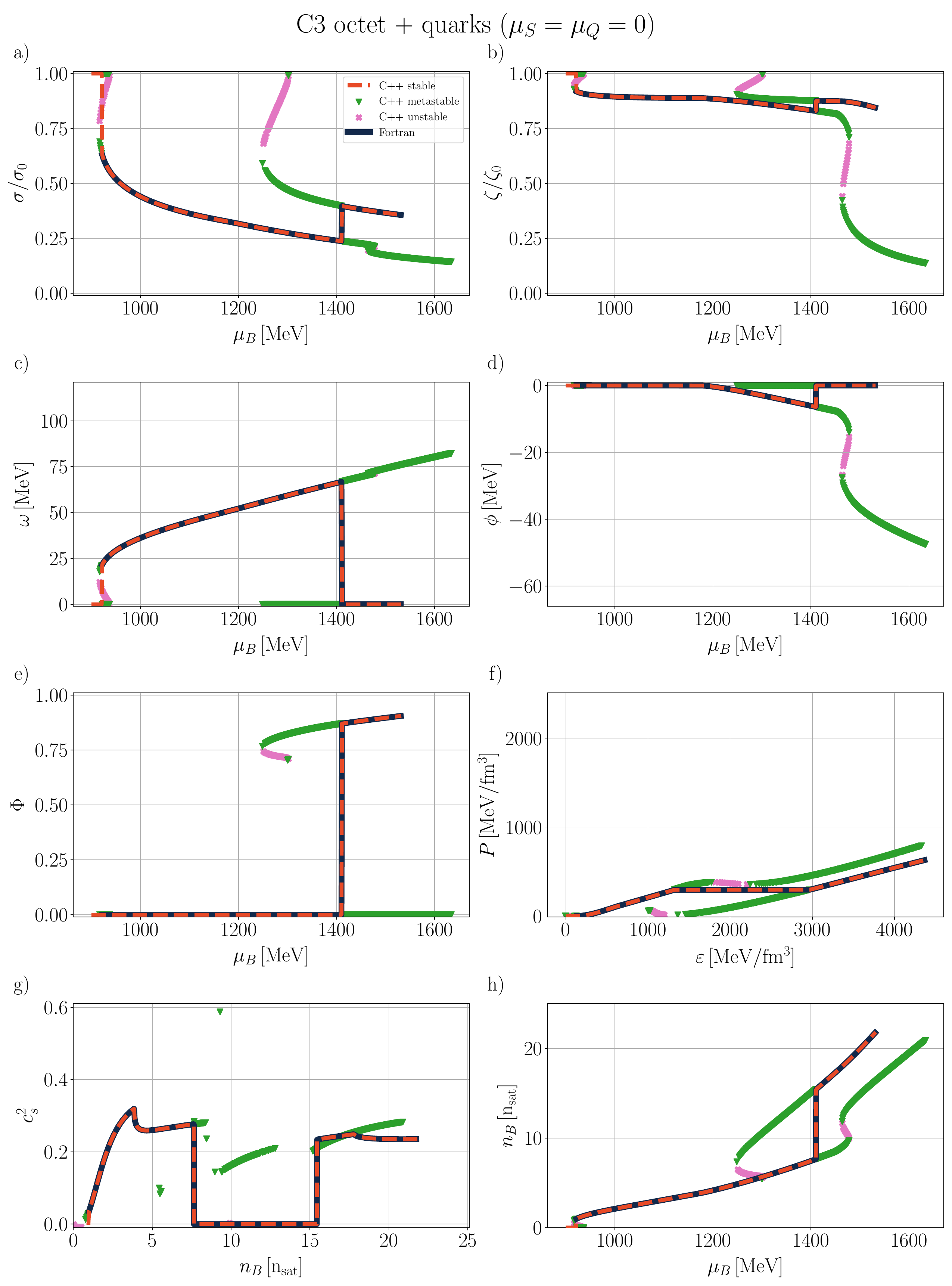}
\caption{C3 ($\mu_S=\mu_Q=0$) octet + quarks: a) b) scalar meson fields (normalized by vacuum values), c) d) vector meson fields, e) and deconfinement field as a function of baryon chemical potential, f) pressure vs energy density, g) speed of sound vs baryon density (in terms of saturation density), h) baryon density (in terms of saturation density) vs baryon chemical potential. Comparison of results from \texttt{Fortran} for stable branches (black solid line) and \texttt{CMF++} for stable (red-orange dashed line), metastable (green upside-down triangles), and unstable (pink x's) branches. 
}
\label{fig:2D_C3_hyper_quarks_muS_0_muQ_0_mf_and_obs}
\end{figure*}

We begin with the C3 coupling in \Cref{fig:2D_C3_hyper_quarks_muS_0_muQ_0_mf_and_obs} in the limit of $\mu_S=\mu_Q=0$.
The mean-field mesons ($\sigma/\sigma_0$, $\zeta/\zeta_0$, $\omega$, and $\phi$) as a function of $\mu_B$ are shown in panels a)-d), respectively.
Additionally, \Cref{fig:2D_C3_hyper_quarks_muS_0_muQ_0_mf_and_obs} contains $\Phi$ in e), the pressure vs the energy density in f), the speed of sound squared (calculated numerically) vs the baryon density in g), and the baryon density vs the baryon chemical potential in h).
Also shown in \Cref{fig:2D_C3_hyper_quarks_muS_0_muQ_0_mf_and_obs} are the different types of solutions obtained in \texttt{CMF++} that we classify as stable, metastable, and unstable solutions.
The legacy  \texttt{Fortran} code only provides stable solutions, which are also shown for comparison. 
The criteria for stability are comprehensively discussed in \Cref{sec:stability}, as well as \Cref{fig:stability}.

We observe a consistent correspondence between the results obtained from \texttt{C++} and \texttt{Fortran} across all mean fields for stable solutions in \Cref{fig:2D_C3_hyper_quarks_muS_0_muQ_0_mf_and_obs} panels a)-d). 
Additionally, $\Phi$ and the thermodynamic quantities are all precisely reproduced in the \texttt{CMF++} version of the code, as one can see in the comparison of the black solid vs red long-dashed lines in panels e)-h). 
In all panels, we see identical results from stable solutions, except for the region with very low $\mu_B$, where \texttt{Fortran} has trouble finding solutions.

\begin{figure}
\centering
\includegraphics[width=.47\textwidth]{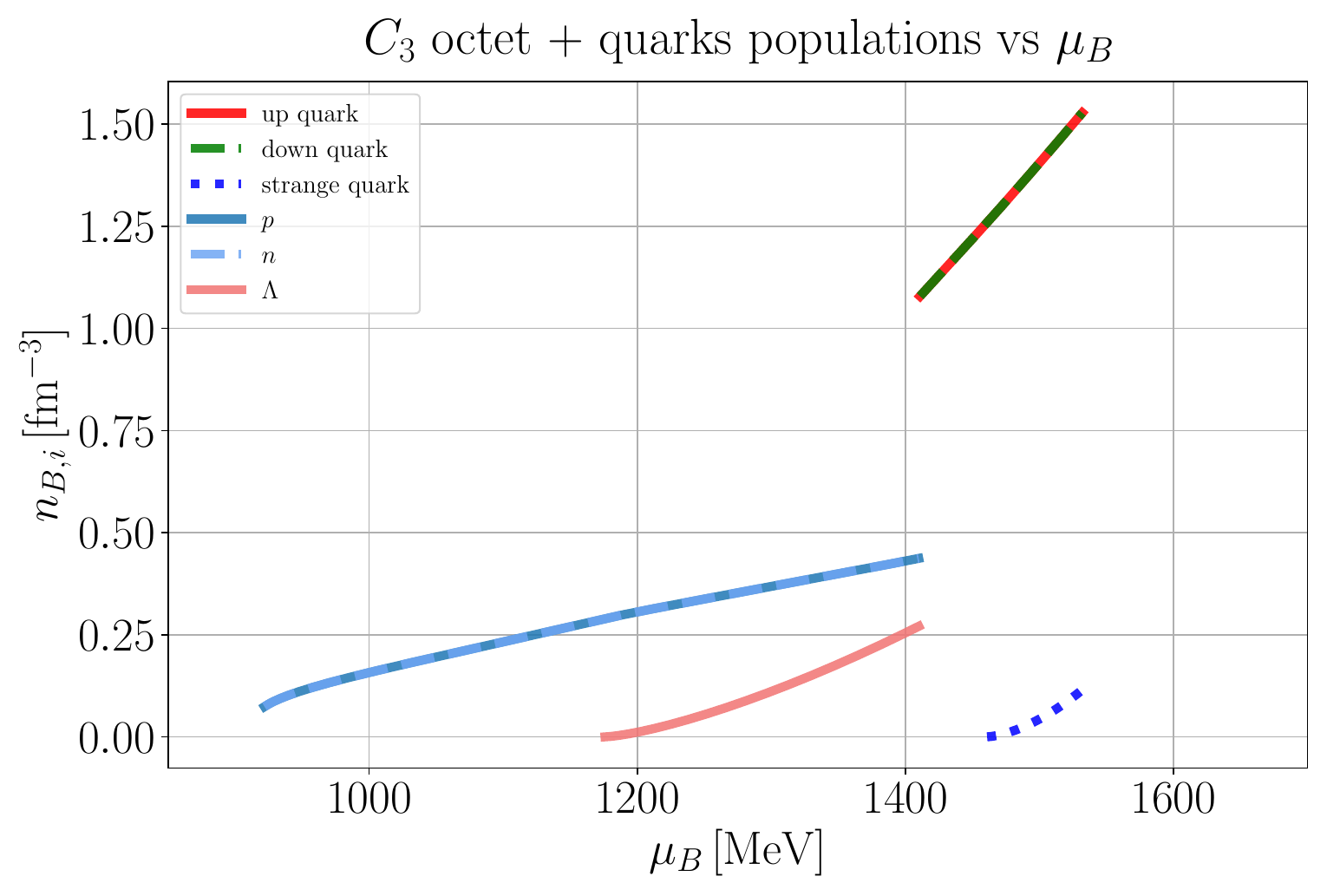}
\caption{C3 ($\mu_S=\mu_Q=0$) octet + quarks: particle populations versus baryon chemical potential using stable solutions from \texttt{CMF++}.}
\label{fig:2D_C3_hyper_quarks_populations_vs_muB}
\end{figure}

Within a given phase of matter (either hadron or quark), the stable solutions for the non-strange scalar $\sigma/\sigma_0$ (panel a)) and strange scalar $\zeta/\zeta_0$ (panel b)) ratios, hereon simply referred to as $\sigma$ and $\zeta$, monotonically decrease with increasing $\mu_B$. 
The decreasing trend in scalar condensates is indicative of chiral symmetry restoration.
Panel b) resembles panel a) but we can see that the $\zeta$ field has a kink just below $\mu_B\sim 1200$ MeV, coinciding with the emergence of $\Lambda$ hyperons (refer to the discussion of hyperon population \Cref{fig:2D_C3_hyper_quarks_populations_vs_muB}). 
Also in the quark phase, we see a kink in $\zeta$ at $\mu_B\sim 1500$ MeV marking the appearance of strange quarks. 
The non-strange vector field $\omega$ (panel c)) exhibits the opposite behavior than (panel a)), while the strange vector field $\phi$ (panel d)) remains zero until the emergence of strange particles (in this case $\Lambda$ hyperon) and then presents a very similar behavior to panel b). 
Post deconfinement phase-transition, both vector fields become zero, as they do not couple to the quarks in our parametrization (see \Cref{here}).
Since we are discussing isospin symmetric matter ($\mu_Q=0$), there is no finite value for isovector mesons. 

As discussed previously in \Cref{sec:stability}, CMF reproduces three distinct phases of matter at $T=0$: vacuum, hadronic phase, and quark phase. 
In \Cref{fig:2D_C3_hyper_quarks_muS_0_muQ_0_mf_and_obs}  the Maxwell constructions of the first-order phase transitions generate vertical lines at a fixed $\mu_B$ for all the mean fields in panels a)-d).
The liquid-gas phase transition (vacuum to hadrons) occurs at low $\mu_B$ and the deconfinement phase transition (hadronic to quark) occurs at intermediate to high $\mu_B$. 
Within the phase transition itself, unstable phases may appear. 

Additionally, within both the hadronic and quark phases, there are differences between phases that only/mainly contain light hadrons or light quarks vs those that contain strange hadrons or strange quarks. 
Separating the light vs strange dominant regimes within a given hadronic or quark phase may also be a phase transition of various order. 
In fact, there can appear sometimes even first-order phase transitions leading into strangeness dominant phases (light hadrons to strange hadron dominated) that occur before the deconfinement phase transition for specific parameter sets of CMF. In the following,
 we discuss the appearance of all possible phases in \Cref{fig:2D_C3_hyper_quarks_muS_0_muQ_0_mf_and_obs} and in subsequent CMF parametrizations across different combinations of $\vec{\mu}$.

In \Cref{fig:2D_C3_hyper_quarks_muS_0_muQ_0_mf_and_obs} the liquid-gas phase transition is shown to take place at $\mu_B=921.5$ MeV.  One can see the telltale vertical line in $\sigma$, $\zeta$, and $\omega$. However, the $\phi$ meson does not experience the liquid-gas phase transition because strangeness is not relevant at such a low $\mu_B$. Similarly, $\Phi$, the order parameter for the deconfinement phase transition remains zero throughout the liquid-gas phase transition.

The phase transition at higher $\mu_B$ is the one related to quark deconfinement. There is a strong relation between $\Phi$ and the meson mean fields. 
The value of $\Phi$ is shown in panel e) of \Cref{fig:2D_C3_hyper_quarks_muS_0_muQ_0_mf_and_obs}. 
The change in its value from $0$ to $\sim1$ at $\mu_B = 1410.5$ MeV signals the change in values for the effective masses of baryons (which become too large for them to be present) and quarks (which become light enough to appear). 

Looking at the $\phi$ meson, we find that strangeness begins to play a role at $\mu_B\sim 1200$ MeV when the $\phi$ meson begins to deviate from 0. 
At the same time, we see that both the $\zeta$ field deviates further from the vacuum at that point as well. 
However, there is not a first-order phase transition as the strangeness begins to play a role because there is no clear vertical line in any of the mean fields between stable phases when the strange mean fields begin to become non-zero. 
We later analyze the susceptibilities in order to determine the order of the strangeness related phase transition.

Even within the first-order phase transition, most of the solutions fulfill the stability criteria (although they do not have the maximum pressure) such that they are labeled metastable (shown in green). 
However, one can see small regions of unstable phases that appear (shown in pink). 
One surprising outcome can be seen fairly clearly in the $\zeta$  plot in panel b) and in the $\phi$  plot in panel d). 
There is a small unstable region around $\mu_B\sim 1500$ MeV that connects two metastable regions, which is indicative of a first-order phase transition that connects two metastable phases. 
The two metastable phases contain strange hadrons (indicated because the $\phi$ and $\zeta$ mean fields mediate the strange interactions). 
Thus, the phase transition (between metastable phases) goes from a hadronic phase with some strangeness into a strangeness-dominated hadronic phase. 
As both chiral condensates, $\sigma$ and $\zeta$ in panels a) and b), decrease across this phase transition, we identify it as a first-order phase transition associated with chiral symmetry.
However, this strangeness-dominated hadronic phase has a lower overall pressure at a fixed $\mu_B$ than the quark phase, such that it is not considered a stable solution.

The EoS is a relationship between the pressure and energy density $P(\varepsilon)$, which is shown in panel f) of \Cref{fig:2D_C3_hyper_quarks_muS_0_muQ_0_mf_and_obs} for C3.
The stable branches with the Maxwell construction show a monotonically increasing EoS with a slight kink around $\varepsilon\sim 500$ MeV/fm$^3$ where the hyperons switch on. 
Then, one can see the first-order phase transition at the plateau where $P$ remains constant while $\varepsilon$ increases. 
In the quark phase, the pressure monotonically increases once again, with a very small kink appearing again when strange quarks become relevant (although that is quite hard to see).

The metastable/unstable solutions are shown in green and pink, respectively, which include both quark and hadronic solutions. 
As discussed previously for the strange mean fields, we can see in panel f) a chiral phase transition within metastable phases (going from a light-dominated hadronic phase into a strangeness-dominated hadronic phase) that appears at higher pressure for a fixed energy density). 
The lower branches that appear at low pressure below the first-order phase transition correspond to light quark phases. Note that if quarks had not been considered, the hadronic phases would be the stable ones throughout and a Maxwell construction would have to be built across the unstable hadronic phase.

Some of the features of the EoS may be difficult to pinpoint in the $P(\varepsilon)$ plot (specifically where strange hadrons or quarks become relevant). 
However, in panel g) showing $c_s^2$ vs $n_B$ these features become clearer (being $c_s^2$ the derivative $dP/d\epsilon$). 
For instance, looking at $c_s^2$ and starting first at low $n_B$, we can see several interesting features. 
First, strange hadrons appear, creating a small peak/kink in $c_s^2$ (denoting a higher-order phase transition)  when the $\Lambda$'s appear. 
Then the next feature is a drop in $c_s^2\rightarrow 0$ that leads to a plateau. The plateau corresponds to the first-order phase transition (as the Maxwell construction was used). 
The metastable and unstable regimes are also shown across the phase transition. They provide a different structure instead of the plateau from the Maxwell construction. 
Finally, after the phase transition, we see that the $c_s^2$ rises again because of the quark phase and there is a small peak in the quark phase that arises from strange quarks appearing. 
Afterwards, $c_s^2$ remains near the QCD conformal limit of $(1/3)$ and is consistent with the pQCD results \cite{Gorda:2021kme},
as well as their constraints following stability and causality \cite{Komoltsev:2021jzg}.

Finally, the $n_B$ vs $\mu_B$ plot is shown in panel h).  
Note that the number density $n_B$ contains an additional non-fermionic contribution from $\Phi$ 
(\Cref{eq:net_energy_pressure_density}).
At the liquid-gas phase transition, there is a small vertical jump at low $\mu_B$, followed by a monotonically increasing $n_B(\mu_B)$ until the deconfinement phase transition is reached just after $\mu_B\sim 1400$ MeV.
Then, in the quark phase, there is a steeper rise in the $n_B(\mu_B)$ compared to what occurred in the hadronic phase (due to the $\Phi$ contribution). 
The metastable/unstable branches are shown in green and pink, respectively.
The branch that is continuous with the hadronic phase at lower $n_B$ is for the hadronic into strangeness-dominated hadronic phases, whereas the upper $n_B(\mu_B)$ branch that connects to the quark phase is the metastable light quark phase. 
We can see that for this specific parametrization, the deconfinement phase transition happens at relatively large $n_B$ (for $\mu_S=\mu_Q=0$).

To analyze the individual contributions of particles to the EoS, we plot the population of particles against $\mu_B$ in \Cref{fig:2D_C3_hyper_quarks_populations_vs_muB}. 
The populations are defined at the baryonic number density of specific species such that if they were all added together we would recover the baryon number density (minus the $\Phi$ contribution in the quark phase) i.e.
\begin{align}
    n_B^{had}&=\sum_i^{had} n_{B,i}=n_p+n_n+n_\Lambda+\dots\label{eqn:POP_had}\\
    n_B^{quark}&=\sum_i^{q} n_{B,i}=\sum_i^{q} B_i n_i=\frac 1 3 n_u+\frac 1 3 n_d+\frac 1 3 n_s\label{eqn:POP_quark}
\end{align}
where \Cref{eqn:POP_had} provides the baryon density in the hadronic phase and \Cref{eqn:POP_quark} provides the baryon density in the quark phase (given the species that appear for this specific C3 coupling at this specific choice in the $\vec{\mu}$ space).
Given that we have $\mu_Q=0$, we are dealing with the symmetric nuclear matter or in other words have isospin symmetry.
Due to isospin symmetry, the in-medium mass and density of protons and neutrons exhibit degeneracy i.e. $n_n=n_p$ and $m_n^*=m_p^*$ at all $\mu_B$. 

The nucleons begin to populate around 
\begin{eqnarray}
    \mu_B^0&=&m_N+B.E \sim 922 \rm{MeV}        
\end{eqnarray}
where $m_N=(m_p+m_n)/2$ is the average nucleon vacuum mass and $B.E$ is the binding energy per nucleon. 
Additionally, at $\mu_B = 1176$  MeV, $\Lambda$ hyperons emerge,  opening the Fermi sea and thereby softening the EoS (this coincides in the first peak in $c_s^2(n_B)$ in  \Cref{fig:2D_C3_hyper_quarks_muS_0_muQ_0_mf_and_obs}). 
Following the deconfinement phase transition around $\mu_B = 1410.5$ MeV, baryons are replaced by quarks, resulting in a growth in the number density of quarks with $\mu_B$. 
The density of $u$ and $d$ quarks are equal across all $\mu_B$. 
Around $\mu_B\sim 1450\;$ MeV, the $s$ quarks switch on, which corresponds to the second peak in $c_s^2(n_B)$ during the quark phase in  \Cref{fig:2D_C3_hyper_quarks_muS_0_muQ_0_mf_and_obs}.
One significant difference in the hadronic vs quark phase is that $\Lambda$ baryon shares a significantly larger fraction of the total baryon number density (with respect to protons and neutrons) than the strange quark does (with respect to up and down quarks) in the regime of densities studied. 

\begin{figure*}
\centering
\includegraphics[width=0.9\textwidth]{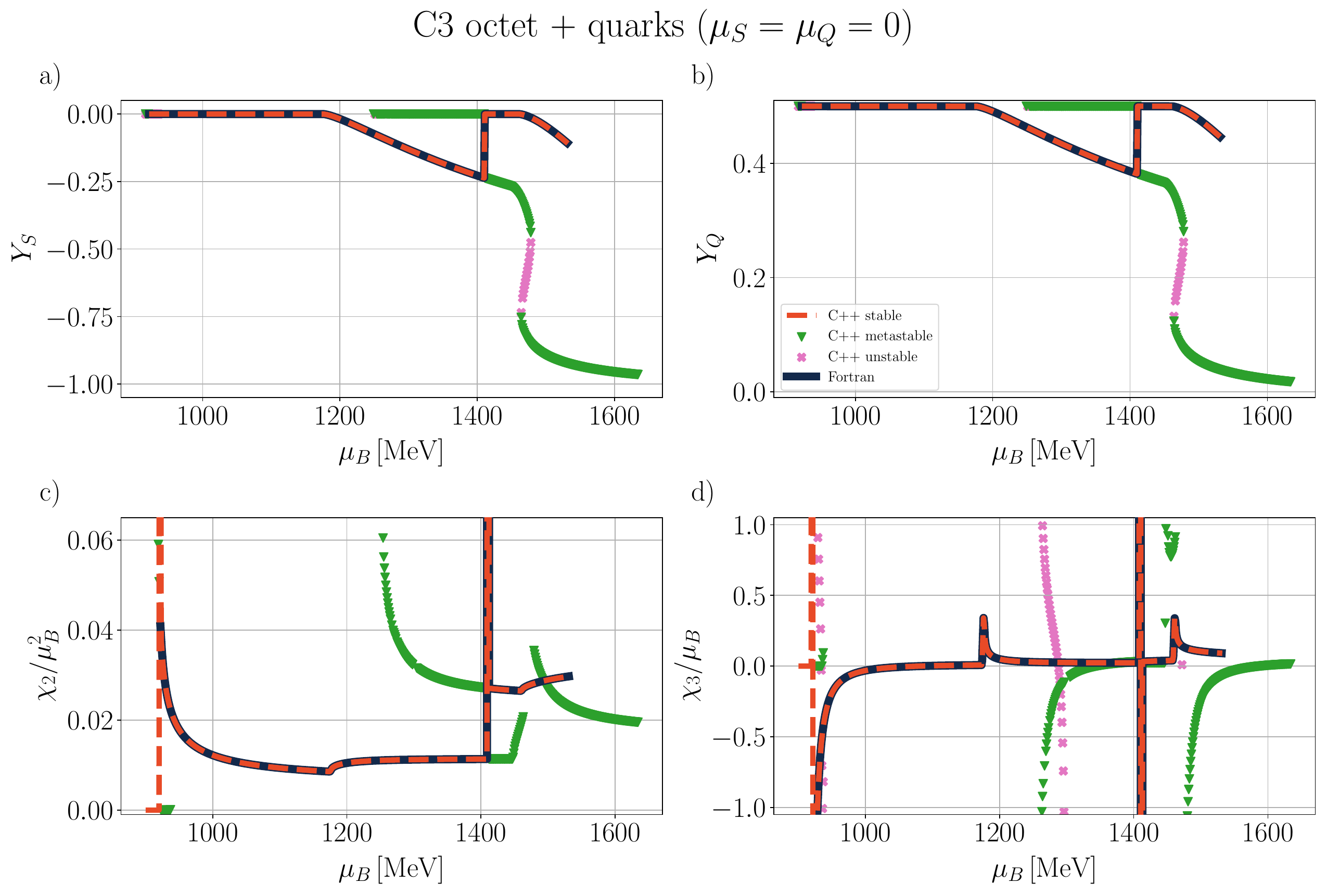}
\caption{C3 ($\mu_S=\mu_Q=0$) octet + quarks: a) strangeness and b) charge fractions vs baryon chemical potential, c) second and d) third order baryon susceptibilities, all versus baryon chemical potential. Comparison of results from \texttt{Fortran} for stable branches (black solid line) and \texttt{CMF++} for stable (red orange-dashed line), metastable (green upside-down triangles), and unstable (pink x's) branches.  
}
\label{fig:2D_C3_hyper_quarks_muS_0_muQ_0_dens_and_sus}
\end{figure*}

In \Cref{fig:2D_C3_hyper_quarks_muS_0_muQ_0_dens_and_sus} we have our last set of figures for C3 and $\mu_S=\mu_Q=0$.  
In panel a) we show the strangeness fraction $Y_S$ vs the baryon chemical potential and in panel b) we show the electric charge fraction $Y_Q$ vs the baryon chemical potential. 
The baryon susceptibilities (calculated numerically) vs the baryon chemical potential are shown for second-order in panel c) and third-order in panel d).

Naively, one may expect that for $\mu_S=\mu_Q=0$, we have $Y_S=0$ and $Y_Q=0.5$. 
However, as we can see in panel a) of \Cref{fig:2D_C3_hyper_quarks_muS_0_muQ_0_dens_and_sus}, it is possible to obtain $Y_S<0$ due to the switching on of the $\Lambda$ baryon in the hadronic phase and later strange quarks in the quark phase (recall that $\Lambda$s and strange quarks carry $S=-1$ strangeness, such that $Y_S<0$).
After the emergence of the $\Lambda$ hyperon (refer to \Cref{fig:2D_C3_hyper_quarks_populations_vs_muB}), the magnitude of $Y_S$ steadily increases until $\mu_B = 1410.5$ MeV, when it jumps to zero at the deconfinement, at which point the strange quarks slowly appear increasing again the strangeness magnitude~(see again \Cref{fig:2D_C3_hyper_quarks_populations_vs_muB}). 
In the limit of isospin symmetric matter, the $\Lambda$ baryon contributes no net-isospin since its quark content is $uds$ and, therefore, is more preferred compared to the $\Sigma^+$ (uus) or $\Sigma^-$ (dds), which would create an isospin imbalance or appear in a pair.
Note the increase in the magnitude of $Y_S$ in the metastable and stable phases at large $\mu_B$ in panel a) of \Cref{fig:2D_C3_hyper_quarks_muS_0_muQ_0_dens_and_sus}. 
This increase in the magnitude of $Y_S$ in the metastable regime arises due to the hadronic phase with many hyperons (i.e. a strangeness-dominated hadronic phase) that would have appeared were quarks not included in the calculation.

In the charge fraction plot in panel b) of \Cref{fig:2D_C3_hyper_quarks_muS_0_muQ_0_dens_and_sus} we find that $Y_Q$ remains at 0.5 (representing an equal amount of positive and neutral nucleons) until the appearance of the neutral $\Lambda$ hyperon. 
After the $\Lambda$ hyperons appear,  $Y_Q$ continues to decrease because they create a further imbalance between charged and neutron hadrons. 
At densities above the deconfinement transition, $Y_Q$ begins to decrease once more with increasing $\mu_B$ due to the  $s$ quark's increasing relevance. 
Note that the isospin fraction $Y_I$ remains zero for all $\mu_B$. 
The relation between the different fractions is related to quantum numbers and can be derived from the Gell-Mann–Nishijima formula, resulting in $Y_Q=Y_I+\frac 1 2 + \frac 1 2 Y_S$ (for negative strangeness). This is discussed in detail in ref. \cite{Aryal:2020ocm}.
We note that the metastable and unstable branches are also shown in $Y_Q(\mu_B)$ and mirror the same qualitative behavior as was already discussed for $Y_S(\mu_B)$.

To gain a clearer insight into the different phase transitions, second and third-order susceptibilities ($\chi_n={d^n P}/{d\mu^n_B}$ with $n=2,3$) are displayed against $\mu_B$ in panels c) and d), respectively. 
The susceptibilities are normalized by different orders of $\mu_B$ to ensure they are dimensionless. 
Recall that the first-order susceptibility is just $n_B$, which was already shown in  \Cref{fig:2D_C3_hyper_quarks_muS_0_muQ_0_mf_and_obs} where we already saw the first-order discontinuities for the liquid-gas phase transition ($\mu_B=921$ MeV) and deconfinement ($\mu_B =1410.5$ MeV).
Then, the two first-order phase transitions are propagated into $\chi_2$ as two divergences and into $\chi_3$ also as two divergences (although they are significantly larger). 

In $\chi_2$ we can see that two kinks appear at $\mu_B=1176$ MeV and $\mu_B=1410.5$ MeV in the stable phases that correspond to the transition where strange baryons and strange quarks appear, respectively. 
Then, in $\chi_3$ the kinks turn into divergences at these same locations, which indicates that the onset of strangeness is a third-order phase transition. 
Thus, we can draw an interesting connection here to $c_s^2(n_B)$ that displays kinks at precisely these locations as well. 
Given that one can show that for a single conserved charge that $c_s^2=n_B/(\mu_B\chi_2)$ \cite{McLerran:2018hbz} and $n_B/\mu_B$ is smooth and continuous, then this kink in $\chi_2$ also appears in $c_s^2(n_B)$ as an inverted kink.

\begin{figure*}
    \centering
    \includegraphics[width=0.8\textwidth]{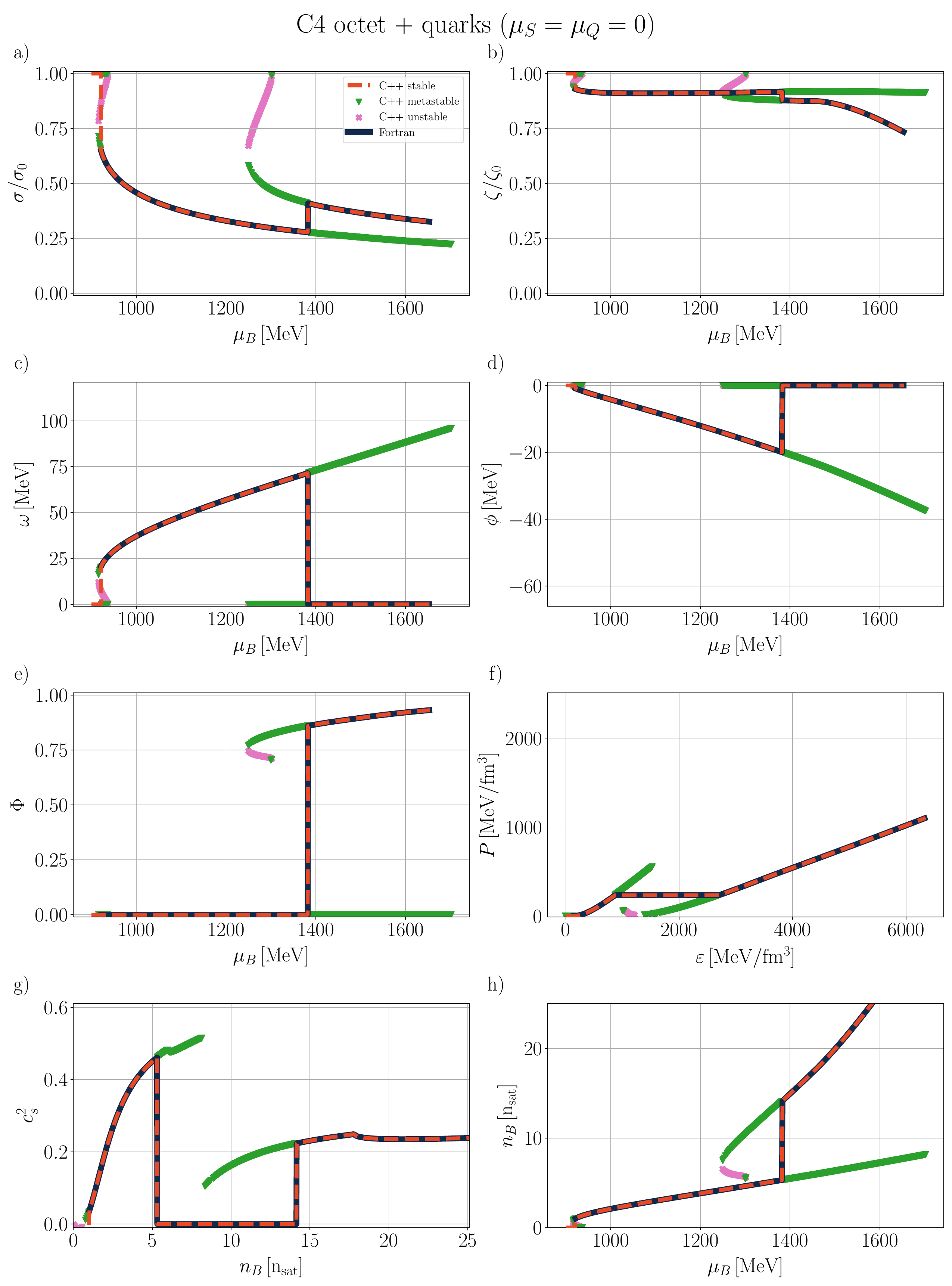}
    \caption{C4 ($\mu_S=\mu_Q=0$) octet + quarks: a) b) scalar meson fields (normalized by vacuum values), c) d) vector meson fields, e) and deconfinement field as a function of baryon chemical potential, f) pressure vs energy density, g) speed of sound vs baryon density, h) baryon density vs baryon chemical potential. Comparison of results from \texttt{Fortran} for stable branches (black solid line) and \texttt{CMF++} for stable (red-orange dashed line), metastable (green upside-down triangles), and unstable (pink x's) branches.
}
\label{fig:2D_C4_hyper_quarks_muS_0_muQ_0_mf_and_obs}
\end{figure*}

\begin{figure}
    \centering
        \includegraphics[width=.47\textwidth]{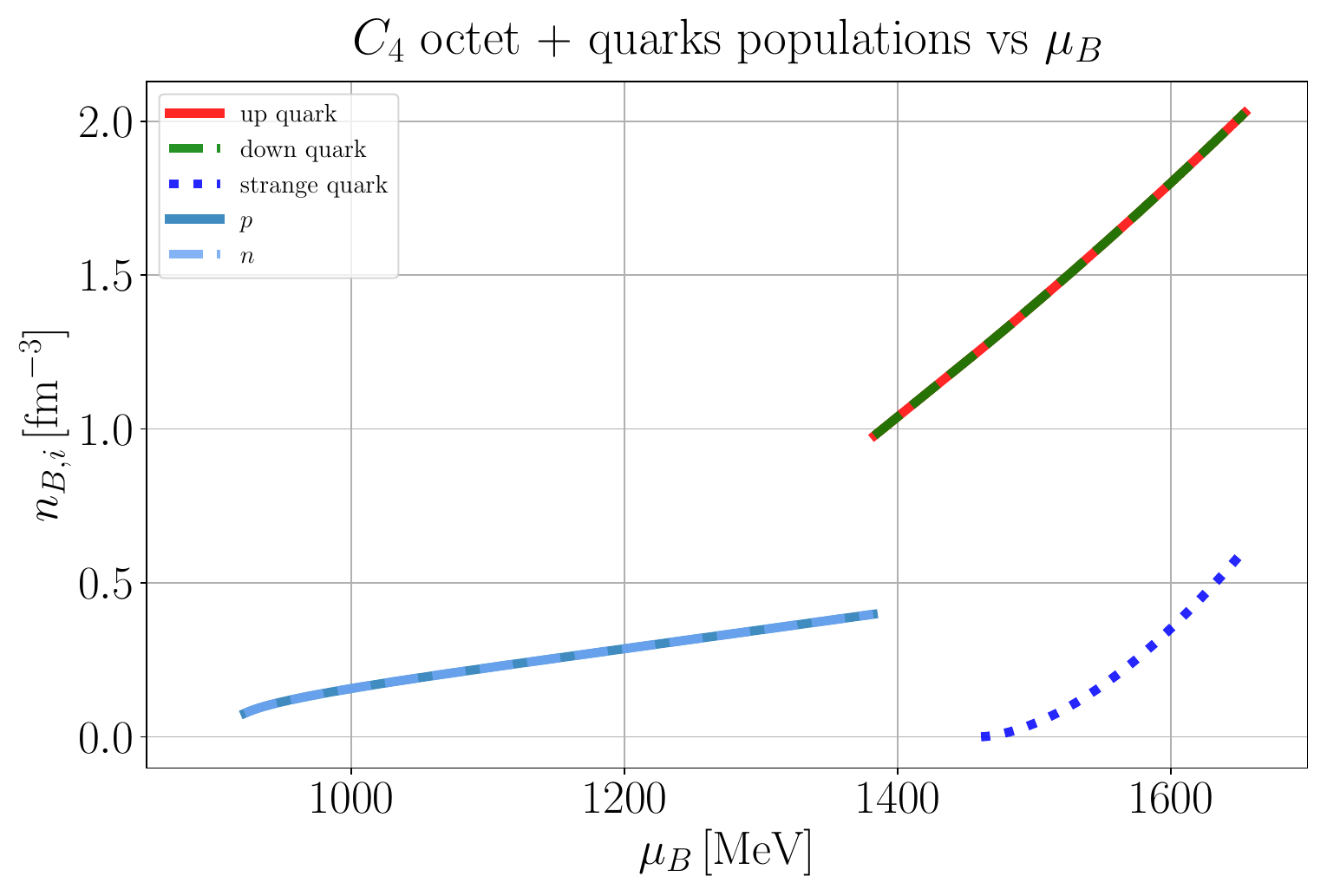}
        \caption{C4 ($\mu_S=\mu_Q=0$) octet + quarks:  particle populations versus baryon chemical potential using stable solutions from \texttt{CMF++}.
        }
\label{fig:2D_C4_hyper_quarks_populations_vs_muB}
\end{figure}

\begin{figure*}
    \centering
    \includegraphics[width=0.9\textwidth]{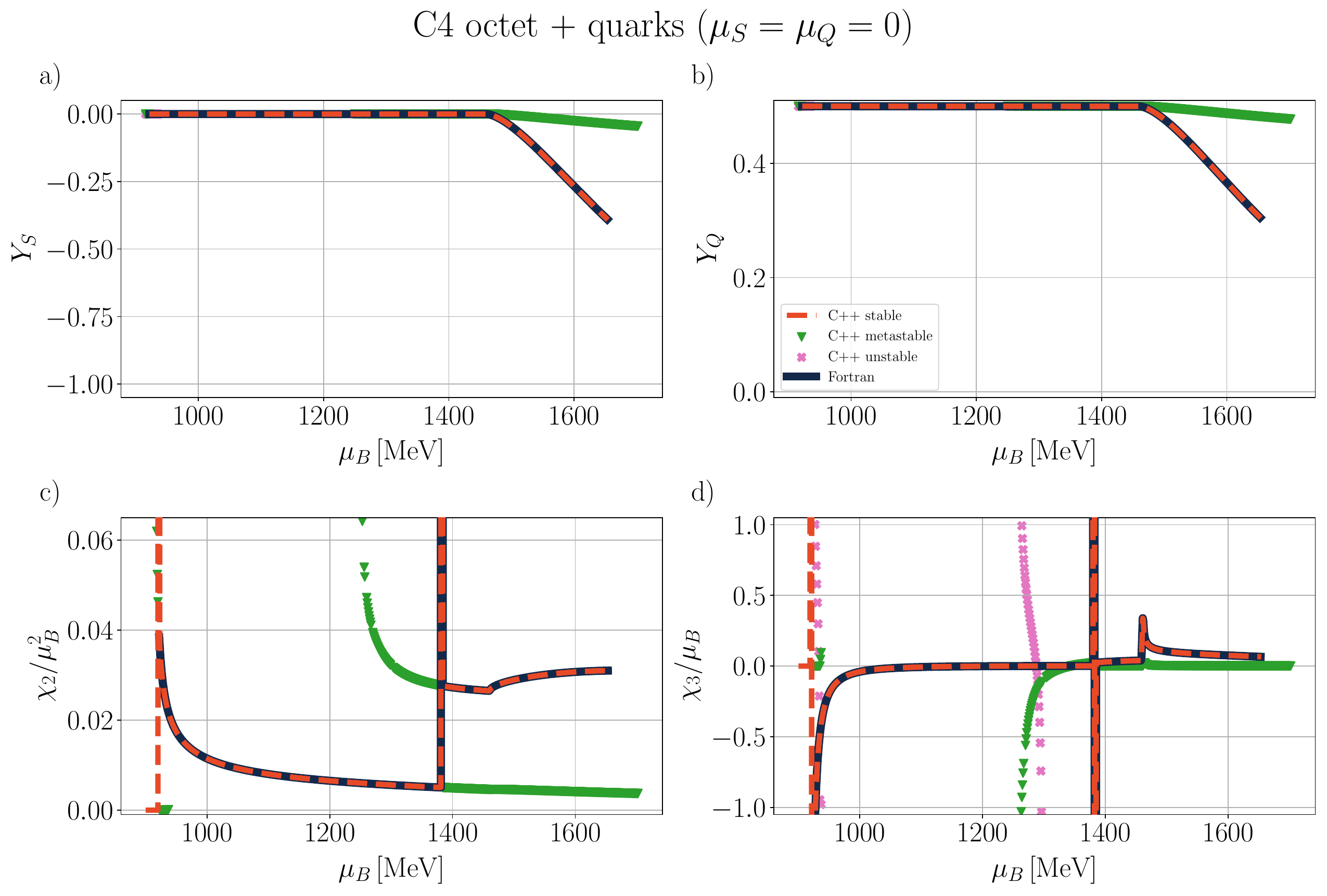}
   \caption{C4 ($\mu_S=\mu_Q=0$) octet + quarks: a) strangeness and b) charge fractions vs baryon chemical potential, c) second and d) third order baryon susceptibilities, all versus baryon chemical potential. Comparison of results from \texttt{Fortran} for stable branches (black solid line) and \texttt{CMF++} for stable (red orange-dashed line), metastable (green upside-down triangles), and unstable (pink x's) branches.  
 }
\label{fig:2D_C4_hyper_quarks_muS_0_muQ_0_dens_and_sus}
\end{figure*}

\Cref{fig:2D_C4_hyper_quarks_muS_0_muQ_0_mf_and_obs} depicts the same set of plots that include mean fields, $\Phi$, and thermodynamic properties as in \Cref{fig:2D_C3_hyper_quarks_muS_0_muQ_0_mf_and_obs}, but for the C4 coupling scheme.
We once again include the information on both the stable, metastable, and unstable phases. 
We also compare the legacy CMF to \texttt{CMF++}. 
We find a strong agreement between the results obtained from \texttt{C++} and \texttt{Fortran} solutions. 

In \Cref{fig:2D_C4_hyper_quarks_populations_vs_muB} the corresponding population plot is shown for the C4 coupling for the combination of octet+quarks.
Finally, in \Cref{fig:2D_C4_hyper_quarks_muS_0_muQ_0_dens_and_sus} the charge and strange fractions are shown as well as the susceptibilities of the pressure. 
In the following, we discuss these three plots and compare and contrast them with the previous C3 coupling that we showed as well. 

The key difference between the C3 and C4 couplings lies in the presence of mixed couplings  $\omega^3 \phi$ and $\omega \phi^3$  in the C4 coupling (\Cref{eq:L_vec}), leading to a different sensitivity to strangeness. 
Due to these changes, there is no stable hadronic phase that includes strange baryons for the C4 couplings at $\mu_S=\mu_Q=0$ (see \Cref{fig:2D_C4_hyper_quarks_populations_vs_muB}).
However, strange quarks do appear in the deconfined quark phase. 

The mean fields associated with strangeness ($\phi$ and $\zeta$) include self-interactions for the $\phi$ meson and the $\zeta$ meson that couples with the non-strange scalar field $\sigma$ (see \Cref{eq:Algebraic_System}), resulting in additional attractive interactions. 
Thus, due to these self-interactions and $\zeta-\sigma$ couplings, even when no strange baryons are present in the hadronic phase, we find that $\phi\neq 0$ and $\zeta/\zeta_0\neq 1$ (see panels b) and d) in \Cref{fig:2D_C4_hyper_quarks_muS_0_muQ_0_mf_and_obs}). 
More specifically due to the $\omega^3 \phi$ and $\omega \phi^3$ coupling terms in C4, $\phi$ has a much larger absolute value in the hadronic phase, see panel d) in \Cref{fig:2D_C4_hyper_quarks_muS_0_muQ_0_mf_and_obs}, compared to C3 in panel d) in \Cref{fig:2D_C3_hyper_quarks_muS_0_muQ_0_mf_and_obs}. 
On the other hand, comparing $\zeta$ in panel b) to C3 in panel b) in \Cref{fig:2D_C3_hyper_quarks_muS_0_muQ_0_mf_and_obs}, we find that $\zeta$ is closer to its vacuum value for C4. 

We only focus on the deconfinement phase transition for C4 because the liquid-gas phase transition has the same properties as C3. 
For the C4 coupling, across panels a) to b) of \Cref{fig:2D_C4_hyper_quarks_muS_0_muQ_0_mf_and_obs}, we observe a jump in the ratio of the scalar mean fields, after which the scalar mean fields continue to decrease. The vector fields become zero at this point as they do not couple to quarks.
Based on the behavior of the mean fields,  we conclude that the phase transition for the C4 coupling shifts to a somewhat lower $\mu_B = 1382.5$  MeV. 
The shift in the phase transition to lower $\mu_B$ is confirmed by the $\Phi$ behavior shown in panel e) where we see the vertical line at $\mu_B = 1382.5$  MeV and the shift is also confirmed later on in the population plot in \Cref{fig:2D_C4_hyper_quarks_populations_vs_muB}. 
The first-order phase transition also appears in panels f) for $P(\varepsilon)$ and g) for $c_s^2(n_B)$ of \Cref{fig:2D_C4_hyper_quarks_muS_0_muQ_0_mf_and_obs}. 
The signatures of the first-order phase transition are similar to C3, even though they appear at a different location.  
C4 has a slightly larger phase transition than C3 (larger jump in $n_B$) because it is a sharper change from protons and neutrons into quarks than if other hadrons had appeared before deconfinement. 

In order to understand strangeness in the coupling C4, let us begin with the population plot in \Cref{fig:2D_C4_hyper_quarks_populations_vs_muB}. 
We see that for the hadronic regime, we only have protons and neutrons (in precisely equal amounts since this is for symmetric nuclear matter) such that there are no strange baryons. 
Thus, in the thermodynamic properties of C4 the lack of strange baryons implies that there is no kink in $c_s^2(n_B)$ in the stable hadron phase (panel g) of \Cref{fig:2D_C4_hyper_quarks_muS_0_muQ_0_mf_and_obs}). 

However, in the quark regime, we see in the population plot in \Cref{fig:2D_C4_hyper_quarks_populations_vs_muB} that strange quarks are present and they play a more significant role compared to the C3 coupling.
Because of the presence of these strange quarks, we see there is a kink in $\zeta$ (panel b) of \Cref{fig:2D_C4_hyper_quarks_muS_0_muQ_0_mf_and_obs})  when the strange quark switches on and then $\zeta$ becomes significantly larger in magnitude in the quark phase as the amount of strangeness increases with $\mu_B$. 
In terms of thermodynamics, we see a kink in $c_s^2(n_B)$ at high $n_B$ when the strange quarks appear (panel g) of \Cref{fig:2D_C4_hyper_quarks_muS_0_muQ_0_mf_and_obs}). 
Additionally, we find that $n_B(\mu_B)$ is significantly steeper in the quark phase (panel h) of \Cref{fig:2D_C4_hyper_quarks_muS_0_muQ_0_mf_and_obs}). 
The influence on $c_s^2$ and $n_B(\mu_B)$ are both consequences of the fact that the strange quarks play a much larger role in the C4 coupling than in the C3 coupling, even though they appear at roughly the same $\mu_B$. 

As discussed above, the C4 coupling does not produce strange hadrons in the hadron phase.
Since strange hadrons soften the EoS, then the C4 coupling has a stiffer EoS at low $n_B$ compared to the C3 coupling (panel f) and g) of \Cref{fig:2D_C4_hyper_quarks_muS_0_muQ_0_mf_and_obs} and \Cref{fig:2D_C3_hyper_quarks_muS_0_muQ_0_mf_and_obs}). 
In fact, the larger $\omega$ value in C4 leads to a stronger repulsive force, that gives us a stiffer $c_s^2$ at low $n_B$. 
The stiffer EoS at low $n_B$ then results in the modeling of more massive neutron stars when finite isospin is included~\cite{Dexheimer:2008ax,Dexheimer:2009hi,Dexheimer:2011pz,Dexheimer:2015qha,Dexheimer:2018dhb,Dexheimer:2020rlp,Dexheimer:2021sxs,Clevinger:2022xzl,Franzon:2015sya,Marquez:2022fzh,Peterson:2023bmr,Hempel:2013tfa,Roark:2018uls,Aryal:2020ocm,Negreiros:2018cho}.
In the quark phase, the $c_s^2(n_B)$ looks very similar between C3 and C4. 
However, because of the stiffer low $n_B$ EoS for C4 (and the fact that $c_s^2$ is a derivative of $P(\varepsilon)$), then that leads to also a stiffer high $n_B$ EoS for C4 as well (even though both of their $c_s^2(n_B)$ look very similar at high $n_B$).

One of the most significant differences between the C4 and C3 (see panels g) of \Cref{fig:2D_C4_hyper_quarks_muS_0_muQ_0_mf_and_obs} and \Cref{fig:2D_C3_hyper_quarks_muS_0_muQ_0_mf_and_obs}) couplings is that the metastable/unstable regimes are significantly different for C3 compared to C4. 
We find that the order parameter $\Phi$ has the same qualitative shape, although the phase transition does take place at a different spot (see panels g) of \Cref{fig:2D_C4_hyper_quarks_muS_0_muQ_0_mf_and_obs} and \Cref{fig:2D_C3_hyper_quarks_muS_0_muQ_0_mf_and_obs}). 
However, when we look at $P(\varepsilon)$ we see that the metastable region is significantly more simplistic for C4, such that there is no longer a chiral phase transition within hadronic metastable phases that existed previously for C3 (see panels f) of \Cref{fig:2D_C4_hyper_quarks_muS_0_muQ_0_mf_and_obs} and \Cref{fig:2D_C3_hyper_quarks_muS_0_muQ_0_mf_and_obs}). 
Thus, C4 does not present a strangeness-dominated metastable hadronic phase (that would be stable if it wasn't for the quarks) as C3 did. 
However, one can see that for the metastable regime in $c_s^2(n_B)$ around $n_B\sim 5 n_{sat}$, there is a small kink, which implies that the strange hadrons would become non-zero if quarks were not considered (see panel g) of \Cref{fig:2D_C4_hyper_quarks_muS_0_muQ_0_mf_and_obs}). 

Due to the absence of hyperons in the baryonic sector (for C4 in the case with $\mu_S = \mu_Q = 0$), both $Y_S$ and $Y_Q$ in panels a) and b) of \Cref{fig:2D_C4_hyper_quarks_muS_0_muQ_0_dens_and_sus} remain $Y_Q=0.5$ and $Y_S=0$ until the strange quarks appear (compare with \Cref{fig:2D_C3_hyper_quarks_muS_0_muQ_0_dens_and_sus}). 
However, once the strange quarks appear after the phase transition, we see that $Y_S$ in the stable phase is significantly larger in overall magnitude for C4 than for C3, demonstrating once again the larger role that strange quarks play in C4.
That effect also leads to a much larger deviation of $Y_Q$ from 0.5 in the quark phase. 
The metastable hadronic phase shown for $Y_S$ and $Y_Q$ also makes it clear what we discussed previously when we looked at the metastable region of $c_s^2(n_B)$ that for the C4 coupling strange baryons appear in the metastable regime.

Finally, we return to the susceptibilities to better understand the order of the phase transitions (panels c) and d) of \Cref{fig:2D_C4_hyper_quarks_muS_0_muQ_0_dens_and_sus}). 
To remind the reader, we already saw jumps in the first-order baryon susceptibility, i.e $n_B$ in \Cref{fig:2D_C4_hyper_quarks_muS_0_muQ_0_mf_and_obs} at $\mu_B = 921.5$ MeV for the liquid-gas phase transition and at $\mu_B = 1382.5$ MeV for the first-order deconfinement phase transition.
Then in the second-order susceptibility, $\chi_2$ we see significantly larger divergences at these points. 
As expected, for $\chi_3$, those divergences are amplified further (higher-order susceptibilities scale with high-orders of the correlation length such that they are more sensitive to phase transitions, see e.g. \cite{Stephanov:2008qz}). 
We can also see the telltale kink in $\chi_2$ when the strange quarks switch on at high $\mu_B$. 
The kink in $\chi_2$ then leads to a divergence in $\chi_3$ such that we also see a third-order phase transition for the C4 coupling within the quark phase. 
This third-order phase transition also has the kink in $c_s^2(n_B)$, providing another example that a third-order phase transition leads to a kink in $c_s^2(n_B)$. $\chi_2$ and $\chi_3$ also show a divergence  for the chiral first-order phase transition  in the hadronic metastable branch.

\subsubsection{C1-C4 with baryon octet + decuplet}

\begin{figure*}
    \centering
    \includegraphics[width=0.8\textwidth]{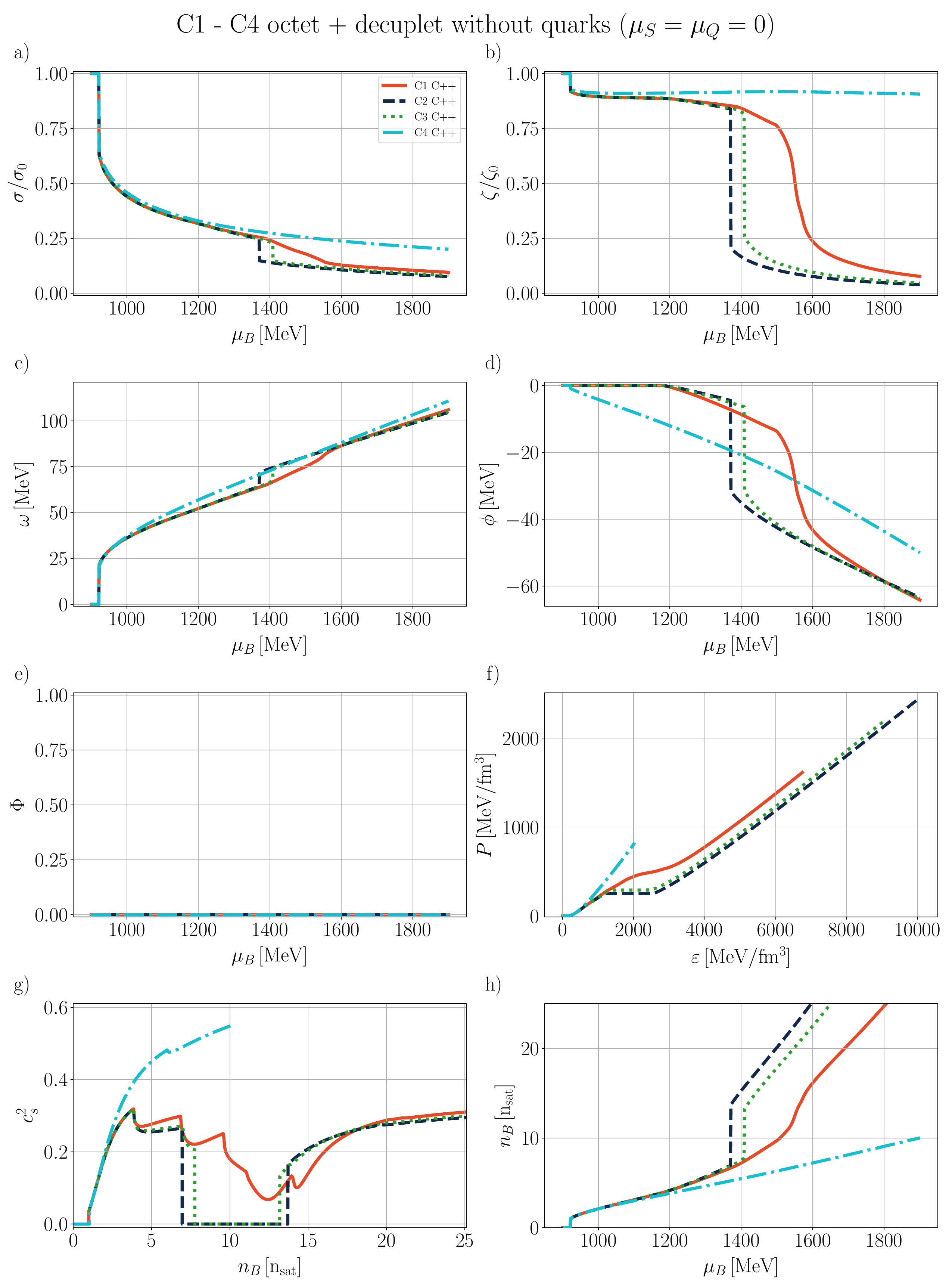}
\caption{C1-C4 ($\mu_S=\mu_Q=0$) octet + decuplet: a) b) scalar meson fields (normalized by vacuum values), c) d) vector meson fields, and e) deconfinement field as a function of baryon chemical potential, f) pressure vs energy density, g) speed of sound vs baryon density, h) baryon density vs baryon chemical potential. Results from \texttt{CMF++} stable solutions for C1 (red-orange solid line), C2 (black dashed line), C3 (green dotted line), and C4 (cyan dash-dotted line).}
\label{fig:2D_all_decuplet_hyper_noquarks_muS_0_muQ_0_mf_and_obs}
\end{figure*}

\begin{figure*}
    \centering
    \begin{subfigure}[b]{0.5\textwidth}
        \centering
        \includegraphics[width=\textwidth]{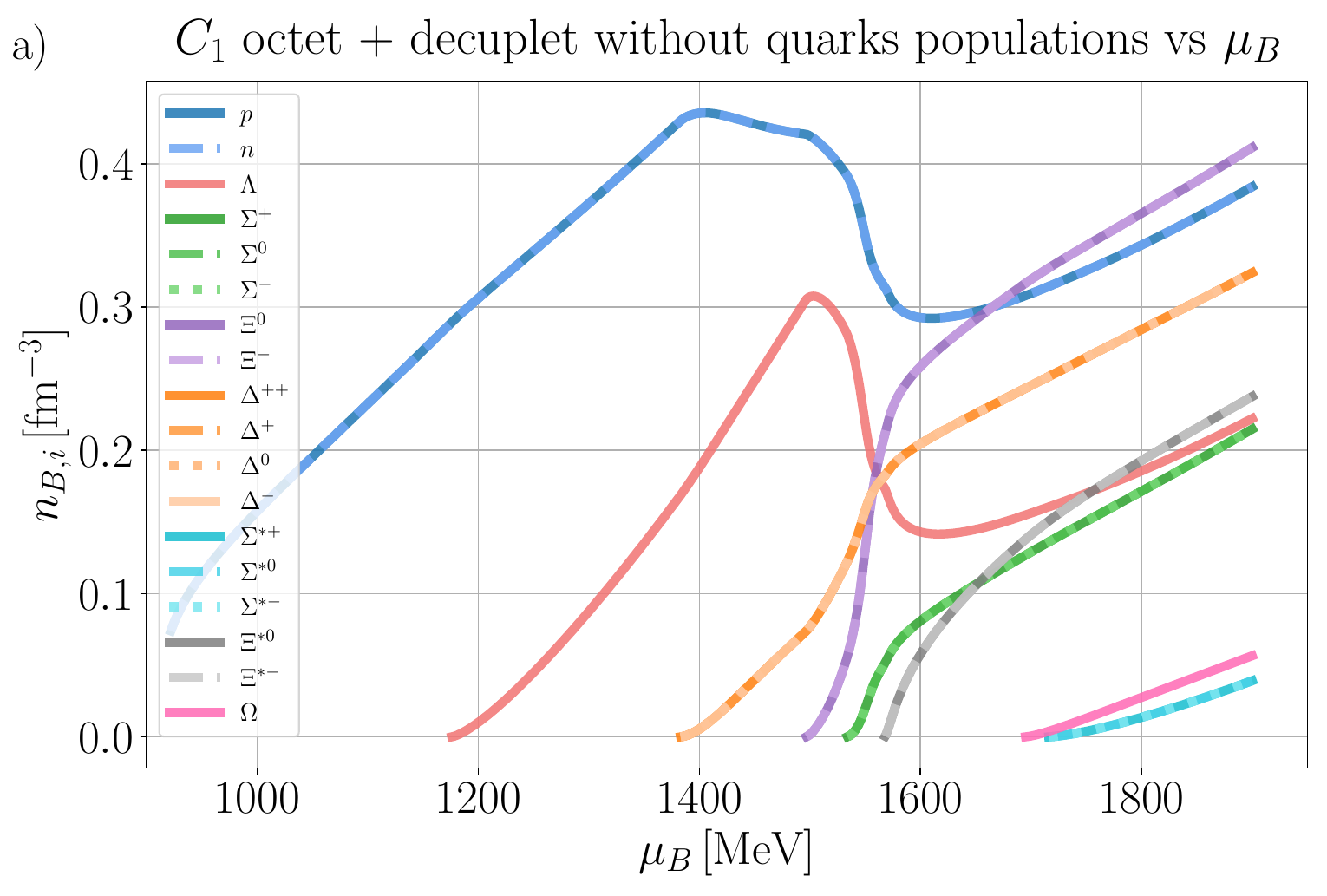}
    \end{subfigure}%
    \begin{subfigure}[b]{0.5\textwidth}
        \centering
        \includegraphics[width=\textwidth]{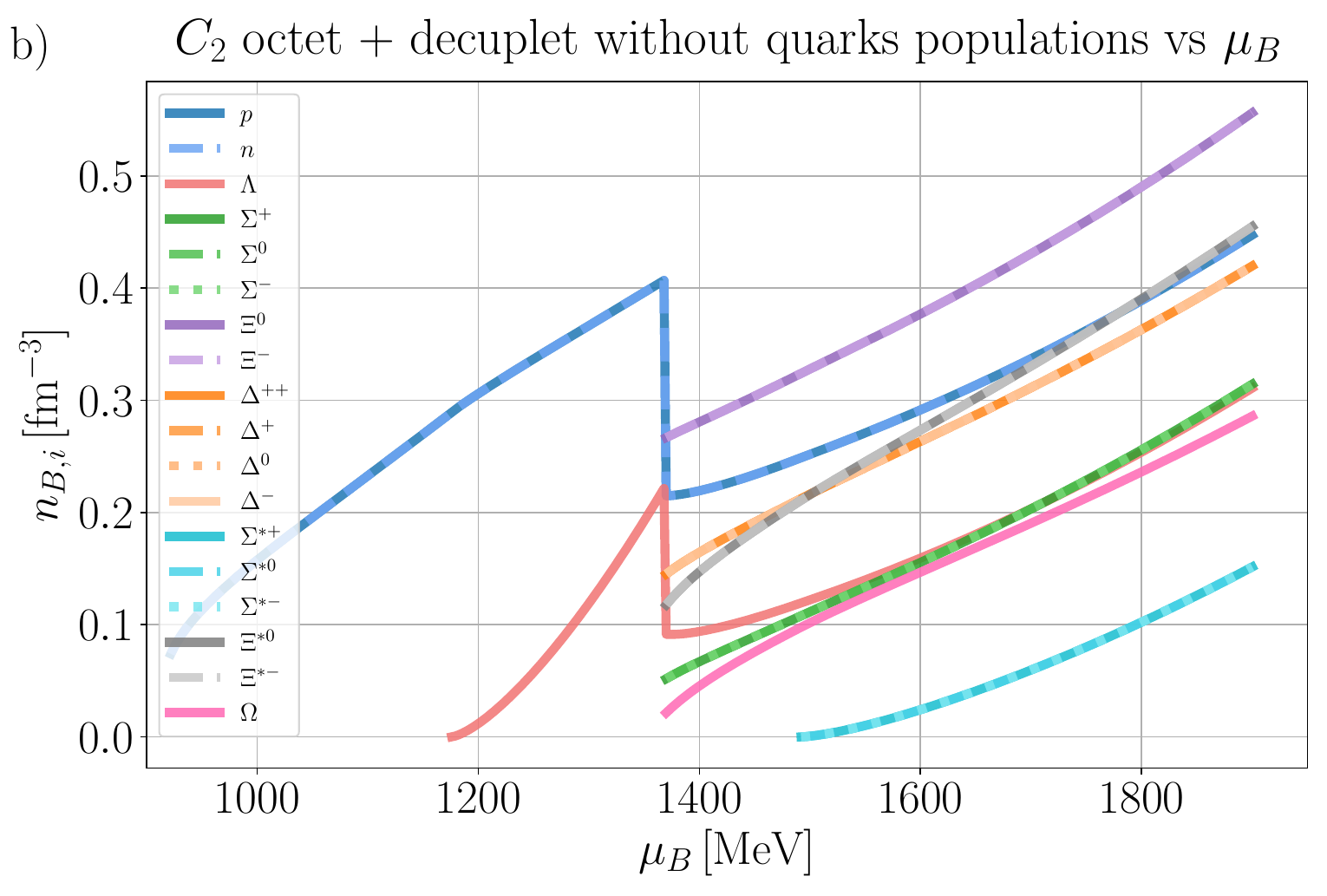}
    \end{subfigure}
    
    \begin{subfigure}[b]{0.5\textwidth}
        \centering
        \includegraphics[width=\textwidth]{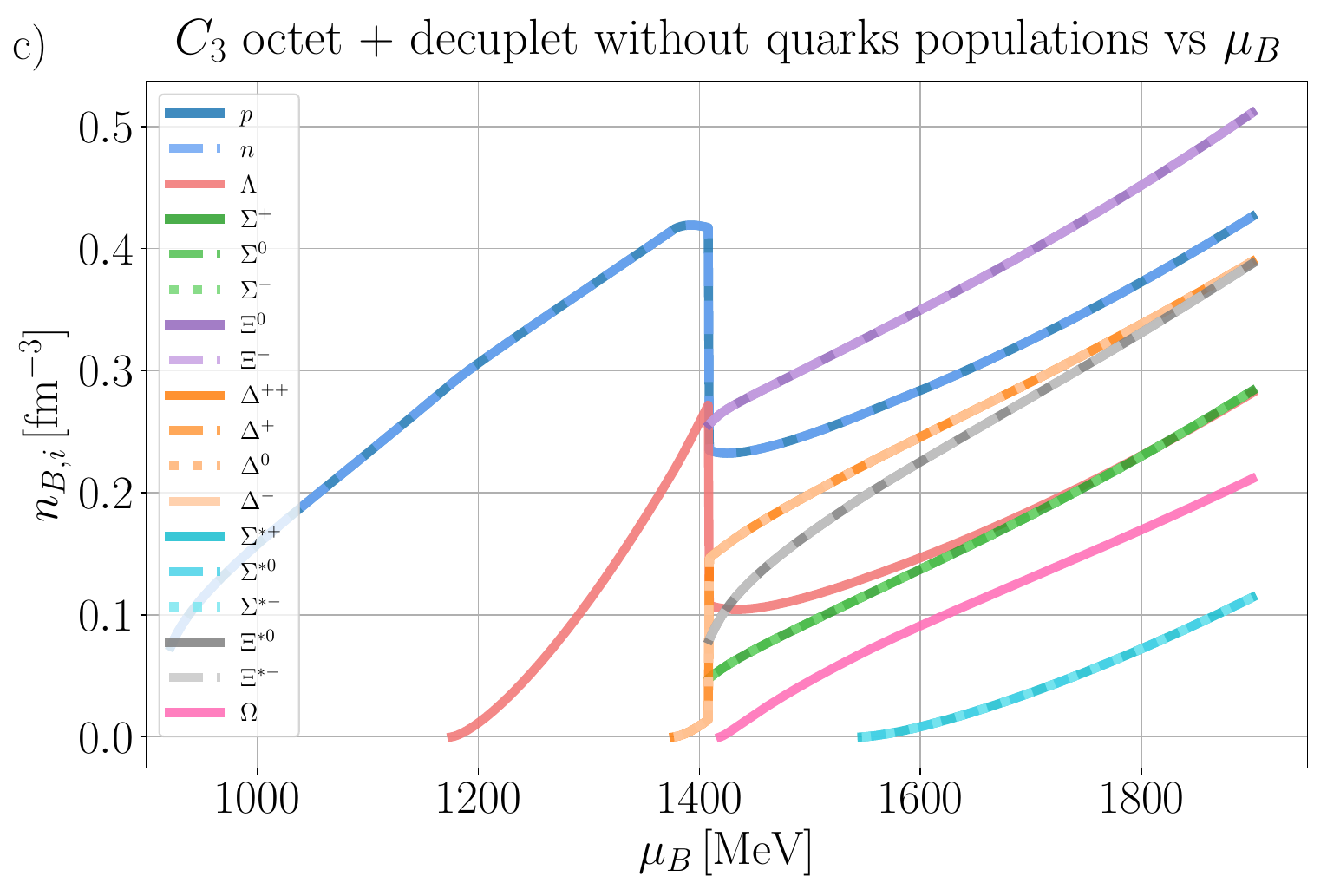}
    \end{subfigure}%
    \begin{subfigure}[b]{0.5\textwidth}
        \centering
        \includegraphics[width=\textwidth]{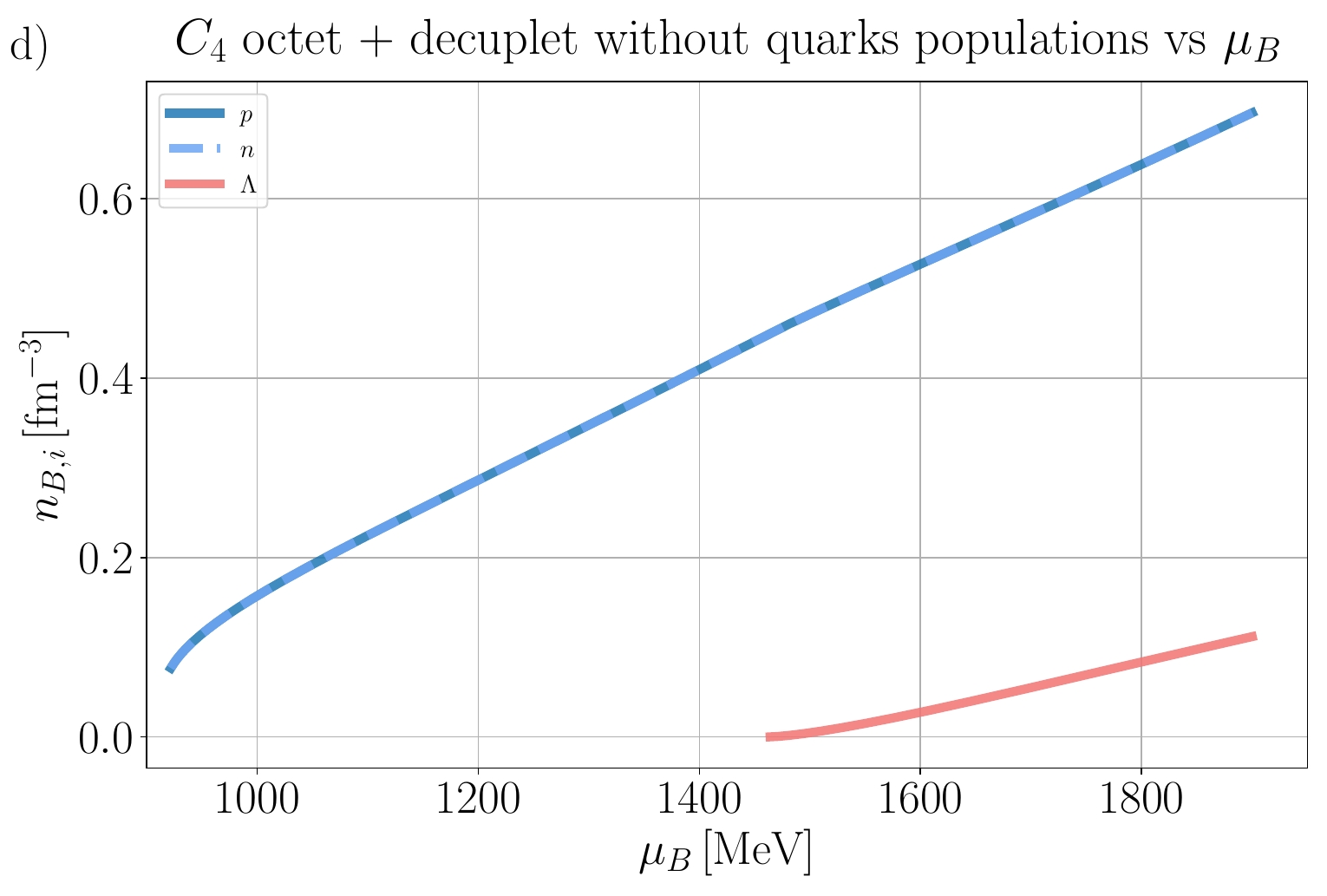}
    \end{subfigure}
    
    \caption{C1-C4 ($\mu_S=\mu_Q=0$) octet + decuplet: particle populations versus baryon chemical potential using stable solutions from \texttt{CMF++}. C1 is shown in panel a), C2 in panel b), C3 in panel c), and C4 in panel d). }
    \label{fig:2D_C1-C4_default_decuplet_noquarks_populations_vs_muB}
\end{figure*}

\begin{figure*}
    \centering
    \includegraphics[width=0.9\textwidth]{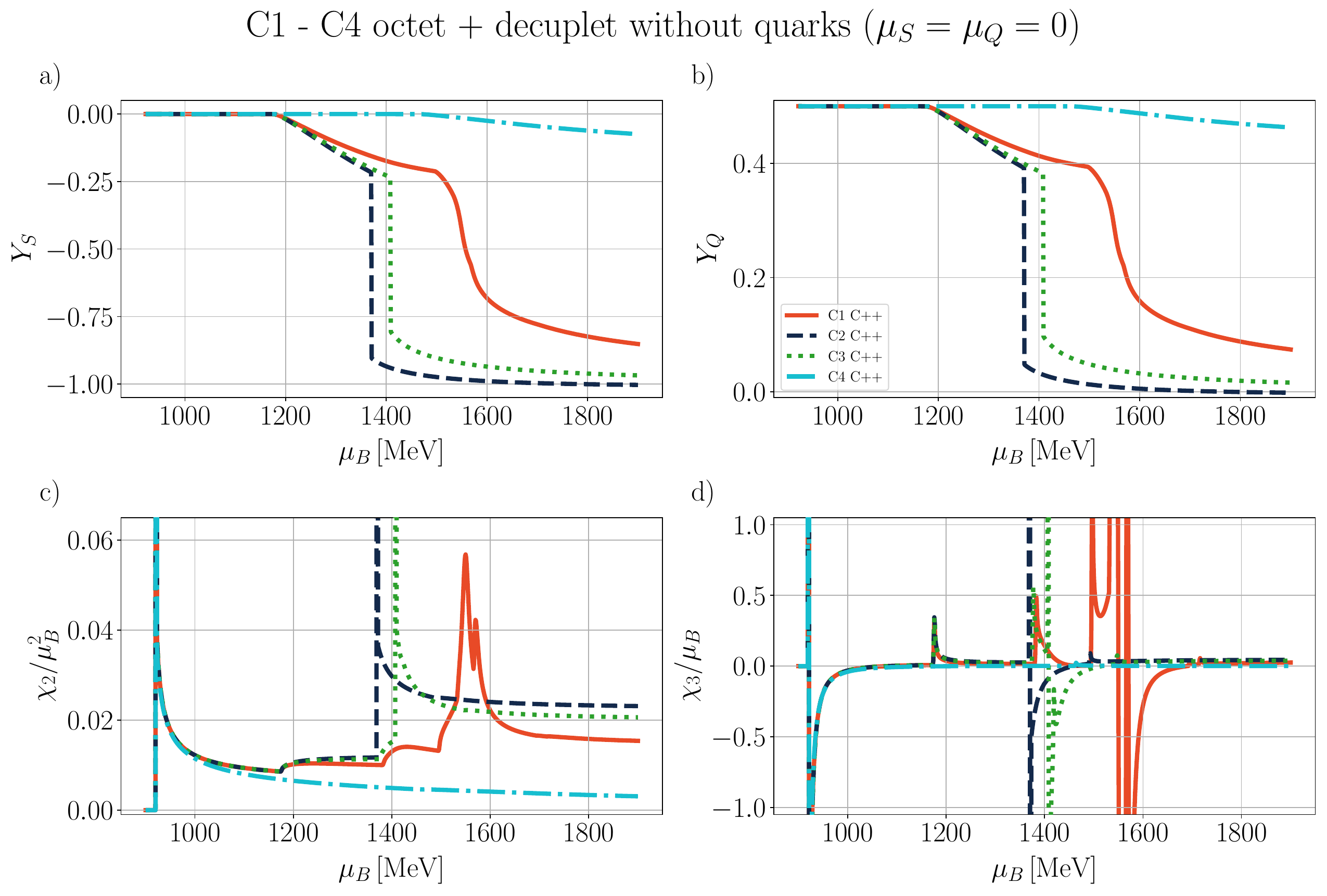}
\caption{C1-C4 ($\mu_S=\mu_Q=0$) octet + decuplet: a) strangeness and b) charge fractions vs baryon chemical potential, c) second and d) third order baryon susceptibilities, all versus baryon chemical potential. Results from \texttt{CMF++} stable solutions for C1 (red-orange solid line), C2 (black dashed line), C3 (green dotted line), and C4 (cyan dash-dotted line).}
\label{fig:2D_all_decuplet_hyper_noquarks_muS_0_muQ_0_dens_and_sus}
\end{figure*}

Up until now, we have considered a combination of the baryon octet and quark phases.  
Here we explore the alternative where no quark phase is present but a larger, more complex hadronic phase is possible wherein both the baryon octet and decuplet are possible. 
Thus, in this scenario, as $\mu_B$ increases we anticipate that a wealth of new baryonic states are switched on. 
Since the decuplet includes $\Delta$'s, the new baryons do not necessarily carry strangeness, but maybe light states as well. 
Note that the appearance of first $\Delta$'s vs hyperons depends on the couplings and parameters in the CMF module. 

At this time, the quark couplings have not yet been fitted while also including the decuplet baryons. 
We have performed some initial testing with the possibility of octet+decuplet+quarks and found that in that case (within the current parametrization) the quarks appear at very large $\mu_B$'s, outside the regime of neutron stars. Thus, for $\mu_Q=\mu_S=0$ we focus on the octet and decuplet without quarks case. We come back later to discuss the case with octet, decuplet, and quarks with $\mu_Q\ne0$ and/or $\mu_S\ne0$.

In \Cref{fig:2D_all_decuplet_hyper_noquarks_muS_0_muQ_0_mf_and_obs} we show the mean fields, $\Phi$, and thermodynamic properties for C1-C4 for the case of the octet+decuplet at $\mu_S=\mu_Q=0$.
While we no longer show the comparison between \texttt{C++} and legacy \texttt{Fortran} versions of CMF, we note that we have checked their consistency for all following plots (see \Cref{sec:C1_C4_baryon_decuplet_muS0_muQ0}). 
However, we do not show the comparisons here to improve the readability of the plots.
Finally, we note that the C4 coupling reproduces for the chosen d.o.f. a much lower density and energy density than the other couplings (for a given $\mu_B$, so it only appears at the beginning of the plots that use those quantities as the x-axis).
In the presence of both the baryon octet and decuplet, the differences between C1-C4 are overall more significant than before. 
Previous work \cite{Dexheimer:2015qha} has shown that C1-C3 produce nearly identical results in the case of just the baryon octet+quarks (although there are small subtle differences in $Y_S$ and $Y_Q$ close to the phase transition from $C1$ vs $C2-C3$). 
Now with the decuplet present, we find that $C1-C4$ all lead to distinct solutions that can be seen quite clearly in both the mean-field mesons and the thermodynamics. 

In panels a)-d) of \Cref{fig:2D_all_decuplet_hyper_noquarks_muS_0_muQ_0_mf_and_obs} we find general trends in the mean-field mesons that are the same as what we saw for the baryon octet+quark configuration i.e. $\sigma$, $\zeta$, and $\phi$ always decrease with $\mu_B$ while $\omega$ increases.
However, their exact behavior within these general trends can differ quite significantly and present jumps/kinks as different particles switch on. 
The C1 and C4 couplings are simpler because they do not contain first-order chiral phase transitions from a light hadronic phase into a strangeness or $\Delta$-dominated hadronic phase. 
Thus, we find that the C1 and C4 coupling schemes change more smoothly as one increases $\mu_B$. 
Specifically, the C4 coupling scheme even with the baryon octet+decuplet, has significantly fewer contributions from hadrons that are not protons and neutrons (see also panel d) of \Cref{fig:2D_C1-C4_default_decuplet_noquarks_populations_vs_muB} for the particle populations).  
Because C4 is dominated by protons and neutrons, it leads to a significantly stiff low $n_B$ EoS, and the mean fields have a smooth behavior across all $\mu_B$.
In contrast, $C1$ has a relatively smooth behavior in $\sigma$ and $\omega$, but there is a more sudden change (not quite first order) in $\zeta$ and $\phi$, where we can see sharp drops in their values around $\mu_B\sim 1500-1600$ MeV (again, in \Cref{fig:2D_all_decuplet_hyper_noquarks_muS_0_muQ_0_mf_and_obs}). 
The increase in the magnitude of $\zeta$ and $\phi$ still occurs across a range of $\mu_B$, such that we would not classify it as a first-order phase transition, but it appears to be a sharper higher-order phase transition (we discuss the exact order in the susceptibilities later on). 

We now come to the C2 and C3 couplings, which demonstrate quite interesting behavior.  For every panel a)-d) in \Cref{fig:2D_all_decuplet_hyper_noquarks_muS_0_muQ_0_mf_and_obs}, we see a clear vertical line around $\mu_B = 1370.5$ MeV for C2 and $\mu_B = 1408.5$ MeV for C3. 
For the mesons, these vertical lines translate to a small jump in $\omega$ and $\sigma$, and a very significant jump in $\zeta$ and $\phi$.
Thus, what we find is that we have a first-order chiral phase transition from a light baryon-dominated regime to a strange and $\Delta$ baryon-dominated regime.
We later explore the implications of what baryons switch on across this phase transition when we explore the population plots. 
This first-order chiral phase transition in the hadronic phase is the same phase transition that we saw previously for C3 couplings when quarks were present in \Cref{fig:2D_C3_hyper_quarks_muS_0_muQ_0_mf_and_obs,fig:2D_C4_hyper_quarks_muS_0_muQ_0_mf_and_obs} that fell in the metastable regime. 

Panel e) in \Cref{fig:2D_all_decuplet_hyper_noquarks_muS_0_muQ_0_mf_and_obs} confirms the absence of quarks, as indicated by the $\Phi$ field remaining zero across the entire $\mu_B$ range.
Thus, the first-order chiral phase transition that we see in \Cref{fig:2D_all_decuplet_hyper_noquarks_muS_0_muQ_0_mf_and_obs} arises entirely from the couplings, effective masses, etc. and is not driven by the explicit order parameter $\Phi$ that we have built into our model. 

Panel f) in \Cref{fig:2D_all_decuplet_hyper_noquarks_muS_0_muQ_0_mf_and_obs} presents the EoS for all coupling cases. 
We see significant differences from C4 compared to the other couplings.  
The C4 couplings lead to a very stiff EoS that steadily rises with $\varepsilon$.  
If we then compare this to the speed of sound in panel g), we see that we also reach the largest $c_s^2$ with C4, such that $c_s^2>0.5$.  
One can see a small kink in $c_s^2$ for C4, which is our first hint that other particles beyond protons and neutrons switch on for C4, but it is a rather subtle effect. 
It is clear then, looking at $n_B(\mu_B)$ in panel h), that the lack of other hadrons that switch on for C4 leads to the lowest corresponding $n_B$ for a given $\mu_B$. 
In other words, the fewer hadronic species possible at a given $\mu_B$ implies a lower $n_B$. 

Let us now return to the EoS in panel f) in \Cref{fig:2D_all_decuplet_hyper_noquarks_muS_0_muQ_0_mf_and_obs} and explore the C1 coupling that appears to fall in between what is seen for C4 and C1-C2. 
Recall that the C1 coupling did not have a first-order chiral phase transition into a strangeness-dominated hadronic phase. 
However, it did demonstrate steep changes in its mean-field mesons, potentially indicating a higher-order phase transition. 
In $P(\varepsilon)$ we can see signs of this higher-order phase transition by the slight flattening in the pressure, but  
$P(\varepsilon)$ never reaches a true plateau like one would anticipate for a first-order phase transition. 
If we then look at the speed of sound in panel g), we can see the quite interesting non-monotonic behavior in C1.  
The kinks in $c_s^2$ correspond to the appearance of new baryons at a given $n_B$. 
Altogether, we find 5 distinct peaks in $c_s^2$ vs $n_B$ for C1, which likely indicates the presence of 5 new hadronic species switching on at certain $n_B$'s.
Additionally, we see a softening where $c_s^2$ drops to $c_s^2\rightarrow 0.1$ but does not hit exactly $0$, as one would expect for a first-order phase transition. 
Finally, this behavior leads to a steeper rise in $n_B(\mu_B)$ in panel h) because of the new hadronic degrees of freedom that switch on. 

Finally, we come to C2-C3 which has shown indications that a first-order phase transition occurred within the mean-field mesons,  even though the deconfinement order parameter is always $\Phi=0$.
At low $n_B$ for C2 and C3 we see a  peak in $c_s^2$ in panel g) of \Cref{fig:2D_all_decuplet_hyper_noquarks_muS_0_muQ_0_mf_and_obs} around $n_B\sim 2.5 n_{sat}$ that occurs at the same location as what is seen for C1. 
Thus, we likely have at least one new particle switching on before the first-order phase transition occurs at higher $n_B$. 
Looking at the $P(\varepsilon)$ relationship in panel f)  we see that there is a plateau consistent with a first-order phase transition that then translates into a region of $c_s^2\rightarrow 0$ in panel g) and a jump in $n_B(\mu_B)$ in panel h) for C2-C3. 
Following the first-order phase transition, we find a very steep increase in $n_B(\mu_B)$ in panel h), which is consistent with what we understood before - new degrees of freedom leads to a larger $n_B$ at a fixed $\mu_B$.

The evidence is quite clear that we have a first-order chiral phase transition, but that the phases of matter are always hadronic (and stable) on both sides of the phase transition.
It is also interesting to compare this chiral phase transition to what we saw previously for the baryon octet into quarks in \Cref{fig:2D_C3_hyper_quarks_muS_0_muQ_0_mf_and_obs} and \Cref{fig:2D_C4_hyper_quarks_muS_0_muQ_0_mf_and_obs}. 
The general behavior of the C2-C3 EoS is very similar for the deconfinement phase transition vs the chiral phase transition. 
In fact, the chiral phase transitions for C2-C3 take place at nearly the same location as the deconfinement phase transition, which is why in \Cref{fig:2D_C3_hyper_quarks_muS_0_muQ_0_mf_and_obs} we saw a phase transition within the metastable regime. But note that in the hadronic case, the first order phase transition is associated with chiral symmetry, since the scalar condensates go down across the phase transition (panel a) of \Cref{fig:2D_all_decuplet_hyper_noquarks_muS_0_muQ_0_mf_and_obs}).
For completeness, in  \Cref{sec:C1_C4_baryon_decuplet_muS0_muQ0}, the \Cref{fig:12_stable+metastable+unstable} shows the stable, metastable, and unstable solutions not shown in \Cref{fig:2D_all_decuplet_hyper_noquarks_muS_0_muQ_0_mf_and_obs}.

Now we discuss the population plots for C1-C4 in \Cref{fig:2D_C1-C4_default_decuplet_noquarks_populations_vs_muB}.
As we discussed, the C4 coupling has the simplest population, so let us begin with that. 
We find that for the C4 coupling the system is heavily dominated by just protons and neutrons, although $\Lambda$ baryons appear at high $\mu_B=1464$ MeV. 
However, their contribution is only a very small fraction of the baryon number. 

Let us now discuss the C1 coupling, for which previously we suspected 5 new particles switching on due to the peaks in $c_s^2$. 
We also previously found that there was no first-order phase transition within the hadronic phase. 
When we look at the population plot in panel a) of \Cref{fig:2D_C1-C4_default_decuplet_noquarks_populations_vs_muB}, we find confirmation of both of these facts. 
The particles slowly turn on (there is no distinct jump in their population numbers) while the influence of the proton/neutron begins to decrease around $\mu_B\sim 1400$ MeV. 
Additionally, we can confirm that new particles (all degeneracies at once) switch on in the (increasing $\mu_B$) order of: first $\Lambda$'s, second $\Delta$'s, third $\Xi^0$ and $\Xi^-$, fourth $\Sigma^{\pm}$ and $\Sigma^0$, fifth $\Xi^{*0}$ and $\Xi^{*-}$.
 Thus, it is true that 5 different species of particles turn on at those peaks, but each peak sometimes includes multiple versions of that particle with different electric charges/isospins.
 We find that, once the $\Delta$ baryons switch on, we reach the peak in protons/neutrons such that they decrease as the contribution of $\Delta$'s increases.
 We also observe that as the $S=-2$ baryons switch on, we see the influence of $\Lambda$'s wane as well. 

For the C2 and C3 coupling (panles b) and c) of \Cref{fig:2D_C1-C4_default_decuplet_noquarks_populations_vs_muB}),  we observe that in the low $n_B$ region the hadron phase consists of predominately protons, neutrons, and $\Lambda$'s that appear a bit before  $\mu_B<1200$ MeV. 
For the C3 coupling, we even see a small contribution of $\Delta$'s just before the first-order phase transition appears into the strangeness and $\Delta$-dominated phase. 
We then see a first-order chiral phase transition at  $\mu_B = 1370.5$ MeV for C2 and at  $\mu_B = 1408.5$ MeV for C3 to a different mixture of hyperons and baryon decuplet (roughly going from top to bottom in density): first $\Xi$, nucleons, $\Delta$, $\Xi^*$, $\Lambda$, and $\Sigma$, then $\Omega$, then $\Sigma^*$. 
C2 and C3 have slightly different mixtures of all these baryons after the first-order phase transition, however, the key result is that $S=-2$ are the most populous baryon in this strangeness-dominated phase and even $S=-3$ $\Omega^-$ baryons are allowed.

Panels a) and b) of \Cref{fig:2D_all_decuplet_hyper_noquarks_muS_0_muQ_0_dens_and_sus} depict the evolution of $Y_S$ and $Y_Q$ with respect to $\mu_B$. 
Because the C4 coupling has the least populous strange baryons, $Y_S$ and $Y_Q$ remain at 0 and 0.5, respectively, until the emergence of the $\Lambda$ hyperon at large $\mu_B$. 
Conversely, for C1, $Y_S$ and $Y_Q$ remain constant until the first strange particle switches on  ($\Lambda$),  where $Y_S$ smoothly decreases together with $Y_Q$. 
As more new hadrons switch on, $Y_S$ and $Y_Q$ both begin to drop more rapidly. 
The largest drop occurs around $\mu_B\sim 1500$ MeV where the $S=-2$ becomes relevant and rapidly increases in importance. 
Finally, for C2-C3 we see a vertical line in $Y_S$ and $Y_Q$, where the chiral first-order phase transition occurs.
The overall magnitude of $Y_S$ is the largest for C2-C3 because even $S=-3$ hadronic states are switched on. 

The susceptibilities in panels c) and d) illustrate the location of the first-order chiral phase transition in C2 and C3 at $\mu_B = 1370.5$ MeV and $\mu_B = 1408.5$ MeV, respectively (in addition to the liquid-gas phase transition). 
For C1 we see an especially surprising effect.
In $\chi_2$ we have two almost second-order phase transitions that appear in the range of $\mu_B\sim 1500-1600$ MeV that relates to $\Sigma$'s and $\Xi$'s rapidly switching on. 
These (nearly) second-order phase transitions lead to the softening in $c_s^2$ that was seen earlier in panel g) of \Cref{fig:2D_all_decuplet_hyper_noquarks_muS_0_muQ_0_mf_and_obs}. 
We specify that these are  ``nearly'' second-order phase transitions because we do not see $c_s^2\rightarrow 0$ and it does not appear that $\chi_2$ is quite diverging, although it looks very close to that behavior. 
$\chi_3$ in panel d) confirms that the other phase transitions that we saw in C1 are all of third-order. 
We do not find any phase transitions in C4 (beyond the liquid-gas phase transition).
For completeness, in \Cref{sec:C1_C4_baryon_decuplet_muS0_muQ0}, \Cref{fig:14_stable+metastable+unstable} shows the stable, metastable, and unstable solutions not shown in \Cref{fig:2D_all_decuplet_hyper_noquarks_muS_0_muQ_0_dens_and_sus}.

\subsection{$\left\{\mu_S \neq 0,\mu_Q=0\right\}$}
\label{sec:muBN0_muQ0_muSN0}

\begin{figure*}
    \centering
    \includegraphics[width=0.9\textwidth]{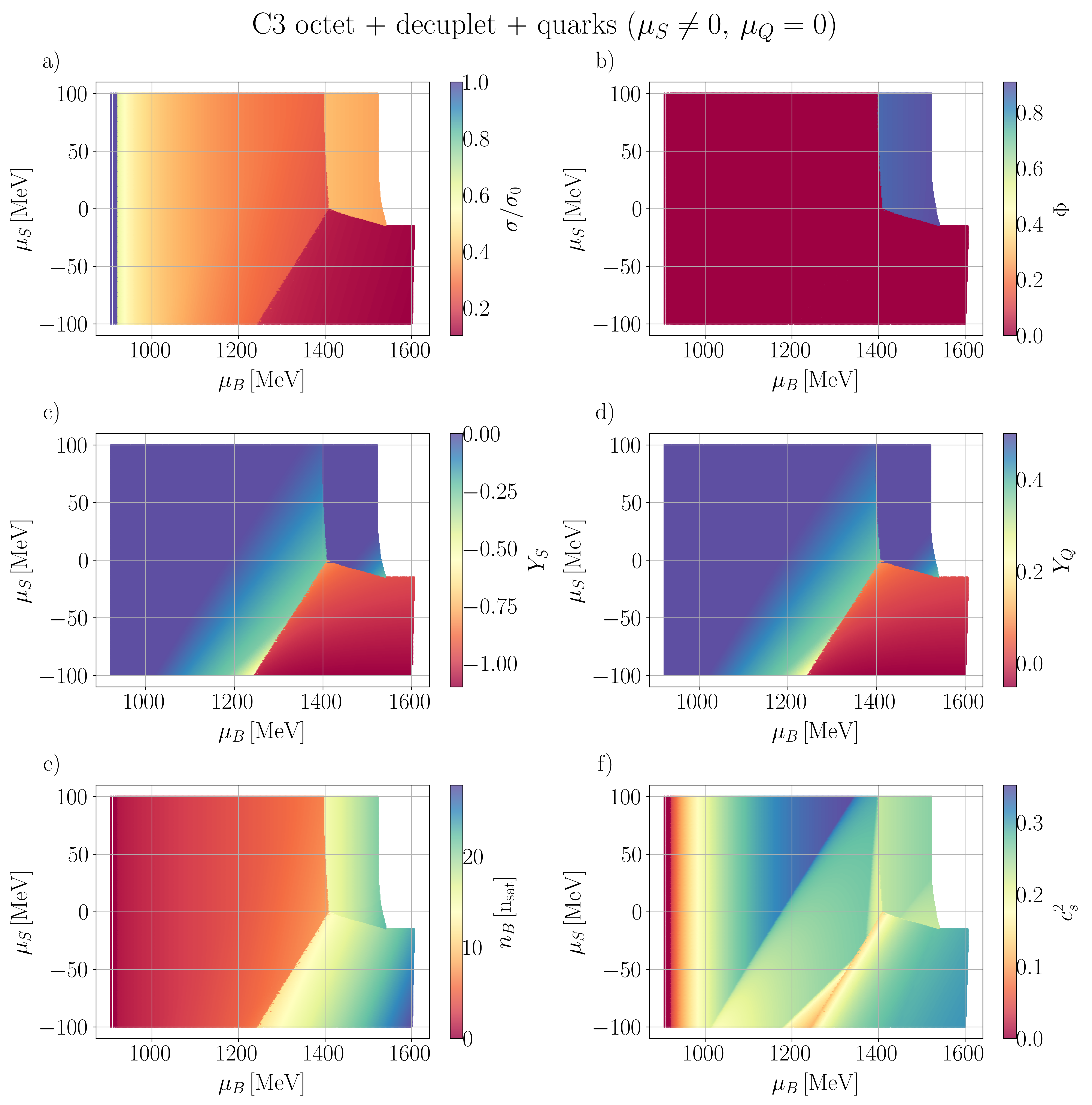}
    \caption{ C3 ($\mu_S \neq 0, \mu_Q=0$) octet + decuplet + quarks: a) scalar meson field $\sigma$ normalized by vacuum value, b) deconfinement field $\Phi$, c) strangeness fraction, d) charge fraction, e) baryon density, and f) speed of sound as functions of baryon and strange chemical potentials.}
\label{fig:3D_C3_decuplet_hyper_quarks_muS_N0_muQ_0_panel}
\end{figure*}

\begin{figure*}
    \centering
    \includegraphics[width=0.9\textwidth]{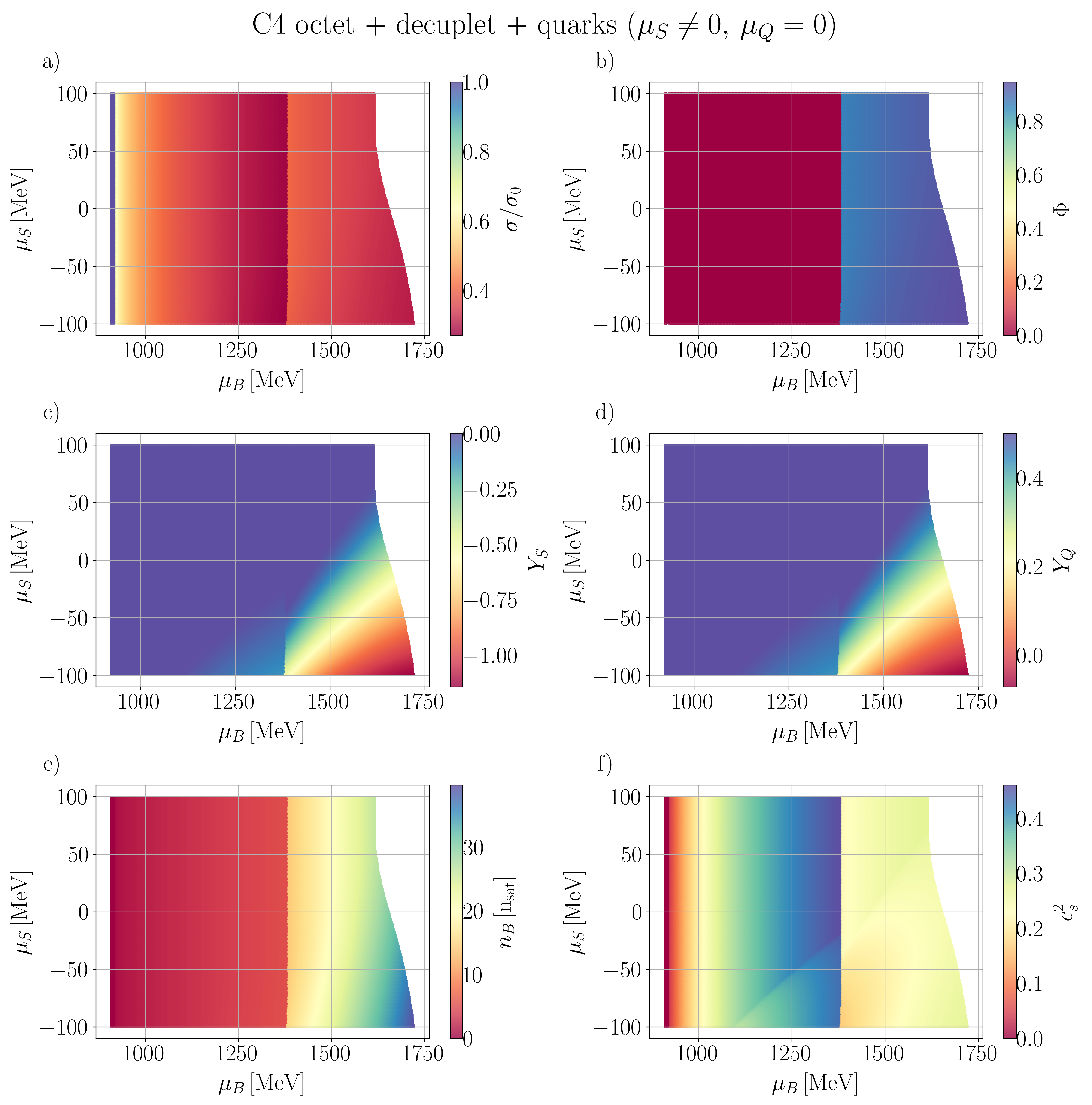}
    \caption{C4 ($\mu_S \neq 0, \mu_Q=0$) octet + decuplet + quarks: a) scalar meson field $\sigma$ normalized by vacuum value, b) deconfinement field $\Phi$, c) strangeness fraction, d) charge fraction, e) baryon density, and f) speed of sound as functions of baryon and strange chemical potentials.}
    \label{fig:3D_C4_decuplet_hyper_quarks_muS_N0_muQ_0_panel}
\end{figure*}

In the following discussion, we analyze the role of the strangeness chemical potential $\mu_S$ while keeping the charge potential fixed, $\mu_Q=0$. 
This scenario is similar to what would be seen in low-energy heavy-ion collisions (although they are typically at finite $T$) because there are local fluctuations of both baryon charge and strangeness such that fluctuations in $\mu_B,\mu_S$ also appear, but the system is nearly isospin symmetric such that $\mu_Q$ (the amount of isospin asymmetry depends on the choice of initial colliding nuclei) is close to zero.

Since the strangeness number is negative (see \Cref{tab:particles}), a positive $\mu_S$ at large $\mu_B$ leads to a smaller magnitude of $Y_S$, and a negative $\mu_S$ leads to a larger magnitude of $Y_S$. 
Nevertheless, a more positive $\mu_S$ could have no effect on the system in the case that there was not enough energy for strange particles to appear in the first place. 
In principle, an extraordinarily large $\mu_S> \mu_B$ would lead to the preference for anti-strange particles, however, this is not a limit that is relevant to neither heavy-ion collisions nor neutron stars, so we do not explore it here. 
In astrophysics, $\mu_S$ is usually set to zero. 
In heavy-ion collisions, the system is globally strangeness neutral and if one is calculating quantities related to the entire system, $\mu_S$ can easily reach up to a third of $\mu_B$ to ensure strangeness neutrality. 
That being said, local fluctuations of strangeness absolutely exist and have been measured experimentally (in fact, strange particles account for roughly $10\%$ of the final particles produced at high energies \cite{ALICE:2016fzo}, it is just that they are exactly balanced by particles that carry anti-strangeness).
Thus, it is possible that one subsection of the fluid experiences $\mu_S<0$ whereas another section experiences $\mu_S>0$ (see e.g.\ \cite{Plumberg:2024leb}).

\subsubsection{C3 and C4 with baryon octet + decuplet + quarks}

We now focus on the C3 and C4 coupling schemes, which are shown using 3D density plots in~\Cref{fig:3D_C3_decuplet_hyper_quarks_muS_N0_muQ_0_panel} and~\Cref{fig:3D_C4_decuplet_hyper_quarks_muS_N0_muQ_0_panel}, respectively, and allow the presence of the baryon octet, decuplet, and quarks. 
In these figures, only the stable solutions of \texttt{C++} results are shown. 
Let us now explain how we can interpret the 3D plots. 
The x-axis corresponds to the baryon chemical potential and the y-axis corresponds to the strange chemical potential. 
Then, each plot uses the color scheme to depict a given variable's value, with its range indicated on the right side of the image. 
Panel a) shows the ratio of the $\sigma$ mean-field to its vacuum value, b) shows the deconfinement order parameter $\Phi$, panels c) and d) show the strangeness and charge fractions, panel e) depicts the ratio of baryon density to the saturation density and panel f) the speed of sound squared in units of the speed of light.

For both C3-C4 in the light hadronic phase (not strangeness or $\Delta$ dominated), we find that the $\sigma$ field is always maximum at low $\mu_B$ and then continuously decreases with $\mu_B$ (regardless of the value in $\mu_S$, as discussed in \Cref{sec:muBN0_muQ0_muS0}. 
At some point, a discontinuity appears that we will discuss later, but then a new phase of matter appears at large $\mu_B$. 
The nature of this new phase of matter at large $\mu_B$ depends on both the coupling and the values of $\mu_S$.

For the C4 coupling, the new phase of matter is a quark phase wherein within the quark phase the $\sigma$ (panel a) of \Cref{fig:3D_C4_decuplet_hyper_quarks_muS_N0_muQ_0_panel})  has a maximum immediately after the phase transition, followed by a monotonic decrease as $\mu_B$ increases. 
Referring to the order parameter $\Phi$ in panel b), we can see clearly that this phase transition corresponds to the deconfinement phase transition, since $\Phi\rightarrow 1$ in the quark phase.
We find that the deconfinement phase transition for C4 does not depend on the $\mu_S$ value, which is because $\Lambda$'s appear significantly later in the C4 coupling such that they do not significantly influence the mean fields/thermodynamics in the hadronic phase. 

However, for the C3 coupling, the intermediate/high $\mu_B$ phase of matter strongly depends on $\mu_S$. 
More positive $\mu_S$ leads to a quark phase for C3 coupling, however, more negative $\mu_S$ leads to a chiral phase transition to a strangeness-dominated hadronic phase.  
We can distinguish between the quark phase by using the order parameter $\Phi$ in panel b) of \Cref{fig:3D_C3_decuplet_hyper_quarks_muS_N0_muQ_0_panel}  that only approaches 1 for most positive values of $\mu_S$ (although some small negative values of $\mu_S$ at large $\mu_B$ do also lead to a quark phase).
Then, it is clear that the other phase of matter must be a strangeness-dominated phase by referring to $Y_S$ in panel c) in \Cref{fig:3D_C3_decuplet_hyper_quarks_muS_N0_muQ_0_panel} where $Y_S\rightarrow -1$ in this phase. 
In the quark phase for C3, $\sigma$ is significantly larger than for the strangeness-dominated phase wherein $\sigma$ approaches zero, which is indicative of the chiral restoration at very high densities.
This highlights the non-trivial relation between $\Phi$ and $\sigma$ at the deconfinement phase transition (see \Cref{fig:2D_all_hyper_quarks_muS_0_muQ_0_mf_and_obs} in \Cref{sec:C1_C4_baryon_octet_quarks}) related through hadronic and quark effective masses (\Cref{eq:effective_mass_hadrons} and \Cref{eq:effective_mass_quarks}).

For C3 (\Cref{fig:3D_C3_decuplet_hyper_quarks_muS_N0_muQ_0_panel}), one very interesting consequence of these two different discontinuities is that it gives rise to a triple point.
In fact, this is a quantum triple point because these calculations are all performed at $T=0$.  
Given that heavy-ion collisions can have fluctuations of $\mu_S$, it may mean that it is possible to experience this point at very low beam energy heavy-ion collisions.  
We have yet to study how the location/existence of this triple point varies with temperature, but we leave that for the next stage of \texttt{CMF++} when the temperature is included. 
To better understand this triple point, we plot susceptibilities for some horizontal lines (fixed $\mu_S$) from \Cref{fig:3D_C3_decuplet_hyper_quarks_muS_N0_muQ_0_panel}. These are shown in \Cref{sec:C3_critical_point_sus}.

In the C3 coupling, as $\mu_S$ decreases and the magnitude of the strangeness fraction $|Y_S|$ increases, the chiral phase transition to the strange-dominated phase rapidly moves to lower values of $\mu_B$ (one can see this clearly in panel c) of \Cref{{fig:3D_C3_decuplet_hyper_quarks_muS_N0_muQ_0_panel}} as the green/yellow colors switch abruptly to red). 
The shift in the phase transition with $\mu_S$ happens because the pressure increases with decreasing $\mu_S$ (for a given $\mu_B$) due to the increasing amount of strange particles, which also increases the baryon density (panel e), filling up the Fermi sea. 
The shift in the phase transition is not related to the stiffening of the EoS (by stiffening we mean an increase in $P$ with respect to $\varepsilon$ because instead hyperons typically soften the EoS).
Meanwhile, in the light hadronic phase, the pressure is not as strongly affected, since only the $\Lambda$ hyperon appears.

We find that for C3 there is already some strangeness (likely the $\Lambda$'s) that switches on during the light hadronic phase. 
One can see that in panel c) in \Cref{fig:3D_C3_decuplet_hyper_quarks_muS_N0_muQ_0_panel} since the blue (consistent with $Y_S\sim 0$) shifts to green/yellow (consistent with $Y_S\sim -0.3$ to $-0.6$).  
Then for $\mu_S>0$ one would go to a light quark phase (strangeness appears to play almost no role in the C3 quark phase for positive $\mu_S$) or into the strangeness-dominated hadronic phase for $\mu_S<0$.

In the strangeness-dominated phase, both the octet hyperons and their corresponding spin 3/2 excitations are present, as well as the $\Delta$s and the $\Omega$ baryons. 
The chiral phase transition was already present in the $\mu_S=0$ case. 
However, at $\mu_S=0$ the chiral phase transition was only observed in the metastable region when quarks were present (see~\Cref{fig:2D_C3_hyper_quarks_muS_0_muQ_0_mf_and_obs}) but in the absence of quarks, the chiral phase transition appeared in the stable region  (\Cref{fig:2D_all_decuplet_hyper_noquarks_muS_0_muQ_0_mf_and_obs}).  

In contrast, for C4 in \Cref{fig:3D_C4_decuplet_hyper_quarks_muS_N0_muQ_0_panel}, we see that $Y_S\sim0$ in panel c) for the entire regime where $\Phi=0$ in panel b).  
However, in the quark phase for C4, a large, negative $\mu_S$ is strongly correlated with a large, negative $Y_S$, such that large negative $\mu_S$ switches on a large number of strange quarks. 
In the $\mu_S>0$ regime in the quark phase for C4, we find almost no strangeness, such that light quarks dominate. 

We have already discussed $Y_S$ quite a bit, but here we briefly discuss the relationship between $Y_S$ and $Y_Q$.
For these results, we have made the assumption that $\mu_Q=0$, which corresponds to isospin symmetric matter, where (as already discussed) $Y_Q=\frac 1 2 + \frac 1 2 Y_S$. 
This clearly holds in panels c) and d) in Figs.\ \Cref{fig:3D_C3_decuplet_hyper_quarks_muS_N0_muQ_0_panel,fig:3D_C4_decuplet_hyper_quarks_muS_N0_muQ_0_panel}.
In fact, this relationship is easy to calculate at the limit of $Y_S=-1$ that corresponds to $Y_Q=0$, which corresponds exactly to what can be seen in the bottom right-hand corner of panels c) and d) in \Cref{fig:3D_C3_decuplet_hyper_quarks_muS_N0_muQ_0_panel,fig:3D_C4_decuplet_hyper_quarks_muS_N0_muQ_0_panel}.

We do not show the $P(\varepsilon)$ relationship directly but rather $c_s^2$ in panel f) because it is easier to see changes in the degrees of freedom and the influence of phase transition in a derivative plot. 
Due to the complexities that appear in $c_s^2$ across the $\mu_B,\mu_S$ plane, we discuss the C3 and C4 couplings separately. 
The blank region at high $\mu_B$, positive $\mu_S$ and following the quark region arises in \Cref{fig:3D_C3_decuplet_hyper_quarks_muS_N0_muQ_0_panel,fig:3D_C4_decuplet_hyper_quarks_muS_N0_muQ_0_panel} because beyond a given $\mu_B$ our quark deconfinement formalism breaks down (for the current parametrization) and hadrons would reappear. This correspond to very high $n_B$s (see panels e)).

The C3 coupling has a number of new particles that switch on as $\mu_B$ increases (this was previously apparent through the population plots at the limit of $\mu_S=0$). 
Here, in the $\mu_B,\mu_S$ plane, we can see new particles switch on through discontinuities in the color spectra, which correspond to kinks in $c_s^2(n_B)$ (panel f) of \Cref{fig:3D_C3_decuplet_hyper_quarks_muS_N0_muQ_0_panel}). 
For C3, the $\Lambda$ particle is the first one to appear, at  $\mu_B=1176$~MeV at $\mu_S=0$. 
The discontinuity in color (or rather kink in $c_s^2$) is correlated with $\mu_S$ as we vary $\mu_B$ such that large, negative values of $\mu_S$ see the kink at low $\mu_B$ whereas large positive values of $\mu_S$ see the kink at large $\mu_B$.  
Additionally, the sharpness of the kink changes with $\mu_S$, negative values of $\mu_S$ have a smoother kink vs positive values of $\mu_S$ that sharpen the kink. 
We can understand the correlated behavior in the location of the kink in $\mu_B^{kink}(\mu_S)$ because $\mu_S<0$ decreases the energy required to produce strange baryons, allowing for them to appear at lower $\mu_B$. 
For positive $\mu_S>0$ there is a very small kink near the deconfinement phase transition, which is a result of the onset of $\Delta$ baryons. 

Following the hadronic phase for C3, there is a first-order phase transition into quarks. 
Because $\mu_S<0$ allows for the possibility of new strange states (at lower $\mu_B$), another kink appears, followed very closely behind by a jump to $c_s^2\rightarrow 0$ (not shown here because the Maxwell construction is not shown) at the first-order phase transition marking the onset of the chiral strangeness-dominated phase. 
In the negative $\mu_S$ region, the kink near the chiral transition is due to the onset of the $\Xi$ particles, which now appear before $\Delta$s (the one at lower $\mu_B$ is due to the $\Lambda$'s).

The C4 coupling results for $c_s^2$ in panel f) of \Cref{fig:3D_C4_decuplet_hyper_quarks_muS_N0_muQ_0_panel} are significantly easier to understand.  
Generally, $c_s^2$ is stiffer than the other couplings and we can see that steady increase up until the quark phase appears.  
At the quark phase, there is a sharp drop in $c_s^2$ that stays close to the conformal limit of $c_s^2\rightarrow 1/3$. 
For $\mu_S<0$ we do see that there is the possibility of $\Lambda$ hyperons switching on in the hadronic phase when a small discontinuity corresponding to a kink in $c_s^2$ can be seen (although the kink appears rather smooth).
Even in the quark phase, there appears a small kink as well, but it is quite difficult to see. 

\subsection{$\left\{\mu_S = 0,\mu_Q \neq 0\right\}$}
\label{sec:muBN0_muQN0_muS0}

\begin{figure*}
    \centering
    \includegraphics[width=0.9\textwidth]{
    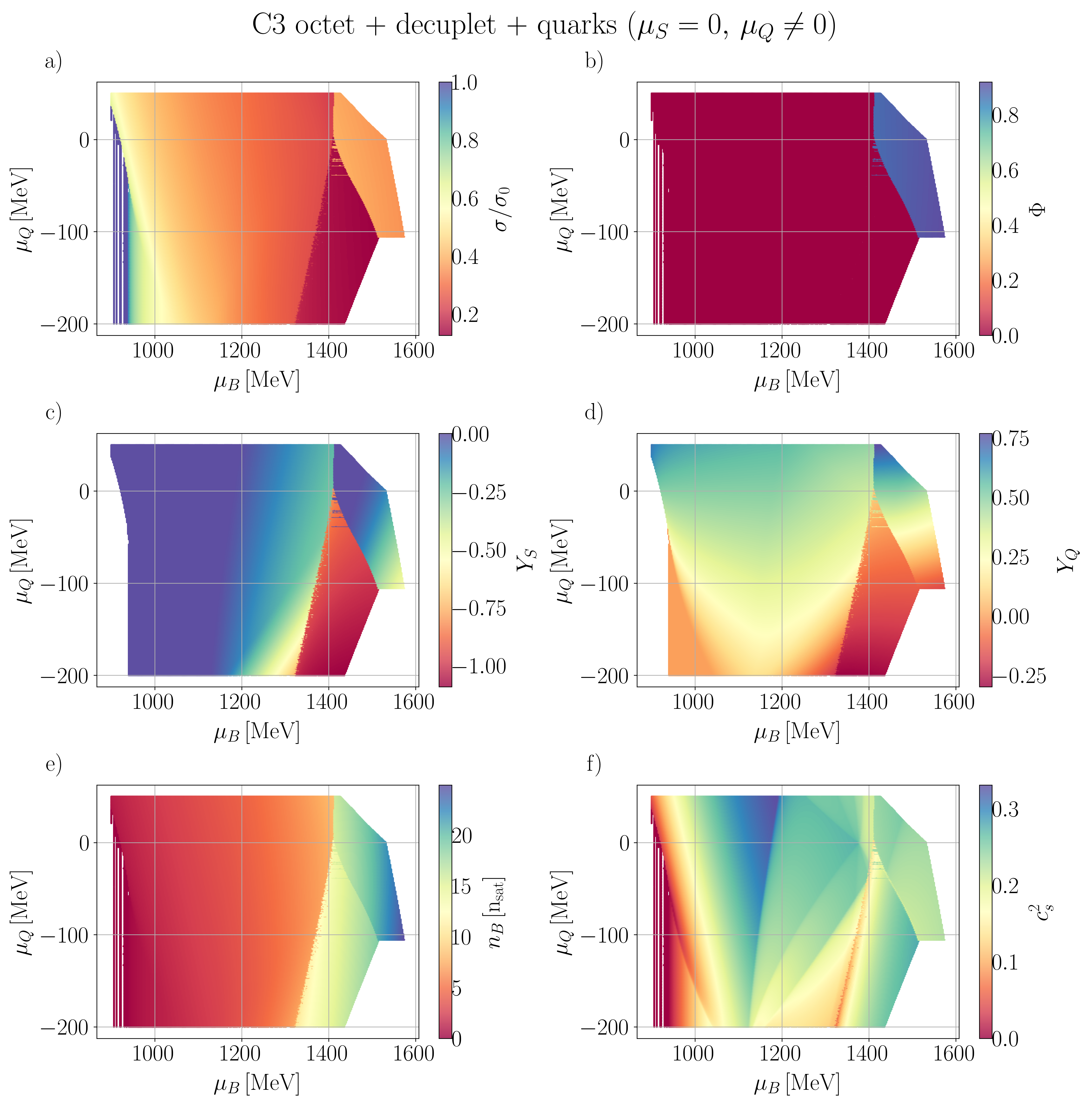}
    \caption{C3 ($\mu_S=0, \mu_Q \neq 0$) octet + decuplet + quarks: a) scalar meson field $\sigma$ normalized by vacuum value, b) deconfinement field $\Phi$, c) strangeness fraction, d) charge fraction, e) baryon density, and f) speed of sound as functions of baryon and charge chemical potentials.}
\label{fig:3D_C3_decuplet_hyper_quarks_muS_0_muQ_N0_panel}
\end{figure*}

\begin{figure*}
    \centering
    \includegraphics[width=0.9\textwidth]{
    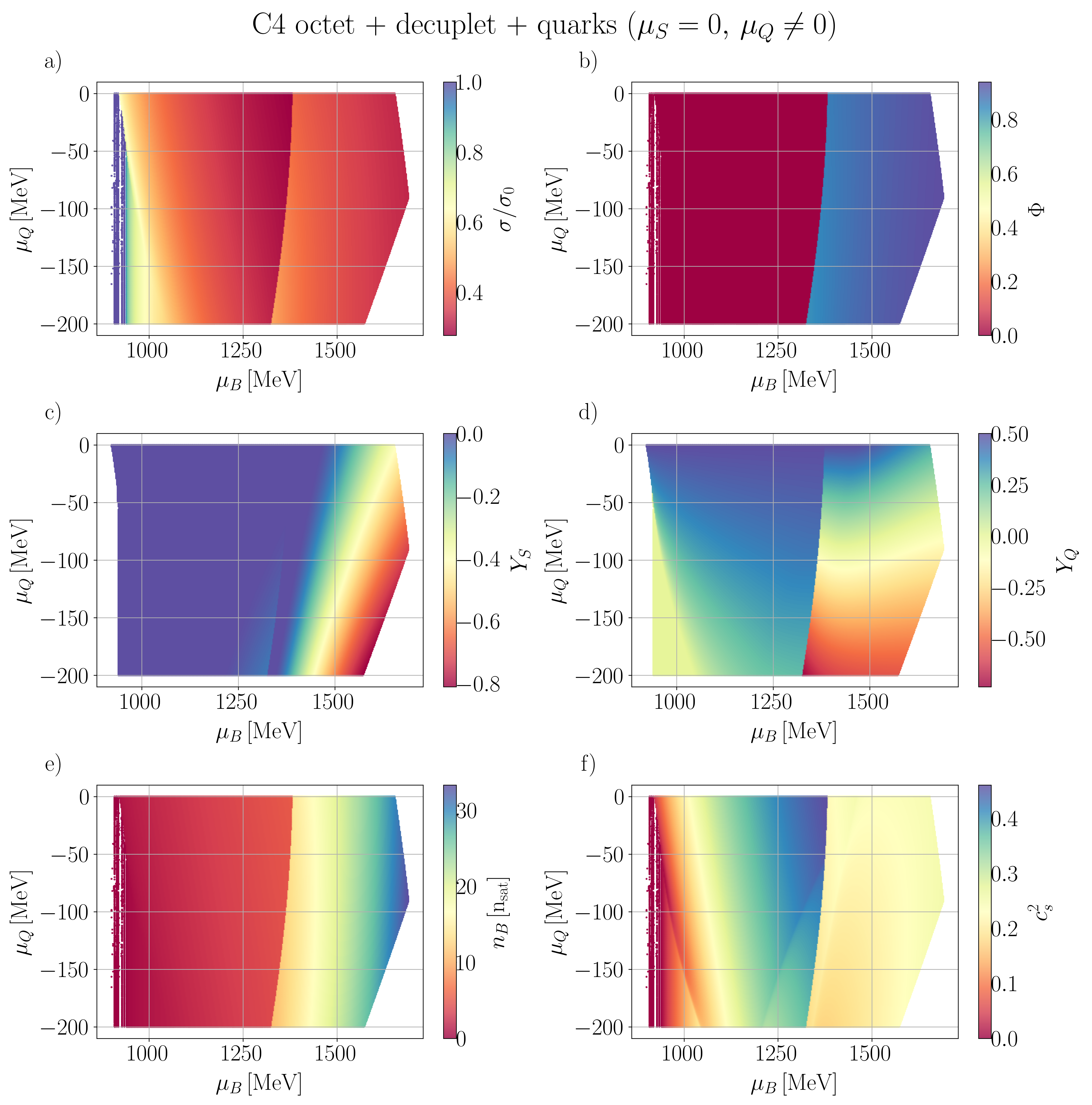}
    \caption{C4 ($\mu_S=0, \mu_Q \neq 0$) octet + decuplet + quarks: a) scalar meson field $\sigma$ normalized by vacuum value, b) deconfinement field $\Phi$, c) strangeness fraction, d) charge fraction, e) baryon density, and f) speed of sound as functions of baryon and charge chemical potentials.}
\label{fig:3D_C4_decuplet_hyper_quarks_muS_0_muQ_N0_panel}
\end{figure*}

Now we analyze the interplay of the charge chemical potential, $\mu_Q$, and $\mu_B$,  while holding $\mu_S=0$. 
The charge chemical potential is directly related to the appearance of electrically charged particles and, unlike the strangeness potential, it breaks the degeneracy between particles of the same family. 
From \Cref{eq:chemical_potential}, we can expect a significantly negative $\mu_Q$ to reduce the overall net positive charge of the system, bringing it farther away from symmetric nuclear matter.  
How this shift away from symmetric nuclear matter depends on the degrees of freedom of the system, their couplings, and interactions. 
For instance, for a system of just neutrons and protons, $\mu_Q<0$ suppresses the $Y_Q$ of the system such that either fewer protons appear or more neutrons appear.
For a system that allows for richer hadron chemistry, a negative $\mu_Q$ suppresses the appearance of positively charged particles and/or enhances the appearance of negatively charged particles. 
We explore only the region of negative $\mu_Q$, since this is the physically relevant region both for heavy-ion collisions where $Z/A\sim 0.4$ (leading to small $\mu_Q<0$) and for asymmetric nuclear matter found in neutron stars (leading to large $\mu_Q<0$), where $\beta$ equilibrium is achieved. 

\subsubsection{C3 and C4 with baryon octet + decuplet + quarks}

The figures follow the same structure as in \Cref{sec:muBN0_muQ0_muSN0}, but where we have changed the y-axis from $\mu_S$ to $\mu_Q$. 
We show the $\sigma $ mean fields a), the deconfinement order parameter b), charge d) and strange c) fractions, baryon density e) and speed of sound squared f) for C3 in ~\Cref{fig:3D_C3_decuplet_hyper_quarks_muS_0_muQ_N0_panel} and for C4 in ~\Cref{fig:3D_C4_decuplet_hyper_quarks_muS_0_muQ_N0_panel}.

At finite $\mu_Q$, the $\sigma$ mean-field (panel a)) still generally decreases with $\mu_B$, within a given phase of matter.
However, we can see that a non-monotonic behavior in $\sigma$ can appear across $\mu_B$. 
In the case of the C4 coupling, we see a non-monotonic behavior in $\sigma$ where there is a minimum in the hadronic phase right before the deconfinement phase transition (see panel b)) wherein $\sigma$ then increases in the quark phase, before decreasing again at high $\mu_B$.
In the case of the C3 coupling, this behavior is more complex and depends on $\mu_Q$ because the existence of the deconfinement phase transition only appears for $\mu_Q\gtrsim -100$ MeV.  
In the C3 coupling, the strangeness-dominated hadronic phase further complicates this non-monotonic behavior of $\sigma$.
For instance, if one fixes $\mu_Q=-50$ MeV, one can see that $\sigma$ steadily decreases until just before $\mu_B\sim 1400$ MeV where a sharp drop appears (going into the strangeness-dominated hadronic phase). Then, going towards higher $\mu_B$, we see a sharp increase in $\sigma$ at the onset of quark deconfinement at around $\mu_B\sim 1450 $ MeV.

At negative $\mu_Q\lesssim -50$ MeV for both the C3 and C4 couplings, a small region with $\sigma/\sigma_0=1$ is seen (dark blue seen at very low $\mu_B$), which indicates the presence of the liquid-gas transition.
When we see $\sigma/\sigma_0=1$, this is an indication that the vacuum solution is the stable solution.
Of course, the actual liquid-gas phase transition should not go to a vacuum phase but rather switch to nuclei (which are not included in our model). For a more in depth discussion on the role the isospin in the liquid-gas phase transition, see~\cite{Ducoin:2005aa,Mallik:2024wzk}.
The blank regions in \Cref{fig:3D_C3_decuplet_hyper_quarks_muS_0_muQ_N0_panel} at high $\mu_B$, below $\mu_Q\sim -100 \, \mathrm{MeV}$ arise because we cannot further find stable hadronic phases that continuously cover $\mu_B$.

As previously discussed, we use the baryon density $n_B$ (panel e)) to determine the existence of a first-order phase transition (a jump in seen in $n_B(\mu_B)$) and/or the order of any given phase transition through the susceptibilities. 
In the C3 coupling scheme, we can immediately see in \Cref{fig:3D_C3_decuplet_hyper_quarks_muS_0_muQ_N0_panel} two first-order lines that appear when $\mu_Q<0$. 
We can then use a combination of the deconfinement order parameter in panel b) (recall that if $\Phi=0$ we are still in a hadronic phase) and the strangeness fraction in panel c) (recall that $Y_S\rightarrow -1$ in the strangeness-dominated phase) to disentangle the deconfinement phase transition vs a first-order chiral phase transition. 
At $\mu_Q\sim0$ these two first-order lines converge into a triple point such that one goes directly from the light hadronic phase into the deconfined quark phase (see that the point where the lines converge also corresponds to an order parameter of $\Phi=1$). 
However, at $\mu_Q<0$ it is clear that as one increases $\mu_B$, then one first reaches a strangeness-dominated phase, and then only at even higher $\mu_B$ is the quark deconfinement phase reached. 
If $\mu_Q\lesssim -110$ MeV, then the quark deconfinement phase transition disappears entirely because there is a region where the code finds no solution. 
The critical $\mu_B$ of the chiral phase transition moves to lower values as $\mu_Q$ decreases since the pressure in the strangeness-dominated phase rises while in light hadronic phase it decreases (both at a given $\mu_B$) due to the larger amount of negatively charged particles in the light hadronic phase, which soften the EoS (pressure vs. energy density). 

For C4 (\Cref{fig:3D_C4_decuplet_hyper_quarks_muS_0_muQ_N0_panel}), there is no chiral phase transition.
Thus, we never reach a strangeness-dominated hadronic phase in the C4 coupling. 
However, we can still use $c_s^2$ in panel f) to identify new particles that have switched on and understand the role that $\mu_Q$ values have in the possibility of opening up these new particles. 
We find that lowering the charge chemical potential to more negative values moves the onset of the $\Lambda$ to lower $\mu_B$, as marked by the strangeness fraction in panel c) and the discontinuity in the speed of sound, shown in panel f), starting around $\mu_B \approx 1370$~MeV and $\mu_Q \approx-65$~MeV. The $\Lambda$ is affected by the charge chemical potential, even though it is not charged, due to the coupling between the $\omega$ and $\phi$ mesons in C4, \Cref{eq:Algebraic_System}, which increases the effective chemical of the $\Lambda$, moving its onset to a lower $\mu_B$ as  $\mu_Q$ decreases. 

In the C3 coupling (\Cref{fig:3D_C3_decuplet_hyper_quarks_muS_0_muQ_N0_panel}), the location of the deconfinement transition in $\mu_B$ is anti-correlated with $\mu_Q$ (in other words more negative $\mu_Q$ leads to a phase transition at a higher $\mu_B$), (see panel b) of \Cref{fig:3D_C3_decuplet_hyper_quarks_muS_0_muQ_N0_panel}).
The anti-correlation of the location of the critical $\mu_B$ for deconfinement with $\mu_Q$ occurs because the quark phase presents a very large density and has a lower pressure for lower $\mu_Q$ (for fixed $\mu_B$).
Within the deconfined quark phase, we find that at low absolute value of $\mu_Q$ there are very few strange quarks. 
However, as the absolute value of $\mu_Q$ increases, it opens up more strange quarks (because they carry electric charge $-1/3$). 
Eventually, at $\mu_Q\sim -130$~MeV, the code no longer finds solutions consistent with a quark phase such that solutions end with the strangeness-dominated hadronic phase. 

For the deconfinement phase transition, we find an opposite effect when it comes to the correlations between the critical $\mu_Q$ and $\mu_Q$ for C4 coupling (\Cref{fig:3D_C4_decuplet_hyper_quarks_muS_0_muQ_N0_panel}). 
At $\mu_Q=0$ we find that the deconfinement phase transition occurs at the maximum $\mu_B$ in panel b) for the order parameter $\Phi$. 
As $\mu_Q$ becomes increasingly negative, then there is a slow shift into the critical $\mu_B$ for deconfinement to lower values. 
Because C4 has a much larger quark phase than C3, we can see more interesting effects that appear in the quark phase at large $\mu_Q$.  
Both the strange and down quarks are preferred for low absolute value of $\mu_Q$ because they carry a negative charge, whereas the positive up quarks are suppressed.  
The consequence of a dominant down/strange quark phase is that $Y_S$ becomes very negative and $Y_Q<0$ as well in the quark phase. This isospin asymmetry causes the pressure to rise as a function of decreasing $\mu_Q$ (for a given $\mu_B$).

While we have discussed how we can use $Y_S$ in panel c) and $Y_Q$ in panel d) to interpret the underlying properties of the phase of matter, here we discuss their general behavior across the $\mu_B,\mu_Q$ phase space. 
Starting with the C3 coupling (\Cref{fig:3D_C3_decuplet_hyper_quarks_muS_0_muQ_N0_panel}), we find that $Y_S$ is mostly 0 in the light hadronic phase. 
Only after $\mu_B\gtrsim 1200$ MeV do we begin to see a slightly negative $Y_S$ due to the appearance of the $\Lambda$.  
However,  $\mu_Q<0$ opens up the strangeness-dominated hadronic phase in the range of $\mu_B\sim 1300-1500$ MeV (depending on $\mu_Q$) wherein $Y_S=-1$ where almost all hadrons carry strangeness (and some carry multiple strangeness). 
Then the deconfined quark phase is significantly less strange and only has some significant contribution of strange quarks when $\mu_Q$ is very negative. 

The $Y_Q$ behavior is quite different than $Y_S$.
The $Y_Q$ plot for C3 in panel d) of \Cref{fig:3D_C3_decuplet_hyper_quarks_muS_0_muQ_N0_panel} has different regimes of interest.
Unsurprisingly, close to $\mu_Q\sim 0$ one is close to the symmetric nuclear matter and $Y_Q\sim 0.5$ (the one exception is in the strangeness-dominated phase where, even at low $\mu_Q$, $Y_Q\sim 0$).
Then at low $\mu_B$ and very negative $\mu_Q$, we find that $Y_Q\rightarrow 0$ because pure neutron matter is reached. 
For $\mu_Q<0$ and intermediate $\mu_B$ (still in the light hadronic phase) we find small, positive values of $Y_Q$. 
In this regime protons have switched on and eventually some other particles like $\Lambda$'s (that are neutral but still suppress $Y_Q$ because they increase $n_B$). 
At very largely negative $\mu_Q$ and large $\mu_B$ but still, in the light hadronic phase, we can even see $Y_Q<0$. 
The strangeness-dominated phase generally has mostly negative values of $Y_Q$ because protons are heavily suppressed while strange baryons with a negative charge are preferred. 
Finally, in the quark phase, the value of $Y_Q$ strongly depends on the value of $\mu_Q$. 
At low absolute value of $\mu_Q$, the quark phase is primarily an even mix of up and down quarks, such that $Y_Q\sim 0.5$, but as $\mu_Q$ becomes more negative then strange and down quarks are preferred, which decreases $Y_Q$. 

For the C4 coupling in \Cref{fig:3D_C4_decuplet_hyper_quarks_muS_0_muQ_N0_panel} we find that $Y_S$ is essentially zero for the light hadronic phase and only because significantly negative once the deconfined quark phase is reached.  Then at large $\mu_B$ and very negative $\mu_Q$, there is a region where strange quarks are preferred such that $Y_Q\rightarrow -0.8$.
%
The $Y_Q$ for C4 reaches 0 for the pure neutron matter region at low $\mu_B$ and  very negative $\mu_Q$. Then at large $\mu_B$ (or smaller $\mu_B$ for large $\mu_Q$) it is close to 0.5 in the hadronic phase. 
For the quark phase, only at $\mu_Q\sim 0$ do we find $Y_Q\sim 0.5$. 
As $\mu_Q$ becomes more negative in the quark phase, we find $Y_Q$ slowly becomes smaller until it eventually becomes negative (and then more and more negative). 

Next, let us discuss the properties of $n_B$ in panel e) for both the C3 and C4 couplings. 
We find that in the light hadronic phase, both have very small values of $n_B$ that slowly increase with $\mu_B$ (regardless of $\mu_Q$). 
For C3 (\Cref{fig:3D_C4_decuplet_hyper_quarks_muS_0_muQ_N0_panel}) in the strangeness-dominated phase, the new hadronic states open up new degrees of freedom, leading to much larger $n_B$. 
For the deconfined quark phase in C3, we find very large values of $n_B$ such that they are much likely well beyond the reach of neutron stars. 
In contrast, C4 (\Cref{fig:3D_C3_decuplet_hyper_quarks_muS_0_muQ_N0_panel}) goes directly to the quark phase at large $\mu_B$. 
In the quark phase, we find a large jump in $n_B$ from hadrons into quarks across the phase transition. 
Then $n_B$ steadily increases with $\mu_B$ in the quark phase, almost independently of $\mu_Q$.
Next, we can use $c_s^2$ to determine when new hadronic species are switching on in the C3 coupling (\Cref{fig:3D_C3_decuplet_hyper_quarks_muS_0_muQ_N0_panel}) due to the kinks/discontinuities that appear. 
From the speed of sound plot (in panel f)), we can identify how the appearance of different baryons changes with  $\mu_B,\mu_Q$. 

For $0<\mu_Q\lesssim -50$ MeV the first discontinuity in $c_s^2(\mu_B)$ as one increases $\mu_B$ is due to the $\Lambda$'s. This is clear because $Y_S$ changes at this point (panel c)), but $Y_Q$ is only mildly affected (panel d)). Then we see a very small kink that appears that causes a $Y_S$ to become more negative and $Y_Q$ to slightly decrease. 
Then, the strangeness-dominated phase appears around $\mu_B\sim 1400$ MeV, which switches on the $\Delta^-$, $\Xi^-$, $\Sigma^-$ baryons, such that $Y_S\rightarrow -1$ and $Y_Q\sim 0$. 
There is then one final discontinuity in $c_s^2$ that indicates the quark deconfinement phase transition. 

For $\mu_Q<-50$ MeV an even richer hadronic phase appears and many new states open up.  
At low $\mu_B$ we have a discontinuity, which occurs as a transition from pure neutron matter into one that includes protons as well (as often seen in neutron star calculations, since the proton chemical potential $\mu_p=\mu_B+\mu_Q$ is less than the neutron one, $\mu_n=\mu_B$). 
Then, the $\Lambda$ baryon switches on at $\mu_B\approx 1170$~MeV, with an associated discontinuity that is only slightly modified by $\mu_Q$ due to the changes in the meson fields. Three more discontinuities appear at sufficiently low $\mu_Q$, associated with the onset of the $\Xi^-$, $\Delta^-$, and $\Sigma^-$ particles, respectively. Then a sharp transition in $c_s^2\rightarrow 0$ appears for the chiral first-order phase transition. 
In summary, the order or appearance is $\Lambda$, $\Delta^-$, $\Xi^-$, $\Sigma^-$, in the hadronic phase, and then, in the strangeness-dominated, the $\Xi$'s become dominant, but all the octet particles appear, as well as the $\Delta$'s. 

Overall, 
we find that $c_s^2$ is pretty dependent on $\mu_Q$ for both C3 and C4.  
At $\mu_Q\sim 0$ for C3 (\Cref{fig:3D_C3_decuplet_hyper_quarks_muS_0_muQ_N0_panel}) we see one kink at around $\mu_B\sim 1200$ MeV and then the triple point at around $\mu_B\sim 1400$ MeV that leads to some kinks with a very brief dip in $c_s^2$ at the phase transition(s) (see \Cref{sec:C3_critical_point_sus} for further discussion). 
As $\mu_Q$ becomes more negative, then $c_s^2$ has a large number of kinks that appear at may see 1-2 regions (depending on the exactly $\mu_Q$) where $c_s\rightarrow 0$. 
For C4 (\Cref{fig:3D_C4_decuplet_hyper_quarks_muS_0_muQ_N0_panel}), we already discussed previously that symmetric nuclear matter has a very stiff EoS with no kinks in the hadronic phase (and only a very tiny one in the quark phase). 
However, as one goes to $\mu_Q<0$ a dip in $c_s^2$ appears when the proton switches one and a kink appears when the $\Lambda$ appears as well, followed by the jump across the first-order deconfinement phase transition. 

\subsection{$\left\{\mu_S \neq 0,\mu_Q \neq 0\right\}$}
\label{sec:muBN0_muQN0_muSN0}

Up until this point, we have always set one or two chemical potentials to zero.  
Due to the limitations of 2-dimensional plots, it is significantly more difficult to vary all 3 chemical potentials at once in a meaningful manner.  
Thus, in the following, we hold either $\mu_S=const$ while varying $\mu_B,\mu_Q$ or hold $\mu_Q=const$, while varying $\mu_B,\mu_S$. 

\begin{figure*}
    \centering
    \includegraphics[width=0.9\textwidth]{
    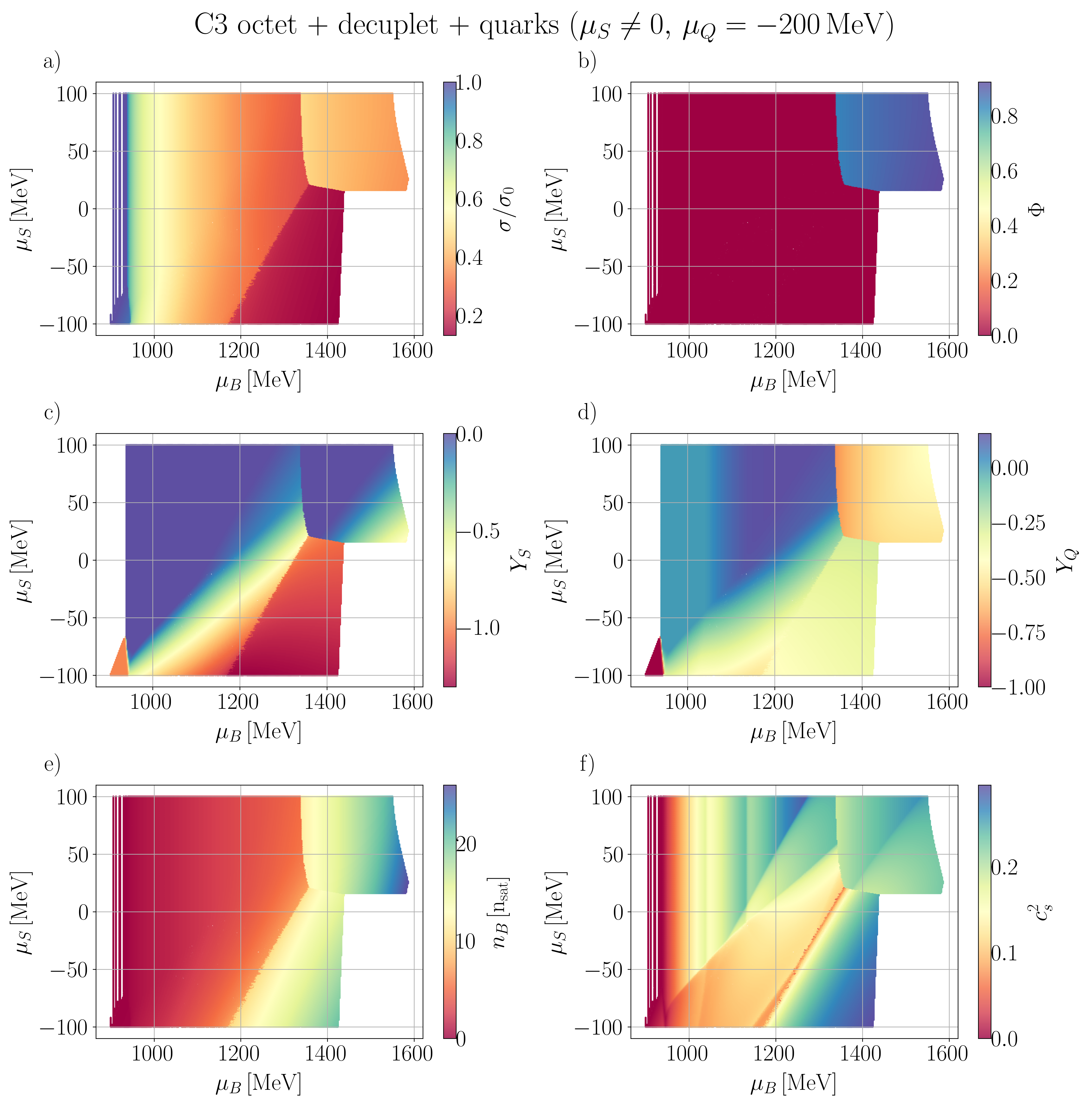}
    \caption{C3 ($\mu_S \neq 0, \mu_Q = -200 \, \mathrm{MeV}$) octet + decuplet + quarks: a) scalar meson field $\sigma$ normalized by vacuum value, b) deconfinement field $\Phi$, c) strangeness fraction, d) charge fraction, e) baryon density, and f) speed of sound as functions of baryon and strange chemical potentials.
    }
\label{fig:3D_C3_decuplet_hyper_quarks_muS_N0_muQ_-200_panel}
\end{figure*}

\begin{figure*}
    \centering
    \includegraphics[width=0.9\textwidth]{
    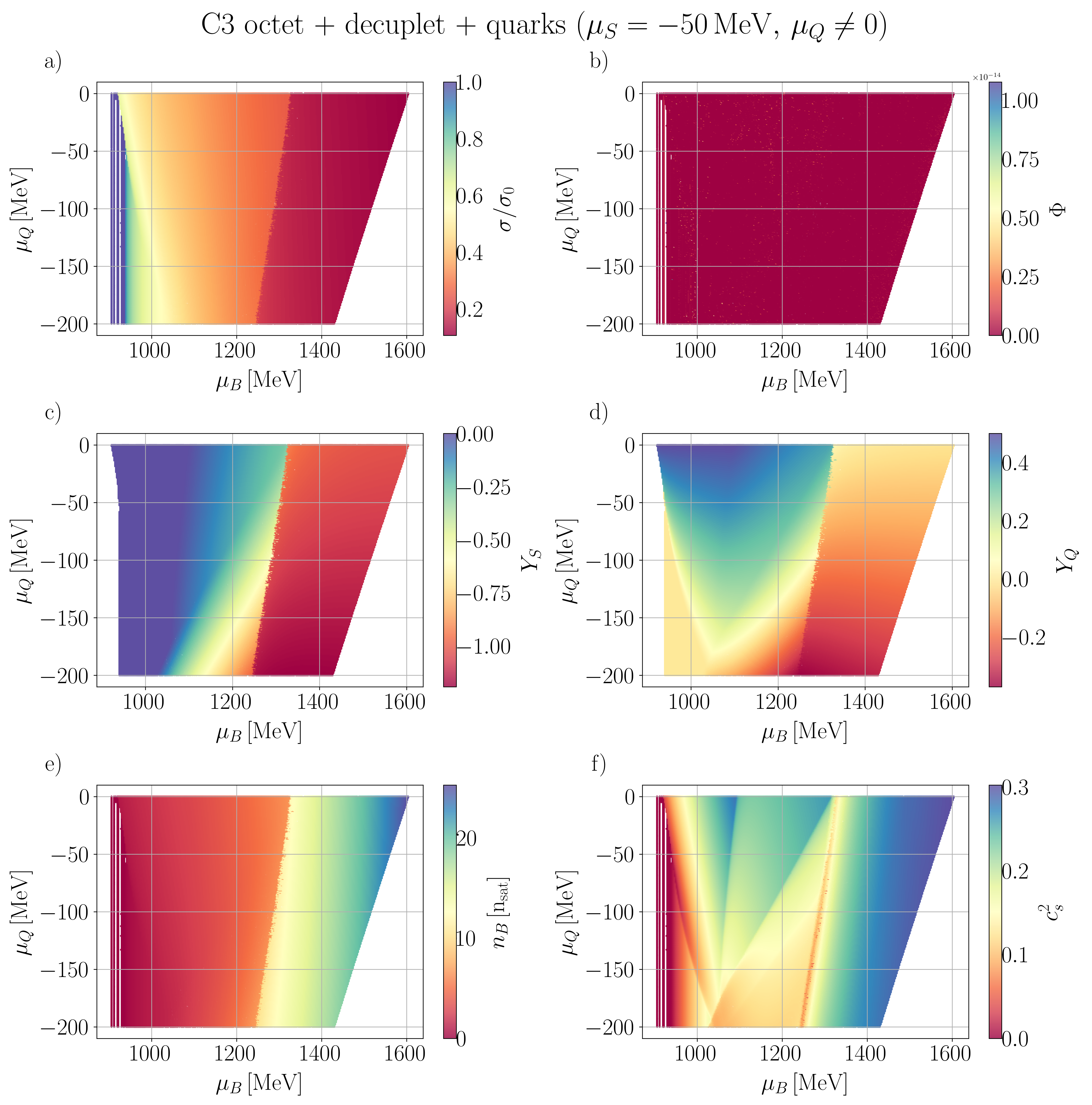}
    \caption{C3 ($\mu_S = -50 \, \mathrm{MeV}, \mu_Q\neq 0$) octet + decuplet + quarks: a) scalar meson field $\sigma$ normalized by vacuum value, b) deconfinement field $\Phi$, c) strangeness fraction, d) charge fraction, e) baryon density, and f) speed of sound as functions of baryon and strange chemical potentials.}
    \label{fig:3D_C3_decuplet_hyper_quarks_muS_-50_muQ_N0_panel}
\end{figure*}

\subsubsection{C3 with baryon octet + decuplet + quarks}

In this section, we consider the baryon octet, decuplet, and quarks to allow for the widest possible range of degrees of freedom while we explore the interplay of $\mu_B,\mu_S,\mu_Q$. 
We focus on the C3 coupling, which has the most distinct features and has a larger variety of particle species that regularly appear in the EoS compared to C4. 
Note that, because heavy particles tend to soften the EoS, the lower amount of heavy particles in the C4 coupling generally has an easier time reproducing astrophysical constraints of neutron star masses and radii~\cite{Dexheimer:2015qha}. 

In~\Cref{fig:3D_C3_decuplet_hyper_quarks_muS_N0_muQ_-200_panel}, the charge chemical potential is kept fixed at $\mu_Q=-200$~MeV, which is a typical value in hadronic neutron star matter, and the strangeness chemical potential is varied. 
Then in~\Cref{fig:3D_C3_decuplet_hyper_quarks_muS_-50_muQ_N0_panel} we  study the opposite scenario wherein the strangeness chemical potential is fixed at $\mu_S=-50$~MeV, and the $\mu_Q$ is varied. 
In each of the panels, we show the $\sigma $ mean-field meson a), the $\Phi$ order parameter b), charge d) and strange c) fractions, baryon density e) and speed of sound squared f).

Let us begin with the fixed $\mu_Q=-200$ MeV in \Cref{fig:3D_C3_decuplet_hyper_quarks_muS_N0_muQ_-200_panel}. 
To explain the $\sigma$, we begin at the low $\mu_B$ end and work our way up to larger $\mu_B$. 
At very low $\mu_B$ and largely negative $\mu_S$ we find a very tiny phase of matter that forms a blue triangle, i.e. $\sigma/\sigma_0\rightarrow 1$.
While one might be tempted to assume that this is a vacuum state since it occurs at low $\mu_B$, we later see that it is dominated by strange baryons (although at very low density).
Then at larger $\mu_B$ the behavior of $\sigma$ has the same qualitative appearance as what was seen in panel a) from \Cref{fig:3D_C3_decuplet_hyper_quarks_muS_N0_muQ_0_panel} (same case but with different $\mu_Q$) and even has similar values for the hadronic phase, the strangeness-dominated phase, and the quark deconfined phase. 

However, the exact location of these phases of matter and the shape of their first-order phase transitions are different at finite $\mu_Q$.
The quark deconfinement in \Cref{fig:3D_C3_decuplet_hyper_quarks_muS_N0_muQ_-200_panel} occurs in a region of positive $\mu_S$ and larger $\mu_B$ and the strangeness-dominated hadronic phase occurs at lower values of the $\mu_S$ compared to what we saw previously in \Cref{fig:3D_C3_decuplet_hyper_quarks_muS_N0_muQ_0_panel}. 
The main role of the negative $\mu_Q$ is to push the triple point from $\mu_S \gtrsim 0$ to $\mu_S \gtrsim 23$~MeV. 
A consequence of the change in the location of the triple point at finite $\mu_Q$ is that the deconfinement phase transition shifts to lower $\mu_B$. 
At, e.g., $\mu_S=-50$~MeV, only the chiral phase transition occurs, as confirmed by the deconfinement field $\Phi$ in panel b).

For $\mu_Q=-200$~MeV  the deconfinement transition shifts to lower $\mu_B$ in the positive $\mu_S$ region (when compared with \Cref{fig:3D_C3_decuplet_hyper_quarks_muS_N0_muQ_0_panel}), due to the higher amount of negatively charged particles at the same $\mu_B$ in the hadronic phase. 
These extra negatively charged strange particles increase the pressure for a given $\mu_B$ (see panels c), d) and e) in \Cref{fig:3D_C3_decuplet_hyper_quarks_muS_N0_muQ_-200_panel}) of the hadronic phase (while softening the EoS).
In contrast, in the quark phase, finite $\mu_Q$ (\Cref{fig:3D_C3_decuplet_hyper_quarks_muS_N0_muQ_-200_panel} vs. \Cref{fig:3D_C3_decuplet_hyper_quarks_muS_N0_muQ_0_panel}) only changes the ratio of up and down quarks, since the phase transition occurs before the onset of strangeness (panel c)).

One conclusion that we can draw from these plots is if C3 correctly describes the matter within neutron stars and heavy-ion collisions, then we could expect that neutron stars have deconfinement phase transition at lower $\mu_B$ than what we expect in heavy-ion collisions (where $\mu_Q$ is negative but normally smaller than what is seen in neutron stars). 
Additionally, because heavy-ion collisions can experience fluctuations in $\mu_S$ due to variations in the local strangeness content, it could be that some regions of the fluid see a first-order phase transition into a strangeness-dominated regime, but other parts of the fluids experience a first-order phase transition into quarks, and yet other parts may fluctuate from quarks into strange baryons (or vice versa). 

When we hold $\mu_S=-50$ MeV, fixed in \Cref{fig:3D_C3_decuplet_hyper_quarks_muS_-50_muQ_N0_panel}, we find that we remove the deconfinement phase transition entirely for C3 (see that $\Phi=0$ in panel b) across all $\mu_B,\mu_Q$).
However, we can see that there is significant variation in $\sigma$ across $\mu_B$ in that we see a small region where $\sigma/\sigma_0\sim 1$ at low $\mu_B$ (panel a)), very negative $\mu_Q$, then a somewhat steadily decreasing values of $\sigma$ followed by a sharp drop for $\sigma\rightarrow 0$. 
Since there is no deconfined phase transition, then this regime only contains hadronic states. 
This sharp drop in $\sigma$ corresponds to a chiral phase transition to the strangeness-dominated hadronic phase (see panel C)). 

In the case of  fixed $\mu_S=-50$~MeV and varying $\mu_Q$ in \Cref{fig:3D_C3_decuplet_hyper_quarks_muS_-50_muQ_N0_panel}, 
we find that the chiral phase transition is at approximately the same location as what was previously seen for $\mu_Q=0$ in \Cref{fig:3D_C3_decuplet_hyper_quarks_muS_0_muQ_N0_panel}, it is just that the strangeness-dominated hadronic phase is now \emph{always} the stable solution regardless of $\mu_Q$ at finite $\mu_S$ (no quark phase). This happens because the additional strange degrees of freedom increase the pressure as a function of $\mu_B$, favoring the chirally restored hadronic phase over the quark one.
Also, the strangeness-dominated phase is able to find solutions out to large $\mu_B$ as well when $\mu_S<0$ as compared to the $\mu_S=0$ case (e.g., ~\Cref{fig:3D_C3_decuplet_hyper_quarks_muS_0_muQ_N0_panel}).  

In both \Cref{fig:3D_C3_decuplet_hyper_quarks_muS_N0_muQ_-200_panel} and \Cref{fig:3D_C3_decuplet_hyper_quarks_muS_-50_muQ_N0_panel}, the $\Xi^-$ appears before the proton. For $\mu_Q \lesssim -170$~MeV when $\mu_S=-50$~MeV and for $\mu_S \lesssim -50$~MeV when $\mu_Q=-200$~MeV.
The strangeness and charge fraction, displayed in panels c) and d) of \Cref{fig:3D_C3_decuplet_hyper_quarks_muS_N0_muQ_-200_panel} and \Cref{fig:3D_C3_decuplet_hyper_quarks_muS_-50_muQ_N0_panel}, show the same tendency as discussed previously: in each phase, both present an overall decrease with $\mu_B$ after the onset of the strangeness and with decreasing $\mu_S$ or $\mu_Q$ at the lowest $\mu$'s analyzed. This feature is mostly due to the substantial contribution of the $\Xi^-$ and $\Sigma^-$ hyperons, which become more relevant than protons at sufficiently low $\mu_S$. We can tell the strong roles of these hyperons by the very negative $Y_S$ in panel c) and the also negative or small $Y_Q$ in panel d) of both figures. In fact, for C3 we find an interesting phenomena at  $\mu_B \lesssim920$~MeV and $\mu_Q \lesssim -70~$MeV,  where there is a phase where only the $\Sigma^-$ particle appears when $\mu_Q=-200$~MeV (\Cref{fig:3D_C3_decuplet_hyper_quarks_muS_N0_muQ_-200_panel}).

The baryon density, shown in panel e) in both \Cref{fig:3D_C3_decuplet_hyper_quarks_muS_N0_muQ_-200_panel} and \Cref{fig:3D_C3_decuplet_hyper_quarks_muS_-50_muQ_N0_panel} increases with $\mu_B$ and with decreasing $\mu_Q$ or $\mu_S$, in all phases. Additionally, the value of the density in the quark and the strangeness-dominated phases are very similar immediately after the transition, with $n_B \gtrsim 10 n_{\rm sat}$.

For $\mu_Q=-200$~MeV, the vertical discontinuities in the speed of sound, displayed in the leftmost part of panel f) in \Cref{fig:3D_C3_decuplet_hyper_quarks_muS_N0_muQ_-200_panel}, indicate the appearance of the proton and the $\Delta^-$. The $\Delta^-$ is then followed by the $\Lambda$'s, $\Xi^-$, and $\Sigma^-$ as one increases $\mu_B$. 
However, $\Xi^-$, and $\Sigma^-$ do not appear above $\mu_S \approx 60$ MeV, but as $\mu_S$ decreases, they eventually overcome protons and neutrons.
While we do not show the population plots in 3D due to their immense complexity, in the hyperonic phase the most abundant particles are in order of abundance $\Xi^-$, $\Xi^0$, $\Xi^{*-}$, n, $\Delta^-$, $\Delta^0$, $\Omega$, $\Sigma^-$, $\Xi^{*0}$, $\Delta^0$, p, $\Delta^+$, $\Lambda$, $\Sigma^0$, $\Delta^++$ and $\Sigma^+$.

For $\mu_S=-50$~MeV (\Cref{fig:3D_C3_decuplet_hyper_quarks_muS_-50_muQ_N0_panel}), the appearance of protons and $\Lambda$ are identified as the first two kinks in panel f) The change on the onset of the $\Lambda$ with $\mu_Q$ is due to the coupling of the vector fields~\Cref{eq:Algebraic_System}. After the $\Lambda$'s appear, the $\Xi^-$ and $\Sigma^-$ appear as one increases $\mu_B$ in the $\mu_Q >-150$~MeV. Below $\mu_Q=-150$~MeV, the $\Xi^-$ eventually becomes more abundant than the $\Lambda$ and the proton.  Across the chiral first-order phase transition, we see that $c_s^2\rightarrow 0$, and the population follows the one in \Cref{fig:3D_C3_decuplet_hyper_quarks_muS_-50_muQ_N0_panel}. 

\section{Final remarks}\label{sec:outlook}

We presented in this paper new results from the Chiral Mean Field (CMF) model at vanishing temperatures. CMF is a relativistic model built to describe bulk dense nuclear matter, describing chiral symmetry restoration and quark deconfinement. Our new calculations use the new \texttt{CMF++} code that was
integrated as a module in the MUSES cyberinfrastructure and is already available to the public as open-source \cite{zenodo_cmf}. 
The runtime improved more
than 4 orders of magnitude in 3 dimensions, when compared to the legacy \texttt{Fortran} code, while showing good agreement in a wide variety of configurations, couplings, thermodynamic quantities, etc. 
In addition to studying (independently) a large range of chemical potentials (baryon $\mu_B$, electric charge $\mu_Q$, and strangeness $\mu_S$),
numerical improvements also allowed us to calculate higher-order susceptibilities for the first time, which allowed us to identify 
a new first-order phase transition associated with chiral symmetry restoration (not related to deconfinement) in CMF, in addition to third-order phase transitions, and a triple point.

For the sake of clarity, we have presented a thorough review of the CMF model, including for the first time a thorough derivation of the formalism focusing on $T=0$. The derivation includes the Lagrangian density, equations of motion, and thermodynamic properties.
We have also derived expressions to calculate thermodynamic stability of the different phases. Together with the numerical improvements, this allowed us to map the unstable and metastable regimes between first-order phase transitions within \texttt{CMF++} and have outlined the complex stability criteria for 3 chemical potentials (BSQ) in the grand canonical ensemble for both $T=0$ (7 stability conditions) and finite $T$ (15 stability conditions).

As QCD cannot be solved in the energy regime we explore in this work, the different CMF C1-C4 vector couplings allow us to explore different vector interactions within our effective model as a way to account for uncertainty in our knowledge of the strong interaction. While exploring these couplings and a large 3-Dimensional parameter space in $\mu_B$, $\mu_S$, $\mu_Q$, we were able to identify different phases of matter that may appear at $T=0$: pure neutron matter, light hadronic matter (protons, neutrons, and sometimes $\Lambda$'s), strangeness and $\Delta$-dominated hadronic phase ($S=-2$ baryons dominate but other baryons may appear such as $\Delta$'s, $\Sigma$'s, and even $\Omega$'s), and deconfined quark matter (although there is a strong flavor dependence of the quark phase such that we have seen quarks phases dominated by light quarks, others by strange quarks, and yet others by down and strange quarks). 

The strangeness and $\Delta$ dominated hadronic phase is especially interesting because often a first-order phase transition appears into this phase of matter such that the EoS may look very similar to one that goes from a light hadronic phase into a deconfined quark phase (although in our model we find that the phase transition into the strangeness-dominated phase tends to be smaller than into quarks). 
We refer to this phase transition as chiral symmetry restoration, as the chiral condensates decrease across it. This phase could be identified, e.g., with a quarkyonic phase \cite{McLerran:2008ua}.
This strangeness-dominated hadronic phase is triggered by large amounts of hyperons or heavier (spin 3/2) baryons appearing in the system.
We found that the strangeness-dominated phase may be a stable solution of the EoS for, e.g., the C3 coupling with negative $\mu_Q$ and/or $\mu_S$ or even hidden within the metastable regime. This shows the importance of exploring different possibilities for the vector couplings, several of which are offered in the \texttt{CMF++} code.

More interestedly, within one of the vector couplings we studied, we found that a triple point appeared in the $\mu_B,\mu_S$ phase space such that at high $\mu_B$ (depending on the $\mu_S$) one could either find a first-order phase transition into deconfined quarks or into a strangeness-dominated hadronic phase (or even a first-order chiral phase transition separating quarks from a strangeness-dominated hadronic phase).
Given that heavy-ion collisions can have local fluctuations of $\mu_S$ due to gluons splitting into quark anti-quark pairs, then it is not unreasonable to think that effects of this triple point could be potentially observed in low-energy heavy-ion collisions. 

Our next step is to develop the finite $T$ version of the  \texttt{CMF++} code, which is already underway.
The finite $T$ version of \texttt{CMF++} will allow direct comparisons to lattice QCD as well as the potential to couple to other finite $T$ codes that reach large $\mu_B$ such as the holography EoS \cite{Hippert:2023bel}.

\acknowledgements

The authors acknowledge useful discussions with Jorge Noronha and Johannes Jahan on the stability constraints. 

All authors of this paper are a part of the MUSES collaboration, which is supported by the NSF under OAC-2103680.
Additional support for the collaboration members includes: RH is supported under OAC-2005572, and OAC-2004879. J.N.H. acknowledges support from the US-DOE Nuclear Science
Grant No. DE-SC0023861, and within the framework of the Saturated Glue (SURGE) Topical Theory Collaboration. R.K., J.P., and V.D were funded by the National Science Foundation under grants PHY1748621 and NP3M PHY2116686, and by the Department of Energy under grant DE-SC0024700.
The work is also supported by the Illinois Campus Cluster, a computing resource that is operated by the Illinois Campus Cluster Program
(ICCP) in conjunction with the National Center for Supercomputing Applications (NCSA),
which is supported by funds from the University of Illinois Urbana Champaign.

Figures in the manuscript were produced using matplotlib~\cite{Hunter:2007,thomas_a_caswell_2023_7697899}.

\bibliography{inspire,not_inspire}

\newpage

\appendix
\setcounter{table}{0}
\renewcommand{\thetable}{A\arabic{table}}

\section{Particle properties and masses}
\label{appendixA}

\Cref{tab:particles} displays the quantum numbers for the degrees of freedom described by the CMF model. \Cref{tab:mass_comparison} displays the masses for these particles, comparing the measured ones with what we currently use in \texttt{CMF++}.

\begin{table}[h!]
\scriptsize
\centering
\def\arraystretch{1.8}
{\begin{tabular}{ccccc}
\hline
\hline
Quark & Symbol & Electric Charge & Isospin$_z$ & Strangeness\vspace{-.2cm}\\
 &  & ($e$) & ($I_{3B}$)  & \\
\hline
up & $u$ & $\frac{2}{3}$ & $\frac{1}{2}$ & $0$\\
down & $d$ & $-\frac{1}{3}$ & $-\frac{1}{2}$ & $0$\\
strange & $s$ & $-\frac{1}{3}$ & $0$ & $-1$\\
\hline
$p$ & $uud$ & $1$ & $\frac{1}{2}$ & $0$ \\
$n$ & $udd$ &  $0$ & $-\frac{1}{2}$ & $0$\\
$\Lambda$ & $uds$ & $0$ & $0$ & $-1$\\
$\Sigma^+$ & $uus$ & $1$ & $1$ & $-1$\\
$\Sigma^0$ & $uds$ & $0$ & $0$ & $-1$\\
$\Sigma^-$ & $dds$  & $-1$ & $-1$ & $-1$\\
$\Xi^0$ & $uss$ &  $0$ & $\frac{1}{2}$ & $-2$\\
$\Xi^-$ & $dss$ &  $-1$ & $-\frac{1}{2}$ & $-2$\\
\hline
$\Delta^{++}$ & $uuu$ &  $2$ & $\frac{3}{2}$ & $0$ \\
$\Delta^+$ & $uud$ &  $1$ & $\frac{1}{2}$ & $0$\\
$\Delta^0$ & $udd$ &  $0$ & $-\frac{1}{2}$ & $0$\\
$\Delta^-$ & $ddd$ &  $-1$ & $-\frac{3}{2}$ & $0$\\
$\Sigma^{*+}$ & $uus$ & $1$ & $1$ & $-1$\\
$\Sigma^{*0}$ & $uds$  & $0$ & 0 & $-1$\\
$\Sigma^{*-}$ & $dds$  & $-1$ & -1 & $-1$\\
$\Xi^{*0}$ & $uss$ &  $0$ & $\frac{1}{2}$ & $-2$\\
$\Xi^{*-}$ & $dss$ & $-1$ & $-\frac{1}{2}$ & $-2$\\
$\Omega^-$ & $sss$ &  $-1$ & 0 & $-3$\\
\hline
\hline
\end{tabular}}
\caption{Quark and baryon quantum numbers \cite{ParticleDataGroup:2024cfk}.}
\label{tab:particles}
\end{table}

\begin{table}[h!]
\centering
{\begin{tabular}{cccc}
\hline\hline
\textbf{Hadron/quark} \ \ \ \ & \textbf{CMF++ mass}  \ \ \ \ & \textbf{PDG mass}\  \\ 
& (MeV) &(MeV) \\ 
\hline \hline
 $u$             &       5      &  $\sim$2.16   \\
 $d$             &      5       &    $\sim$4.70   \\
 $s$             &      150       &    $\sim$93.5  \\
\hline
$\sigma$        &   597.76          &  $\sim$ 500    \\
$\zeta$        &      972.86      &   $\sim$ 980   \\
$\delta$        &      953.53       &  $\sim$ 980    \\
\hline
$\pi$           &   139         &  139 .57  \\
$K$             &        498   &    $\sim$ 498  \\
$\eta$          &     574.74        &   547.86   \\
$\eta^\prime$   &     958         &  957.78   \\
\hline
$\omega$          &    780.65        &   782.66    \\
$\rho$          &        761.06    &   775.26    \\
$\phi$          &       1019     &    1019.461   \\
\hline
$n$             &      937.24       &   939.57   \\
$p$             &      937.24       &   938.27   \\
$\Lambda$             &    1115         &   1115.7   \\
$\Sigma^+$             &     1202        &    1189.4  \\
$\Sigma^0$             &    1202      &  1192.6    \\
$\Sigma^-$             &    1202         &   1197.4   \\
$\Xi^0$   &         1336.26    &   1314.9  \\
$\Xi^-$ &             1336.26          &   1321.7      \\
\hline
$\Delta^{++}$             &    1232         &  1232.0    \\
$\Delta^+$             &      1232       &  1232.0   \\
$\Delta^0$             &  1232           &  1232.0   \\
$\Delta^-$             &     1232        &  1232.0   \\
$\Sigma^{*^{+}}$             &    1385        & 1382.83     \\
$\Sigma^{*^{0}}$               &     1385        &   1382.7   \\
$\Sigma^{*^{-}}$                &    1385         &  1387.2    \\
$\Xi^{*^{0}}$             &       1538      &  1531.80    \\
$\Xi^{*^{-}}$             &         1538    &  1535.0  \\
$\Omega^-$             &     1691        &   1672.4   \\
\hline \hline
\end{tabular}}
\caption{Comparison of quark, meson, and baryon masses: \texttt{CMF++} vs. PDG 2024 values~\cite{ParticleDataGroup:2024cfk}.}
\label{tab:mass_comparison}
\end{table}

\section{Particle multiplets}
\label{app:multiplets}

The following baryon and meson matrices are constructed from the triplet tensor products ${\bf{3}} \otimes {\bf{3}} \otimes {\bf{3}}$ and ${\bf{3}} \otimes \bar{\bf{3}}$, respectively, where ${\bf{3}}=(u, d, s)^T$~\cite{Lie_group_lecture_notes}

\begin{itemize}
\item Baryon Matrix
\begin{equation}
B=\begin{pmatrix}
\frac{\Sigma^{0}}{\sqrt{2}}+\frac{\Lambda}{\sqrt{6}} & \Sigma^{+} & p \\
\Sigma^{-} & \frac{-\Sigma^{0}}{\sqrt{2}}+\frac{\Lambda}{\sqrt{6}} & n \\
\Xi^{-} & \Xi^{0} & -2\frac{\Lambda}{\sqrt{6}}
\end{pmatrix}\,.
\label{Bmatrix}
\end{equation}

\item Scalar-Meson Matrix
\begin{equation}
X=\begin{pmatrix}
\frac{\delta^{0}+\sigma}{\sqrt{2}} & \delta^{+} & \kappa^{+} \\
\delta^{-} & \frac{-\delta^{0}+\sigma}{\sqrt{2}} & \kappa^{0} \\
\kappa^{-} & \bar{\kappa}^{0} & \zeta
\end{pmatrix}\,.
\label{Xmatrix}
\end{equation}

\item Vector-Meson Matrix
\begin{equation}
V_\mu=\begin{pmatrix}
\frac{\rho_\mu^{0}+\omega_\mu}{\sqrt{2}} & \rho_\mu^{+} & K_\mu^{*+} \\
\rho_\mu^{-} & \frac{-\rho_\mu^{0}+\omega_\mu}{\sqrt{2}} & K_\mu^{*0} \\
K_\mu^{*-} & \bar{K}_\mu^{*0} & \phi_\mu
\end{pmatrix}\,.
\label{Vmatrix}
\end{equation}

\item Pseudoscalar-Meson Matrix
\begin{align}
P=\begin{pmatrix}
\frac{1}{\sqrt{2}}\left(\pi^{0}+\frac{\eta^{8}}{\sqrt{1+2w^{2}}}\right) & \pi^{+} & 2\frac{K^{+}}{w+1} \\
\pi^{-} & \frac{1}{\sqrt{2}}\left(-\pi^{0}+\frac{\eta^{8}}{\sqrt{1+2w^{2}}}\right) & 2\frac{K^{0}}{w+1} \\
2\frac{K^{-}}{w+1} & 2\frac{\bar{K^{0}}}{w+1} & -\sqrt{\frac{2}{1+2w^{2}}}\eta^{8}
\end{pmatrix}\,,\nonumber\\
\label{Pmatrix}
\end{align}
where $w=\sqrt{2}\zeta_{0}/\sigma_{0}$. 

\item Pseudoscalar-Meson Singlet Matrix
\begin{equation}
Y=\sqrt{\frac{1}{3}}\eta^{0}\begin{pmatrix}
1 & 0 & 0 \\
0 & 1 & 0 \\
0 & 0 & 1
\end{pmatrix}\,.
\label{Ymatrix}
\end{equation}

\end{itemize}

\begin{widetext}

\section{Lagrangian density calculations}

\subsection{The self-interaction term for scalar mesons}
\label{app:CMFdetails}

In this study, we consider a term with the form
\begin{equation}
\begin{split}
\mathcal{L}_{\rm scal}= & -\dfrac{1}{2}k_{0}\chi^2I_{2}+k_{1}I_2^{2}+k_{2}I_{4} + 2 k_3 \chi I_0+k_{3N}\chi I_3+\dfrac{\epsilon}{3}\chi^4\ln\dfrac{I_{0}}{\mathrm{det}\langle X_0\rangle} -k_4\chi^4+ \frac{\chi^4}{4}\ln \bigg( \frac{\chi^4}{\chi_0^4}\bigg)\,.    
\end{split}
\end{equation}
In the mean-field approximation, ~\Cref{Xmatrix} becomes
\begin{equation}
X=\begin{pmatrix}\dfrac{\delta+\sigma}{\sqrt{2}} & 0 & 0\\
0 & \dfrac{-\delta+\sigma}{\sqrt{2}} & 0\\
0 & 0 & \zeta
\end{pmatrix}\,,
\end{equation}
and from \Cref{10},
\begin{equation}
\begin{split}
I_{0}= &~\mathrm{det}(X) =\bigg(\dfrac{\delta+\sigma}{\sqrt{2}}\bigg)\bigg(\dfrac{-\delta+\sigma}{\sqrt{2}}\bigg)\bigg(\zeta\bigg)=\left( \dfrac{\sigma^2-\delta^2}{2}\right)\zeta\,.
\end{split}
\end{equation}
For the vacuum expectation value of $X$, $\langle X_0\rangle$, the isovector meson $\delta$ does not contribute
\begin{equation}
\mathrm{det}\langle X_0\rangle=\dfrac{\sigma_0^2\zeta_0}{2}\,.
\end{equation}
For the other terms in the Lagrangian, we use
\begin{equation}
\mathrm{Tr}(X^n)=\bigg(\dfrac{\delta+\sigma}{\sqrt{2}}\bigg)^n+\bigg(\dfrac{-\delta+\sigma}{\sqrt{2}}\bigg)^n+\bigg(\zeta\bigg)^n\,,
\end{equation}
such that from \Cref{10,11,12} becomes
\begin{equation}
I_2= \mathrm{Tr}(X^2)=  \sigma^2+\zeta^2+\delta^2,\quad 
I_{3}=\mathrm{Tr}(X^{3})=\dfrac{\sigma^{3}+3\sigma\delta^{2}}{\sqrt{2}}+\zeta^{3}\,,
\quad
I_{4} =\mathrm{Tr}(X^{4})=\dfrac{\sigma^{4}+6\sigma^{2}\delta^{2}+\delta^{4}}{2}+\zeta^{4}\,.
\end{equation}

The scalar Lagrangian is then given by
\begin{align}
\mathcal{L}_{\rm scal}=&- \dfrac{1}{2}k_{0}\chi^2\left(\delta^{2}+\sigma^{2}+\zeta^{2}\right)+k_{1}\left(\delta^{2}+\sigma^{2}+\zeta^{2}\right)^{2} +k_{2}\left[\dfrac{\delta^{4}}{2}+3\delta^{2}\sigma^{2}+\dfrac{\sigma^{4}}{2}+\zeta^{4}\right] +k_3\chi\left(\dfrac{\sigma^2-\delta^2}{2}\right)\zeta \\
&+k_{3N}\chi \left(\dfrac{\sigma^{3}+3\sigma\delta^{2}}{\sqrt{2}}+\zeta^{3} \right)+\dfrac{\epsilon}{3}\chi^4\ln\dfrac{-\delta^{2}\zeta+\sigma^{2}\zeta}{\sigma_{0}^{2}\zeta_{0}} - k_4\chi^4 + \frac{1}{4}\ln \bigg( \frac{\chi^4}{\chi_0^4}\bigg)\,.
\end{align}

\subsection{The baryon-meson interaction term}
\label{app:interactions} 

To calculate \Cref{symcouple,antisymcouple}, we write explicitly the $\bar{B}OBM$ matrix

\begin{align}
\bar{B}OBM=
\begin{pmatrix}\bar{B}OBM_{(1,1)} & \bar{B}OBM_{(1,2)} & \bar{B}OBM_{(1,3)}\\ \bar{B}OBM_{(2,1)}&\bar{B}OBM_{(2,2)}&
\bar{B}OBM_{(2,3)}&\\ \bar{B}OBM_{(3,1)}&\bar{B}OBM_{(3,2)}&\bar{B}OBM_{(3,3)}
\end{pmatrix}\,,\nonumber\\
\end{align}
having for a diagonal $M$ 
\begin{align}
\bar{B}OBM_{(1,1)} =& \Bigg[ \left(\dfrac{\bar{\Sigma}^{0}O}{\sqrt{2}}+\dfrac{\bar{\Lambda}^{0}O}{\sqrt{6}}\right)\left(\dfrac{\Sigma^{0}}{\sqrt{2}}+\dfrac{\Lambda}{\sqrt{6}}\right)+\bar{\Sigma}^{-}O\Sigma^{-}+\bar{\Xi}^{-}O\Xi^{-}\Bigg] \bigg(M_{11}\bigg)\,,
\end{align}
\begin{align}
\bar{B}OBM_{(2,2)} =& \Bigg[ \bar{\Sigma}^{+}O\Sigma^{+}+\left(\dfrac{-\bar{\Sigma}^{0}O}{\sqrt{2}}+\dfrac{\bar{\Lambda}^{0}O}{\sqrt{6}}\right)\left(\dfrac{-\Sigma^{0}}{\sqrt{2}}+\dfrac{\Lambda}{\sqrt{6}}\right) +\bar{\Xi}^{0}O\Xi^{0}\Bigg]\bigg(M_{22}\bigg)\,,
\end{align}
\begin{align}
\bar{B}OBM_{(3,3)} =& \Bigg[ \bar{p}Op+\bar{n}On+\dfrac{2}{3}\bar{\Lambda}^{0}O\Lambda\Bigg]\bigg(M_{33}\bigg)\,,
\end{align}
with trace
\begin{align}
\mathrm{Tr}(\bar{B}OBM) = & \Bigg[ \left(\dfrac{\bar{\Sigma}^{0}O}{\sqrt{2}}+\dfrac{\bar{\Lambda}^{0}O}{\sqrt{6}}\right)\left(\dfrac{\Sigma^{0}}{\sqrt{2}}+\dfrac{\Lambda}{\sqrt{6}}\right)+\bar{\Sigma}^{-}O\Sigma^{-}+\bar{\Xi}^{-}O\Xi^{-}\Bigg]\bigg(M_{11}\bigg) +\Bigg[ \bar{\Sigma}^{+}O\Sigma^{+}+\left(\dfrac{-\bar{\Sigma}^{0}O}{\sqrt{2}}+\dfrac{\bar{\Lambda}^{0}O}{\sqrt{6}}\right)\nonumber \\
&\left(\dfrac{-\Sigma^{0}}{\sqrt{2}}+\dfrac{\Lambda}{\sqrt{6}}\right)+\bar{\Xi^{0}}O\Xi^{0}\Bigg]\bigg(M_{22}\bigg) +\Bigg[ \bar{p}Op+\bar{n}On+\dfrac{2}{3}\bar{\Lambda}^{0}O\Lambda\Bigg]\bigg(M_{33}\bigg)\,.
\end{align}

On the other hand,
\begin{align}
\bar{B}OMB=
\begin{pmatrix}\bar{B}OMB_{(1,1)} & \bar{B}OMB_{(1,2)} & \bar{B}OMB_{(1,3)}\\ \bar{B}OMB_{(2,1)}&\bar{B}OMB_{(2,2)}&
\bar{B}OMB_{(2,3)}&\\ \bar{B}OMB_{(3,1)}&\bar{B}OMB_{(3,2)}&\bar{B}OMB_{(3,3)}
\end{pmatrix}\,,\nonumber \\
\end{align}
and again, explicitly for a diagonal $M$,
\begin{align}
\bar{B}OMB_{(1,1)} =&\bigg( \dfrac{\bar{\Sigma}^{0}O}{\sqrt{2}}+\dfrac{\bar{\Lambda}^{0}O}{\sqrt{6}}\bigg)\bigg(\dfrac{\Sigma^{0}}{\sqrt{2}}+\dfrac{\Lambda}{\sqrt{6}}\bigg)M_{11}+\bar{\Sigma}^{-}O\Sigma^{-}M_{22}+\bar{\Xi}^{-}O\Xi^{-}M_{33}\,,
\end{align}
\begin{align}
\bar{B}OMB_{(2,2)} =&\bar{\Sigma}^{+}O\Sigma^{+}M_{11} + \bigg(\dfrac{-\bar{\Sigma}^{0}O}{\sqrt{2}}+\dfrac{\bar{\Lambda}^{0}O}{\sqrt{6}}\bigg)\bigg(\dfrac{-\Sigma^{0}}{\sqrt{2}}+\dfrac{\Lambda}{\sqrt{6}}\bigg)M_{22} + \bar{\Xi}^{0}O\Xi^{0}M_{33}\,,
\end{align}
\begin{align}
\bar{B}OMB_{(3,3)} =&\bar{p}Op~M_{11} + \bar{n}On~M_{22} + \dfrac{2}{3}\bar{\Lambda}^{0}O\Lambda^0M_{33}\,,
\end{align}
with trace
\begin{align}
\mathrm{Tr}(\bar{B}OMB) = &\bigg[\bigg(\dfrac{\bar{\Sigma}^{0}O}{\sqrt{2}}+\dfrac{\bar{\Lambda}^{0}O}{\sqrt{6}}\bigg)\bigg(\dfrac{\Sigma^{0}}{\sqrt{2}}+\dfrac{\Lambda}{\sqrt{6}}\bigg)+\bar{\Sigma}^{+}O\Sigma^{+}+\bar{p}Op\bigg]\bigg(M_{11}\bigg) +\bigg[\bar{\Sigma}^{-}O\Sigma^{-}+\bigg(\dfrac{-\bar{\Sigma}^{0}O}{\sqrt{2}}+\dfrac{\bar{\Lambda}^{0}O}{\sqrt{6}}\bigg)\nonumber\\
&\bigg(\dfrac{-\Sigma^{0}}{\sqrt{2}}+\dfrac{\Lambda}{\sqrt{6}}\bigg)+\bar{n}On\bigg]\bigg(M_{22}\bigg)
+\bigg[\bar{\Xi}^{-}O\Xi^{-}+\bar{\Xi}^{0}O\Xi^{0}+\dfrac{2}{3}\bar{\Lambda}^{0}O\Lambda\bigg]\bigg(M_{33}\bigg)\,.
\end{align}
Also,
\begin{equation}
\mathrm{Tr}(\bar{B}OB) = \sum_{i\in B}\bar{\psi}_iO\psi_i\,,
\end{equation}
and
\begin{align}
\mathrm{Tr}(\bar{B}OB)\mathrm{Tr}(M) & = \bigg(\bar{p}Op+\bar{n}On+\bar{\Lambda}^{0}O\Lambda+\bar{\Sigma}^{+}O\Sigma^{+}+\bar{\Sigma}^{0}O\Sigma^{0}+\bar{\Sigma}^{-}O\Sigma^{-}+\bar{\Xi}^{0}O\Xi^{0}+\bar{\Xi}^{-}O\Xi^{-}\bigg)\bigg(M_{11}+M_{22}+M_{33}\bigg)\,.
\label{eq:rrr1}
\end{align}
Combining those into \Cref{symcouple,antisymcouple}, we obtain
\begin{align}
[\bar{B}OBM]_{\rm AS}=& \bigg[\bar{p}Op+\bar{\Sigma}^{+}O\Sigma^{+}-\bar{\Sigma}^{-}O\Sigma^{-}-\bar{\Xi}^{-}O\Xi^{-}\bigg]\bigg(M_{11}\bigg)\nonumber\\ +&\bigg[\bar{n}On+\bar{\Sigma}^{-}O\Sigma^{-}-\bar{\Sigma}^{+}O\Sigma^{+}-\bar{\Xi^{0}}O\Xi^{0}\bigg]\bigg(M_{22}\bigg)\nonumber\\  +&\bigg[-\bar{p}Op-\bar{n}On+\bar{\Xi}^{-}O\Xi^{-}+\bar{\Xi}^{0}O\Xi^{0}\bigg]\bigg(M_{33}\bigg),
\label{eq:rrr2}
\end{align}
\begin{align}
[\bar{B}OBM]_{\rm S} & =\dfrac{1}{3}\bigg(\bar{\Sigma}^{0}O\Sigma^{0}-\bar{\Lambda}^{0}O\Lambda+\bar{\Sigma}^{+}O\Sigma^{+}+\bar{p}Op+\bar{\Sigma}^{-}O\Sigma^{-}+\bar{\Xi}^{-}O\Xi^{-}-2\bar{\Xi}^{0}O\Xi^{0}-2\bar{n}On\bigg)\bigg(M_{11}\bigg)\nonumber\\&+\dfrac{1}{3}\bigg(\bar{\Sigma}^{0}O\Sigma^{0}-\bar{\Lambda}^{0}O\Lambda+\bar{\Sigma}^{+}O\Sigma^{+}+\bar{n}On+\bar{\Sigma}^{-}O\Sigma^{-}+\bar{\Xi}^{0}O\Xi^{0}-2\bar{\Xi}^{-}O\Xi^{-}-2\bar{p}Op\bigg)\bigg(M_{22}\bigg)\nonumber\\&+\dfrac{1}{3}\bigg(\bar{p}Op+\bar{n}On+\bar{\Xi}^{0}O\Xi^{0}+\bar{\Xi}^{-}O\Xi^{-}+2\bar{\Lambda}^{0}O\Lambda-2\bar{\Sigma}^{+}O\Sigma^{+}-2\bar{\Sigma}^{0}O\Sigma^{0}-2\bar{\Sigma}^{-}O\Sigma^{-}\bigg)\bigg(M_{33}\bigg)\,,
\label{eq:rrr3}
\end{align}
where the $\Sigma\Lambda$ terms are not considered as they do not contribute to the mean-field approximation.
\\

\textbf{Case 1)} For $M=X$ (the scalar-meson matrix in the mean-field approximation)
\begin{align*}
X = \mathrm{diag}\bigg(\dfrac{\delta+\sigma}{\sqrt{2}},\dfrac{-\delta+\sigma}{\sqrt{2}},\zeta\bigg)\,,\quad \quad
\mathrm{Tr}(X)=\sqrt2\sigma+\zeta\,,
\end{align*}
\begin{align*}
\sigma=\dfrac{X_{11}+X_{22}}{\sqrt2}\,,
\quad
\delta=\dfrac{X_{11}-X_{22}}{\sqrt2}\,,
\quad
\zeta=X_{33}\,.
\end{align*}
To illustrate how to find the coupling for the baryons, we take the particular case of the proton. Its couplings to the scalar mesons $\sigma$, $\delta$ and $\zeta$ are found by replacing~\Cref{eq:rrr1},~\Cref{eq:rrr2}, and~\Cref{eq:rrr3} in~\Cref{eq:L_BM}, such that
\begin{align}
\mathcal{L}_{{\rm int},Xp} & = -\sqrt2g_8^X\bigg[\alpha_X\bigg(\bar{p}p\big(X_{11}-X_{33}\big)\bigg) +(1-\alpha_X)\bigg(\dfrac{1}{3}\bar{p}p(X_{11}-2X_{22}+X_{33})\bigg)\bigg]
-\dfrac{1}{\sqrt3}g_1^X\bigg(\bar{p}p(X_{11}+X_{22}+X_{33})\bigg)\,.
\end{align}
From the mass term of Dirac Lagrangian for fermions, we find
\begin{align}
m^*_p=-\frac{\mathcal{L}_{{\rm int},X,p}}{{\bar{p}p}}=&\sqrt2g_8^X\bigg(\alpha_X\left[(\dfrac{\sigma+\delta}{\sqrt2}-\zeta)\right]+(1-\alpha_X)\left[\dfrac{1}{3}(\sqrt2\delta-\dfrac{\sigma-\delta}{\sqrt2}+\zeta)\right]\bigg)+\dfrac{1}{\sqrt3}g_1^X\left[(\sqrt2\sigma+\zeta)\right]\,\nonumber 
\\
=& g_8^X\bigg(\alpha_X\left[\sigma+\delta-\sqrt2\zeta-\dfrac{1}{3}(3\delta-\sigma+\sqrt2\zeta)\right]+\dfrac{1}{3}(3\delta-\sigma+\sqrt2\zeta)\bigg)+\dfrac{1}{\sqrt3}g_1^X\left[(\sqrt2\sigma+\zeta)\right]\,\nonumber 
\\
=& g_8^X\bigg(\alpha_X\left[\dfrac{4}{3}(\sigma-\sqrt2\zeta)\right]-\dfrac{1}{3}(\sigma-\sqrt2\zeta)+\delta\bigg)+\dfrac{1}{\sqrt3}g_1^X\left[(\sqrt2\sigma+\zeta)\right]\, \nonumber 
\\
=&\dfrac{1}{\sqrt3}g_1^X(\sqrt2\sigma+\zeta)+\dfrac{g_8^X}{3}(4\alpha_X-1)(\sigma-\sqrt2\zeta)+g_8^X\delta\,,\label{eq:pp_lagr}
\end{align}
which is a term appearing in the total effective mass of the proton in~\Cref{eq:effective_masses}.  By rearranging the terms for a particular scalar meson, we get
\begin{align}
m^*_p=&  \left[\left( \frac{g_8^X}{3}(4 \alpha_X - 1)  + \sqrt{\frac{2}{3}} g_1^X \right) \sigma +
 \left( -\frac{\sqrt{2}}{3}g_8^X(4 \alpha_X - 1)  + \frac{1}{\sqrt{3}} g_1^X \right)  \zeta  + g_8^X \delta \right]{\bar{p}p}\,.
\end{align}

The neutron has the same coupling to the $\sigma$ and $\zeta$ mesons, but it couples to the $\delta$ with opposite sign, so we can write
\begin{flalign}
    g_{N\sigma}&= \frac{g_8^X}{3}(4 \alpha_X - 1)  + \sqrt{\frac{2}{3}} g_1^X, \nonumber \\
    g_{N\zeta}&= -\frac{\sqrt{2}}{3}g_8^X(4 \alpha_X - 1)  + \frac{1}{\sqrt{3}} g_1^X, \\
    g_{p\delta}&=g_8^X, \qquad g_{n\delta}=-g_8^X\,. \nonumber 
\end{flalign}
Indeed, this holds for all baryons in the octet: the $\sigma$ and $\zeta$ couplings are equal for the baryon families, while the $\delta$ coupling differentiates them due to isospin. Additionally, for the hyperons, the addition of the symmetry-breaking term to fix their potentials,~\Cref{eq:hyperon_potential_lagrangian}, adds an additional contribution proportional to $m^{HO}_3$ in the $\sigma$ and $\zeta$ couplings. For the $\Lambda$ hyperon:
\begin{equation}
    \begin{split}
        g_{\sigma \Lambda} &=\frac{2}{3} g_8^X (\alpha_X-1) + \sqrt{\frac{2}{3}} g_1^X + \sqrt{2} m^{HO}_3\,, \\
        g_{\zeta \Lambda} &= -\frac{2\sqrt{2}}{3} g_8^X (\alpha_X-1) + \frac{1}{\sqrt{3}} g_1^X + m^{HO}_3\,, \\
        g_{\delta \Lambda}&= 0\,.
    \end{split}
\end{equation}
For the $\Sigma$'s: 
\begin{equation}
    \begin{split}
        g_{\sigma \Sigma} &= -\frac{2}{3} g_8^X (\alpha_X-1) + \sqrt{\frac{2}{3}} g_1^X + \sqrt{2} m^{HO}_3\,, \\
        g_{\zeta \Sigma} &= \frac{2\sqrt{2}}{3} g_8^X (\alpha_X-1) + \frac{1}{\sqrt{3}} g_1^X + m^{HO}_3\,, \\
        g_{\delta \Sigma^+}&= 2 g_8^X\alpha_X, \qquad  g_{\delta \Sigma^0}= 0, \qquad g_{\delta \Sigma^-}= -2 g_8^X\alpha_X\,. 
    \end{split}
\end{equation}
And for the $\Xi$'s:
\begin{equation}
    \begin{split}
        g_{\sigma \Xi} &= -\frac{1}{3} g_8^X (2\alpha_X+1) + \sqrt{\frac{2}{3}} g_1^X + \sqrt{2} m^{HO}_3\,, \\
        g_{\zeta \Xi} &= \frac{\sqrt{2}}{3} g_8^X (2\alpha_X+1) + \frac{1}{\sqrt{3}} g_1^X + m^{HO}_3\,, \\
        g_{\delta \Xi^0} &= g_8^X (2\alpha_X-1), \qquad g_{\delta \Xi^-}= -g_8^X (2\alpha_X-1)\,.
    \end{split}
\end{equation}

From this discussion, we can identify the effective masses, written explicitly in terms of the original couplings. In~\Cref{eq:pp_lagr}, the singlet term $m_B^*=\dfrac{1}{\sqrt3}g_1^X(\sqrt2\sigma+\zeta)$  is identical for all baryons. The second term exists for both nucleons $m_N^*\equiv\dfrac{g_8^X}{3}(4\alpha_X-1)(\sigma-\sqrt2\zeta)$, and the third $\delta$ term differentiates the nucleons due to isospin. We can repeat this for all hyperons as well, the terms that are identical for the hyperon multiplets are
\begin{align}
\rm{second\ term\ of}\ m_\Lambda^*=&-\dfrac{2}{3}g_8^X(\alpha_X-1)(\sqrt2\zeta-\sigma) +m^{HO}_3(\sqrt{2}\sigma + \zeta)\,,\nonumber\\
\rm{second\ term\ of}\ m_\Sigma^*=&\dfrac{2}{3}g_8^X(\alpha_X-1)(\sqrt2\zeta-\sigma)+m^{HO}_3(\sqrt{2}\sigma + \zeta)\,,\nonumber\\
\rm{second\ term\ of}\ m_\Xi^*=&\dfrac{1}{3}g_8^X(2\alpha_X+1)(\sqrt2\zeta-\sigma)+m^{HO}_3(\sqrt{2}\sigma + \zeta)\,.
\end{align}
The full effective mass expressions, including the constant mass term $\Delta m_i$, are 
\begin{align}
m_p^*= &\Delta m_N+ m_B^*+m_N^*+g_8^X\delta\,,\nonumber\\
m_n^*=&\Delta m_N+ m_B^*+m_N^*-g_8^X\delta\,,\nonumber\\
m_{\Lambda}^*=& \Delta m_\Lambda+ m_B^*+m_\Lambda^*\,,\nonumber\\
m_{\Sigma^+}^*=&\Delta m_\Sigma+ m_B^*+m_\Sigma^*+2g_8^X\alpha_X\delta\,,\nonumber\\
m_{\Sigma^0}^*=&\Delta m_\Sigma +m_B^*+m_\Sigma^*\,,\nonumber\\
m_{\Sigma^-}^*=& \Delta m_\Sigma + m_B^*+m_\Sigma^*-2g_8^X\alpha_X\delta\,,\nonumber\\
m_{\Xi^0}^*=& \Delta m_\Xi + m_B^* +m_\Xi^*+g_8^X(2\alpha_X-1)\delta\,,\nonumber\\  
m_{\Xi^-}^*=&\Delta m_\Xi + m_B^*+m_\Xi^*-g_8^X(2\alpha_X-1)\delta\,.
\end{align}\\

\textbf{Case 2)} For $M=V$ (the vector-meson matrix in the mean-field approximation)
\begin{align}
V = \mathrm{diag}\Big(\dfrac{\rho+\omega}{\sqrt{2}},\dfrac{-\rho+\omega}{\sqrt{2}},\phi\Big)\,,
\quad \quad
\mathrm{Tr}(V)=\sqrt2\omega+\phi\,,
\end{align}
\begin{align}
\omega=\dfrac{V_{11}+V_{22}}{\sqrt2}\,,
\quad
\rho=\dfrac{V_{11}-V_{22}}{\sqrt2}\,,
\quad
\phi=V_{33}\,.
\end{align}
For the vector mesons, an additional complication arises: the $\omega$ and $\phi$ fields are defined as a combination of the singlet and octet mesons ($v^{1}$ and $v^{8}$) as
\begin{equation}
    \begin{split}
        \omega &= \cos\theta_V v^{1} + \sin\theta_V v^{8}\,, \\
        \phi &= -\sin\theta_V v^{1} + \cos \theta_V v^{8}\,, \\
    \end{split}
\end{equation}
which adds a dependence on $\theta_V$ in the baryon-meson couplings. In principle, this is also the case for the $\sigma$ and $\zeta$ mesons, but for them, the mixing angle is $\theta_S=0$, such that $\sigma$ is purely singlet and $\zeta$ is purely octet, and there is no angular contribution. For the vectors on the other hand, it is customary to take the ideal mixing angle, $\tan\theta_V=\frac{1}{\sqrt{2}}$, with the $\alpha_V=1$ condition (see \Cref{sec:the_baryon_meson_interaction_term} for more). See~\cite{Dover:1985ba} for a more complete discussion.
 
\end{widetext}

\section{Equations of motion and thermodynamics for a free Fermi gas}
\label{app:idealgas}

For a given fermion (or antifermion) $i$ of mass $m_i$, which must obey Fermi-Dirac statistics, the distribution function as a function of energy level, temperature, and chemical potential reads
\begin{equation}\label{eq:fermidirac}
f_{i\pm}(E_i,T,\mu_i)=\dfrac{1}{e^{(E_i\mp\mu_i)/T}+1}\,.
\end{equation}
The Dirac Lagrangian density for spin 1/2 fermions is given by
\begin{equation}
\mathcal{L}=\bar{\psi}_i(i\gamma_\mu\partial^\mu-m_i)\psi_i\,,
\end{equation}
from which applying the Euler-Lagrange equations 
\begin{equation}
\dfrac{\partial\mathcal{L}}{\partial\psi_{i}}-\partial_{\mu}\left(\dfrac{\partial\mathcal{L}}{\partial\left(\partial_{\mu}\psi_{i}\right)}\right)=0\,,\ \ \ \ 
\dfrac{\partial\mathcal{L}}{\partial\bar{\psi}_{i}}-\partial_{\mu}\left(\dfrac{\partial\mathcal{L}}{\partial\left(\partial_{\mu}\bar{\psi}_{i}\right)}\right)=0\,,
\end{equation}
for each particle (or antiparticle) resulting in the equations of motion
\begin{equation}
i\partial_{\mu}\bar{\psi_i}\gamma^{\mu}+m_i\bar{\psi_i}=0\,,\ \ \ \ 
i\gamma^{\mu}\partial_{\mu}\psi_i-m_i\psi_i=0\,.
\end{equation}

These are both linear and first-order, indicating a plane-wave solution of the form
\begin{equation}\label{eq:psi_solution}
\psi_i(t,\vec{x})=\Psi(\vec{k},s)e^{-i(E_i t-\vec{k}\cdot\vec{x})}\,,
\end{equation}
where $\Psi$ is a four-vector spinor for Fermi momentum $\vec{k}$ and spin $s$. If $\psi_i$ satisfies the equations of motion, then the Lagrangian is zero. We can use this in the energy-momentum tensor together with the $3+1$ dimensional Minkowski metric 
$g_{\mu \nu} = \mathrm{diag(+,-,-,-)}$
to obtain the energy-momentum tensor
\begin{equation}
T_{\mu\nu}=-\mathcal{L}g_{\mu\nu}+\dfrac{\partial\mathcal{L}}{\partial (\partial_{\mu}\psi_{i})}\partial_{\nu}\psi_{i}\,,
\label{Tfull}
\end{equation}
yielding
\begin{equation}
T_{\mu\nu}=i\bar{\psi_i}\gamma^\mu\partial_\nu\psi_i\,.
\end{equation}

The energy density and pressure are then obtained either directly from the Lagrangian density using finite-$T$ Quantum Field Theory and thermodynamical relations, or using the ideal fluid approximation (where there is no dissipation and all non-diagonal terms vanish), resulting in the latter case in
\begin{equation}
\varepsilon_i=T_{00}\,,\ \ \ \ 
P_i=\dfrac{1}{3}\sum_{j=1}^3 T_{jj}\,,
\end{equation}
giving
\begin{equation}\label{eq:energy_and_pressure}
\varepsilon_i=i\bar{\psi_i}\gamma^0\partial_0\psi_i\,, \ \ \ \ 
P_i=-\dfrac{i}{3}\bar{\psi_i}\vec{\gamma}\cdot\vec{\nabla}\psi_i\,. 
\end{equation}
Additionally, we assume there is rotational symmetry, which is broken by the presence of magnetic fields, that would to different pressures in the directions longitudinal and perpendicular to the local direction of the field.
Applying plane-wave $\psi$ and periodic boundary conditions it can be shown that
\begin{equation}
\varepsilon_i=\dfrac{\gamma_i}{2\pi^2}\int_0^\infty dk E_i k^2(f_{i+}+f_{i-})\,,
\label{epsT}
\end{equation}
and
\begin{equation}
P_i=\dfrac{1}{3}\dfrac{\gamma_i}{2\pi^2}\int_0^\infty dk\dfrac{k^4}{E_i}(f_{i+}+f_{i-})\,,
\end{equation}
where $\gamma_i=2$ for baryons and leptons and $\gamma_i=6$ for quarks is the particle degeneracy. $E_i=\sqrt{k^2+m_i^2}$ are particle energy levels. Likewise, the number density and entropy density can be calculated as
\begin{equation}
n_i=\dfrac{\gamma_i}{2\pi^2}\int_0^\infty dkk^2(f_{i+}-f_{i-})\,,
\end{equation}
and
\begin{align}
s_i=&\dfrac{\gamma_i}{2\pi^2}\int_0^\infty dk k^2\Bigg[f_{i+}\ln\left(\dfrac{1}{f_{i+}}\right)
+f_{i-}\ln\left(\dfrac{1}{f_{i-}}\right)\nonumber\\
&+(1-f_{i+})\ln\left(\dfrac{1}{1-f_{i+}}\right)+(1-f_{i-})\ln\left(\dfrac{1}{1-f_{i-}}\right)\Bigg]\,.
\label{entT}
\end{align}

Additionally, in the presence of interactions, the scalar (number) density $n_{sc,i}$ (or the source for scalar fields) is
\begin{equation}\label{eq:scalar_densityT}
 n_{sc,i}=\dfrac{\gamma_i}{2 \pi^2}\int_0^\infty dk \dfrac{k^2m_i}{E_i}\left(f_{i+}+f_{i-}\right)\,.
\end{equation}

In the limit of zero temperature, the Fermi-Dirac distribution for fermions, $f_{i+}$, becomes unity between $k=0$ and $k=k_{Fi} $ and zero for higher $k$'s. The Fermi-Dirac distribution for antifermions, $f_{i-}$, becomes zero. As a consequence, the direct integration of eqs.~\Cref{epsT}--~\Cref{eq:scalar_densityT} yield eqs.~\Cref{eps0}--~\Cref{eq:scalar_density} and $s_i=0$. 

\section{How to use the software}
\label{sec:Run_Software}

There are multiple ways to run the CMF solver software to calculate equations of state:
\begin{itemize}
  \item To run a calculation using a standalone script, download the source code package from the associated software publication \cite{zenodo_cmf}, where you can also find instructions detailing how to compile and execute the code along with directions to the MUSES support community.
  \item You may also run the CMF solver as a MUSES module in a processing workflow executed by the MUSES Calculation Engine. This method allows you to optionally include other MUSES modules in your workflows to perform more complex data processing. The Calculation Engine is also free and open-source software, available both for download and as an online service offered by the MUSES collaboration. Although a local installation requires Docker Compose, the use of containerization means you can run the software without installing the complex set of specific dependencies required by the CMF module. See the MUSES project website to learn more \cite{MUSES}.
\end{itemize}

\section{Thermodynamic stability in the different ensembles}
\label{Ensembles}

Given a multi-variable function dependent on $3$ parameters F(a,b,c), we can ensure it possesses a minimum by showing that it has an extremum 
\begin{equation}
dF(a,b,c)=0\,,
\end{equation}
and that it has positive concavity. The latter follows from the determinant of the Hessian matrix 
\begin{equation}
    M=\begin{bmatrix}
    \frac{\partial^2 F}{\partial a^2}\Big|_{b,c}&
    \frac{\partial^2 F}{\partial a\partial b}\Big|_{c} & \frac{\partial^2 F}{\partial a\partial c}\Big|_{b} \\
    \frac{\partial^2 F}{\partial b\partial a}\Big|_{c}&
    \frac{\partial^2 F}{\partial b^2}\Big|_{a,c} & \frac{\partial^2 F}{\partial b\partial c}\Big|_{a} \\
    \frac{\partial^2 F}{\partial c\partial a}\Big|_{b}&
    \frac{\partial^2 F}{\partial c\partial b}\Big|_{a} & \frac{\partial^2 F}{\partial c^2}\Big|_{a,b} \\
    \end{bmatrix}\,,
\end{equation}
and its submatrices being $\ge0$. In the case of, e.g., $\frac{\partial^2 F}{\partial a\partial b}\Big|_{c}$, it is implied that this means $\frac{\partial }{\partial b}\Big|_{a,c}\frac{\partial F}{\partial a}\Big|_{b,c}$. The order of derivatives does not matter, as this matrix is symmetric, but permutations of the variables must be included, as there is no physical justification for any particular order.

\subsection{Microcanonical ensemble}

In this ensemble, based on the conservation of energy $E=-PV+TS+N_x\mu_x$, the fixed variables are a number of $x$ particles $N_x$, volume $V$, and energy $E$. Minimization of energy implies that the differential 
\begin{equation}
dE=-PdV+TdS+\mu_xdN_x=0\,,
\end{equation}
and det$M\ge0$ with ($a\rightarrow V$, $b\rightarrow S$, and $c\rightarrow N_x$)
\begin{equation}
    M=\begin{bmatrix}
    -\frac{\partial P}{\partial V}\Big|_{S,N_x} & -\frac{\partial P}{\partial S}\Big|_{V,N_x} & -\frac{\partial P}{\partial N_x}\Big|_{V,S} \\
    \frac{\partial T}{\partial V}\Big|_{S,N_x} & 
    \frac{\partial T}{\partial S}\Big|_{V,N_x} & \frac{\partial T}{\partial N_x}\Big|_{S,V}  \\
   \frac{\partial \mu_x}{\partial V}\Big|_{N_x,S} & \frac{\partial \mu_x}{\partial S}\Big|_{N_x,V} & \frac{\partial \mu_x}{\partial N_x}\Big|_{V,S} 
    \end{bmatrix}\,,
\end{equation}
where we used that 
$dE/dV=-P$, $dE/dS=T$, and $dE/dN_x=\mu_x$.

In the zero-temperature limit, $M$ reduces to
\begin{equation}
    M=\begin{bmatrix}
    -\frac{\partial P}{\partial V}\Big|_{N_x} & -\frac{\partial P}{\partial N_x}\Big|_{V}  \\
   \frac{\partial \mu_x}{\partial V}\Big|_{N_x} &
   \frac{\partial \mu_x}{\partial N_x}\Big|_{V} 
    \end{bmatrix}\,,
    \label{micro}
\end{equation}
and stability requires
\begin{align}
-\frac{\partial P}{\partial V}\Big|_{N_x}&\ge 0\,,\\
\frac{\partial \mu_x}{\partial N_x}\Big|_{V}&\ge 0 \,,
\label{leChatelier}
\end{align}
and
\begin{equation}
-\frac{\partial P}{\partial V}\Big|_{N_x}\frac{\partial \mu_x}{\partial N_x}\Big|_{V}+\frac{\partial P}{\partial N_x}\Big|_{V} \frac{\partial \mu_x}{\partial V}\Big|_{N_x}
\ge0\,.
\end{equation}
Using the Maxwell relation $-\frac{\partial P}{\partial N}\Big|_{V}=\frac{\partial \mu_x}{\partial V}\Big|_{N}$, we obtain
\begin{equation}
-\frac{\partial P}{\partial V}\Big|_{N_x}\frac{\partial \mu_x}{\partial N_x}\Big|_{V}\ge\left(\frac{\partial P}{\partial N_x}\Big|_{V}\right)^2\,.
\end{equation}

Using \Cref{leChatelier} and the definition of $n_x=N_x/V$, one can also write
\begin{align}
\frac{n_x^2}{N_x}\frac{\partial P}{\partial n_x}\Big|_{N_x}&\ge 0\,,\\
\frac{\partial P}{\partial n_x}\Big|_{N_x}&\ge 0\,.
\label{leChatelier2}
\end{align}

\subsection{Canonical ensemble}
\label{ensembles}

In this ensemble, based on the conservation of (Helmholtz) free energy $F=E-ST$, the fixed variables are a number of particles $N$, volume $V$, and energy $T$. Minimization of free energy implies that the differential 
\begin{equation}
dF=-PdV-SdT+\mu_xdN=0\,,
\end{equation}
and det$M\ge0$ with ($a\rightarrow V$, $b\rightarrow T$, and $c\rightarrow N_x$)
\begin{equation}
    M=\begin{bmatrix}
    -\frac{\partial P}{\partial V}\Big|_{T,N_x} & 
    -\frac{\partial P}{\partial T}\Big|_{V,N_x} & 
    -\frac{\partial P}{\partial N_x}\Big|_{V,T} \\
    -\frac{\partial S}{\partial V}\Big|_{T,N_x} & 
    -\frac{\partial S}{\partial T}\Big|_{V,N_x} & 
    -\frac{\partial S}{\partial N_x}\Big|_{T,V}  \\
   \frac{\partial \mu_x}{\partial V}\Big|_{N_x,T} & \frac{\partial \mu_x}{\partial T}\Big|_{N_x,V} & \frac{\partial \mu_x}{\partial N_x}\Big|_{V,T} 
    \end{bmatrix}\,,
\end{equation}
where we used that 
$dF/dV=-P$, $dF/dT=-S$, and $dF/dN_x=\mu_x$.

In the zero-temperature limit, $M$ reduces once more to \Cref{micro}.

\subsection{Grand canonical ensemble}

In this ensemble, based on the conservation of grand potential $\Omega=E-TS-\mu_xN$, the fixed variables are chemical potentials $\mu_x$, volume $V$, and energy $T$. Minimization of grand potential implies that the differential 
\begin{equation}
d\Omega=-PdV-SdT-N_xd\mu_x=0\,,
\end{equation}
and det$M\ge0$ with ($a\rightarrow V$, $b\rightarrow T$, and $c\rightarrow \mu_x$)
\begin{equation}
    M=\begin{bmatrix}
    -\frac{\partial P}{\partial V}\Big|_{T,\mu_x} & 
    -\frac{\partial P}{\partial T}\Big|_{V,\mu_x} & 
    -\frac{\partial P}{\partial \mu_x}\Big|_{V,T} \\
    -\frac{\partial S}{\partial V}\Big|_{T,\mu_x} & 
    -\frac{\partial S}{\partial T}\Big|_{V,\mu_x} & 
    -\frac{\partial S}{\partial \mu_x}\Big|_{T,V}  \\
   -\frac{\partial N_x}{\partial V}\Big|_{\mu_x,T} & 
   -\frac{\partial N_x}{\partial T}\Big|_{\mu_x,V} & 
   -\frac{\partial N_x}{\partial \mu_x}\Big|_{V,T} 
    \end{bmatrix}\,,
\end{equation}
where we used that 
$d\Omega/dV=-P$, $d\Omega/dT=-S$, and $d\Omega/d\mu_x=-N_x$.

In the zero-temperature limit, $M$ reduces to
\begin{equation}
    M=\begin{bmatrix}
    -\frac{\partial P}{\partial V}\Big|_{\mu_x} & 
    -\frac{\partial P}{\partial \mu_x}\Big|_{V}\\
   -\frac{\partial N_x}{\partial V}\Big|_{\mu_x} & 
   -\frac{\partial N_x}{\partial \mu_x}\Big|_{V} 
    \end{bmatrix}\,,
\end{equation}
and stability requires
\begin{align}
    -\frac{\partial P}{\partial V}\Big|_{\mu_x}&\ge 0\,,\\
    -\frac{\partial N_x}{\partial \mu_x}\Big|_{V} &\ge 0\,,
\end{align}
and
\begin{equation}
\frac{\partial P}{\partial V}\Big|_{\mu_x}\frac{\partial N_x}{\partial \mu_x}\Big|_{V}-\frac{\partial P}{\partial \mu_x}\Big|_{V} \frac{\partial N_x}{\partial V}\Big|_{\mu_x}
\ge0\,.
\end{equation}
Using the Maxwell relation $\frac{\partial P}{\partial \mu_x}\Big|_{V}=\frac{\partial N}{\partial V}\Big|_{\mu_x}$, we obtain
\begin{equation}
\frac{\partial P}{\partial V}\Big|_{\mu_x}\frac{\partial N_x}{\partial \mu_x}\Big|_{V}\ge\left(\frac{\partial P}{\partial \mu_x}\Big|_{V}\right)^2\,.
\end{equation}

\subsection{Infinite volume limit}

For bulk matter, it is convenient to divide our grand potential with respect to the volume to get $P=-\Omega/V=-\epsilon-Ts-\mu_xn$, the fixed variables now being only $\mu_x$ and $T$. Maximization of the pressure implies that the differential 
\begin{equation}
dP=sdT+n_xd\mu_x=0\,,
\end{equation}
and det$M\ge0$ with ($a\rightarrow \infty$, $b\rightarrow T$, and $c\rightarrow \mu_x$)
\begin{equation}
    M=\begin{bmatrix} 
    \frac{\partial s}{\partial T}\Big|_{\mu_x} & 
    \frac{\partial s}{\partial \mu_x}\Big|_{T}  \\
   \frac{\partial n_x}{\partial T}\Big|_{\mu_x} & 
   \frac{\partial n_x}{\partial \mu_x}\Big|_{T} 
    \end{bmatrix}\,,
\label{D20}\end{equation}
\vspace{0.8cm}
where we used that 
$dP/dT=s$, and $dP/d\mu_x=n_x$.

In the zero-temperature limit, $M$ reduces to
\begin{equation}
    M=\begin{bmatrix} 
   \frac{\partial n_x}{\partial \mu_x} 
    \end{bmatrix}\,,
\end{equation}
and stability requires
\begin{align}
    \frac{\partial n_x}{\partial \mu_x} &\ge 0\,.
    \label{result}
\end{align}
We can also write for our particular case
\begin{align}
\frac{\partial n_x}{\partial P}\frac{\partial P}{\partial \mu_x} &\ge 0\,,\\
\frac{\partial n_x}{\partial P}n_x &\ge 0\,,
\end{align}
which using \Cref{result} means
\begin{align}
\frac{\partial P}{\partial n_x}\ge 0\quad \rm{for}\quad n_x\ge0\,.
\end{align}
This is the case for baryon number, $x=B$, at $T=0$. Note that this is similar to~\Cref{leChatelier2}, one of the stability conditions in the microcanonical or canonical ensembles.

\begin{widetext}
\subsection{Multiple chemical potentials}
\label{Ensembles5}

Expanding~\Cref{D20} to the case of $3$ chemical potentials, $\mu_B$, $\mu_S$, and $\mu_Q$
\begin{equation}\label{eqn:M_matrix_finiteT_appen}
   M=\begin{bmatrix}
    \frac{\partial s}{\partial T}\big|_{\vec{\mu}} & \frac{\partial s}{\partial \mu_B}\big|_{T,\mu_S,\mu_Q} & \frac{\partial s}{\partial \mu_S}\big|_{T,\mu_B,\mu_Q} & \frac{\partial s}{\partial \mu_Q}\big|_{T,\mu_B,\mu_S} \\
    \frac{\partial n_B}{\partial T}\big|_{\vec{\mu}} & \frac{\partial n_B}{\partial \mu_B}\Big|_{T,\mu_S,\mu_Q} & \frac{\partial n_B}{\partial \mu_S}\Big|_{T,\mu_B,\mu_Q} & \frac{\partial n_B}{\partial \mu_Q}\Big|_{T,\mu_B,\mu_S} \\
   \frac{\partial n_S}{\partial T}\big|_{\vec{\mu}} &   \frac{\partial n_S}{\partial \mu_B}\Big|_{T,\mu_S,\mu_Q} & \frac{\partial n_S}{\partial \mu_S}\Big|_{T,\mu_B,\mu_Q} & \frac{\partial n_S}{\partial \mu_Q}\Big|_{T,\mu_B,\mu_S} \\
     \frac{\partial n_Q}{\partial T}\big|_{\vec{\mu}} & \frac{\partial n_Q}{\partial \mu_B}\Big|_{T,\mu_S,\mu_Q} & \frac{\partial n_Q}{\partial \mu_S}\Big|_{T,\mu_B,\mu_Q}  & \frac{\partial n_Q}{\partial \mu_Q}\Big|_{T,\mu_B,\mu_S}
    \end{bmatrix}\,,
\end{equation}
or using~\Cref{72,75}.
\begin{equation}\label{eqn:M_matrix_finiteT_appen_chi}
    M=\begin{bmatrix}
    \frac{\partial s}{\partial T}\big|_{\vec{\mu}} & \frac{\partial s}{\partial \mu_B}\big|_{T,\mu_S,\mu_Q} & \frac{\partial s}{\partial \mu_S}\big|_{T,\mu_B,\mu_Q} & \frac{\partial s}{\partial \mu_Q}\big|_{T,\mu_B,\mu_S} \\
    \frac{\partial n_B}{\partial T}\big|_{\vec{\mu}} & \chi_2^B & \chi_{11}^{BS} & \chi_{11}^{BQ} \\
   \frac{\partial n_S}{\partial T}\big|_{\vec{\mu}} &   \chi_{11}^{SB} & \chi_2^S & \chi_{11}^{SQ} \\
     \frac{\partial n_Q}{\partial T}\big|_{\vec{\mu}} & \chi_{11}^{QB} & \chi_{11}^{QS}  & \chi_2^Q
    \end{bmatrix}\,.
\end{equation}
Then, to find the stability constraints, one must ensure that the determinants of all submatrices are positive. 
Thus, one can show that at finite $T$ and infinite $V$ the constraints are:
\begin{align}
    \chi_2^B\geq0\,, \quad \chi_2^S&\geq 0\,, \quad \chi_2^Q\geq0\,,\\
\chi_2^B\chi_2^S&\geq \left(\chi_{11}^{BS}\right)^2\,,  \\
\chi_2^S\chi_2^Q&\geq \left(\chi_{11}^{SQ}\right)^2\,, \\
\chi_2^B\chi_2^Q&\geq \left(\chi_{11}^{BQ}\right)^2\,, 
\end{align}
\begin{align}
\chi_2^B\chi_2^S\chi_2^Q+2\left(\chi_{11}^{BS}\chi_{11}^{BQ}\chi_{11}^{SQ}\right)&\geq \chi_2^B\left(\chi_{11}^{SQ}\right)^2+\chi_2^S\left(\chi_{11}^{BQ}\right)^2+\chi_2^Q\left(\chi_{11}^{BS}\right)^2\,,
\end{align}
\begin{align}
\frac{\partial s}{\partial T}\bigg|_{\vec{\mu}} &\geq  0\,, \\
\chi_2^B \frac{\partial s}{\partial T}\bigg|_{\vec{\mu}}&\geq  \left[\frac{\partial s}{\partial \mu_B}\bigg|_{T,\mu_S,\mu_Q}\right]^2\,, \\
\chi_2^S \frac{\partial s}{\partial T}\bigg|_{\vec{\mu}}&\geq  \left[\frac{\partial s}{\partial \mu_S}\bigg|_{T,\mu_B,\mu_Q}\right]^2\,, \\
\chi_2^Q \frac{\partial s}{\partial T}\bigg|_{\vec{\mu}}&\geq  \left[\frac{\partial s}{\partial \mu_Q}\big|_{T,\mu_B,\mu_S}\right]^2\,, 
\end{align}
\begin{align}
    \frac{\partial s}{\partial T}\bigg|_{\vec{\mu}}\left[\chi_2^B\chi_2^S-\left(\chi_{11}^{BS}\right)^2\right] +2\chi_{11}^{BS}\frac{\partial s}{\partial \mu_B}\bigg|_{T,\mu_S,\mu_Q} \frac{\partial s}{\partial \mu_S}\bigg|_{T,\mu_B,\mu_Q} &\geq & \chi_2^B \left(\frac{\partial s}{\partial \mu_S}\bigg|_{T,\mu_B,\mu_Q}\right)^2+\chi_2^S \left(\frac{\partial s}{\partial \mu_B}\bigg|_{T,\mu_S,\mu_Q}\right)^2\,,\\
    \frac{\partial s}{\partial T}\bigg|_{\vec{\mu}}\left[\chi_2^B\chi_2^Q-\left(\chi_{11}^{BQ}\right)^2\right] +2\chi_{11}^{BQ}\frac{\partial s}{\partial \mu_B}\bigg|_{T,\mu_S,\mu_Q} \frac{\partial s}{\partial \mu_Q}\bigg|_{T,\mu_B,\mu_S} &\geq & \chi_2^B \left(\frac{\partial s}{\partial \mu_Q}\bigg|_{T,\mu_B,\mu_S}\right)^2+\chi_2^Q \left(\frac{\partial s}{\partial \mu_B}\bigg|_{T,\mu_S,\mu_Q}\right)^2\,,\\
    \frac{\partial s}{\partial T}\bigg|_{\vec{\mu}}\left[\chi_2^S\chi_2^Q-\left(\chi_{11}^{SQ}\right)^2\right] 
    +2\chi_{11}^{SQ}\frac{\partial s}{\partial \mu_S}\bigg|_{T,\mu_B,\mu_Q} \frac{\partial s}{\partial \mu_Q}\bigg|_{T,\mu_B,\mu_S} 
    &\geq & 
    \chi_2^S \left(\frac{\partial s}{\partial \mu_Q}\bigg|_{T,\mu_B,\mu_S}\right)^2
    +\chi_2^Q \left(\frac{\partial s}{\partial \mu_S}\bigg|_{T,\mu_B,\mu_Q}\right)^2\,,
\end{align}
\begin{align}
    \frac{\partial s}{\partial T}\bigg|_{\vec{\mu}}\left[\chi_2^B\chi_2^S\chi_2^Q+2\chi_{11}^{BS}\chi_{11}^{SQ}\chi_{11}^{BQ}-\sum_{j=B,S,Q}\chi_2^j
    \left(\chi_{11}^{j+1,j+2}\right)^2\right]
    +\sum_{j=B,S,Q}\left[\left(\chi_{11}^{j,j+1}\right)^2-\chi_2^j\chi_2^{j+1}\right]\left(\frac{\partial s}{\partial \mu_{j+2}}\bigg|_{T,\mu_{\neq j+2}}\right)^2& &\nonumber\\
+2\sum_{j=B,S,Q}\left[ \chi_2^j\chi_{11}^{j+1,j+2}-\chi_{11}^{j,j+1}\chi_{11}^{j,j+2} \right]\frac{\partial s}{\partial \mu_{j+1}}\bigg|_{T,\mu_{\neq j+1}}\frac{\partial s}{\partial \mu_{j+2}}\bigg|_{T,\mu_{\neq j+2}}&<&0\,.\nonumber\\
& &
\end{align}
\end{widetext}

\section{Additional results: C1-C4 with baryon octet + quarks}
\label{sec:C1_C4_baryon_octet_quarks}

\renewcommand\thefigure{G\arabic{figure}}  
\setcounter{figure}{0}

\begin{figure*}
    \centering
    \includegraphics[width=0.8\textwidth]{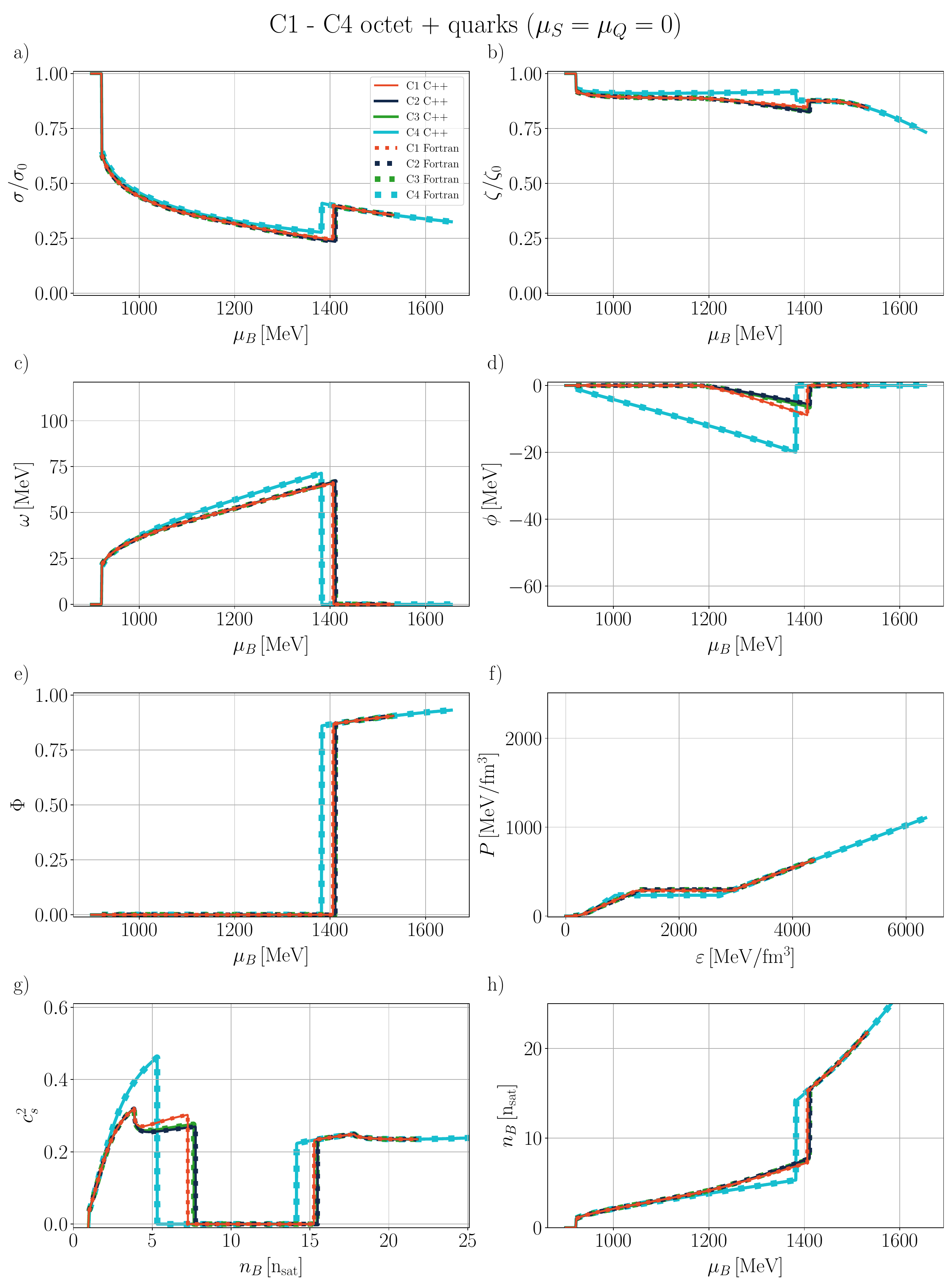}
    \caption{C1-C4 ($\mu_S=\mu_Q=0$) octet + quarks: a) b) scalar meson fields (normalized by vacuum values), c) d) vector meson fields, and e) deconfinement field as a function of baryon chemical potential, f) pressure vs energy density, g) speed of sound vs baryon density, h) baryon density vs baryon chemical potential. Comparison of results from \texttt{Fortran} for stable branch (dashed lines) and \texttt{CMF++} for stable branch (solid lines) for C1 (red-orange), C2 (black), C3 (green), and C4 (cyan).
    }
\label{fig:2D_all_hyper_quarks_muS_0_muQ_0_mf_and_obs}
\end{figure*}

\begin{figure*}
    \centering
    \includegraphics[width=0.8\textwidth]{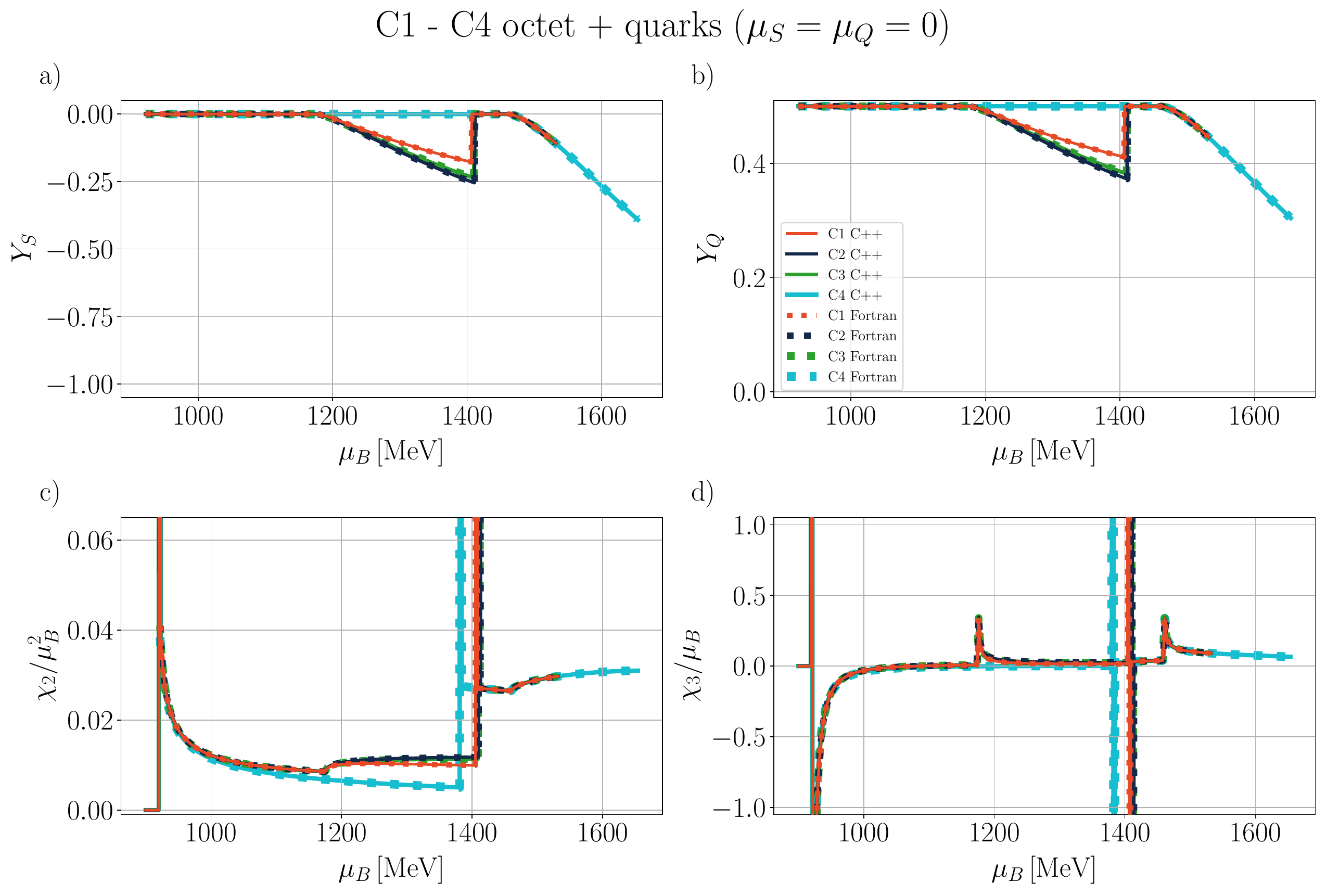}
 \caption{C1-C4 ($\mu_S=\mu_Q=0$) octet + quarks: a) strangeness and b) charge fractions vs baryon chemical potential, c) second and d) third order baryon susceptibilities, all versus baryon chemical potential. Comparison of results from \texttt{Fortran} for stable branch (dashed lines) and \texttt{CMF++} for stable branch (solid lines) for C1 (red-orange), C2 (black), C3 (green), and C4 (cyan).}
\label{fig:2D_all_hyper_quarks_muS_0_muQ_0_dens_and_sus}
\end{figure*}

Below we make a direct comparison between all four couplings C1-C4 for only the stable solutions for the baryon octet+quarks at $\mu_S=\mu_Q=0$. 
%
%
Within panels a)-d) of \Cref{fig:2D_all_hyper_quarks_muS_0_muQ_0_mf_and_obs}, we depict mean-field mesons against $\mu_B$. As the previous discussions have covered in detail the C3 and C4 coupling schemes, we proceed to examine C1 and C2 in detail here.
We note that the stable solutions of \texttt{C++} and legacy \texttt{Fortran} solutions match for these couplings, except for the liquid-gas first-order phase transition, where \texttt{Fortran} presents no points. 
The $\sigma$ and $\omega$ mean fields (panel a) and b)) follow a trend similar to C3 and C4.
The C1 and C2 coupling schemes
showcase a pronounced deconfinement first-order phase transition (panel e)) around $\mu_B = 1406.5$ MeV, and $\mu_B = 1411.5$ MeV, respectively.

Concerning the strange mean-field mesons, for C1 and C2, they decrease with increasing $\mu_B$, particularly upon the emergence of hyperons, until the deconfinement phase transition. Above the deconfinement phase transition, the strange field $\zeta$ decreases in value, while, the $\phi$ field drops to zero in the quark phase (due to its lack of coupling with quarks, see \Cref{tab:quark_param}), exactly as C3 and C4. 


For C1 and C2, $\Phi$ (panel e)) behaves just like C3. For the EoS (panel f)), all couplings exhibit first-order deconfinement phase transition. In the hadronic phase, the wiggle observed in  C1 and C2, just like C3, is due to the emergence of $\Lambda$ hyperons, indicating a higher-order phase transition as confirmed by the speed of sound plot in panel g). In the quark phase, all coupling schemes overlap because vector fields do not couple with quarks. Panel h) shows that the density of C1 and C2 aligns with C3.

In panels a) and b) of \Cref{fig:2D_all_hyper_quarks_muS_0_muQ_0_dens_and_sus}, we observe $Y_S$ and $Y_Q$, with C1 and C2 resembling C3. Lastly, panels c) and d) depict susceptibilities for all coupling cases, with the discontinuities marking the first-order phase transitions (marked by discontinuities in $\chi_2$). The discontinuities in $\chi_3$ indicate that the onset of
strangeness in the stable phases (both hadronic and quark) is a third-order phase transition.

\section{Additional results: stable, metastable, and unstable solutions for C1-C4 with baryon octet + decuplet ($\mu_S=\mu_Q=0$)}
\label{sec:C1_C4_baryon_decuplet_muS0_muQ0}
\renewcommand\thefigure{H\arabic{figure}}  
\setcounter{figure}{0}

\begin{figure*}
    \centering
    \includegraphics[width=0.8\textwidth]{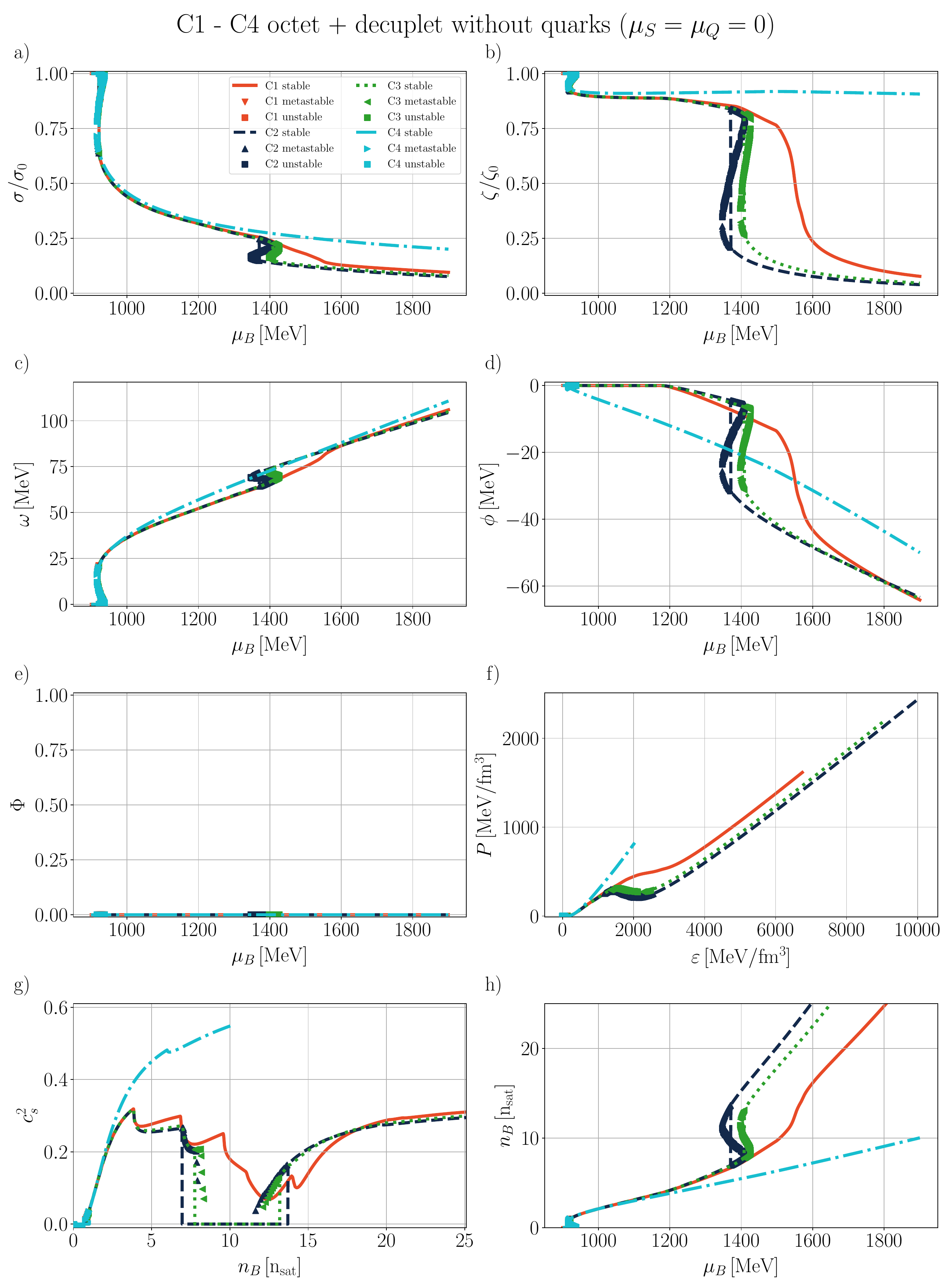}
\caption{C1-C4 ($\mu_S=\mu_Q=0$) octet + decuplet: a) b) scalar meson fields (normalized by vacuum values), c) d) vector meson fields, and e) deconfinement field as a function of baryon chemical potential, f) pressure vs energy density, g) speed of sound vs baryon density, h) baryon density vs baryon chemical potential. Results from \texttt{CMF++} solutions for C1 (red-orange), C2 (black), C3 (green), and C4 (cyan) for stable (full lines), metastable (upside-down triangles), and unstable (squares) branches. }
\label{fig:12_stable+metastable+unstable}
\end{figure*}

\Cref{fig:12_stable+metastable+unstable} shows the same results as in \Cref{fig:2D_all_decuplet_hyper_noquarks_muS_0_muQ_0_mf_and_obs} including the octet and the decuplet of baryons for parametrizations C1-C4 but including also metastable and unstable phases. One can see that for the parametrizations C2 and C3 that present a first-order chiral symmetry phase transition, there is an entire unstable branch that connect both sides of the Maxwell construction for the hadronic (chiral) phase transition. This is easier seen for the speed of sound in panel g). Note that this is different from the deconfinement phase transition, where in that case there are metastable branches instead around the Maxwell construction. For the liquid-gas phase transition at low $\mu_B$, all parametrizations show unstable, then metastable branches.

\begin{figure*}
    \centering
    \includegraphics[width=0.9\textwidth]{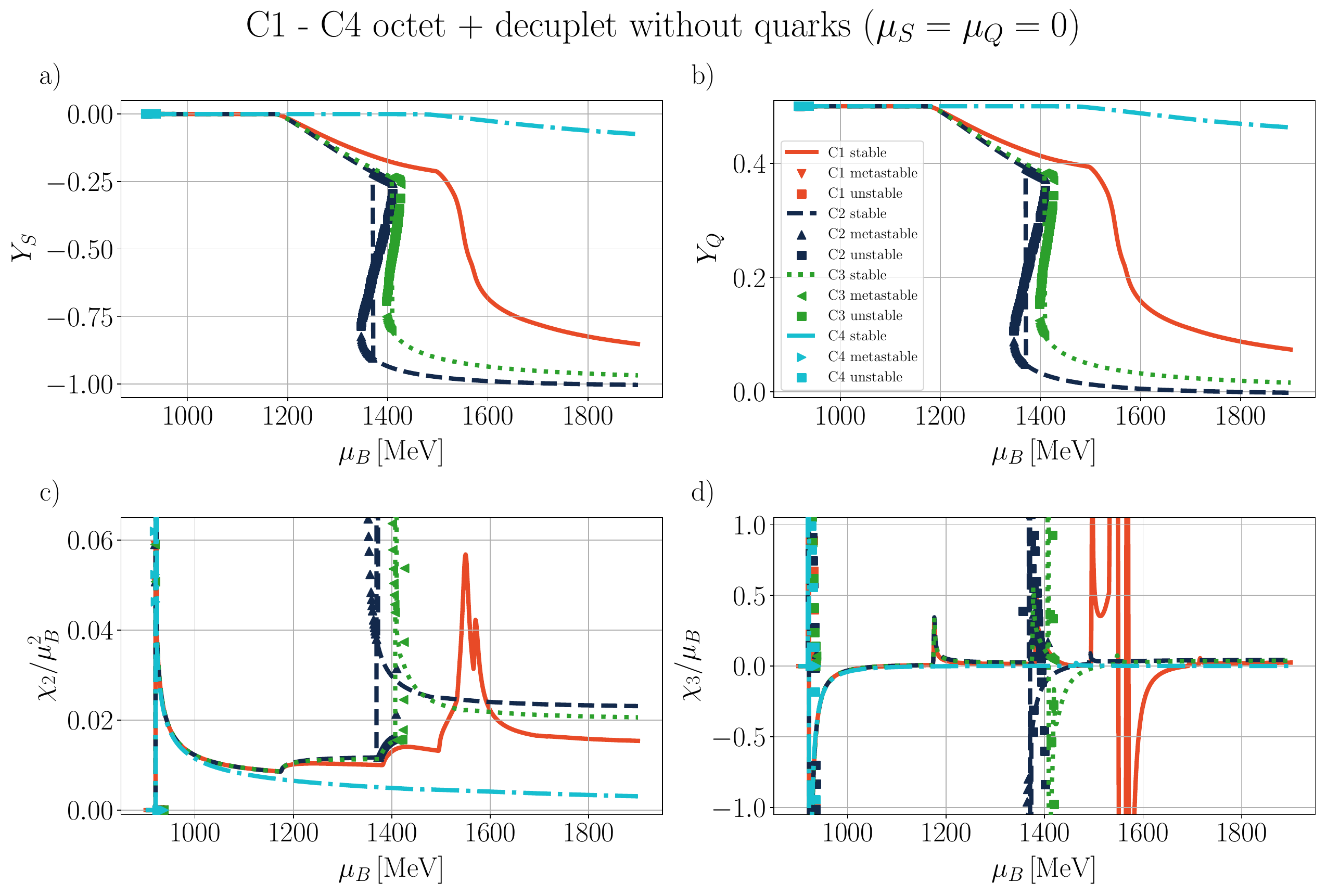}
\caption{C1-C4 ($\mu_S=\mu_Q=0$) octet + decuplet: a) strangeness and b) charge fractions vs baryon chemical potential, c) second and d) third order baryon susceptibilities, all versus baryon chemical potential. Results from \texttt{CMF++} solutions for C1 (red-orange), C2 (black), C3 (green), and C4 (cyan) for stable (full lines), metastable (upside-down triangles), and unstable (squares) branches. }
\label{fig:14_stable+metastable+unstable}
\end{figure*}

Likewise, \Cref{fig:14_stable+metastable+unstable} shows the same results as in \Cref{fig:2D_all_decuplet_hyper_noquarks_muS_0_muQ_0_dens_and_sus} including the octet and the decuplet of baryons for parametrizations C1-C4 but including also metastable and unstable phases. Metastable and unstable phases can be seen again for the  chiral phase transition for C3 and C4 in all panels, as well as the liquid-gas phase transition for al couplings. In the bottom panels, the susceptibilities show metastable and unstable phases only for the phase transitions of first order, not, e.g., for the chiral symmetry restoration for C1.

\section{Susceptibilities for C3 parametrization with baryon octet + decuplet + quarks}
\label{sec:C3_critical_point_sus}
\renewcommand\thefigure{I\arabic{figure}}  
\setcounter{figure}{0}

\begin{figure*}
    \centering
    \includegraphics[width=0.9\linewidth]{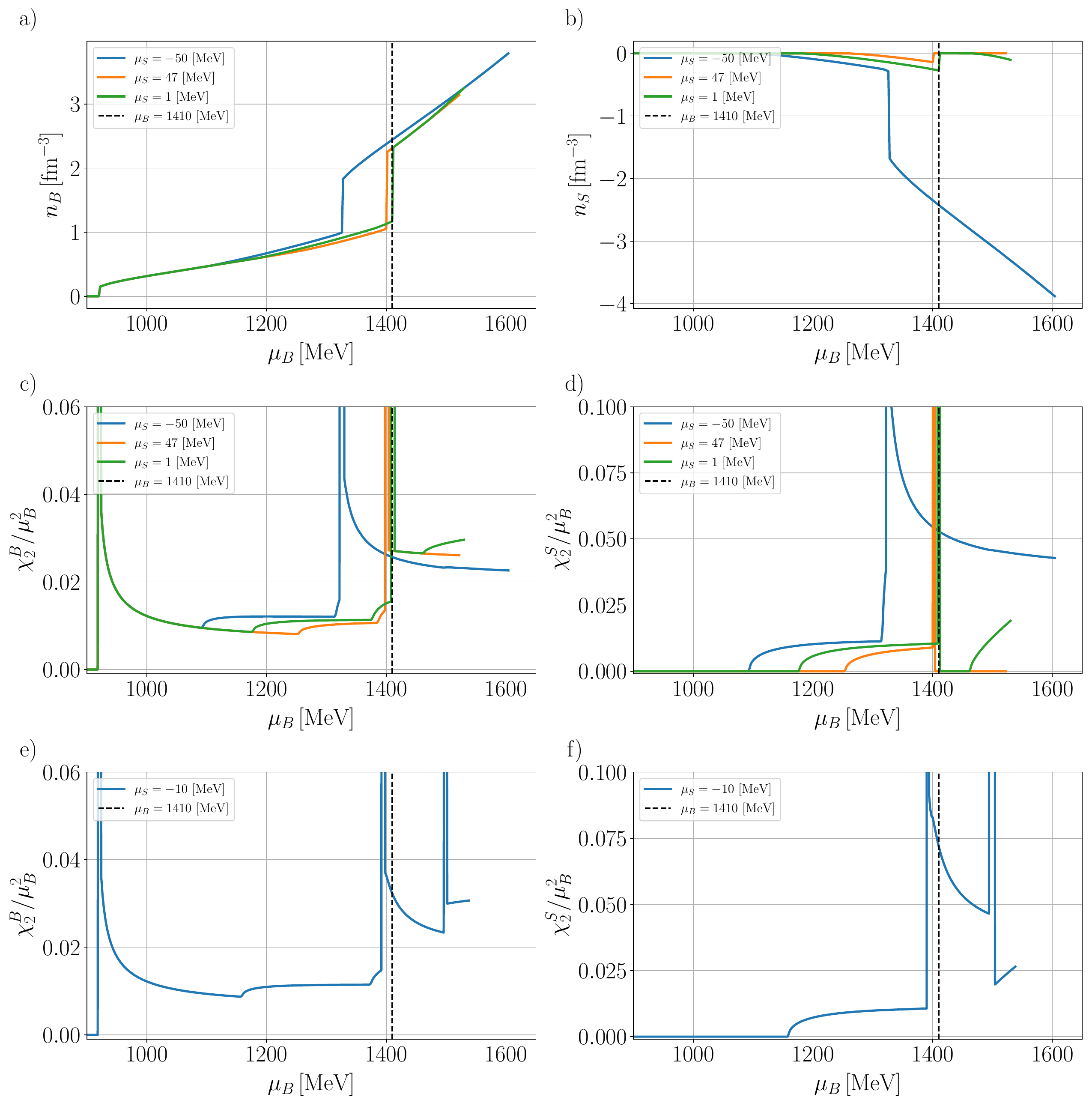}
\caption{C3 ($\mu_Q=0$) octet + decuplet + quarks: 
a) baryon density, b) strangeness density, c) and e) second-order baryon susceptibilities, and d) and f) second-order strangeness susceptibilities, all as a function of baryon chemical potential. 
The top rows demonstrate first-order phase transitions into the quark phase (orange), into the strangeness-dominated hadronic phase (blue), and passing by the triple point (green). 
The black dashed line indicates the exact $\mu_B$ where the triple point is located in \Cref{fig:3D_C3_decuplet_hyper_quarks_muS_N0_muQ_0_panel} (at $\mu_B=1410$, $\mu_S=1$, $\mu_Q=0$ MeV). 
The bottom row displays the regime where one crosses from the light hadronic phase at low $\mu_B$, into the strangeness-dominated phase, followed by the quark phase at high $\mu_B$ (both phase transitions are first-order). The left panels also show the liquid-gas first-order phase transition. Results shown for stable branches from \texttt{C++}. }
\label{fig:2D_C3_muS_N0_muQ_0_sus_comparison}
\end{figure*}

\begin{figure*}
    \centering
    \includegraphics[width=0.9\textwidth]{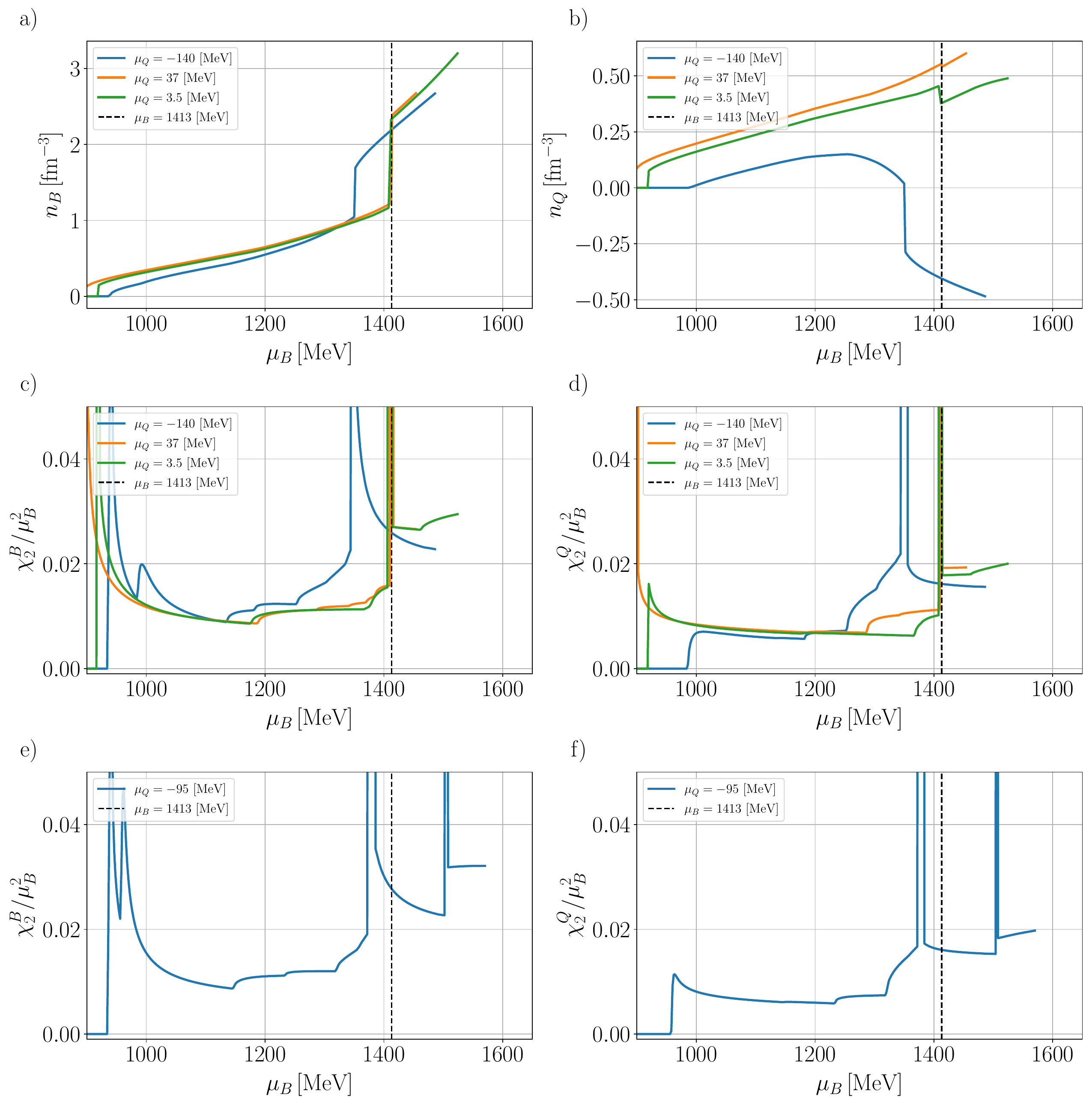}
\caption{C3 ($\mu_S=0$) octet + decuplet + quarks: 
a) baryon density, b) charge density, c) and e) second-order baryon susceptibilities, and d) and f) second-order charge susceptibilities, all as a function of baryon chemical potential. 
The top rows demonstrate first-order phase transitions into the quark phase (orange), into the strangeness-dominated hadronic phase (blue), and passing by the triple point (green). 
The black dashed line indicates the exact $\mu_B$ where the triple point is located in \Cref{fig:3D_C3_decuplet_hyper_quarks_muS_0_muQ_N0_panel} (at $\mu_B=1413$, $\mu_S=0$, $\mu_Q=3.5$ MeV). 
The bottom row displays the regime where one crosses from the light hadronic phase al low $\mu_B$, into the strangeness-dominated phase, followed by the quark phase at high $\mu_B$ (both phase transitions are first-order). The left panels and some curves in the right panels also show the liquid-gas first-order phase transition. Results shown for stable branches from \texttt{C++}. 
}
\label{fig:2D_C3_muS_0_muQ_N0_sus_comparison}
\end{figure*}

\begin{figure*}
    \centering
    \includegraphics[width=0.9\textwidth]{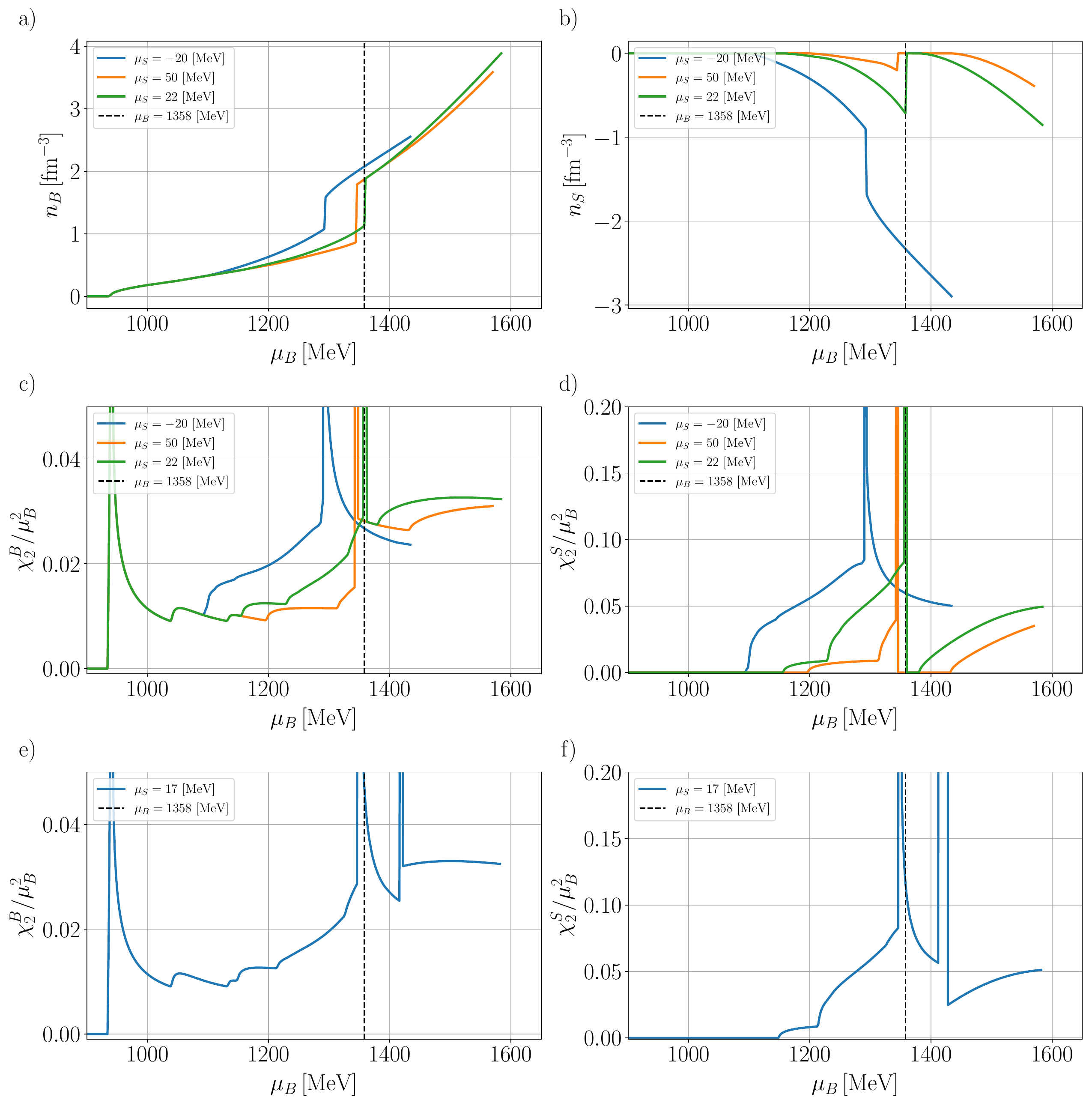}
\caption{C3 ($\mu_Q=-200$) octet + decuplet + quarks: 
a) baryon density, b) strangeness density, c) and e) second-order baryon susceptibilities, and d) and f) second-order strangeness susceptibilities, all as a function of baryon chemical potential. 
The top rows demonstrate first-order phase transitions into the quark phase (orange), into the strangeness-dominated hadronic phase (blue), and passing by the triple point (green). 
The black dashed line indicates the exact $\mu_B$ where the triple point is located in \Cref{fig:3D_C3_decuplet_hyper_quarks_muS_N0_muQ_-200_panel} (at $\mu_B=1358$, $\mu_s=22$, $\mu_Q=-200$ MeV). 
The bottom row displays the regime where one crosses from the light hadronic phase at low $\mu_B$, into the strangeness-dominated phase, followed by the quark phase at high $\mu_B$ (both phase transitions are first-order). The left panels also show the liquid-gas first-order phase transition. Results shown for stable branches from \texttt{C++}.}
\label{fig:2D_C3_muS_N0_muQ_-200_sus_comparison}
\end{figure*}

Here we show in detail the susceptibilities for the C3 parametrization including baryons from the octet and decuplet plus quarks for different combinations of $\mu_Q$ and $\mu_S$. In \Cref{fig:2D_C3_muS_N0_muQ_0_sus_comparison,fig:2D_C3_muS_0_muQ_N0_sus_comparison,fig:2D_C3_muS_N0_muQ_-200_sus_comparison}, we present the baryon number density $n_B$ in panel a), the strangeness $n_S$ or charge $n_Q$ number density in panel b), the dimensionless second-order susceptibilities for baryon number $\chi_2^B/\mu_B^2$ in panels c) and e), and the dimensionless second-order susceptibilities for strangeness $\chi_2^S/\mu_B^2$ or charge $\chi_2^Q/\mu_B^2$ in panels d) and f), all vs. $\mu_B$. We normalize all second-order susceptibilities by $\mu_B^2$ to make them dimensionless. 
The choice of using the baryon chemical potential to normalize $\chi^S$ and $\chi^Q$ is somewhat arbitrary but it allows for easier comparisons between the strangeness and baryon susceptibilities in terms of order of magnitude. 

In  \Cref{fig:2D_C3_muS_N0_muQ_0_sus_comparison} we choose various horizontal cuts of \Cref{fig:3D_C3_decuplet_hyper_quarks_muS_N0_muQ_0_panel}, where $\mu_S$ is held constant (the second and third rows are distinguished only by the choices in $\mu_S$ values and separated for the sake of visibility). 
We choose four specific slices in $\mu_S$ to compare: 
\begin{itemize}
    \item The green solid line in panels a), b), c) and d) passes through the triple point at $\mu_B=1410$, $\mu_S=1$, $\mu_Q=0$ MeV
    \item The orange solid line in panels a), b), c) and d) at $\mu_S=47$ MeV passes above the triple point from a light hadronic phase  into a quark phase (with a first-order phase transition) such that is occurs due to deconfinement. 
    \item The blue solid line in panels a), b), c) and d) at $\mu_S=-50$ MeV
    passes below the triple point from a light hadronic phase  into a strangeness-dominated phase (with a chiral first-order phase transition). 
    \item The blue solid line in the bottom row in panels e) and f) at $\mu_S=-10$ MeV passes from a light hadronic phase  into a strangeness-dominated phase (with a chiral first-order phase transition). Then it also has another first-order phase transition from the strangeness-dominated phase into the quark phase (as one increases $\mu_B$).
\end{itemize}
Then we can compare how the densities and susceptibilities are affected for strangeness vs baryon numbers, depending on the phases of matter. 

From the susceptibilities we can determine the type of phase transitions that occur in \Cref{fig:3D_C3_decuplet_hyper_quarks_muS_N0_muQ_0_panel}. 
All fixed $\mu_S$ slices in the EOS shown here present the liquid-gas phase transition at low $\mu_B$ (indicated by a small jump in $n_B$, panel a), and a divergence in $\chi_2^B$, panel c)). The liquid-gas phase transition only affects the baryon number, so we do not see a any changes in the strangeness susceptibilities. 
In panels a) and c) at larger $\mu_B$ we can clearly see that each slice of $\mu_S$ passes across another first-order phase transition, such that the jump in $n_B$ and divergence in $\chi_2^B$ (indicating the first-order phase transition) change location depending on $\mu_S$. 
Then, in panels b) and  d) at the same location in $\mu_B$ (for each respective line) we see a jump in $n_S$ and divergence in $\chi_2^S$ as well.

It is interesting to note that the difference in  $\Delta n_B=n_{B,2}-n_{B,1}$ for the phases across the first-order phase transition line is nearly the same, regardless if one goes into a strangeness-dominated phase or into a quark phase (blue vs. orange line in panel a)). 
However, we note that the difference in strangeness i.e. $\Delta n_S=n_{S,2}-n_{S,1}$ (blue vs. orange line in panel b)) is significantly different, depending on which phase of matter it transitions into. 
Additionally, the signs of $\Delta n_S$ are opposite since the quark phase actually only consists of up and down quarks close to the phase transition (strangeness only appears at higher $\mu_B)$ such that strange particles disappear, whereas the strangeness-dominated phase increases the number of hyperons such that strange particles dominate and $n_S$ becomes more negative.
The implications of this opposite sign in $\Delta n_S$  is that the $\chi_2^S$ susceptibilities are very different for the strangeness-dominated phase transition vs the deconfinement one. 
Regardless, both blue lines, orange lines, and green lines diverge (panels c) and d)), indicating that both B and S conserved charges are relevant for these phase transitions.

The green lines in panels a)-d) pass by the point where the three phases converge (light hadronic, strangeness-dominated, and quark phase). 
This point is indicated by a black dashed line at $\mu_B=1410$, $\mu_S=1$, $\mu_Q=0$ MeV. 
Looking at panels a) and b) we can determine that this is a triple point (not a tricritical point) since the order of the phase transition is still first-order (jumps appear in both $n_B$ and $n_S$, leading to a divergence at the triple point in both $\chi_2^B$ and $\chi_2^S$ as well). 
At $\mu_B$ above the triple point for a fixed $\mu_S$ then one reaches the quark phase (note that given different trajectories in $\mu_B,\mu_S$ crossing the triple point, one could instead end up in the strangeness-dominated phase).

We can also compare the small regime of phase space at slightly negative $\mu_S$ where one passes from a light hadronic phase, into a strangeness-dominated phase, followed by a quark phase (as one increases $\mu_B$). 
Thus, in panel e) from \Cref{fig:2D_C3_muS_N0_muQ_0_sus_comparison}, three phase transitions are observed: the first-order liquid-gas phase transition at low $\mu_B$, the first-order chiral phase transition from a light-hadron-dominated phase to a strangeness-dominated phase (below $\mu_B=1400$ MeV), and the first-order phase transition marking the deconfinement transition (near $\mu_B=1500$ MeV).
As they are first-order phase transitions, they lead to divergences in the second-order susceptibilities (and would lead to further divergences in higher-order susceptibilities as well). 
From the results seen in panel d), it can be noted that $\chi_2^S$ diverges from the light into the strangeness-dominated phase, then again due to deconfinement, so this transition occurs both in B and S. 

In  \Cref{fig:2D_C3_muS_0_muQ_N0_sus_comparison} we choose various similar horizontal cuts but now of \Cref{fig:3D_C3_decuplet_hyper_quarks_muS_0_muQ_N0_panel}, where $\mu_Q$ is held constant (the second and third row are once more distinguished only by the choices in $\mu_Q$ values and separated for the sake of visibility). 
We choose again four specific slices in $\mu_Q$ to compare: 
\begin{itemize}
    \item The green solid line in panels a), b), c) and d) passes through the triple point at $\mu_B=1413$, $\mu_S=0$, $\mu_Q=3.5$ MeV.
    \item The orange solid line in panels a), b), c) and d) at $\mu_Q=37$ MeV passes above the triple point from a light hadronic phase  into a quark phase (with a first-order chiral phase transition) such that is occurs due to deconfinement. 
    \item The blue solid line in panels a), b), c) and d) at $\mu_Q=-140$ MeV
    passes below the triple point from a light hadronic phase  into a strangeness-dominated phase (with a first-order chiral phase transition). 
    \item The blue solid line in the bottom row in panels e) and f) at $\mu_Q=-95$ MeV passes from a light hadronic phase  into a strangeness-dominated phase (with a chiral first-order phase transition), and then has a another first-order phase transition into the quark phase at higher $\mu_B$. 
\end{itemize}
Then we can compare again how the densities and susceptibilities are affected for electric charge vs baryon numbers, depending on the phases of matter. 

From the susceptibilities we can determine the order of phase transitions that occur in \Cref{fig:3D_C3_decuplet_hyper_quarks_muS_0_muQ_N0_panel}. 
All fixed $\mu_Q$ slices in the EOS shown here have the liquid-gas phase transition at low $\mu_B$ (indicated by a small jump in $n_B$ in panel a) and a divergence in $\chi_2^B$ in panel c) and e) ).  Unlike in \Cref{fig:2D_C3_muS_N0_muQ_0_sus_comparison}, we see a more complicated structure around the liquid-gas phase transition, which occurs for a couple of reasons. 
To start with, when we vary $\mu_Q$, we are actually also varying the location of the liquid gas phase transition in $\mu_B$ (negative $\mu_Q$ shifts it to higher $\mu_B$, whereas positive $\mu_Q$ shifts it to lower $\mu_B$). 
Furthermore, changing $\mu_Q$ also changes the critical $\mu_B$ where protons can appear. Recall that neutrons are unaffected by $\mu_Q$ because their effective chemical potentials are $\mu_n=\mu_B$ whereas the effective chemical potential for protons are $\mu_p=\mu_B+\mu_Q$. 
Then, we could think about a critical $\mu_B^c$ for protons to appear, which is $\mu_B^c=\mu_p-\mu_Q$ such that large, positive values of $\mu_Q$ allow the protons to appear earlier whereas as large, negative values of $\mu_Q$ lead to the protons switching on at higher $\mu_B$. 
We see exactly this effect occur in panel c), where the solid blue line presents a second peak in $\chi_2^B$ at low $\mu_B$ (not a divergence though), which is where the protons switch on just above the liquid-gas phase transition and that coincides with the $\chi_2^Q$ becoming non-zero. 
Looking at low $\mu_B$ in panels b) and d), one sees that for positive $\mu_Q$ there is also a first-order liquid-gas phase transition in Q, seen both as a jump in $n_Q$ and a divergence in $\chi_2^Q$.

Our phase structure in $\mu_B,\mu_Q$ space has the same phases of matter that we saw previously (light hadronic, strangeness-dominated, and quark phase) but the shape of the phase diagram is different such that there is a much larger region where one transitions between all three dense-matter phases (see \Cref{fig:3D_C3_decuplet_hyper_quarks_muS_0_muQ_N0_panel}). 
As it happened for $\mu_S$, when one varies $\mu_Q$ it opens up the possibility of reaching a regime where negatively charged baryons dominate in the strangeness-dominated regime.

In panels a) and c) at larger $\mu_B$ we can again clearly see that each slice of $\mu_Q$ passes across another first-order phase transition, such that the jump in $n_B$ and divergence in $\chi_2^B$ (indicating the first-order phase transition) change location depending on $\mu_Q$. 
Then, in panels b) and  d) at the same location in $\mu_B$ (for each respective line) we see a jump in $n_Q$ and divergence in $\chi_2^Q$ as well, indicating that both B and Q conserved charges are relevant for these phase transitions. The green lines pass by the triple point indicating that the phase transition there is still first order. At $\mu_B$ above the triple point for a fixed $\mu_Q$ then one reaches the quark phase, while at $\mu_B$ below the triple point one reaches the strangeness-dominated (now with all dense matter first order phase transitions showing the same sign in $\Delta n_Q$).

In panel e) at larger $\mu_B$ we can clearly see that the slice of  negative $\mu_Q$ (with smaller magnitude) passes across two first-order phase transitions at high $\mu_B$, into the strangeness-dominated phase, followed by into the quark phase. A consequence of negative $\mu_Q$ is the appearance of kinks in $\chi_2^B$ as well as in $\chi_2^Q$ (panels c)-f)), which indicate the appearance of new particles (beyond neutrons and protons) switching on. At least some of these particles are hyperons evidenced by the  negative $Y_S$ that appears in panel c) of \Cref{fig:3D_C3_decuplet_hyper_quarks_muS_0_muQ_N0_panel} at very negative $\mu_Q$.
As discussed previously in the main text, these kinks lead to third-order phase transitions, which would have divergences in $\chi_3$ (not shown here).

In \Cref{fig:2D_C3_muS_N0_muQ_-200_sus_comparison}, we choose various horizontal cuts of \Cref{fig:3D_C3_decuplet_hyper_quarks_muS_N0_muQ_-200_panel}, where $\mu_S$ is held constant. Note that \Cref{fig:3D_C3_decuplet_hyper_quarks_muS_N0_muQ_-200_panel} is very similar to the phase diagram in \Cref{fig:3D_C3_decuplet_hyper_quarks_muS_N0_muQ_0_panel} where $\mu_Q=0$, except that we now show the consequence of $\mu_Q=-200$ MeV that is more realistic for neutron stars. 
In fact, the main consequence of using a fixed $\mu_Q=-200$ MeV is that the triple point shifts into the positive $\mu_S$ regime where it is located at $\mu_B=1358$ MeV, $\mu_S=22$ MeV, $\mu_Q=-200$ MeV (previously for $\mu_Q=0$ the triple point was near $\mu_S\sim 0$). 

In the top rows, the green line indicates the trajectory that passes across the triple point (that spikes at the triple point, which is shown with a black dashed line). 
Then we also show two trajectories that appear at $\mu_S=50$ MeV and $\mu_S=-20$ MeV that demonstrate the chiral transition across into deconfined quarks or demonstrates the transition into a strangeness-dominated regime. 
Here we once again see that both the deconfinement phase transition and the chiral phase transition into a strangeness-dominated phase lead to spikes in both $\chi_2^B$ and $\chi_2^S$ (but now with the same sign in $\Delta n_S$, panels a) and b)). The liquid-gas first order phase transition only appears in the left panels.
Then in the lower row we see the spikes in $\chi_2^B$ that indicate the first-order phase transitions for light to strangeness-dominated at intermediate $\mu_B$ (near $\mu_B=1350$ MeV) and from strangeness-dominated into quarks at higher $\mu_B$ (above $\mu_B=1400$ MeV). 
The later two phase transitions also have spikes in $\chi_2^S$ because of the role that strangeness plays in these phase transitions. 
Finally, we point out that at $\mu_Q=-200$ MeV we find even more significant kinks at intermediate $\mu_B$ where new particles are switching on, which indicates negative values of $\mu_S$ and $\mu_Q$ make it easier for new hyperons to appear. 

\end{document}